\documentclass[5p]{elsarticle}       
\usepackage[utf8]{inputenc}
\usepackage{amsfonts}
\usepackage{amsmath}
\usepackage{multicol}
\usepackage{amssymb}
\usepackage{graphicx}
\usepackage{mathtools}
\usepackage{xcolor}
\usepackage{blindtext}
\usepackage{multicol}

\usepackage{colortbl}
\usepackage{arydshln}
\usepackage{booktabs}
\usepackage{hhline}

\usepackage{breqn}
\usepackage{enumitem}
\usepackage{xr}
\newcommand{\subscript}[2]{$#1 _ #2$}

 \usepackage{breqn} 
\graphicspath{{./}{./Figures/}}

\usepackage{units}

\newcounter{theExample}
\newenvironment{Example}{
\par\smallskip\small\refstepcounter{theExample}%
\noindent\hspace*{-0em}\textbf{Example~\arabic{theExample}}:~%
\leftskip0em\ignorespaces%
}{
\par\smallskip
}
\newcounter{theRemark}
\newenvironment{Remark}{
\par\smallskip\refstepcounter{theRemark}%
\noindent%
\textbf{Remark~\arabic{theRemark}}:~%
\ignorespaces%
}{
\par\smallskip
}

\makeatletter
\newcommand{\subalign}[1]{%
	\vcenter{%
		\Let@ \restore@math@cr \default@tag
		\baselineskip\fontdimen10 \scriptfont\tw@
		\advance\baselineskip\fontdimen12 \scriptfont\tw@
		\lineskip\thr@@\fontdimen8 \scriptfont\thr@@
		\lineskiplimit\lineskip
		\ialign{\hfil$\m@th\scriptstyle##$&$\m@th\scriptstyle{}##$\crcr
			#1\crcr
		}%
	}
}
\makeatother

\begin{document}

\title{Graph Signal Processing -- Part II: Processing and Analyzing Signals on Graphs}

\author[etf]{\!Ljubi\v{s}a \!Stankovi\'{c}}
\ead{ljubisa@ucg.ac.me}
\author[ic]{\!Danilo \!Mandic}
\ead{d.mandic@imperial.ac.uk}
\author[etf]{\!Milo\v{s} \!Dakovi\'{c}} 
\ead{milos@ucg.ac.me} 
\author[etf]{\!Milo\v{s} \!Brajovi\'{c}}
\ead{milosb@ucg.ac.me}
%\author[pit]{\!Ervin \!Sejdi\'{c}}
%\ead{esejdic@pitt.edu}
\author[ic]{\!Bruno \!Scalzo}
\ead{bruno.scalzo-dees12@imperial.ac.uk}
\author[ic]{\!Anthony G. \!Constantinides}
\ead{a.constantinides@imperial.ac.uk}

\address[etf]{University of Montenegro, Podgorica, Montenegro }
\address[ic]{Imperial College London, London, United Kingdom }
%\address[pit]{University of Pittsburg, Pittsburg, PA, USA }

\date{Received: date / Accepted: date}

\begin{abstract}  Data analytics on graphs deals with information processing of data acquired on irregular but structured graph domains. The focus of Part I of this monograph has been on both the fundamental and higher-order graph properties, graph topologies, and spectral representations of graphs. Part I also establishes  rigorous frameworks 
	for vertex clustering and graph segmentation, and illustrates the power of graphs in various data association tasks. Part II embarks on these concepts to address the algorithmic and practical issues centered round data/signal processing on graphs, that is, the focus is on the analysis and estimation of both deterministic and random data on graphs.  The fundamental ideas related to graph signals are introduced through a simple and intuitive, yet illustrative and general enough case study of multisensor temperature field estimation. 
	 The concept of systems on graph is defined using graph signal shift operators, which generalize the corresponding principles from traditional learning systems. At the core of the spectral domain representation of graph signals and systems is the Graph Discrete Fourier Transform (GDFT), which is defined based on the eigendecomposition of both the adjacency matrix and the graph Laplacian. The spectral domain representations are then used as the basis to introduce graph signal filtering concepts and address their design, including Chebyshev polynomial approximation series.  Ideas related to the sampling of graph signals, and in particular the challenging topic of  data dimensionality reduction through graph subsampling, are presented and further linked with  compressive sensing. The principles of time-varying signals on graphs and  basic definitions related to random graph signals are next reviewed. Localized graph signal analysis in the joint vertex-spectral domain is referred to as the vertex-frequency analysis, since it can be considered as an extension of classical time-frequency analysis to the graph domain of a signal. Important topics related to the local graph Fourier transform (LGFT) are covered, together with its various forms including the graph spectral and vertex domain windows and the inversion conditions and relations. A  link between the LGFT with spectral varying window and the spectral graph wavelet transform (SGWT) is also established. Realizations of the LGFT and SGWT using polynomial (Chebyshev) approximations of the spectral functions are further considered and supported by examples. Finally, energy versions of the vertex-frequency representations are introduced, along with their relations with classical time-frequency analysis, including a vertex-frequency distribution that can satisfy the marginal properties. The material is supported by numerous examples.    
\end{abstract}

\maketitle

\setcounter{tocdepth}{3}

\tableofcontents

\section{Introduction}

Graphs are irregular structures which naturally account for data integrity, however, traditional approaches have been established outside Machine Learning and Signal Processing, and largely focus on analyzing the underlying graphs rather than dealing with signals on graphs. 
On the other hand, given the rapidly increasing availability of multisensor and multinode measurements, likely recorded on irregular or ad-hoc grids, it would be extremely advantageous to analyze such structured data as ``signals on graphs'' and thus benefit from the ability of graphs to account for spatial sensing awareness, physical intuition and sensor importance, together with the inherent \textquotedblleft local versus global\textquotedblright\ sensor association.  The aim of Part II of our monograph is therefore to establish a common language between graph signals which are observed on irregular signal domains, and some of the most fundamental paradigms in Learning Systems, Signal Processing and Data Analytics, such as spectral analysis, system transfer function, digital filter design, parameter estimation, and optimal denoising.  

 In classical Data Analytics and Signal Processing, the signal domain is determined by equidistant time instants or by a set of spatial sensing points on a uniform grid. However, increasingly the actual data sensing domain may not even be related to the physical dimensions of time and/or  space, and it typically does exhibit various forms of irregularity, as, for example, in social or web-related networks, where the sensing points and their connectivity pertain to specific objects/nodes and ad-hoc topology of their links. It should be noted that  even for the data acquired on well defined time and space domains,  the introduction of new relations between the signal samples, through graphs, may yield new insights into the analysis and provide enhanced data processing (for example, based on local similarity, through neighborhoods). We therefore set out to show that the advantage of graphs over classical data domains is that graphs account naturally and comprehensively for irregular data relations in the problem definition, together with the corresponding data connectivity in the analysis \citep{mouraaa2018graph,VetterliBook,sandryhaila2013discrete,ekambaram2014graph,shuman2013emerging,hamon2016extraction,chen2014signal,gavili2017shift}. 

 To build up the intuition behind the fundamental ideas of signals/data on graph,  a simple yet general example of multisensor temperature estimation is first considered in Section \ref{sec2}. Basic concepts regarding the signals and systems on graphs are presented in Section \ref{sec3}, including basic definitions, operations and transforms, which generalize the foundations of traditional signal processing. Systems on graphs are interpreted starting from a comprehensive account of the existing and the introduction of a novel, isometric,  graph signal shift operator.  Further, graph Fourier transform is defined based on both the adjacency matrix and the graph Laplacian and it serves  as the basis to introduce graph signal filtering concepts. Various ideas related to the sampling of graph signals, and particularly, the challenging topic of their subsampling, are reviewed in Section \ref{sec4}. Sections \ref{sec5} and \ref{sec6} present the concepts of time-varying signals on graphs and introduce basic definitions related to random graph signals. Localized graph signal behavior can be simultaneously characterized in the vertex-frequency domain, which is discussed in Section \ref{sec7}. This Section also covers the important topics of local graph Fourier transform, various forms of its inversion, relations with the frames framework and links with the graph wavelet transform.  Energy versions of the vertex-frequency representations are also considered, along with their relations with  classical time-frequency analysis.

\section{Problem Statement: An Illustrative Example}
\label{sec2}
\begin{figure}[t]
	
	\includegraphics[]{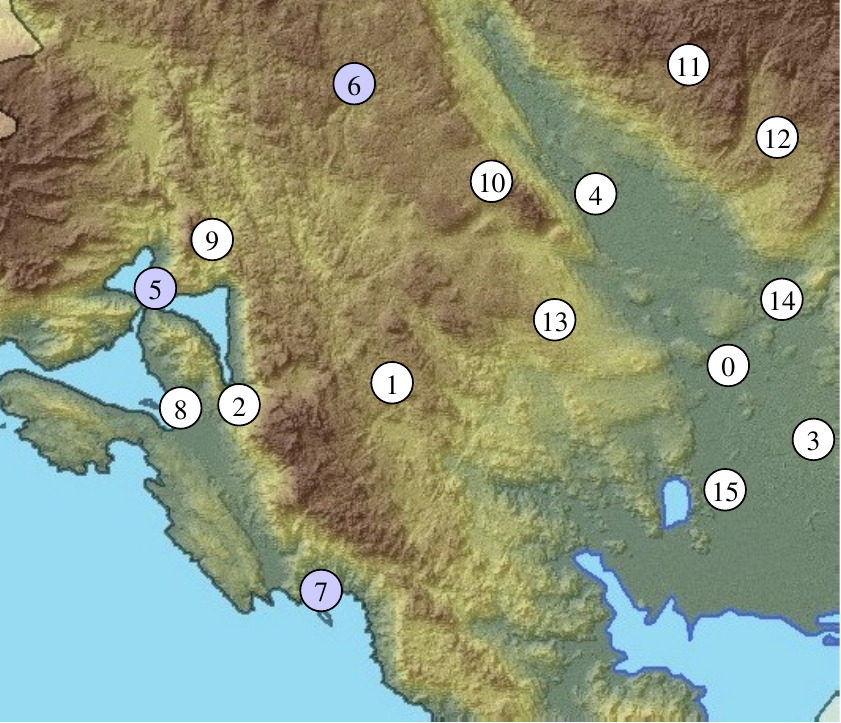}(a)
	
	\vspace{5mm}
	
	\raggedleft
	\includegraphics[scale=0.95]{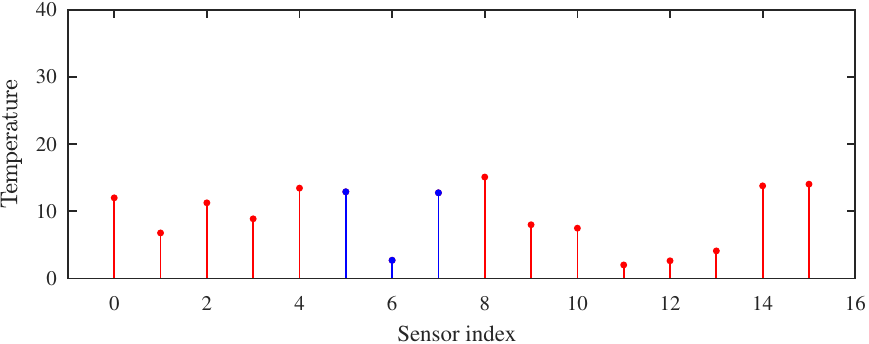}(b)
	
	\vspace{5mm}
	
	\raggedleft
	\includegraphics[scale=0.95]{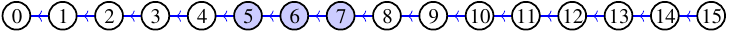} \hspace{3mm}(c)

	\caption{Temperature sensing as a classic data analytics problem.
		(a) Sensing locations in a geographic region along the Adriatic sea.  (b) Temperatures measured at $N=16$ sensing locations. In standard data estimation, the spatial sensor index is used for the horizontal axis and serves as the data domain. This domain can be interpreted as a directed 	path graph structure, shown in the bottom panel (c). Observe that the consecutive samples (vertices) on this 	path graph offer no physical intuition or interpretation, as in this \textquotedblleft brute force\textquotedblright \, arrangement, for example,  vertex $6$ is located on a high mountain, whereas its neighboring vertices $5$ and $7$ are located along the sea; despite the consecutive index numbers these sensors are physically distant, as indicated by their very different temperature measurements.    
	}
	\label{fig:MNE_fig_a}
\end{figure}

Consider a multi-sensor setup for measuring a temperature field in a region of interest. The temperature sensing locations are chosen according to the significance of a particular geographic area to local users, with $N=16$ sensing points in total, as shown in Fig. \ref{fig:MNE_fig_a}(a). The temperature field is denoted by $\{x(n)\}$, with $n$ as the sensor index, while a snapshot of its values is given in Fig. \ref{fig:MNE_fig_a}(b). 
Each measured sensor signal can then be mathematically expressed as \begin{equation}
x(n)=s(n)+\varepsilon(n), \,\,\,\,\,\, n=0,1,\dots,15, \label{nsig}
\end{equation}
where $s(n)$ is the true temperature  that would have been obtained in  ideal measuring conditions and $\varepsilon(n)$ comprises the adverse effects of the local environment on sensor readings or faulty sensor activity, and is referred to as ``noise" in the sequel. For illustrative purposes, in our study each $\varepsilon(n)$ was modeled as a realization of white, zero-mean, Gaussian process, with standard deviation $\sigma_\varepsilon=2$, that is, $\varepsilon(n) \in \mathcal{N}(0,4)$. It was added to the signal, $s(n)$, to yield the signal-to-noise ratio in $x(n)$ of $SNR_{in}=14.2$ dB.  

\begin{figure}[thtb]
	\centering
	\includegraphics[]{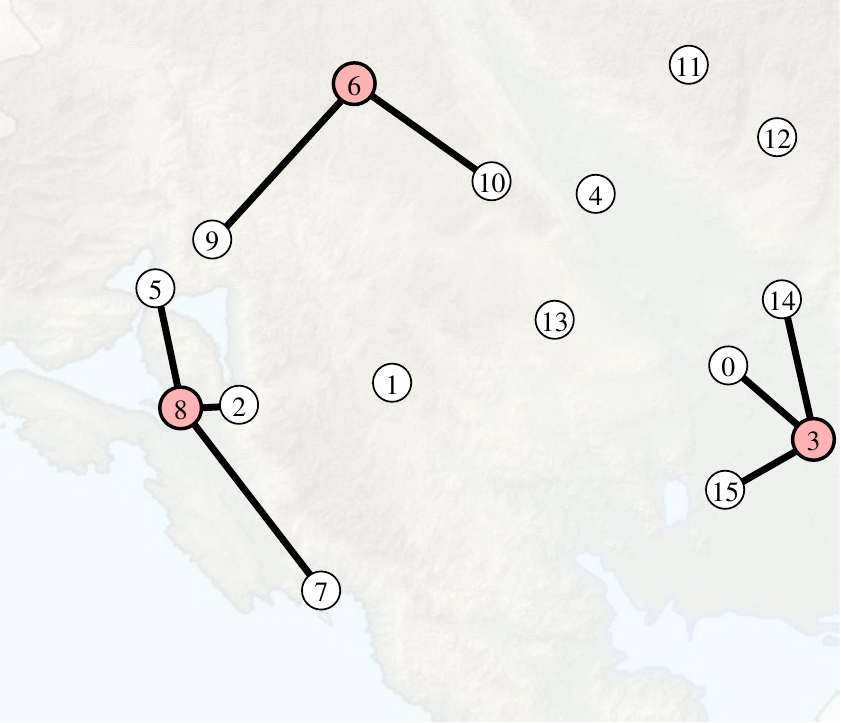}(a)
	
	\vspace{5mm}
	
	\includegraphics[]{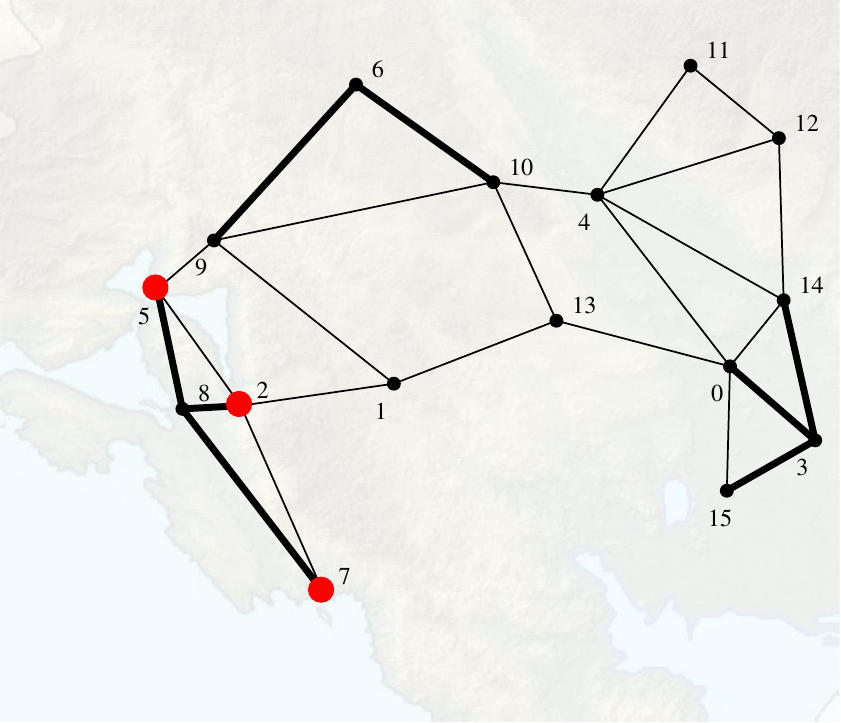}(b)
	
	\caption{Temperature sensing setup as a graph signal estimation problem. (a) Local neighborhood for the sensing points  $n=3$, $6$, and $8$. These neighborhoods are chosen using ``domain knowledge" dictated by the local terrain and by taking into account the sensor distance and altitude.  Neighboring sensors for each of these sensing locations (vertices) are chosen in a physically meaningful way and their relation is indicated by the connectivity lines, that is, graph edges. (b) Local neighborhoods for all sensing vertices, presented in a graph form.}
	\label{fig:MNE_fig_b}
\end{figure}

\begin{Remark}
Classical data analytics require a  rearrangement of the quintessentially irregular spatial temperature sensing arrangement in Fig. \ref{fig:MNE_fig_a}(a) into a linear structure shown in Fig. \ref{fig:MNE_fig_a}(b). Obviously, such ``lexicographic" ordering is not amenable to exploiting the information related to the actual sensor locations, which is inherently dictated by the terrain. This renders classical analyses of this multisensor temperature field inapplicable (or at best suboptimal), as the performance critically depends on the chosen sensor ordering scheme. This exemplifies that even a most routine multisensor measurement  setup requires a more complex estimation structure than the standard linear one corresponding to the classical signal processing framework, shown in Fig. \ref{fig:MNE_fig_a}(b).  
\end{Remark}

To introduce a ``situation-aware" noise reduction scheme for the temperature field in Fig. \ref{fig:MNE_fig_a}, we proceed to explore  a graph-theoretic framework to this problem, starting from a local signal average operator. In classical analysis, this may be achieved through a moving average operator, e.g., by averaging across the neighboring data samples, or equivalently neighboring sensors in the linear data setup in Fig. \ref{fig:MNE_fig_a}(b), and  for each sensing point. Physically, such local neighborhood should include close neighboring sensing points but only those which also exhibit similar meteorological properties defined by the sensor distance, altitude difference, and other terrain specific properties. In other words, since the sensor network in Fig. \ref{fig:MNE_fig_a} measures a set of related temperatures from irregularly spaced sensors, an effective estimation strategy should include domain knowledge -- not possible to achieve with standard methods (linear path graph).   

To illustrate the advantages of approaches based on local information (neighborhood based) , consider the neighborhoods for the sensing points $n=3$ (low land), $n=6$ (mountains), and $8$ (coast), shown in Fig. \ref{fig:MNE_fig_b}(a). The cumulative temperature for each sensing point is then given by
$$y(n)=\sum_{m \text{ at  and around } n  } x(m),$$ 
so that the local average temperature for a sensing point $n$ may be obtained by dividing the cumulative temperature, $y(n)$, with the number of included sensing points (size of local neighborhood).
For example, for the sensing points $n=3$ and $n=6$, presented in Fig. \ref{fig:MNE_fig_b}(a), the ``domain knowledge aware" local estimation takes the form
\begin{align}
y(3) & =x(3)+x(0)+x(14)+x(15) \label{20VA}\\
y(6) & = x(6)+x(9)+x(10). \label{37VA}
\end{align}
For convenience, the full set of relations among the sensing points can now be arranged into a matrix form, to give
\begin{equation}
\mathbf{y}=\mathbf{x}+\mathbf{A}\mathbf{x}, \label{1}
\end{equation} 
where the \textbf{adjacency matrix} $\mathbf{A}$, given in (\ref{A16full}), indicates the connectivity structure of the sensing locations; this local connectivity structure should be involved in the calculation of each $y(n)$.
\begin{figure*}[tbp]
\begin{align}
\mathbf{A}= 
{	\color{blue}
	\begin{matrix}
	\text{\footnotesize 0}\\
	\text{\footnotesize 1}\\
	\text{\footnotesize 2}\\
	\text{\footnotesize 3}\\
	\text{\footnotesize 4}\\
	\text{\footnotesize 5}\\
	\text{\footnotesize 6}\\
	\text{\footnotesize 7}\\
	\text{\footnotesize 8}\\
	\text{\footnotesize 9}\\
	\text{\footnotesize 10}\\
	\text{\footnotesize 11}\\
	\text{\footnotesize 12}\\
	\text{\footnotesize 13}\\
	\text{\footnotesize 14}\\
	\text{\footnotesize 15}
	\end{matrix}
} 
\left[
\begin{array}{*{16}c}
0 & 0 & 0 & 1 & 1 & 0 & 0 & 0 & 0 & 0 & 0 & 0 & 0 & 1 & 1 & 1  \\
0 & 0 & 1 & 0 & 0 & 0 & 0 & 0 & 0 & 1 & 0 & 0 & 0 & 1 & 0 & 0  \\
0 & 1 & 0 & 0 & 0 & 1 & 0 & 1 & 1 & 0 & 0 & 0 & 0 & 0 & 0 & 0  \\
1 & 0 & 0 & 0 & 0 & 0 & 0 & 0 & 0 & 0 & 0 & 0 & 0 & 0 & 1 & 1  \\
1 & 0 & 0 & 0 & 0 & 0 & 0 & 0 & 0 & 0 & 1 & 1 & 1 & 0 & 1 & 0  \\
0 & 0 & 1 & 0 & 0 & 0 & 0 & 0 & 1 & 1 & 0 & 0 & 0 & 0 & 0 & 0  \\
0 & 0 & 0 & 0 & 0 & 0 & 0 & 0 & 0 & 1 & 1 & 0 & 0 & 0 & 0 & 0  \\
0 & 0 & 1 & 0 & 0 & 0 & 0 & 0 & 1 & 0 & 0 & 0 & 0 & 0 & 0 & 0  \\
0 & 0 & 1 & 0 & 0 & 1 & 0 & 1 & 0 & 0 & 0 & 0 & 0 & 0 & 0 & 0  \\
0 & 1 & 0 & 0 & 0 & 1 & 1 & 0 & 0 & 0 & 1 & 0 & 0 & 0 & 0 & 0  \\
0 & 0 & 0 & 0 & 1 & 0 & 1 & 0 & 0 & 1 & 0 & 0 & 0 & 1 & 0 & 0  \\
0 & 0 & 0 & 0 & 1 & 0 & 0 & 0 & 0 & 0 & 0 & 0 & 1 & 0 & 0 & 0  \\
0 & 0 & 0 & 0 & 1 & 0 & 0 & 0 & 0 & 0 & 0 & 1 & 0 & 0 & 1 & 0  \\
1 & 1 & 0 & 0 & 0 & 0 & 0 & 0 & 0 & 0 & 1 & 0 & 0 & 0 & 0 & 0  \\
1 & 0 & 0 & 1 & 1 & 0 & 0 & 0 & 0 & 0 & 0 & 0 & 1 & 0 & 0 & 0  \\
1 & 0 & 0 & 1 & 0 & 0 & 0 & 0 & 0 & 0 & 0 & 0 & 0 & 0 & 0 & 0 
\end{array}\label{A16full}
\right] \\
	{	\color{blue}
		\begin{array}{*{16}c}
		\text{\footnotesize 0 \hspace{-1.8mm} } &
		\text{\footnotesize 1 \hspace{-1.8mm} }  &
		\text{\footnotesize 2 \hspace{-1.8mm}  }  &
		\text{\footnotesize 3 \hspace{-1.8mm} }  &
		\text{\footnotesize 4 \hspace{-1.8mm}  }  &
		\text{\footnotesize 5 \hspace{-1.8mm}  }  &
		\text{\footnotesize 6 \hspace{-1.8mm} }  &
		\text{\footnotesize 7 \hspace{-1.8mm} }  &
		\text{\footnotesize 8 \hspace{-1.8mm} }  &
		\text{\footnotesize 9 \hspace{-1.8mm} }  &
		\text{\footnotesize 10 \hspace{-3.3mm} }  &
		\text{\footnotesize 11 \hspace{-3.3mm} }  &
		\text{\footnotesize 12 \hspace{-3.3mm} }  &
		\text{\footnotesize 13 \hspace{-3.3mm} }  &
		\text{\footnotesize 14 \hspace{-3.3mm} }  &
		\text{\footnotesize 15 \hspace{0mm} }
		\end{array}\nonumber
	}
\end{align}
\vspace{-5mm}
\end{figure*}

\begin{figure*}[tbp]
	\begin{align}
	\mathbf{W}=
	{	\color{blue}
		\begin{matrix}
		\text{\footnotesize   0}\\
		\text{\footnotesize   1}\\
		\text{\footnotesize   2}\\
		\text{\footnotesize   3}\\
		\text{\footnotesize   4}\\
		\text{\footnotesize   5}\\
		\text{\footnotesize   6}\\
		\text{\footnotesize   7}\\
		\text{\footnotesize   8}\\
		\text{\footnotesize   9}\\
		\text{\footnotesize   10}\\
		\text{\footnotesize   11}\\
		\text{\footnotesize   12}\\
		\text{\footnotesize   13}\\
		\text{\footnotesize   14}\\
		\text{\footnotesize   15}
		\end{matrix}
	} 
	\left[
	\begin{array}{*{16}r}
	0    & 0    & 0    & 0.97 & 0.91 & 0    & 0    & 0    & 0    & 0    & 0    & 0    & 0    & 0.05 & 0.90 & 0.94  \\
	0    & 0    & 0.03 & 0    & 0    & 0    & 0    & 0    & 0    & 0.37 & 0    & 0    & 0    & 0.78 & 0    & 0     \\
	0    & 0.03 & 0    & 0    & 0    & 0.96 & 0    & 0.95 & 0.98 & 0    & 0    & 0    & 0    & 0    & 0    & 0     \\
	0.97 & 0    & 0    & 0    & 0    & 0    & 0    & 0    & 0    & 0    & 0    & 0    & 0    & 0    & 0.88 & 0.96  \\
	0.91 & 0    & 0    & 0    & 0    & 0    & 0    & 0    & 0    & 0    & 0.06 & 0.01 & 0.01 & 0    & 0.94 & 0     \\
	0    & 0    & 0.96 & 0    & 0    & 0    & 0    & 0    & 0.97 & 0.01 & 0    & 0    & 0    & 0    & 0    & 0     \\
	0    & 0    & 0    & 0    & 0    & 0    & 0    & 0    & 0    & 0.62 & 0.40 & 0    & 0    & 0    & 0    & 0     \\
	0    & 0    & 0.95 & 0    & 0    & 0    & 0    & 0    & 0.94 & 0    & 0    & 0    & 0    & 0    & 0    & 0     \\
	0    & 0    & 0.98 & 0    & 0    & 0.97 & 0    & 0.94 & 0    & 0    & 0    & 0    & 0    & 0    & 0    & 0     \\
	0    & 0.37 & 0    & 0    & 0    & 0.01 & 0.62 & 0    & 0    & 0    & 0.25 & 0    & 0    & 0    & 0    & 0     \\
	0    & 0    & 0    & 0    & 0.06 & 0    & 0.40 & 0    & 0    & 0.25 & 0    & 0    & 0    & 0.85 & 0    & 0     \\
	0    & 0    & 0    & 0    & 0.01 & 0    & 0    & 0    & 0    & 0    & 0    & 0    & 0.92 & 0    & 0    & 0     \\
	0    & 0    & 0    & 0    & 0.01 & 0    & 0    & 0    & 0    & 0    & 0    & 0.92 & 0    & 0    & 0.01 & 0     \\
	0.05 & 0.78 & 0    & 0    & 0    & 0    & 0    & 0    & 0    & 0    & 0.85 & 0    & 0    & 0    & 0    & 0     \\
	0.90 & 0    & 0    & 0.88 & 0.94 & 0    & 0    & 0    & 0    & 0    & 0    & 0    & 0.01 & 0    & 0    & 0     \\
	0.94 & 0    & 0    & 0.96 & 0    & 0    & 0    & 0    & 0    & 0    & 0    & 0    & 0    & 0    & 0    & 0   
	\end{array}\label{W16full}
	\right]
	\\
	{	\color{blue}
		\begin{array}{*{16}c}
		\text{\footnotesize 0 \hspace{2.7mm} } &
		\text{\footnotesize 1 \hspace{2.7mm} }  &
		\text{\footnotesize 2 \hspace{2.7mm}  }  &
		\text{\footnotesize 3 \hspace{2.7mm} }  &
		\text{\footnotesize 4 \hspace{2.7mm}  }  &
		\text{\footnotesize 5 \hspace{2.7mm}  }  &
		\text{\footnotesize 6 \hspace{2.7mm} }  &
		\text{\footnotesize 7 \hspace{2.7mm} }  &
		\text{\footnotesize 8 \hspace{2.7mm} }  &
		\text{\footnotesize 9 \hspace{1.8mm} }  &
		\text{\footnotesize 10 \hspace{1.3mm} }  &
		\text{\footnotesize 11 \hspace{1.3mm} }  &
		\text{\footnotesize 12 \hspace{1.3mm} }  &
		\text{\footnotesize 13 \hspace{1.3mm} }  &
		\text{\footnotesize 14 \hspace{1.3mm} }  &
		\text{\footnotesize 15 \hspace{0mm} }
		\end{array} \nonumber
	}
	\end{align}
	\vspace{-5mm}
\end{figure*}
\begin{figure*}[hbtp]
	\begin{align}
	\setlength{\arraycolsep}{1.9pt}
	\mathbf{L}\!=\!\!
	\left[
	\begin{array}{*{16}r}
	3.77 &  0    &  0    & -0.97 & -0.91 &  0    &  0    &  0    &  0    &  0    &  0    &  0    &  0    & -0.05 & -0.90 & -0.94  \\
	0    &  1.19 & -0.03 &  0    &  0    &  0    &  0    &  0    &  0    & -0.37 &  0    &  0    &  0    & -0.78 &  0    &  0     \\
	0    & -0.03 &  2.93 &  0    &  0    & -0.96 &  0    & -0.95 & -0.98 &  0    &  0    &  0    &  0    &  0    &  0    &  0     \\
	-0.97 &  0    &  0    &  2.81 &  0    &  0    &  0    &  0    &  0    &  0    &  0    &  0    &  0    &  0    & -0.88 & -0.96  \\
	-0.91 &  0    &  0    &  0    &  1.94 &  0    &  0    &  0    &  0    &  0    & -0.06 & -0.01 & -0.01 &  0    & -0.94 &  0     \\
	0    &  0    & -0.96 &  0    &  0    &  1.94 &  0    &  0    & -0.97 & -0.01 &  0    &  0    &  0    &  0    &  0    &  0     \\
	0    &  0    &  0    &  0    &  0    &  0    &  1.02 &  0    &  0    & -0.62 & -0.40 &  0    &  0    &  0    &  0    &  0     \\
	0    &  0    & -0.95 &  0    &  0    &  0    &  0    &  1.89 & -0.94 &  0    &  0    &  0    &  0    &  0    &  0    &  0     \\
	0    &  0    & -0.98 &  0    &  0    & -0.97 &  0    & -0.94 &  2.89 &  0    &  0    &  0    &  0    &  0    &  0    &  0     \\
	0    & -0.37 &  0    &  0    &  0    & -0.01 & -0.62 &  0    &  0    &  1.24 & -0.25 &  0    &  0    &  0    &  0    &  0     \\
	0    &  0    &  0    &  0    & -0.06 &  0    & -0.40 &  0    &  0    & -0.25 &  1.56 &  0    &  0    & -0.85 &  0    &  0     \\
	0    &  0    &  0    &  0    & -0.01 &  0    &  0    &  0    &  0    &  0    &  0    &  0.93 & -0.92 &  0    &  0    &  0     \\
	0    &  0    &  0    &  0    & -0.01 &  0    &  0    &  0    &  0    &  0    &  0    & -0.92 &  0.94 &  0    & -0.01 &  0     \\
	-0.05 & -0.78 &  0    &  0    &  0    &  0    &  0    &  0    &  0    &  0    & -0.85 &  0    &  0    &  1.68 &  0    &  0     \\
	-0.90 &  0    &  0    & -0.88 & -0.94 &  0    &  0    &  0    &  0    &  0    &  0    &  0    & -0.01 &  0    &  2.74 &  0     \\
	-0.94 &  0    &  0    & -0.96 &  0    &  0    &  0    &  0    &  0    &  0    &  0    &  0    &  0    &  0    &  0    &  1.91 
	\end{array}\label{LL16full}
	\right]
	\end{align} 
\end{figure*}

\begin{figure}[thpb]
	\centering
	\includegraphics[]{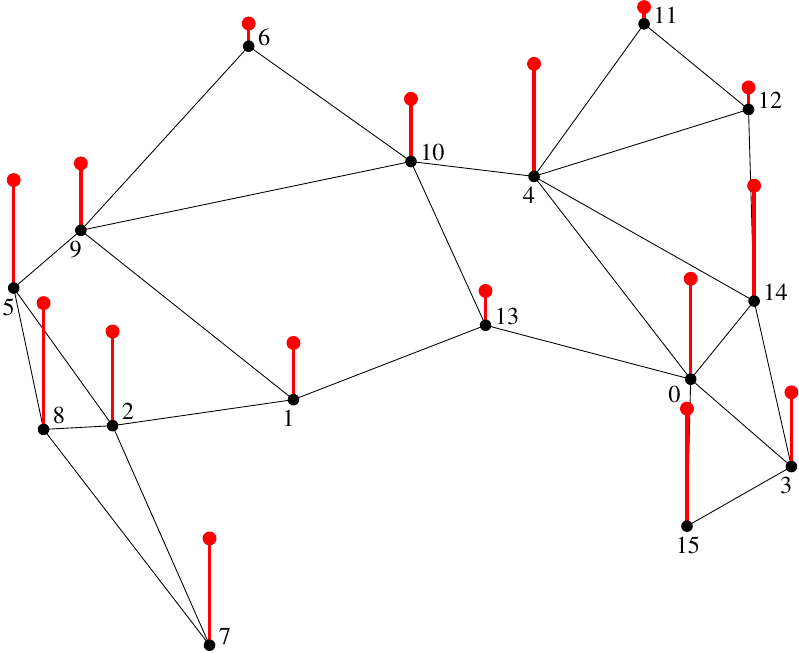}
	
	\vspace{5mm}
	
	 \includegraphics[]{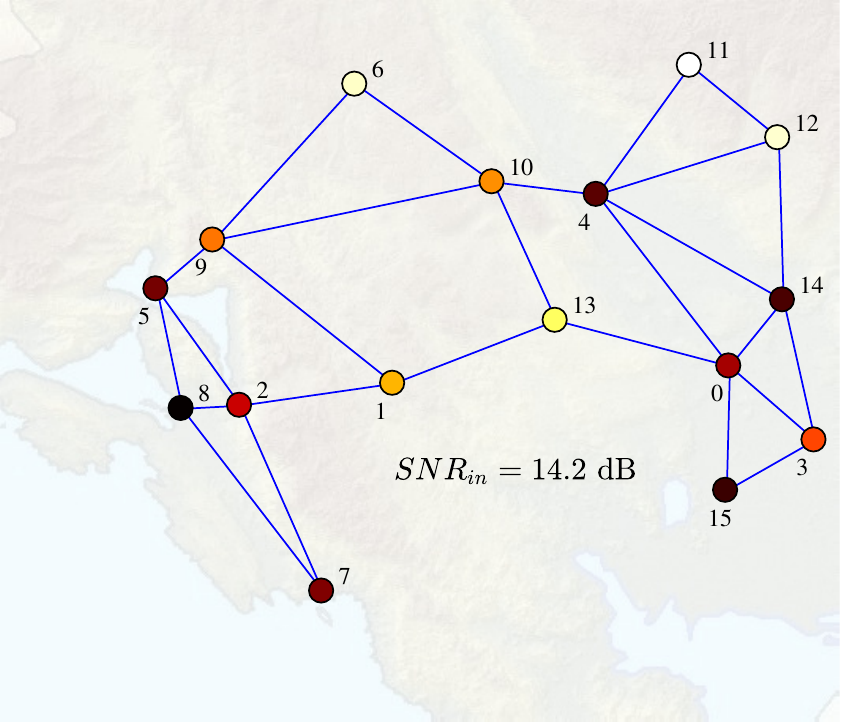}
	\caption{From a multi-sensor temperature measurement to a graph signal.  The temperature field is represented on a graph that combines the spatially unaware measurements in Fig. \ref{fig:MNE_fig_a}(b) and the physically relevant graph topology in Fig. \ref{fig:MNE_fig_b}(b). The graph signal values are represented in two ways: (top) by vertical lines for which the length is proportional to the signal values, and (bottom) by using a ``hot'' colormap to designate the signal values at the  vertices. }
	\label{fig:MNE_fig_e}
\end{figure}

This simple real-world example can be interpreted within the graph signal processing framework as follows: 
\begin{itemize}
	\item Sensing points where the signal is measured are designated as the \textbf{graph vertices}, as in Fig. \ref{fig:MNE_fig_a},
	\item Vertex-to-vertex lines which indicate physically meaningful connectivity among the sensing points become the \textbf{graph edges}, as in Fig. \ref{fig:MNE_fig_b}(a),
	\item The vertices and edges form \textbf{a graph}, as in Fig. \ref{fig:MNE_fig_b}(b), a new very structurally rich signal domain, 
	\item  The graph, rather than a standard vector of sensing points, is then used for analyzing and processing data, as it exhibits both spatial and physical domain awareness,  
	\item The measured temperatures are now interpreted as \textbf{signal samples on graph}, as shown in Fig. \ref{fig:MNE_fig_e}, 	
	\item   Similar to traditional signal processing, this new \textbf{graph signal} may have many realizations on the same graph and may comprise noise,
	\item Through relation (\ref{1}), we have therefore introduced a simple \textbf{system on a graph} for physically and spatially aware signal averaging (a linear first-order system on a graph).
\end{itemize}

To emphasize our trust in a particular sensor (i.e., to model sensor  relevance),  a weighting scheme may be imposed, in the form
\begin{equation}
y(n)=x(n)+\sum_{m\ne n} W_{nm}x(m), \label{2}
\end{equation}
where $W_{nm}$ are the elements of the weighting matrix, $\mathbf{W}$.  
 
 There are three classes of approaches to the definition of graph edges and their corresponding weights,  $ W_{nm}$: 
 \begin{itemize}
 	\item
  Already physically well defined edges and weights,
  \item Definition of edges and weights based on the geometry of vertex positions, 
  \item 
  Data similarity based methods for learning the underlying graph topology. 
  \end{itemize}
  All three approaches to define the edge weights are covered in detail in Part III of this monograph.

  Since in our case of geographic temperature measurements, the graph weights do not belong to the class of obvious and physically well defined edges and weights, we will employ  the “geometry of the vertices” based approach for the definition of the edges and weights. In this way, the weight elements,  $ W_{nm}$, for the neighboring vertices are calculated based on the horizontal vertex distance, $r_{mn}$, and the altitude difference, $h_{mn}$, as 
  \begin{equation}W_{mn}=e^{-\alpha r_{mn}-\beta h_{mn}}, \label{weigthsdef}
  \end{equation} 
  where $\alpha$ and $\beta$ are suitable constants. The so obtained weight matrix, $\mathbf{W}$, is given in (\ref{W16full}).

Based on (\ref{1}), a weighted  graph signal estimator of cumulative temperature now becomes 
\begin{equation}
\mathbf{y}=\mathbf{x}+\mathbf{W}\mathbf{x}. \label{3}
\end{equation}
In order to produce unbiased estimates, instead of the cumulative sums in (\ref{1}) and (\ref{2}), the weighting coefficients within the estimate for each $y(n)$ should sum up to unity. This can be achieved through a  normalized form of (\ref{3}), given by
\begin{equation}
\mathbf{y}=\frac{1}{2}(\mathbf{x}+{\mathbf{D}^{-1}}\mathbf{W}\mathbf{x}),\label{AWjed}
\end{equation}
where the elements of the diagonal normalization matrix, $\mathbf{D}$, are equal to the \textbf{the degree matrix} elements, $D_{nn}=\sum_m W_{nm}$, while  ${\mathbf{D}^{-1}}\mathbf{W}$ is a \textbf{random walk (diffusion) shift operator} \cite{LNDM,stankovic2017LLLvertex}.

\begin{figure}[thpb]
	\centering
	\includegraphics[]{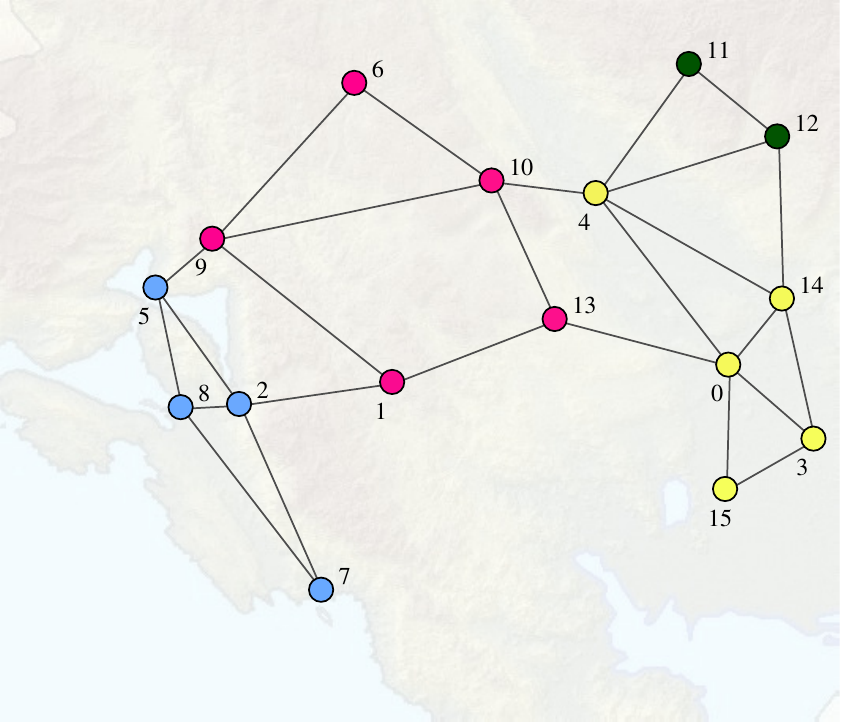}
	\caption{Clustering of the graph from Fig. \ref{fig:MNE_fig_b}(b) based on the graph Laplacian eigenvectors, $\mathbf{u}_1$, $\mathbf{u}_2$, and $\mathbf{u}_3$. Observe the correct clustering of the graph into the clusters that belong to the seaside area (blue), low mountains (red), low land (yellow), and high mountains (green).}
	\label{MNE_fig_h}
\end{figure}

Now that we have defined the graph vertices and edge weights we may resort to the data-agnostic clustering approaches, given in {\color{blue}Part I - Section  \ref{I-data_clustering_section}}, to cluster the vertices of this graph based on the graph topology. Fig. \ref{MNE_fig_h}   shows the clustering result obtained based on  the three smoothest eigenvectors, $\mathbf{u}_1$, $\mathbf{u}_2$, and $\mathbf{u}_3$ (excluding the constant eigenvector, $\mathbf{u}_0$), of the graph Laplacian matrix, $\mathbf{L}=\mathbf{D}-\mathbf{W}$,  of which the values are given in (\ref{LL16full}).
Notice that even such a simple graph clustering scheme was capable of identifying different physically meaningful geographic regions. This also means that temperature estimation can roughly be performed within each cluster, which may even be treated as an independent graph (see graph segmentation and graph cuts in {\color{blue}Part I, Section \ref{I-section_vertex_clustering_and_mapping}}), rather than over the whole sensor network.

The above-introduced graph data estimation framework is quite general and admits application to many different  scenarios where, after identifying a suitable graph topology, we desire to perform estimation on data acquired on such graphs, the subject of this part of the monograph.

\section{Signals and Systems on Graphs} 
\label{sec3}

\begin{figure}
\centering
\includegraphics[]{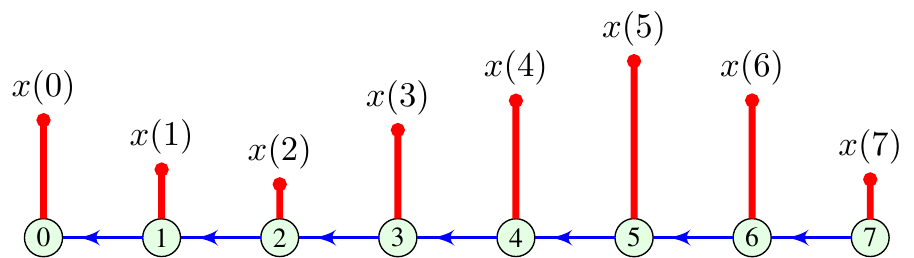}
\caption{Directed 	path graph representation of a classical time-domain signal defined on an equidistant discrete-time grid.}
\label{fig:sig-time}
\end{figure}

In classical data analytics, a signal is sampled at successive, equally spaced, time instants.  This then dictates the ordering of signal samples, with $x(n)$ being preceded by $x(n-1)$ and succeeded by $x(n+1)$. The \textquotedblleft time  distance\textquotedblright \, between data samples is therefore an inherent parameter in standard data processing algorithms.  
The relation between sampling instants can also be represented in a graph form, whereby the vertices that correspond to the instants when the signal is sampled and the corresponding edges define the linear sampling (vertex) ordering. The equally spaced nature of sampling instants in classical scenarios can then be represented with equal weights for all edges (for example, normalized to $1$), as shown in Fig. \ref{fig:sig-time}.

\begin{figure}
\centering
\includegraphics[]{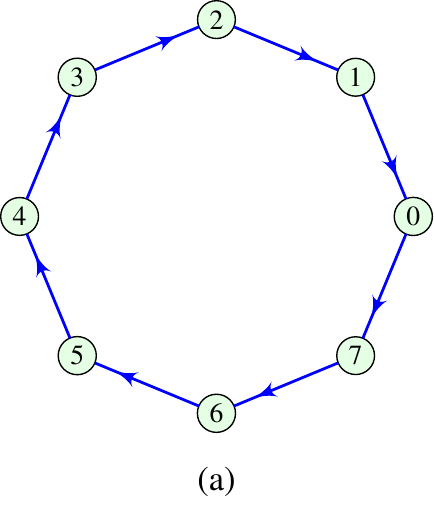}
\hspace{5mm}
\includegraphics[]{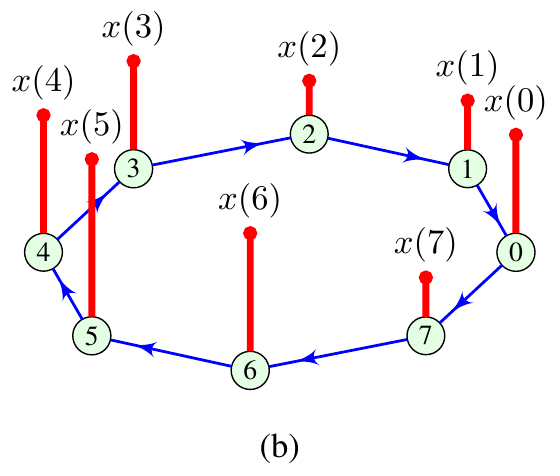}
\caption{Graph representation of periodic data. (a) A directed circular graph. (b) A periodic signal measured on a circular graph. Signal values, $x(n)$, are designated by vertical lines at the corresponding vertex, $n$.}
\label{fig:sig-on-graph}
\end{figure}

\begin{figure}
\centering
\includegraphics[]{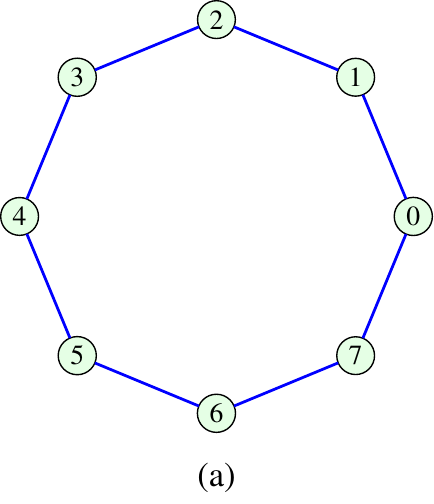}
\hfill
\includegraphics[]{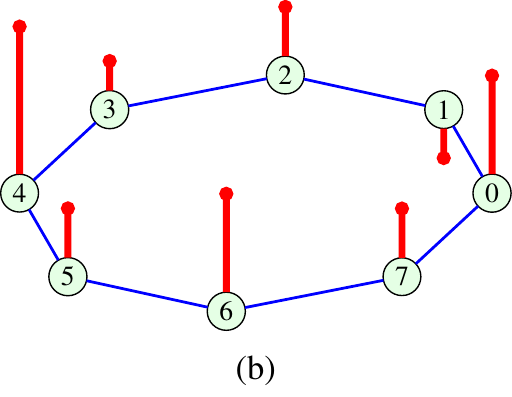}
\caption{Undirected circular graph (a) and signal on the graph (b). Signal values, $x(n)$, are presented as vertical lines at the corresponding vertex, $n$.}
\label{fig:sig-circ-graph}
\end{figure}

\begin{figure}
	\centering
	\includegraphics[]{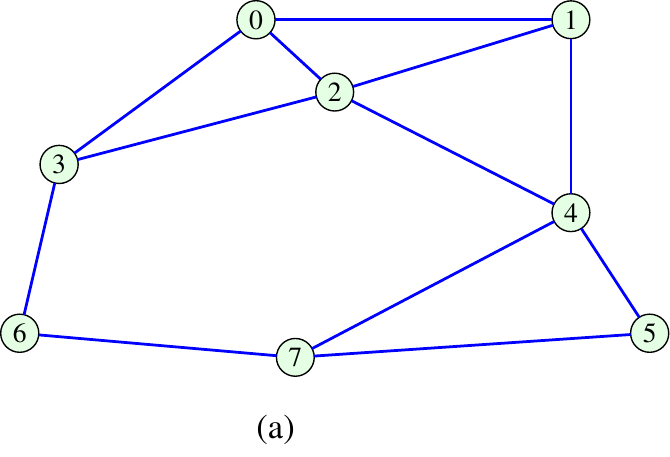}
	\hfill
	\includegraphics[]{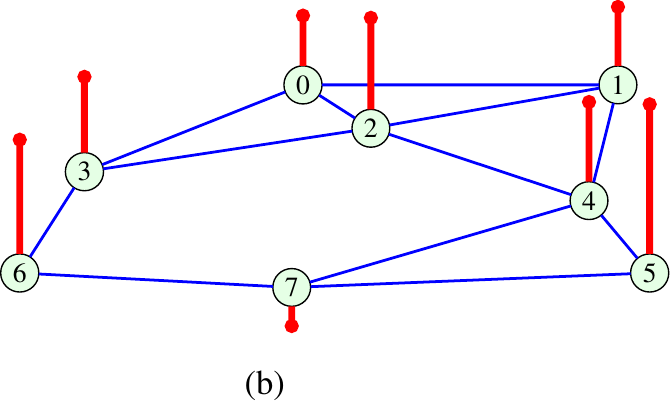}
%	\caption{Signals on general undirected graphs. (a) A signal on an undirected circular graph. (b) A signal on an arbitrary undirected  graph. Signal values are presented as vertical lines coming out of the corresponding vertices.}
	\caption{Arbitrary undirected graph (a) and signal on graph (b). Signal values, $x(n)$, are presented as vertical lines at the corresponding vertex, $n$.}
	\label{fig:sig-arb-graph}
\end{figure}

Algorithms defined in discrete time (like, for example, those based on the DFT or other similar data transforms), usually assume periodicity of the analyzed signals, which means that sample $x(N-1)$ is succeeded by sample $x(0)$, in a perpetual sequence. Notice that this case corresponds to the circular graph, shown in Fig.~\ref{fig:sig-on-graph}, which allows us to use this model in many standard data transforms, such as the DFT, DCT, wavelets, and to define graph-counterparts of other processing algorithms, based on these transforms.

A signal on general (including also circular) undirected graph is defined by associating real (or complex) data values, $x(n)$, to each vertex, as shown in Fig. \ref{fig:sig-circ-graph} and Fig. \ref{fig:sig-arb-graph}. Such signal values can be arranged in a vector form $$\mathbf{x}=[x(0),x(1),\ldots,x(N-1)]^T,$$
so that a graph may be considered as a generalized signal domain.

This allows, in general, for any linear processing scheme for a graph signal observed at a vertex, $n$, to be defined as a linear combination of the signal value, $x(n)$, at this vertex and the signal samples, $x(m)$, at the neighboring vertices, that is    
\begin{equation}
y(n)=x(n)h(n,n)+\sum_{m \in \mathcal{V} _n}x(m)h(m,n), \label{genGRSIS}
\end{equation}
where $\mathcal{V} _n$ is the set of vertices in the neighborhood of vertex $n$, and $h(m,n)$ the scaling coefficients.  

\begin{Remark} The estimation form in (\ref{genGRSIS}) is highly vertex-dependent; it is vertex-invariant only in a very specific case of regular graphs, where $\mathcal{V} _n$ is a $K$-neighborhood of the vertex $n$,  with $h(n,m)=h(n-m)$.
\end{Remark}

We now proceed to define various forms of vertex-invariant filtering functions, using shifts on a graph. These will then be  used to introduce efficient graph signal processing methods \cite{sandryhaila2014discrete,sandryhaila2014big,venkitaraman2016hilbert,Agaskar,7986982,wang2016local,segarra2016stability}.

\subsection{Adjacency Matrix and Graph Signal Shift} \label{graphshiftsec}
Consider a graph signal, $\mathbf{x}$, for which $x(n)$ is the observed sample at a vertex $n$. A signal shift on a graph can be defined as movement of the signal sample, $x(n)$, from its original vertex, $n$, along all walks of length one, that is $K=1$, that start at vertex $n$. If the signal shifted in this way is denoted by  $\mathbf{x}_1$, then its values can be defined using the graph adjacency matrix, $\mathbf{A}$, as
\begin{equation}
\mathbf{x}_1=\mathbf{A}\mathbf{x}.
\end{equation}

\begin{figure}[t]
\centering
\includegraphics[]{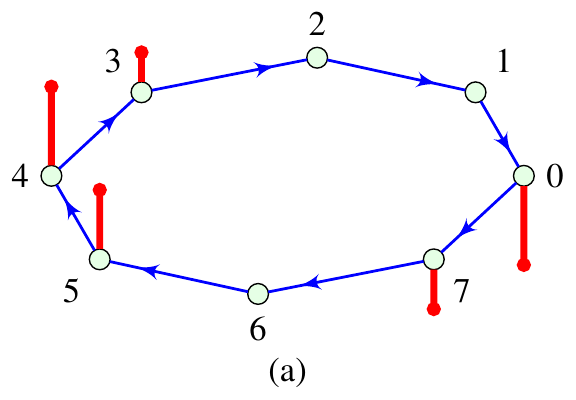}
\hfill
\includegraphics[]{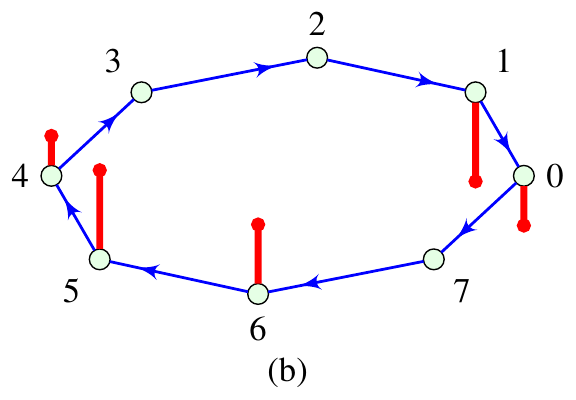}
\caption{Graph shift operator  on a directed graph (classical circular shift). (a) Elements of a signal, $\mathbf{x}$, shown as red lines on a directed circular graph. (b) The shifted version, $\mathbf{Ax}$, of the graph signal from (a). The adjacency matrix of for this graph is given in (\ref{I-AdjMtxFirsDir}) in Part I.}
\label{fig:sig-fourier}
\end{figure}

\begin{figure}[t]
\centering
\includegraphics[]{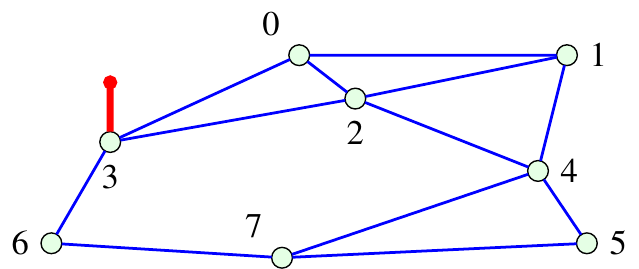} (a)

\vspace{3mm}
\includegraphics[]{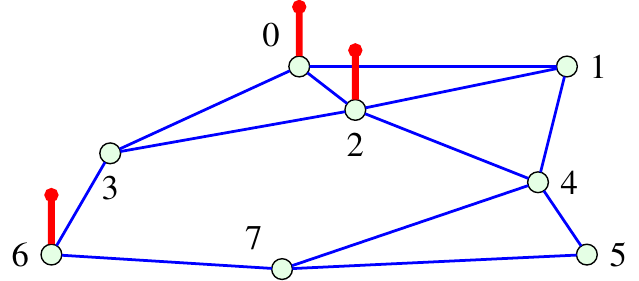}(b)

\caption{Graph signal shift on an undirected graph. (a) A simple signal, $\mathbf{x}$, on an undirected graph. (b) Shifted version, $\mathbf{Ax}$, of the graph signal from (a).}
\label{fig:sig-gr1}
\end{figure}

\begin{Example} As an illustration of a graph signal and its shifted  version, consider the signal on a circular graph from Fig. \ref{fig:sig-on-graph}(a). The original signal, $\mathbf{x}$,  is shown in Fig. \ref{fig:sig-fourier}(a), and its shifted version, $\mathbf{x}_1$, in  Fig. \ref{fig:sig-fourier}(b). Another simple signal on the undirected graph from Fig. \ref{fig:sig-arb-graph} (a) is presented in Fig. \ref{fig:sig-gr1}(a), with its shifted version, $\mathbf{x}_1=\mathbf{A}\mathbf{x}$, shown in Fig. \ref{fig:sig-gr1}(b).
\end{Example}

A signal shifted by two graph shifts is obtained by further shifting $\mathbf{x}_1=\mathbf{A}\mathbf{x}$ by one shift. The resulting, twice shifted, graph signal is then given by $$\mathbf{x}_2=\mathbf{A}\mathbf{x}_1=\mathbf{A}(\mathbf{A}\,\mathbf{x})=\mathbf{A}^2\,\mathbf{x}.$$ 

Therefore, in general, an $m$ times shifted signal on graph is given by $$\mathbf{x}_m=\mathbf{A}\mathbf{x}_{m-1}=\mathbf{A}^m\,\mathbf{x}.$$

\begin{Remark} Like the standard shift operator, the second order shift of a graph signal is obtained by shifting the  already once shifted signal.  The role of the shift operator is assumed by the adjacency matrix, $\mathbf{A}$.
\end{Remark}
	
\subsection{Systems Based on Graph Shifted Signals}\label{Sec:SysA}
Very much like in standard linear shift-based systems, a system on a graph can be implemented as a linear combination of a graph signal, $\mathbf{x}$,  and its graph shifted versions,   $\mathbf{A}^m\, \mathbf{x}$, $m=1,2,\dots,M-1$. The output signal from a system on a graph can then be written as
\begin{equation}
\mathbf{y}=h_0 \mathbf{A}^0\, \mathbf{x}+h_1 \mathbf{A}^1\, \mathbf{x}+\dots+h_{M-1} \mathbf{A}^{M-1}\, \mathbf{x}=\sum_{m=0}^{M-1}h_m \mathbf{A}^m\, \mathbf{x}
\label{eq:filter-time}
\end{equation}
where $\mathbf{A}^0=\mathbf{I}$, by definition, and  $h_0$, $h_1$, \ldots, $h_{M-1}$ are the system coefficients. For a circular (classical linear system) graph, this relation reduces to the well known Finite Impulse Response (FIR) filter, given by,
\begin{equation}
y(n)=h_0x(n)+h_1x(n-1)+\cdots+h_{M-1}x(n-M+1).\label{ClassFIR}
\end{equation}

Keeping in mind that the matrix $\mathbf{A}^m$ describes walks of the length $K=m$ in a graph (see Property $M_2$ in Part 1), the output graph signal, $y(n)$, is calculated as a linear combination of the input graph signal values and the signal values observed at vertices belonging to the $(M-1)$-neighborhood  of the considered vertex $n$.

\begin{Remark}  When the minimal and characteristic polynomial are of the same degree, a physically meaningful system order $(M-1)$ should be lower than the number of vertices $N$, that is is, $M\le N$. The corresponding condition in classical signal analysis would that the number, $M$, of the system impulse response coefficients, $h_m$,  in (\ref{ClassFIR}) should be lower or equal to the total number of signal samples, $N$ (for the graph in Fig. \ref{fig:sig-fourier} it means that the meaningful graph signal shifts are $m=0,1,2,\dots,N-1$, since the shift for $m=N$ reduce to the shift for $m=0$, the shift for $m=N+1$ is equivalent to the shift for $m=1$, and so on). Therefore, in general, the system order $(M-1)$ should be lower than the degree $N_m$ of the minimal polynomial of the adjacency matrix $\mathbf{A}$. For more detail see {\color{blue}Part I, Section \ref{I-section_eigenvalue_decomposition_adjacency}}.
\end{Remark}

\begin{Remark}   Any system of order $M-1\ge N_m$ can be reduced to a system of order $N_m-1$. 
\end{Remark}

\begin{Remark}  If the system order is greater than or equal to the degree of the minimal polynomial, $M-1\ge N_m$, then there exist more than one system producing the same output signal for a given input signal. All such systems on a graph are called equivalent. 
\end{Remark}

The statements in the last three remarks will be addressed in more detail in Section \ref{SpecDFD}, with their proofs  also  provided.

\begin{Example} Consider a signal on graph from Fig. \ref{fig:sig-arb-graph}(a), given in Fig.~\ref{fig:filt2}(a), and a linear system which operates on this graph, defined by the coefficients $h_0=1$, $h_1=0.5$. Observe that this system on a graph corresponds to a simple classical first-order weighted moving average system.  The output graph signal  then represents a weighted average of the signal value at a vertex $n$ and the signal values at its $K=1$ neighborhood. The output graph signal is shown in Fig.~\ref{fig:filt2}(b). 

\begin{figure}[t]
\centering
\includegraphics[]{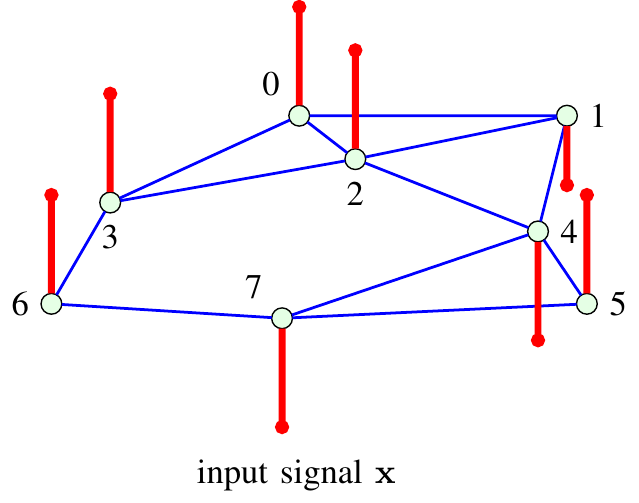}
(a)
\hfill
\includegraphics[]{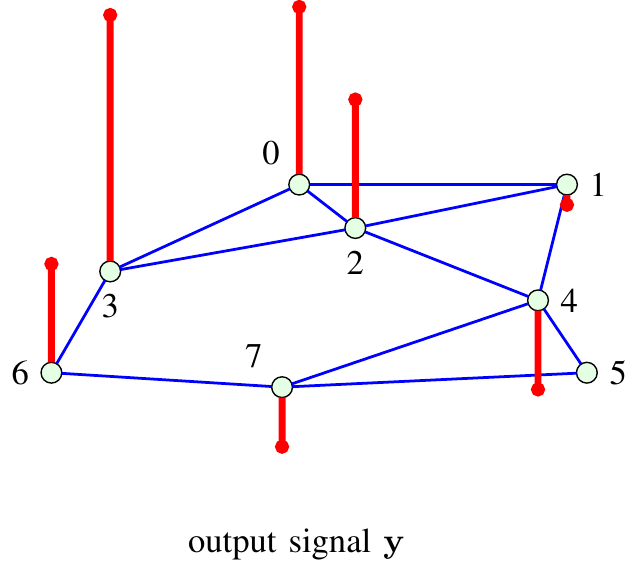}
(b)
\caption{Example of vertex domain signal filtering. (a) An arbitrary  graph signal. (b) The output signal obtained  through a first-oder (averaging) system on a graph, defined as $\mathbf{y}=\mathbf{x} +0.5\; \mathbf{A} \mathbf{x}$.}
\label{fig:filt2}
\end{figure} 
\end{Example}

\noindent\textbf{General system on graph.} A system on a graph may be defined in the vertex domain as     
\begin{equation}
\mathbf{y}=H(\mathbf{A})\mathbf{x},
\end{equation}
where $H(\mathbf{A})$ is a vertex domain system (filter) function. A system on a graph is then linear and shift invariant if it satisfies the following properties of:
\begin{enumerate}  
	\item Linearity
$$H(\mathbf{A})(a_1\mathbf{x}_1+a_2\mathbf{x}_2)=a_1\mathbf{y}_1+a_2\mathbf{y}_2.$$ 
\item Shift invariance
$$H(\mathbf{A})[\mathbf{A}\mathbf{x}]=\mathbf{A}[H(\mathbf{A})\mathbf{x}]=\mathbf{Ay}.$$
\end{enumerate}

\begin{Remark} A system on a graph defined by 
\begin{equation}
H(\mathbf{A})=h_0 \mathbf{A}^0+h_1 \mathbf{A}^1+\dots+h_{M-1} \mathbf{A}^{M-1} \label{Vetrx_DF}
\end{equation}
is linear and shift invariant since $\mathbf{A}\mathbf{A}^m=\mathbf{A}^m\mathbf{A}$.  
\end{Remark}

\subsection{Graph discrete Fourier transform (GDFT), adjacency matrix based definition}

Classical exploratory data analysis often employs estimation of signals in the spectral (Fourier) domain; this has led to a number of simple and efficient algorithms. While standard spectral analysis employs  an equidistant grid in both time and frequency, following the ideas of a system on a graph, we next show that spectral domain representations of graph signals are naturally based on spectral decompositions of the  adjacency matrix or graph Laplacian.
 
\noindent\textbf{The graph Fourier transform} of a signal, $\mathbf{x}$, is defined as
\begin{equation}
\mathbf{X}=\mathbf{U}^{-1}\mathbf{x}
\label{eq:gft}
\end{equation}
where $\mathbf{X}$ denotes a vector of the GDFT coefficients, and $\mathbf{U}$ is a matrix whose columns represent the eigenvectors of the adjacency matrix, $\mathbf{A}$. 
Denote the elements of the vector $\mathbf{X}$ by $X(k)$, for $k=0,1,\ldots,N-1$, and recall that for undirected graphs, the adjacency matrix is symmetric, that is, $\mathbf{A}^T=\mathbf{A}$, and that the eigenmatrices of a symmetric matrix satisfy the property $$\mathbf{U}^{-1}=\mathbf{U}^T.$$
 The element, $X(k)$, of the graph Fourier transform vector, $\mathbf{X}$, therefore represents a projection of the considered graph signal, $x(n)$, onto the $k$-th eigenvector of $\mathbf{A}$ (a basis function), given by 
\begin{gather}
X(k)=\sum_{n=0}^{N-1}x(n)u_k(n). \label{GFTDEF}
\end{gather}
In this way, the graph discrete Fourier transform can be interpreted as a set of projections (signal decomposition) onto the set of eigenvectors, $\mathbf{u}_0, \mathbf{u}_1,\dots,\mathbf{u}_{N-1} $, which serve as orthonormal basis functions.
 
The inverse graph discrete Fourier transform is then straightforwardly  obtained from (\ref{eq:gft}) as
\begin{equation}
\mathbf{x}=\mathbf{U}\,\mathbf{X},
\label{eq:igft} 
\end{equation}
or element-wise
\begin{gather}
x(n)=\sum_{k=0}^{N-1}X(k)u_k(n). \label{IGFTDEF}
\end{gather}

Observe that, for example,  for a circular graph from Fig.~\ref{fig:sig-on-graph}, the graph discrete Fourier transform pair in (\ref{GFTDEF}) and (\ref{IGFTDEF}) reduces to the standard discrete Fourier transform (DFT) pair. For this reason, the transform in (\ref{GFTDEF}) and its inverse in (\ref{IGFTDEF}) are referred to as the \textit{graph discrete Fourier transform} (GDFT) and the \textit{inverse graph discrete Fourier transform} (IGDFT).

\subsection{System on a graph in the GDFT domain}
Consider a general system on a graph defined in (\ref{Vetrx_DF}),
\begin{gather}
\mathbf{y}=H(\mathbf{A})\mathbf{x} 
=\Big(h_0 \mathbf{A}^0+h_1 \mathbf{A}^1+\dots+h_{M-1} \mathbf{A}^{M-1}\Big)\mathbf{x}. \label{Vetrx_DF_Sys}
\end{gather}
 Upon employing the spectral representation of the adjacency matrix, $\mathbf{A}=\mathbf{U}\mathbf{\Lambda}\mathbf{U}^{-1}$, we have
\begin{gather}
\mathbf{y}=\Big(h_0 \mathbf{U}\mathbf{\Lambda}^0\mathbf{U}^{-1}+h_1 \mathbf{U}\mathbf{\Lambda}^1\mathbf{U}^{-1}+\dots+h_{M-1} \mathbf{U}\mathbf{\Lambda}^{M-1}\mathbf{U}^{-1}\Big) \mathbf{x} \notag \\
=\mathbf{U} \big( h_0 \mathbf{\Lambda}^0+h_1 \mathbf{\Lambda}^1+\dots+h_{M-1} \mathbf{\Lambda}^{M-1}\big)\mathbf{U}^{-1}\, \mathbf{x}\notag\\
=
\mathbf{U}\, H(\mathbf{\Lambda})\mathbf{U}^{-1}\, \mathbf{x},
\end{gather}
with the system on a graph transfer function
\begin{gather}
H(\mathbf{\Lambda})
=h_0 \mathbf{\mathbf{\Lambda}}^0+h_1 \mathbf{\mathbf{\Lambda}}^1+\dots+h_{M-1} \mathbf{\mathbf{\Lambda}}^{M-1}, \label{Vetrx_DF_SysScep}
\end{gather}
where $\mathbf{\Lambda}$ is the matrix of eigenvalues of $\mathbf{A}$.

A pre-multiplication of this relation with $\mathbf{U}^{-1}$, yields
\begin{equation}
\mathbf{U}^{-1}\mathbf{y}=
 H(\mathbf{\Lambda})\mathbf{U}^{-1}\, \mathbf{x}
\end{equation}
From (\ref{eq:gft}), the terms $\mathbf{U}^{-1}\, \mathbf{y}$ and $\mathbf{U}^{-1}\, \mathbf{x}$ are respectively the GDFTs of the output graph signal, $\mathbf{y}$, and the input graph signal, $\mathbf{x}$, so that the spectral domain system on a graph relation becomes
\begin{equation}
\mathbf{Y}=
 H(\mathbf{\Lambda})\, \mathbf{X}, \label{sisteonGRAPH}
\end{equation} 
The output graph signal in the vertex domain can then be calculated as
\begin{equation}
\mathbf{y}=H(\mathbf{A})\mathbf{x}=\mathrm{IGDFT}\{
H(\mathbf{\Lambda})\, \mathbf{X}\}. \label{sisteonGRAPH_VS}
\end{equation}
The element-wise form of the system on a graph  in (\ref{sisteonGRAPH}) is of the form
$$Y(k)=(h_0+h_1 \lambda_k+\dots+h_{M-1}\lambda^{M-1}_k)X(k),$$
where $\lambda_k$ denotes the $k$th eigenvalue of the adjacency matrix, $\mathbf{A}$. From (\ref{Vetrx_DF_SysScep}) and the above equation, we can now define the \textit{transfer function of a system on a graph}  in the form
\begin{equation}H(\lambda_k)=\frac{Y(k)}{X(k)}=h_0+h_1 \lambda_k+\dots+h_{M-1}\lambda^{M-1}_k. \label{H_in_the_GDFT}
\end{equation}

\begin{Remark} The classical linear system  in (\ref{ClassFIR}) can be obtained directly from its graph counterpart in  (\ref{H_in_the_GDFT}) when the graph is directed and circular. This is because the adjacency matrix of a directed circular graph has eigenvalues $\lambda_k=e^{-j2\pi k /N}$ ({\color{blue}see Part I, Section \ref{I-Section_DFT_basis_functions}}  for more detail on directed circular graphs), which are identical to the samples on the unit circle in classical DFT. 
\end{Remark}

Similar to the $z$-transform in classical signal processing, for systems on graphs we can also introduce the system transfer function in the $z$-domain .

\noindent\textbf{The $z$-domain transfer function} of a system on a graph is defined as
\begin{equation}H(z^{-1})=\mathcal{Z}\{h_n\}=h_0+h_1 z^{-1} +\dots+h_{M-1}z^{-(M-1)}, \label{H_in_the_GzT}
\end{equation}
for $n=0,1,\dots,M-1$. Obviously, from (\ref{H_in_the_GDFT}), we have
$$H(\lambda_k)=H(z^{-1})\big\vert_{z^{-1}=\lambda_k}.$$
However, the definition of the $z$-transform for arbitrary graph signals, $x(n)$ and $y(n)$, that would satisfy the relation $Y(z^{-1})=H(z^{-1})X(z^{-1})$ is not straightforward, which limits the application of the $z$-transform on graphs. This will be discussed in more detail in Section \ref{GZT}. 

\subsection{Graph Signal Filtering in the  Spectral Domain of the Adjacency Matrix}

The energy of a graph shifted signal is given by 
$$\left\Vert \mathbf{x}_1\right\Vert _{2}^{2}=\left\Vert \mathbf{Ax}\right\Vert _{2}^{2}.$$
 However, as shown in Fig. \ref{fig:sig-gr1}, in general, the energy of a shifted signal is not the same as the energy of the original signal,  that is
 $$\left\Vert \mathbf{Ax}\right\Vert _{2}^{2} \ne \left\Vert \mathbf{x}\right\Vert _{2}^{2}.$$
 On the other hand, in graph signal processing it is often desirable that a graph shift does not increase signal energy. One such graph shift operator is introduced bellow.  

\begin{Remark} Using the matrix two-norm it is straightforward to show that the ratio of energies of the graph shifted signal, $\mathbf{Ax}$, and the original graph signal, $\mathbf{x}$, satisfies the relation 
\begin{gather}
\max \{\frac{\left\Vert \mathbf{Ax}\right\Vert _{2}^{2}}{\left\Vert
	\mathbf{x}\right\Vert _{2}^{2}}\}=\max \{\frac{\mathbf{x}^{T}\mathbf{A}^{T}\mathbf{Ax}%
}{\left\Vert \mathbf{x}\right\Vert _{2}^{2}} \}= \lambda_{\max}^2. \label{Mat_Nor_limit}
\end{gather}
where $\lambda_{\max}=\max_k{|\lambda_k|},~k=0,1,\dots,N-1$.
\end{Remark}

\subsubsection{Normalization of the Adjacency Matrix}
From (\ref{Mat_Nor_limit}), for the energy of a graph shifted signal, $\left\Vert \mathbf{Ax}\right\Vert _{2}^{2}$, not to exceed the energy of the original graph signal, $\left\Vert \mathbf{x}\right\Vert _{2}^{2}$, we may employ \textit{the normalized adjacency matrix}, defined as
\begin{equation}
\mathbf{A}_{norm}=\frac{1}{\lambda_{\max}}\mathbf{A} \label{NormAdA}
\end{equation}
as a graph shift operator within any system on a graph. While this kind of normalization still does not make the shift on a graph isometric, the energy of the signal shifted in this way is guaranteed not to be bigger than the energy of the original graph signal, since
$$\left\Vert \mathbf{A}_{norm}\mathbf{x}\right\Vert _{2}^{2} \leq \left\Vert \mathbf{x}\right\Vert _{2}^{2}.$$
  The equality holds only for a very specific signal which is proportional to the eigenvector that corresponds to $\lambda_{\max}$. 

The basic shift on a graph, system on a graph, and graph spectral domain representations can be implemented with the normalized adjacency matrix in (\ref{NormAdA}) in the same way as with the original adjacency matrix. \textit{An important property which does not apply to standard adjacency matrices is that the normalization of adjacency matrix yields a simpler eigenvector and eigenvalue ordering scheme, as shown next.} 

\subsubsection{Spectral Ordering of Eigenvectors of the Adjacency Matrix}\label{AMord}
For physically meaningful low-pass and high-pass filtering on a graph, we need to establish the notion of spectral order. This, in turn, requires a criterion to classify the eigenvectors (corresponding to the GDFT basis functions) into the slow-varying and fast-varying ones. 

\begin{Remark}
In classical Fourier analysis, the basis functions are ordered according to their frequency, whereby, for example, low-pass (slow varying) basis functions are harmonic functions characterized by low frequencies. On the other hand, the notion of frequency of the eigenvectors of the graph adjacency matrix, which serve as a basis for for signal decomposition, is not defined and we have to find another criterion to classify or rank order the eigenvectors. Again, we draw the inspiration from classical Fourier analysis which suggests that the energy of the “signal change” can be used  instead of frequency to indicate the rate of change of an eigenvector along time. 
\end{Remark}

\noindent\textbf{Energy of signal change.} The \textit{first graph difference} can be defined for graph signals as a difference of the original graph signal and its graph shift, that is,
$$\mathbf{\Delta x}=\mathbf{x}-\mathbf{x}_1=\mathbf{x}-\mathbf{A}_{norm}\mathbf{x}.$$
In analogy to classical analysis, the energy of signal change can then be defined as the energy of the first difference of a graph signal $\mathbf{x}$, and takes the form
\begin{gather}
E_{\Delta x}=\left\Vert \mathbf{x}-\mathbf{A}_{norm}\mathbf{x}\right\Vert_2^2
=\left\Vert \mathbf{x}-\frac{1}{\lambda_{\max}}\mathbf{A}\mathbf{x}\right\Vert_2^2.\notag
\end{gather}

When the graph signal assumes a specific form of an eigenvector, $\mathbf{x}=\mathbf{u}$, of the adjacency matrix, $\mathbf{A}$,  the energy of this eigenvector change is equal to 
\begin{gather}
E_{\Delta u}=\left\Vert \mathbf{u}-\frac{1}{\lambda_{\max}}\lambda\mathbf{u}\right\Vert_2^2=
\left|1-\frac{\lambda}{\lambda_{\max}}\right|^2, \label{AgraphORD}
\end{gather}
whereby the normalized adjacency matrix, $\mathbf{A}_{norm}$, is used to bound the energy of the shifted graph signal. In the derivation we have also used $\mathbf{A} \mathbf{u} = \lambda \mathbf{u}$ and $\left\Vert\mathbf{u}\right\Vert_2^2=1$. 

Now, the lower values of $E_{\Delta u}$ indicate that $\mathbf{u}$ is slow-varying, $E_{\Delta u}=0$ indicates  that the signal is constant, while larger values of $E_{\Delta u}$ are associated with fast changes of $\mathbf{u}$ in time. The form in (\ref{AgraphORD}) is also referred to as \textit{the two-norm total variation of a basis function/eigenvector}. Therefore, if the change in a basis function, $\mathbf{u}$, has a large energy, then the eigenvector, $\mathbf{u}$, can be considered to belong to the higher spectral content of the graph signal.

\begin{Remark}
	From (\ref{AgraphORD}), the energy of the rate of change of a graph signal is minimal for $\lambda=\lambda_{\max}$ and it increases as $\lambda$ decreases ({\color{blue}see Fig. \ref{I-GSPb_spectrum2a} in Part 1}). %Fig. \ref{GSPb_spectrum2a}. 
\end{Remark}

Now that we have established a criterion for the ordering of eigenvectors, based on the corresponding eigenvalues, we shall proceed to define an \textit{ideal low-pass filter on a graph}. The intuition behind low-pass filtering in the graph domain is that such a filter should  pass unchanged all signal components (eigenvectors of $\mathbf{A}$) for which the rates of change are slower than that defined by the cut-off eigenvalue,  $\lambda_{c}$ (\textit{cf.} cut-off frequency), while all signal components (eigenvectors) which exhibit variations which are faster than that defined by the cut-off eigenvalue, $\lambda_{c}$, should be suppressed. The ideal low-pass filter in the graph domain is therefore defined as 
$$
f(\lambda)= \begin{cases}
1, & \text{ for } \lambda > \lambda_{c}, \\
0, & \text{ for other } \lambda.
\end{cases}
$$

\begin{Example} Consider again the undirected graph from Fig. \ref{fig:sig-arb-graph}(a) on which we observe a graph signal shown in {\color{blue}Fig.~\ref{fig:filtering}(a)}, which is constructed as a linear combination of two of the eigenvectors of the adjacency matrix of this graph to give  $\mathbf{x}=3.2\mathbf{u}_7+2\mathbf{u}_6$ (eigenvectors of the adjacency matrix  of the considered graph are presented in {\color{blue} Part I, Fig. \ref{I-GSPb_spectrum2a}}). The signal is corrupted by additive white Gaussian noise, $\boldsymbol{\varepsilon}$, at the signal-to-noise (SNR) ratio of $SNR_{in}=2.7$dB and the noisy graph signal, $\mathbf{x}_{\varepsilon}=\mathbf{x}+\boldsymbol{\varepsilon}$, is shown in Fig.~\ref{fig:filtering}(b). This noisy signal is next filtered using an ideal spectral domain graph filter with a cut-off eigenvalue of $\lambda_c=1$. The output signal, $\mathbf{x}_f$, is shown in Fig.~\ref{fig:filtering}(c). With $SNR_{out}=18.8$dB, an increase in signal quality of $16.1$dB is achieved with this type of filtering.  

\begin{figure}
\centering
\includegraphics[]{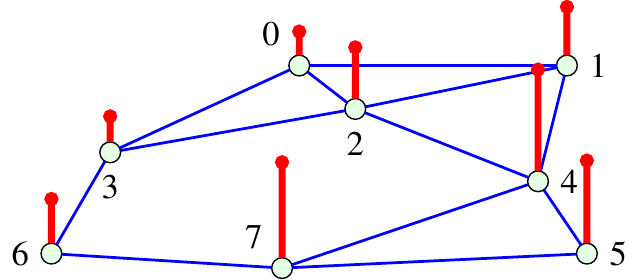}
\includegraphics[]{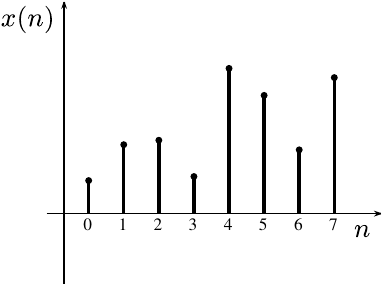}

(a) original signal, $\mathbf{x}=3.2\mathbf{u}_7+2\mathbf{u}_6$ \\[3mm] 

\includegraphics[]{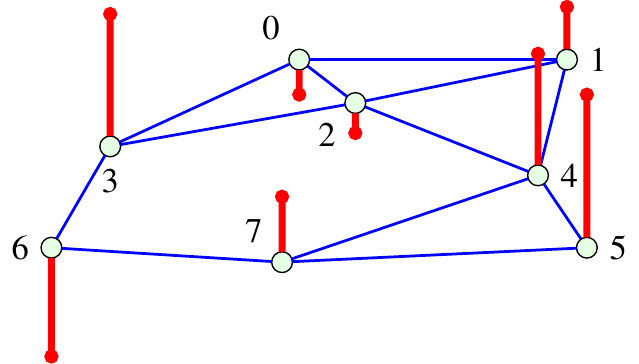}
\includegraphics[]{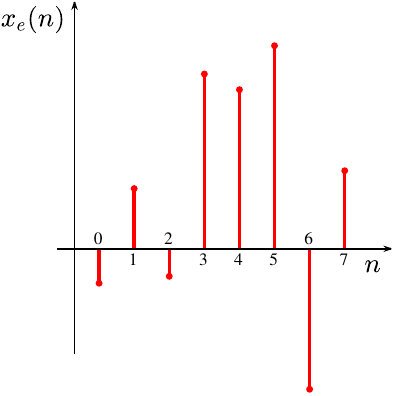}

(b) noisy signal, $\mathbf{x}_{\varepsilon}=\mathbf{x}+\boldsymbol{\varepsilon}$\\[3mm] 

\includegraphics[]{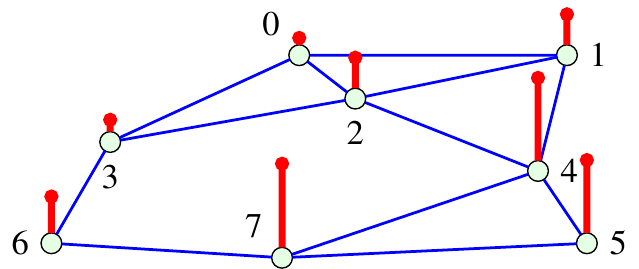}
\includegraphics[]{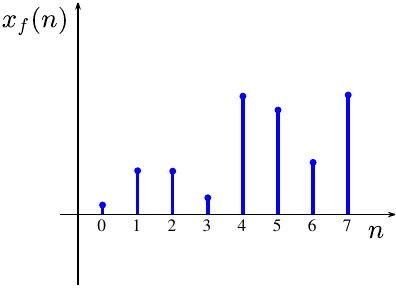}

(c) filtered signal

\caption{A low-pass graph signal filtering example. (a) Original signal, $\mathbf{x}=3.2\mathbf{u}_7+2\mathbf{u}_6$. (b) Noisy signal, $\mathbf{x}_{\varepsilon}=\mathbf{x}+\boldsymbol{\varepsilon}$, at an $SNR=2.7$dB. (c) Filtered signal, at an $SNR=18.8$dB. . Ideal low-pass filtering based on the two highest eigenvalues in the pass-band was applied.}
\label{fig:filtering} 

\end{figure}
\end{Example}

 \begin{Remark} The energy of the rate of change of an eigenvector  is consistent with the classical DFT based filtering when $\lambda_k=\exp(-j2\pi k/N)$ and $\lambda_{\max}=1$. 
\end{Remark}
 
\subsubsection{Spectral Domain Filter Design}\label{SpecDFD}

We shall denote by $G(\mathbf{\Lambda})$  the desired  graph transfer function of a system defined on a graph.  Then, a system with this transfer function can be implemented either in the spectral domain or in the vertex domain. 

In the spectral domain, the implementation is straightforward and can be performed in the following three steps: 
\begin{enumerate}
\item 
Calculate the GDFT of the input graph signal, $\mathbf{X}=\mathbf{U}^{-1}\mathbf{x}$, 
\item 
Multiply the GDFT of the input graph signal by the graph transfer function, $G(\mathbf{\Lambda})$, to obtain the output spectral form,  $\mathbf{Y}=G(\mathbf{\Lambda})\mathbf{X}$, and 
\item 
Calculate the output graph signal as the
inverse GDFT of $\mathbf{Y}$ in Step 2, that is, $\mathbf{y}=\mathbf{U}\mathbf{Y}$. 
\end{enumerate}

This procedure may be computationally very demanding for large graphs, where it may be more convenient to implement the desired filter (or its close approximation) directly in the vertex domain.

For the implementation in the vertex domain, the task is to find the coefficients (\textit{cf.} standard impulse response) $h_0,h_1,\dots,h_{M-1}$ in (\ref{eq:filter-time}), such that their spectral representation, $H(\mathbf{\Lambda})$,  is equal (or approximately equal) to the desired $G(\mathbf{\Lambda})$. This is performed in the following way. The transfer function of the vertex domain system is given by (\ref{H_in_the_GDFT}) as
$H(\lambda_k)=h_0 +h_1 \lambda_k^1+\cdots + h_{M-1} \lambda_k^{M-1}$ and should be equal to the desired transfer function, $G(\lambda_k)$, for $k=0,1,\dots,N-1$. This condition leads to a system of linear equations 
\begin{gather}
	h_0 +h_1 \lambda_0^1+\cdots h_{M-1} \lambda_0^{M-1}=G(\lambda_0) \nonumber \\
	h_0 +h_1 \lambda_1^1+\cdots + h_{M-1} \lambda_1^{M-1}=G(\lambda_1) \nonumber
	\nonumber \\
	\vdots \nonumber \\
	h_0 +h_1 \lambda_{N-1}^1+\cdots + h_{M-1} \lambda_{N-1}^{M-1}=G(\lambda_{N-1}).
	\label{eq:sistem-filt}
\end{gather}
The matrix form of this system is then
\begin{equation} 
\mathbf{V}\!_\lambda \, \mathbf{h}=\mathbf{g}, \label{eq:sistem-filtV}
\end{equation}
where $\mathbf{V}\!_\lambda$ is the Vandermonde matrix form of the eigenvalues $\lambda_k$, given by
\begin{equation}
 \mathbf{V}\!_\lambda=
\begin{bmatrix} 
	1 & \lambda_0^1 & \cdots & \lambda_0^{M-1}\\
	1 & \lambda_1^1 & \cdots & \lambda_1^{M-1} \\
	\vdots & \vdots & \ddots  & \vdots  \\
	1 & \lambda_{N-1}^1 & \cdots & \lambda_{N-1}^{M-1}
\end{bmatrix} \label{VandMatr}	
\end{equation}
and 
\begin{equation}\mathbf{h}=[h_0, h_1,\dots,h_{M-1}]^T\end{equation}
is the vector of  system coefficients which need to be calculated to obtain the desired
\begin{equation}
\mathbf{g}=[G(\lambda_0), G(\lambda_1),\dots,G(\lambda_{N-1})]^T=\textrm{diag}(G(\mathbf{\Lambda})). \label{DesSyst}
\end{equation}

\subsubsection*{Comments on the solution in (\ref{eq:sistem-filt}):}
\begin{enumerate}
\bigskip \item 
Consider the case with $N$ vertices and with all distinct eigenvalues of the adjacency matrix (in other words, the minimal polynomial is equal to the characteristic polynomial, $P_{min}(\lambda)=P(\lambda)$).
\begin{enumerate}
\bigskip \item 
If the filter order, $M$, is such that $M=N$, then the solution to (\ref{eq:sistem-filt}) is unique, since the determinant of the Vandermonde matrix is always nonzero.
\bigskip \item 
If the filter order, $M$, is such that $M<N$, then the system in (\ref{eq:sistem-filt}) is overdetermined. Therefore, the solution to (\ref{eq:sistem-filt}) can only be obtained in the least squares sense (as described later in this section).
\end{enumerate}
\bigskip \item
If some of the eigenvalues are of a degree higher than one (minimal polynomial order, $N_m$, is lower than the number of vertices, $N$) the system in  (\ref{eq:sistem-filt}) reduces to a system of $N_m$ linear equations (by removing multiple equations which correspond to the repeated eigenvalues $\lambda$). 
\begin{enumerate}
\bigskip \item 
If the filter order, $M$, is such that $N_m <M \le N $, the system in (\ref{eq:sistem-filt}) is underdetermined. In that case $(M-N_m)$ filter coefficients are free variables and the system has an infinite number of solutions, while all so obtained \textit{filters are equivalent}.

\bigskip \item 
If the filter order is such that $M=N_m$, the solution to the system in (\ref{eq:sistem-filt}) is unique.

\bigskip \item 
If the filter order is such that $M<N_m$, the system in (\ref{eq:sistem-filt}) is overdetermined and the solution is obtained in the least squares sense.

\end{enumerate}
\bigskip \item Any filter of an order $M>N_m$ has a \textit{unique equivalent filter} of order $N_m$. This equivalent filter can be obtained by setting the free variables to zero, that is, $h_i=0$ for $i=N_m,N_m+1,\ldots,N-1$. 

\end{enumerate}

\subsubsection*{Finding the system coefficients}

\noindent\textbf{Exact solution:} For $M=N=N_m$,  that is, when the filter order is equal to the number of vertices and the order of minimal polynomial, the solution to the system  in (\ref{eq:sistem-filt}) or (\ref{eq:sistem-filtV}) is unique and is obtained from 
$$ \mathbf{h}=\mathbf{V}_{\!\lambda}^{-1}\mathbf{g}.$$

\noindent\textbf{Least-squares solution.} For the overdetermined case, when $M< N_m$, the mean-square approximation  of $\mathbf{h}=[h_0, h_1,\dots,h_{M-1}]^T$ in $ \mathbf{V}\!_\lambda \mathbf{h}=\mathbf{g}$ is obtained by minimizing the squared error 
$$ e=\left\Vert \mathbf{V}\!_\lambda \mathbf{h}-\mathbf{g}\right\Vert _2 ^2.$$
From $\partial e /\partial \mathbf{h}^T=\mathbf{0} $ we then have 
$$ \mathbf{\hat h}=(\mathbf{V}_{\!\lambda}^T \mathbf{V}_{\!\lambda})^{-1}\mathbf{V}_{\!\lambda}^T \mathbf{g}=\text{pinv}(\mathbf{V}\!_\lambda)\mathbf{g}.$$
Since $M<N_m$, the obtained solution, $\mathbf{\hat h}$, is the least-squares approximation for $ \mathbf{V}\!_\lambda \mathbf{h}=\mathbf{g}$. Given that this solution may not satisfy $ \mathbf{V}\!_\lambda \mathbf{h}=\mathbf{g}$,  the designed coefficient vector, $\mathbf{\hat g}$ (its spectrum $\hat{G}(\mathbf{\Lambda})$), obey
$$ \mathbf{V}\!_\lambda \mathbf{\hat{h}}=\mathbf{\hat{g}}$$
which, in general, differs from the desired system coefficients, $\mathbf{g}$ (their spectrum $G(\mathbf{\Lambda})$).

\begin{Example}\label{GRDesignEx} Consider the unweighted graph from {\color{blue}Fig.~\ref{fig:sig-arb-graph}(a)} and the task of the synthesis of a desired filter for which the frequency response is described by
$$\mathbf{g}=[0, 0, 0, 0, 0, 0.5, 1, 1]^T.$$
This filter was designed for various filter orders $M=1,2,4,6,$ using (\ref{eq:sistem-filt}) and the results are shown in Fig.~\ref{FIR_FR}. For clarity, analytically, the vertex domain realization of the filter with $M=4$ is given by  
\begin{equation}
\mathbf{y}=0.1734 \mathbf{A}^0\mathbf{x} + 0.3532\mathbf{A}^1\mathbf{x} +0.0800\mathbf{A}^{2}\mathbf{x} -0.0336\mathbf{A}^3\mathbf{x}, \notag
\end{equation}
however, the exact frequency response $\mathbf{\hat{g}}= \mathbf{g}$ is only obtained with $M=N=8$.

\begin{figure}
\centering
\includegraphics[]{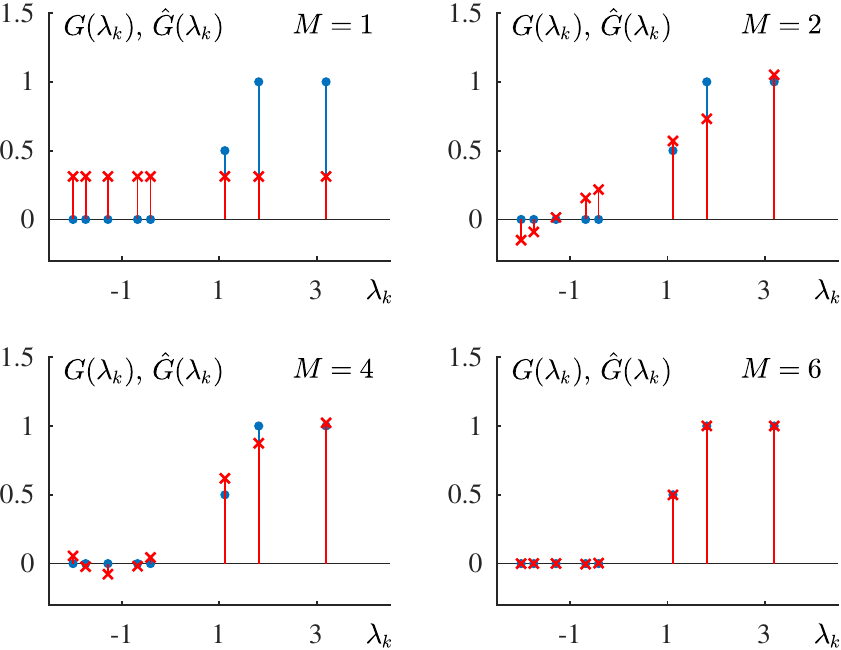}
\caption{Design of a graph filter with a specified transfer function in the spectral domain (\textit{cf.} standard frequency response). The desired spectral response, $G(\lambda_k)$, is denoted by blue circles. Red asterisks designate the spectral response of the filter designed in Example \ref{GRDesignEx}, denoted by $\hat{G}(\lambda_k)$,  obtained with $M$ filter coefficients, $h_0$, $h_1$, \ldots, $h_{M-1}$, in the vertex domain. }
\label{FIR_FR}
\end{figure}

\end{Example}

\subsubsection{Polynomial (Chebyshev) Approximation of the System on a Graph Transfer Function}\label{Plyappsystem}

Without loss of generality, it can be considered that the desired transfer function, $\mathbf{g}=[G(\lambda_0), G(\lambda_1),\allowbreak \dots, \allowbreak G(\lambda_{N-1})]^T$, consists of samples taken from a continuous function of $\lambda$ within the interval $\lambda_{\min} \le \lambda \le \lambda_{\max}$, where $\lambda_{\min}$ and $\lambda_{\max}$ denote the minimum and maximum value of $\{\lambda_0,\lambda_1,\dots, \allowbreak \lambda_{N-1}\}$, respectively. The variable $\lambda$ of the desired transfer function, $G(\lambda)$, is continuous, and the system on graph uses only the values at discrete points $\lambda \in \{\lambda_0,\lambda_1,\dots,\lambda_{N-1}\}$. Therefore, for a polynomial approximation, $P(\lambda)$, of the desired transfer function, $G(\lambda)$, it is important that the error at the points within the considered interval, $\lambda_{\min} \le \lambda \le \lambda_{\max}$, is bounded and sufficiently small. 

This problem is known in algebra as the min-max approximation, and its goal is to find an approximating polynomial that has the smallest maximum absolute error from the desired function value. The min-max polynomials can be approximated by the truncated Chebyshev polynomials, $P(\lambda)$, which yield approximations of the desired function having almost min-max behavior. 

For this the reason, the approximation of the desired transfer function, $G(\lambda)$, may be performed using the truncated Chebyshev polynomial
\begin{equation}
P_{M-1}(z)=\frac{c_0}{2}+\sum_{m=1}^{M-1}c_mT_m(z),
\end{equation} 
where $T_m(z)$ are the Chebyshev polynomials defined as
\begin{gather}
T_0(z)=1, \nonumber \\
T_1(z)=z, \nonumber \\
T_2(z)=2z^2-1, \nonumber \\
T_3(z)=4z^3-3z, \nonumber \\
\vdots \nonumber \\
T_m(z)=2zT_{m-1}(z)-T_{m-2}(z), \label{recurChebPol}
\end{gather}
with the variable $\lambda$ being centered and normalized as
\begin{equation}z=\frac{2\lambda-(\lambda_{\max}+\lambda_{\min})}{\lambda_{\max}-\lambda_{\min}}, \label{MappLam2z}
\end{equation}
 such that $-1\le z \le 1$ (required by the Chebyshev polynomial definition). The inverse mapping, from $z$ to $\lambda$, is given by
 $$\lambda=\frac{1}{2}\Big(z(\lambda_{\max}-\lambda_{\min})+\lambda_{\max}+\lambda_{\min}\Big).$$
 
 Since the  Chebyshev polynomials are orthogonal, with measure $dz/\sqrt{1-z^2}$, the Chebyshev coefficients, $c_m$, for an expansion of the desired function, $G(z)$, into the polynomial series, $P_{M-1}(z)$, are easily obtained as 
 \begin{gather*}
 c_m=\frac{2}{\pi} \int_{-1}^{1}G(z)T_m(z)\frac{dz}{\sqrt{1-z^2}} \\
 =\frac{2}{\pi} \int_{0}^{\pi}cos(m\theta)G(cos(\theta))d\theta.
 \end{gather*} 

\begin{Example}\label{Cheb_GRDesignEx}Consider  the unweighted graph from {\color{blue}Fig.~\ref{fig:sig-arb-graph}(a)} with the desired transfer function
	$$G(\lambda)=\frac{1+\mathrm{sign}(\lambda-\lambda_5)}{2}. $$
The samples of $G(\lambda)$ at  the discrete points 
	$$ \lambda_k \in \{-2,   -1.74,   -1.28,   -0.68,   -0.41,    1.11,    1.81,    3.19\},$$
	  correspond to the values of $G(\lambda_k)$ in Example \ref{GRDesignEx}, Fig. \ref{FIR_FR}. Since the minimum and maximum eigenvalues are $\lambda_{\min}=-2$ and $\lambda_{\max}=3.19$, this yields the desired transfer function with a variable $z$ within a normalized interval, $-1\le z \le 1$,  
	$$G(z)=\frac{1+\mathrm{sign}(z-z_5)}{2}, $$
where $z_5$ is defined by (\ref{MappLam2z}) as
$$z_5=\frac{2\lambda_5-(\lambda_{7}+\lambda_{0})}{\lambda_{7}-\lambda_{0}}=0.2.$$	
%	  The Chebyshev coefficients $c_0$, $c_1$,  $c_2$, and $c_3$, are ...
%	The transfer function is 
%	$$P_{M-1}(z)= $$
%	and it shown in Fig. \ref{FIR_FR_Chebyshev}
	
	\begin{figure}
		\centering
		\includegraphics{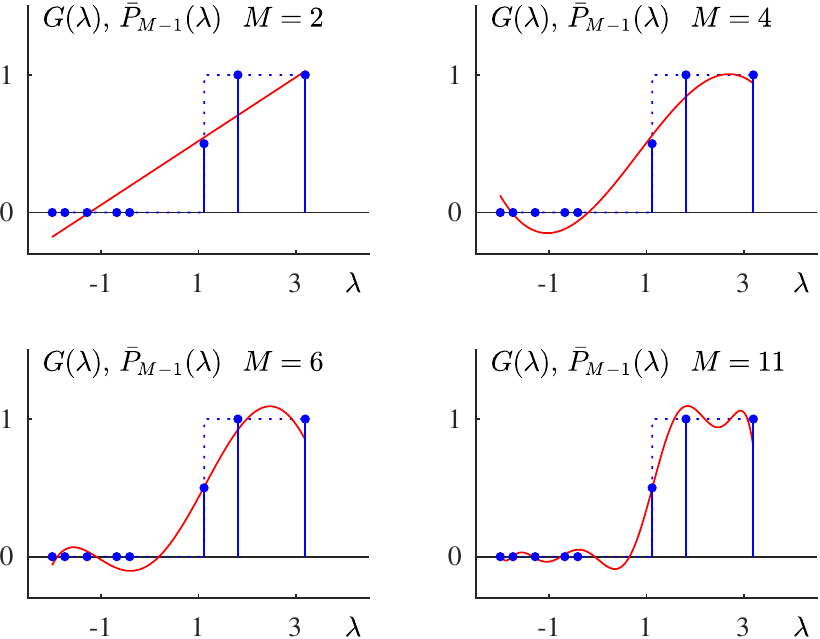}
		\caption{Design of a graph filter with a specified transfer function in the spectral domain using the Chebyshev polynomial approximation of order $(M-1)$ with $M$ terms, $T_0(z), \, T_1(z), \dots, T_{M-1}(z)$. The desired spectral response, $G(\lambda)$, is denoted by blue dashed line and blue dots. Red lines designate the spectral response of the  designed Chebyshev approximation. }
		\label{FIR_FR_Chebyshev}
	\end{figure}

\begin{figure}
	\centering
 	\includegraphics[]{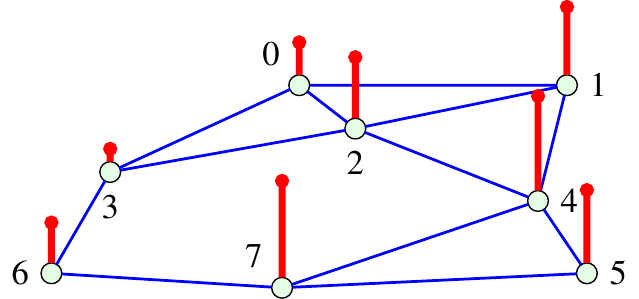}
	\includegraphics[]{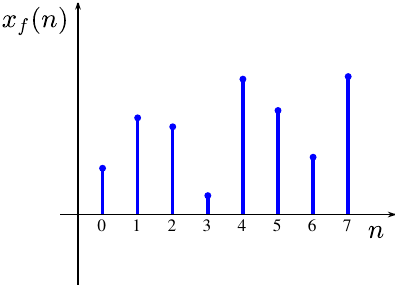}
	
	\caption{Vertex-domain filtering result for the noisy signal from Fig. \ref{fig:filtering}, using the Chebyshev approximation of the desired transfer function from Fig. \ref{FIR_FR_Chebyshev} with $M=4$.}
	\label{fig:signal_filtering_cebyshev} 
	
\end{figure}
	
	The Chebyshev series for $(M-1)=3$ is given by
		\begin{gather*}
		P_{M-1}(z)=0.43+0.62T_1(z)+0.12T_2(z)-0.18T_3(z) \\
		=0.31+1.16z+0.24z^2-0.72z^3.
%			=0.43+0.08z+24z^3-0.72z^3
		\end{gather*}
	Upon the change of variables, $z \rightarrow \lambda$, we obtain the form 
	 $$\bar{P}_{M-1}(\lambda)=0.07+0.36\lambda+0.11\lambda^2-0.04\lambda^3. $$
	
	 Graph signal filtering can now be performed in the vertex domain using
	 $$\mathbf{y}=\bar{P}_{M-1}(\mathbf{A})\mathbf{x},$$
	 where  $$\bar{P}_{M-1}(\mathbf{A})=0.07+0.36\mathbf{A}+0.11\mathbf{A}^2-0.04\mathbf{A}^3.$$ The result of the vertex domain filtering using $\bar{P}_{M-1}(\mathbf{A})$ is shown in Fig. \ref{fig:signal_filtering_cebyshev} for the noisy signal from Fig. \ref{fig:filtering}, with the SNR improvement of 16.76 dB. 
	\end{Example}
	 
	 \noindent\textbf{Calculation complexity.} If the number of nonzero elements in the adjacency matrix, $\mathbf{A}$, is $N_{\mathbf{A}}$, then the number of arithmetic operations (additions) to calculate $\mathbf{Ax}$ is of $N_{\mathbf{A}}$ order. The same number of operations is required to calculate $\mathbf{A}^2\mathbf{x}=\mathbf{A}(\mathbf{A}\mathbf{x})$  using the available $\mathbf{Ax}$. This means that the total number arithmetic operations (additions) to calculate all $\mathbf{Ax}$, $\mathbf{A}^2\mathbf{x}$,...,$\mathbf{A}^{M-1}\mathbf{x}$ is of order $MN_{\mathbf{A}}$. Adding these terms requires additional $MN_{\mathbf{A}}$ arithmetic operations (additions), while the calculation of all terms of the form $c_m\mathbf{A}^{m}\mathbf{x}$ requires an order of $MN_{\mathbf{A}}$ multiplications by constants $c_m$, $m=0,1,\dots,M-1$. Therefore, to calculate the output graph signal,  $\mathbf{y}=\bar{P}_{M-1}(\mathbf{A})\mathbf{x}$, an order of $2MN_{\mathbf{A}}$ additions and $MN_{\mathbf{A}}$ multiplications is needed. Notice that the eigenanalysis of the adjacency matrix, $\mathbf{A}$, requires an order of $N^3$ arithmetic operations. For large graphs, the advantage in calculation complexity of the vertex domain realization with the polynomial transfer function approximation, $\mathbf{y}=\bar{P}_{M-1}(\mathbf{A})\mathbf{x}$, is obvious.     
	 
	 As  is common place in standard filter design theory, the transition intervals of the approximated transfer function, $G(\lambda)$, can be appropriately smoothed, to improve the approximation.

In general, the mapping in (\ref{MappLam2z}) from $\lambda$ to $z$ can be written as $z=a\lambda+b$, where $a=2/(\lambda_{\max}-\lambda_{\min})$ and $b=-(\lambda_{\max}+\lambda_{\min})/(\lambda_{\max}-\lambda_{\min})$. The Chebyshev polynomials series in $\lambda$ is then of the form
\begin{equation}
\bar{P}_{M-1}(\lambda)=\frac{c_0}{2}+\sum_{m=1}^{M-1}c_m\bar{T}_m(\lambda),
\end{equation} 
with $\bar{T}_0(\lambda)=1$, $\bar{T}_1(\lambda)=a\lambda+b$, and $$\bar{T}_m(\lambda)=2(a\lambda+b)\bar{T}_{m-1}(\lambda)-\bar{T}_{m-2}(\lambda),$$ 
for $m \ge 2$.

 The same relations hold for
\begin{equation}
\bar{P}_{M-1}(\mathbf{A})=\frac{c_0}{2}+\sum_{m=1}^{M-1}c_m\bar{T}_m(\mathbf{A}),
\end{equation} 
This change of variables admits recursive calculation, as in (\ref{recurChebPol}).

\subsubsection{Inverse System on a Graph}\label{inversesystem}

A system on a graph, $H(\mathbf{\Lambda})$, which represents an inverse of the system on a graph, given by $G(\mathbf{\Lambda})$, can be obtained from their generic relationship $$H(\mathbf{\Lambda})G(\mathbf{\Lambda})\mathbf{X}=\mathbf{X}.$$
According to (\ref{DesSyst}), this in turn means that if all $G(\lambda_k) \ne 0$ and  $P(\lambda)=P_{min}(\lambda)$, then $H(\lambda_k)=1/G(\lambda_k)$ for each $k$. 

\subsection{Graph Fourier Transform Based on the Laplacian}

Similar to the graph graph discrete Fourier transform based on the adjacency matrix, spectral representation of a graph signal can  be alternatively based on eigenvalue decomposition of the graph Laplacian, given by
$$\mathbf{L}=\mathbf{U}\mathbf{\Lambda}\mathbf{U}^{-1}$$
or $\mathbf{L}\mathbf{U}=\mathbf{U}\mathbf{\Lambda}$. 

\textit{Although the analysis can be conducted in a unified way for spectral decompositions based on both the adjacency matrix and the graph Laplacian, due to their different behavior and scope of application, these will be considered separately. 
}

The graph Fourier transform of a signal, $\mathbf{x}$, which employs the graph Laplacian eigenvalue decomposition to define its basis functions, is given by
\begin{equation}
\mathbf{X}=\mathbf{U}^{-1}\mathbf{x},
\end{equation}
where the matrix $\mathbf{U}$ comprises in its columns the eigenvectors of the graph Laplacian. The inverse graph Fourier transform then follows immediately in the form
\begin{equation}
\mathbf{x}=\mathbf{U}\,\mathbf{X}.
\end{equation}

% Fig. \ref{I-fig:spec-graph}(e) in Part I
In the case of undirected circular unweighted graph, such as the graph in {\color{blue} Fig. \ref{fig:sig-circ-graph}}(a),  this Laplacian based spectral analysis reduces to the standard Fourier transform, but with real-valued basis functions (eigenvectors), as shown in {\color{blue}Part I, equation (\ref{I-GFT_lap})}. 

\subsection{Ordering and Filtering in the Laplacian Spectral Domain}
\label{sec:OrderinLaplacian}

As shown in Section \ref{AMord}, \textit{the graph shift and the adjacency matrix are related to the first finite difference of eigenvectors in the vertex domain, while the rate of the eigenvector change is related to its corresponding eigenvalue (cf. standard frequency). } A similar approach can be used for the Laplacian based eigendecomposition. 
We have seen that for standard time domain signals, the Laplacian of a circle graph represents the second order finite difference, $y(n)$, of a signal $u(n)$, that is 
$$y(n)=-u(n-1)+2u(n)-u(n+1),$$ 
as shown in {\color{blue}Section \ref{I-section_gft_laplacian_spectrum} in Part I}. A compact expression for all elements of the Laplacian can then be written in a matrix form as $\mathbf{y=Lu}$. It is now obvious that the eigenvectors, $\mathbf{u}$, which exhibit small variations should also have a small cumulative energy of the second order difference 
$$E_u=\sum_n \Big[\Big(u(n)-u(n-1)\Big)^2+\Big(u(n)-u(n+1)\Big)^2\Big]/2.$$ 
Recall that this expression corresponds to the quadratic form of the eigenvector, $\mathbf{u}$, defined by   $E_u=\mathbf{u}^T\mathbf{Lu}$.  

The above reasoning for the Laplacian quadratic form can also be used for graph signals. As a default case for the Laplacian analysis we will consider undirected weighted  graphs, where by definition
 $$\mathbf{Lu}=\lambda \mathbf{u}, \qquad \mathbf{u}^T\mathbf{u}=1$$
 or
 $$\mathbf{u}^T\mathbf{Lu}=\lambda \mathbf{u}^T\mathbf{u}=\lambda=E_u.$$
This means that the quadratic form of an eigenvector, $\mathbf{u}_k$, is equal to its corresponding eigenvalue. This is elaborated in detail in {\color{blue}  Section \ref{I-section_smoothness_of_eigenvectors} in Part I}, where we have shown that
\begin{gather} 
   \mathbf{u}_k^T \mathbf{L} \mathbf{u}_k = \lambda_k=
  \frac{1}{2} \sum_{n=0}^{N-1}\ \sum_{m=0}^{N-1} W_{nm}\Big(u_k(n) - u_k(m)\Big)^2 \ge 0.
\label{eq:energijaLaplaciana2}
  \end{gather}
Obviously, a small $\mathbf{u}_k^T\mathbf{Lu}_k=\lambda_k$ implies a small variation of  $W_{nm}(u_k(n) - u_k(m))^2$ in the eigenvector $\mathbf{u}_k$, and for each vertex $n$. Consequently, the eigenvectors corresponding to small $\lambda_k$ correspond to the low-pass part of a graph signal.  In other words, the smaller the smoothness index (curvature), $\mathbf{u}_k^T\mathbf{Lu}_k=\lambda_k$,  the smoother the eigenvector, $\mathbf{u}_k$.     
 
An \textit{ideal low-pass filter} in the Laplacian spectrum domain, with a cut-off eigenvalue $\lambda_{c}$, can be therefore defined as 
$$
f(\lambda)= \begin{cases}
1, & \text{ for } \lambda < \lambda_{c} \\
0, & \text{ for other } \lambda.
\end{cases}
$$

\begin{figure}[thpb]
	\centering
	\includegraphics[]{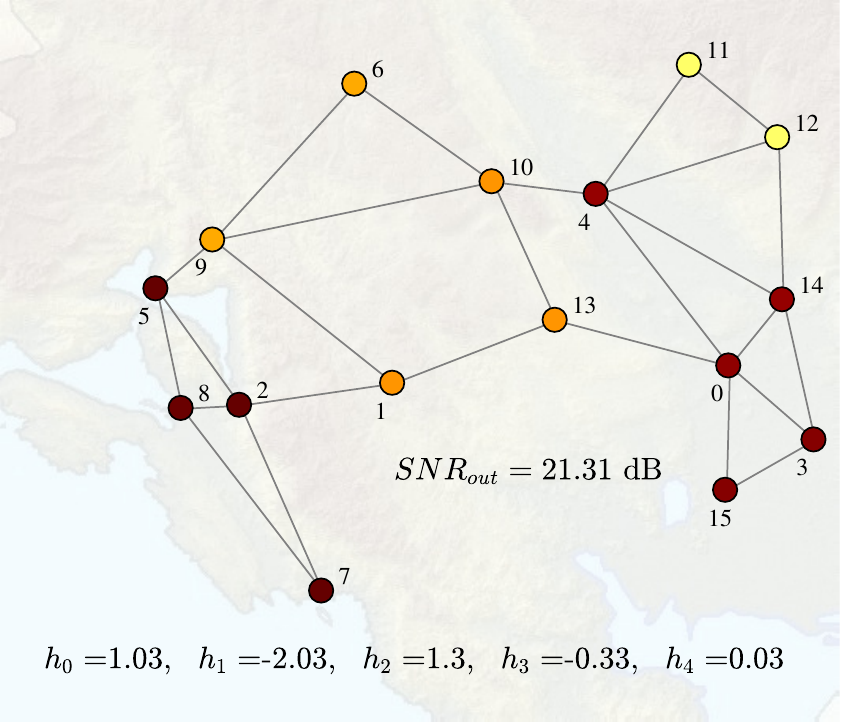}
	\caption{Graph filtering of a noisy signal from Fig. \ref{fig:MNE_fig_e}, using a  fourth-order system given by $\mathbf{y}=h_0 \mathbf{L}^0\, \mathbf{x}+h_1 \mathbf{L}^1\, \mathbf{x}+h_{2} \mathbf{L}^{2}+h_{3} \mathbf{L}^{3}+h_{4} \mathbf{L}^{4}$.}
	\label{MNE_fig_VanderM_filt}
\end{figure}

\begin{Example} Consider a signal on the undirected graph from Fig. \ref{fig:sig-arb-graph}(a),  shown in Fig.~\ref{fig:filtering_Laplacian}(a). This graph signal is generated as a linear combination of two Laplacian eigenvectors (which correspond to the slow-varying signal part), to give $\mathbf{x}=2\mathbf{u}_0+1.5\mathbf{u}_1$. The Laplacian eigenvectors of the considered graph are presented in {\color{blue} Part I,  Fig. \ref{I-GSPb_spectrum3a}},
while the considered graph signal is shown in Fig. \ref{fig:filtering_Laplacian}(a). The original signal, $\mathbf{x}$, was then corrupted by white Gaussian noise at the signal-to-noise ratio of $SNR_{in}=-1.76$ dB, and shown in Fig. \ref{fig:filtering_Laplacian}(b). Next, this noisy graph signal was filtered using an ideal spectral domain graph filter, with a cut-off eigenvalue $\lambda_c=2$, to obtain the filtered signal, $\mathbf{x}_f$, shown in Fig. \ref{fig:filtering_Laplacian}(c). The so achieved output SNR was $SNR_{out}=21.29$ dB, that is, despite its simplicity, the graph filter achieved a gain in SNR of $23.05$ dB, as compared to the noisy signal in Fig. \ref{fig:filtering_Laplacian}(b).

To further illustrate the principle of graph filtering, the noisy signal from Fig. \ref{fig:MNE_fig_e} was filtered using a  filter with the spectral cut-off at $\lambda_c=0.25$ and the result is shown in Fig. \ref{MNE_fig_ideal_filt}. The same signal was also filtered using a polynomial approximation to the low-pass system, as illustrated in  Fig. \ref{MNE_fig_VanderM_filt}.

\begin{figure}
	\centering
	\includegraphics[]{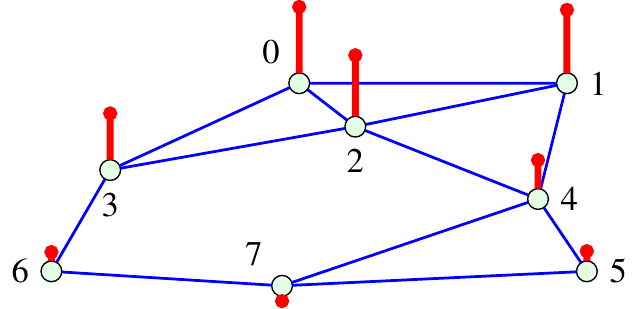}
	\includegraphics[]{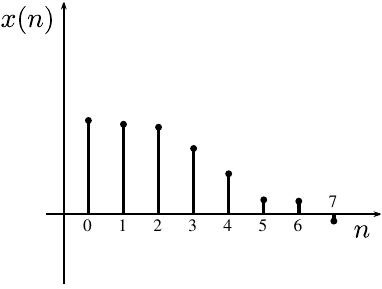}
	
	(a) original signal\\[3mm] 
	
	\includegraphics[]{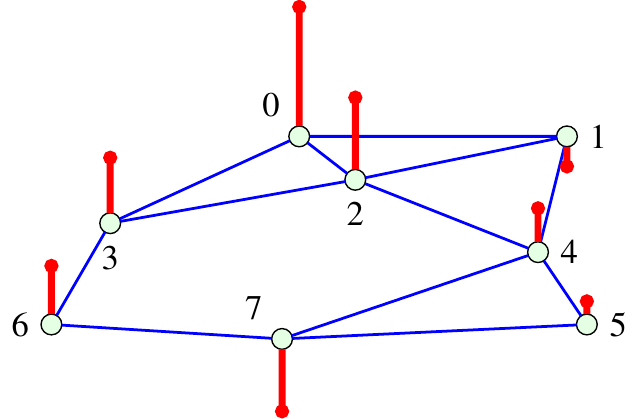}
	\includegraphics[]{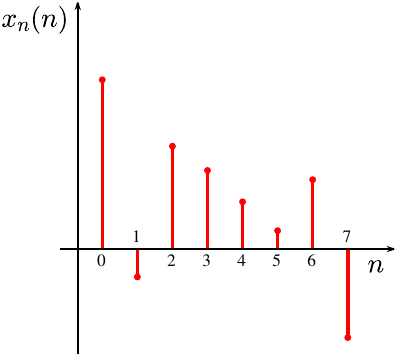}
	
	(b) noisy signal\\[3mm] 
	
	\includegraphics[]{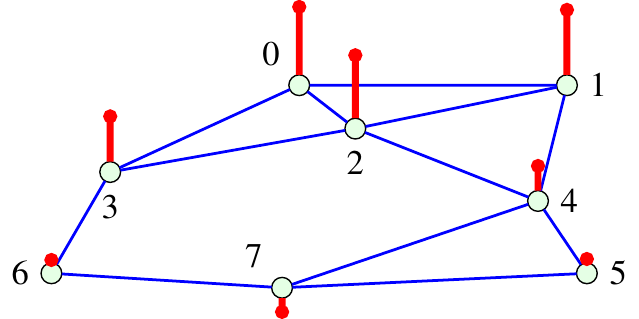}
	\includegraphics[]{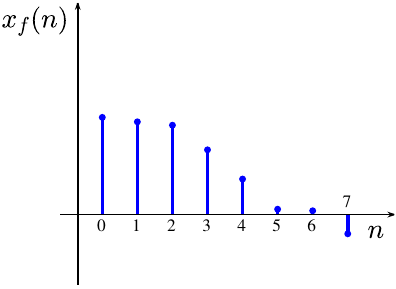}
	
	(c) filtered signal
	
	\caption{Graph signal filtering example. (a) Original signal. (b) Noisy signal.  (c) Filtered signal. Low pass filtering was performed based on  the two lowest eigenvalues of the graph Laplacian.}
	\label{fig:filtering_Laplacian} 
	
\end{figure}

\begin{figure}[thpb]
	\centering
	\includegraphics{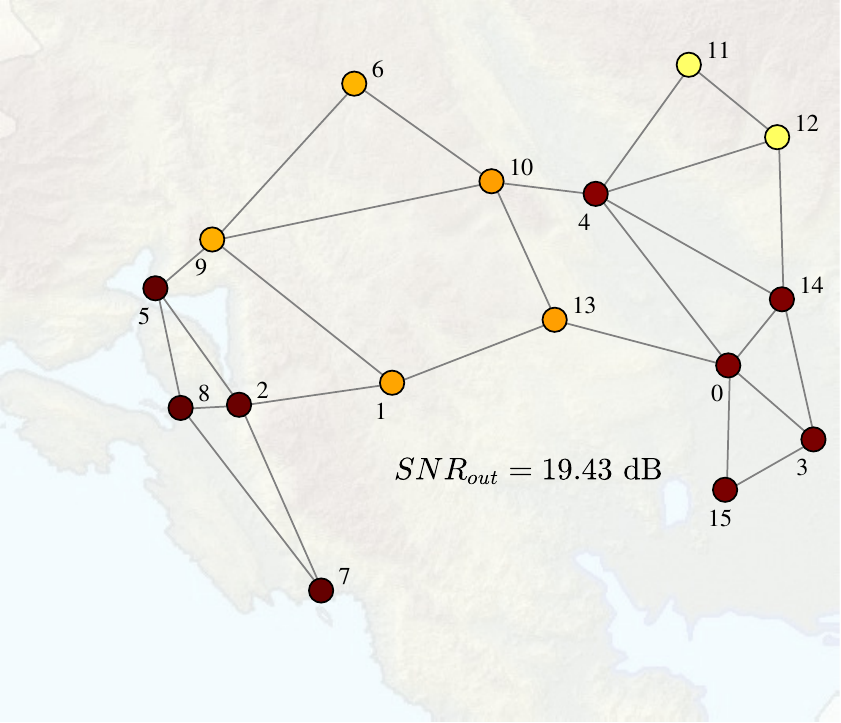}
	\caption{Denoising results for the noisy signal from Fig. \ref{fig:MNE_fig_e}, which was filtered using a low-pass graph filter with $\lambda_c=0.25$.}
	\label{MNE_fig_ideal_filt}
\end{figure}

\end{Example}

\noindent\textbf{Laplacian versus adjacency-based GDFT for regular graphs.} A direct relation between the adjacency-based and Laplacian-based  spectral decomposition can be established for $\mathcal{J}$-regular unweighted graphs {\color{blue} (see  (\ref{I-regulGGGG}) in Part I)}, for which
$$\mathbf{L}=\mathcal{J}\mathbf{I}-\mathbf{A}$$
to yield
$$\lambda^{(A)}_k=\mathcal{J}-\lambda^{(L)}_k,$$
where the eigenvalues of the adjacency matrix and the graph Laplacian are respectively denoted by   $\lambda^{(A)}_k$ and $\lambda^{(L)}_k$, while they share the same eigenvectors. 

\begin{Remark} Rank-ordering of the eigenvectors, $\mathbf{u}_k$, from low-pass to high-pass, which is based on the respective eigenvalues,  $\lambda^{(A)}_k$ and $\lambda^{(L)}_k$, yields exactly opposite ordering for these two graph spectral decompositions. For example, the smoothest eigenvector is obtained for $\min_k{\lambda^{(L)}_k}=\lambda^{(L)}_0=0$ or for $\max_k{\lambda^{(A)}_k}=\lambda_{\max}=\mathcal{J}-\lambda_0^{(L)}=\mathcal{J}.$ 
\end{Remark}
 
\subsection{Systems on a Graph Defined Using the Graph Laplacian}

Following on the discussion in Section \ref{Sec:SysA} and equation (\ref{eq:filter-time}), a system on a graph, defined using the graph Laplacian, has the form
\begin{align}
\mathbf{y}&=h_0 \mathbf{L}^0\, \mathbf{x}+h_1 \mathbf{L}^1\, \mathbf{x}+\dots+h_{M-1} \mathbf{L}^{M-1}\, \mathbf{x}\notag\\&=\sum_{m=0}^{M-1}h_m \mathbf{L}^m\, \mathbf{x}.
\label{eq:filterL-time}
\end{align}

\textit{For an unweighted graph, this definition of a system on a graph can be related to the corresponding adjacency matrix form as $\mathbf{L}=\mathbf{D}-\mathbf{A}$.}

The \textbf{spectral domain description of a system on a graph} is then obtained through the Laplacian eigenvalue decomposition, to yield
\begin{align}
\mathbf{y}&=\mathbf{U} \mathbf{Y}=\sum_{m=0}^{M-1}h_m \mathbf{L}^m\, \mathbf{x}=H(\mathbf{L}) \mathbf{x}\notag\\
&=\mathbf{U}H(\mathbf{\Lambda}) \mathbf{U}^T  \mathbf{x}=\mathbf{U}H(\mathbf{\Lambda}) \mathbf{X},
\label{eq:filterLF-time}
\end{align}
where we used the property of the eigendecomposition of matrix polynomial, 
\begin{align}
H(\mathbf{L})=\sum_{m=0}^{M-1}h_m \mathbf{L}^m=\sum_{m=0}^{M-1}h_m \mathbf{U}\mathbf{\Lambda}^m \mathbf{U}^T= \mathbf{U}H(\mathbf{\Lambda}) \mathbf{U}^T 
\end{align}
described in {\color{blue} Section \ref{I-Section_decomposition_of_matrix_powers} in Part I}, and the notation
\begin{align}
H(\mathbf{\Lambda})=\sum_{m=0}^{M-1}h_m \mathbf{\Lambda}^m
\end{align}
 to obtain 
$$\mathbf{Y}=H(\mathbf{\Lambda}) \mathbf{X} $$
or in an element-wise form
$$ Y(k)=H(\lambda_k)X(k), \,\, k=0,1,\dots,N-1.$$

In the vertex domain, the $n$-th element of the  output signal, $\mathbf{y}=\mathbf{U}H(\mathbf{\Lambda}) \mathbf{U}^T  \mathbf{x}$, of a system on a graph is given by
\begin{equation}
y(n)=\sum_{k=0}^{N-1}\sum_{i=0}^{N-1}x(i)u_k(i)H(\lambda_k)u_k(n)=
\sum_{i=0}^{N-1}x(i)h_n(i), \label{conv_Hk}
\end{equation}
for which the transfer function is defined by
\begin{equation}
H(\lambda_k)=h_0+h_1\lambda_k+\dots+h_{M-1}\lambda_k^{M-1} \label{Huge_LConv}
\end{equation}
and the \textit{graph impulse response} is
\begin{equation}
h_n(i)=\sum_{k=0}^{N-1}H(\lambda_k)u_k(n)u_k(i)=\mathcal{T}_n\{h(i)\}. \label{GGGCON}
\end{equation}
 
\begin{Remark}The expression for $y(n)$ in (\ref{conv_Hk}) can be interpreted as a\textit{ generalized convolution on graphs}, which is performed using a generalized graph shift of the impulse response, $h_n(i)$, in the vertex domain. 
\end{Remark}

We next proceed to describe the generalized convolution on graphs through responses to the unit delta pulses. 
For illustration, consider the delta function located at a graph vertex $m$,  given by
\begin{equation}
\delta_m(n)=\begin{cases}
1, & \text{ for } m=n \\
0, & \text{ for } m\ne n
\end{cases} \label{DeltaDEF}
\end{equation}
with the corresponding GDFT
\begin{equation}
\Delta(k)=\sum_{n=0}^{N-1} \delta_m(n) u_k(n) =u_k(m), \label{DeltaGDFT}
\end{equation}
which is defined based on graph Laplacian eigenvectors.

Observe that, similar to the standard time domain, any graph signal can be written as a sum of delta functions at the graph vertices, that is
$$ 
x(n)=\sum_{i=0}^{N-1}x(i)\delta_n(i)
$$
or in a vector form
$$ 
\mathbf{x}=\sum_{i=0}^{N-1}x(i)  \boldsymbol{\delta}_i,
$$
where $\boldsymbol{\delta}_i$ is a vector with elements $\delta(n-i)$, as in (\ref{DeltaDEF}). Then,  the system output, $\mathbf{y}$, takes the form 
\begin{align*}
	\mathbf{y}&=\sum_{m=0}^{M-1}h_m \mathbf{L}^m\, \mathbf{x}=\mathbf{U}H(\mathbf{\Lambda}) \mathbf{U}^T \mathbf{x}\\
	&=\sum_{i=0}^{N-1}x(i) \mathbf{U}H(\mathbf{\Lambda}) \mathbf{U}^T \boldsymbol{\delta}_i
\end{align*}
and its elements are obtained as
$$
y(n)=\sum_{i=0}^{N-1}x(i)\sum_{k=0}^{N-1}u_k(n)H(\lambda_k)u_k(i)=\sum_{i=0}^{N-1}x(i)h_n(i),
$$
where $h_n(i)$ are related to $H(\lambda_k)$ as in  (\ref{GGGCON}).

\begin{Remark} According to (\ref{eq:filterLF-time}), the form of the graph convolution operator for a vertex $n$, given in (\ref{conv_Hk}), is localized within the $(M-1)$-neighborhood of vertex $n$. This localization property is even more important for large graphs. 
\end{Remark}	

A generalized convolution for two arbitrary graph signals will be addressed next.

\subsection{Convolution of Signals on a Graph}\label{Sec:ConvSigGRR}

Consider two graph signals, $x(n)$ and $h(n)$. A generalized convolution operator for these two signals on a graph is defined using their graph Laplacian spectra \citep{shuman2016vertex}, 
based on the assumption that the spectrum of a convolution on a graph 
$$y(n)=x(n)*h(n)$$
is equal to the product of the corresponding  spectra of graph signals, $x(n)$ and $h(n)$, that is
\begin{equation}
Y(k)=X(k)H(k),  \label{GraphConcDEF}    
\end{equation}
in the element-wise form. The output of the generalized graph convolution operation,
$x(n)*h(n)$, is then equal to the inverse GDFT of the spectral product $Y(k)$ in (\ref{GraphConcDEF}),  that is
\begin{align*}
y(n)&=x(n)*h(n)\\&=\sum_{k=0}^{N-1}Y(k)u_k(n) 
=\sum_{k=0}^{N-1}X(k)H(k)u_k(n), \nonumber
\end{align*}
where 
\begin{equation}H(k)=\sum_{n=0}^{N-1}h(n)u_k(n). \label{GGGIDF}
\end{equation}
Notice the difference between the definition of $H(k)$ in (\ref{GGGIDF}) and $H(\lambda_k)$ in (\ref{Huge_LConv}). Both these forms will be discussed in more detail in the next section. 

\noindent\textbf{Shift on a graph -- an alternative definition.} The above framework of generalized graph convolution can also serve as a  basis for a convenient definition of a shift on a graph. Consider the graph signal,
$h(n)$, and the delta function located at a vertex $m$. Here, we will use $h_m(n)$
to denote the shifted version of the graph signal, $h(n)$, \textquotedblleft toward\textquotedblright \, a vertex $m$. This kind of shifted signal will be defined following the reasoning  in classical signal processing where the shifted signal is obtained as a convolution of the original signal and an appropriately shifted delta function. Therefore, a graph shifted signal is here defined through a generalized graph convolution, $h(n)*\delta_m(n)$, whose GDFT is equal to $H(k)u_k(m)$, according to (\ref{DeltaGDFT}) and (\ref{GraphConcDEF}). The graph shifted signal is then the IGDFT of $H(k)u_k(m)$, that is
\begin{equation}
h_m(n)=h(n)*\delta_m(n)=\sum_{k=0}^{N-1}H(k)u_k(m)u_k(n). \label{signShiftOngr}
\end{equation}
The same relation follows when calculating the inverse GDFT of $X(k)H(k)$, to yield
\begin{align}y(n)&=\sum_{k=0}^{N-1}X(k)H(k)u_k(n) \nonumber \\&=\sum_{k=0}^{N-1}\sum_{m=0}^{N-1}x(m)u_k(m)H(k)u_k(n) \nonumber \\
&=\sum_{m=0}^{N-1}x(m)h_m(n)=x(n)*h(n), \label{Huge_LConv22}
\end{align}
where 
\begin{align}
	h_m(n)=\sum_{k=0}^{N-1}H(k)u_k(m)u_k(n)=T_m\{h(n)\} \label{GSHGSH}
\end{align}
 is another version of graph shifted signal. Since the definition of $H(k)$ as a GDFT of a signal $h(n)$ differs from that in (\ref{Huge_LConv}), these produce different shift operations, which are respectively denoted by $T_m\{h(n)\}$ and $\mathcal{T}_m\{h(n)\}$. 
 
 \begin{Remark}
 	Note that neither of the two shift operations, (\ref{GGGCON}) or (\ref{GSHGSH}), satisfy the property that a shift by $0$ is equal to the original signal, $h_0(n) \ne h(n)$.  
 \end{Remark}

\begin{figure}
\centering
\includegraphics[]{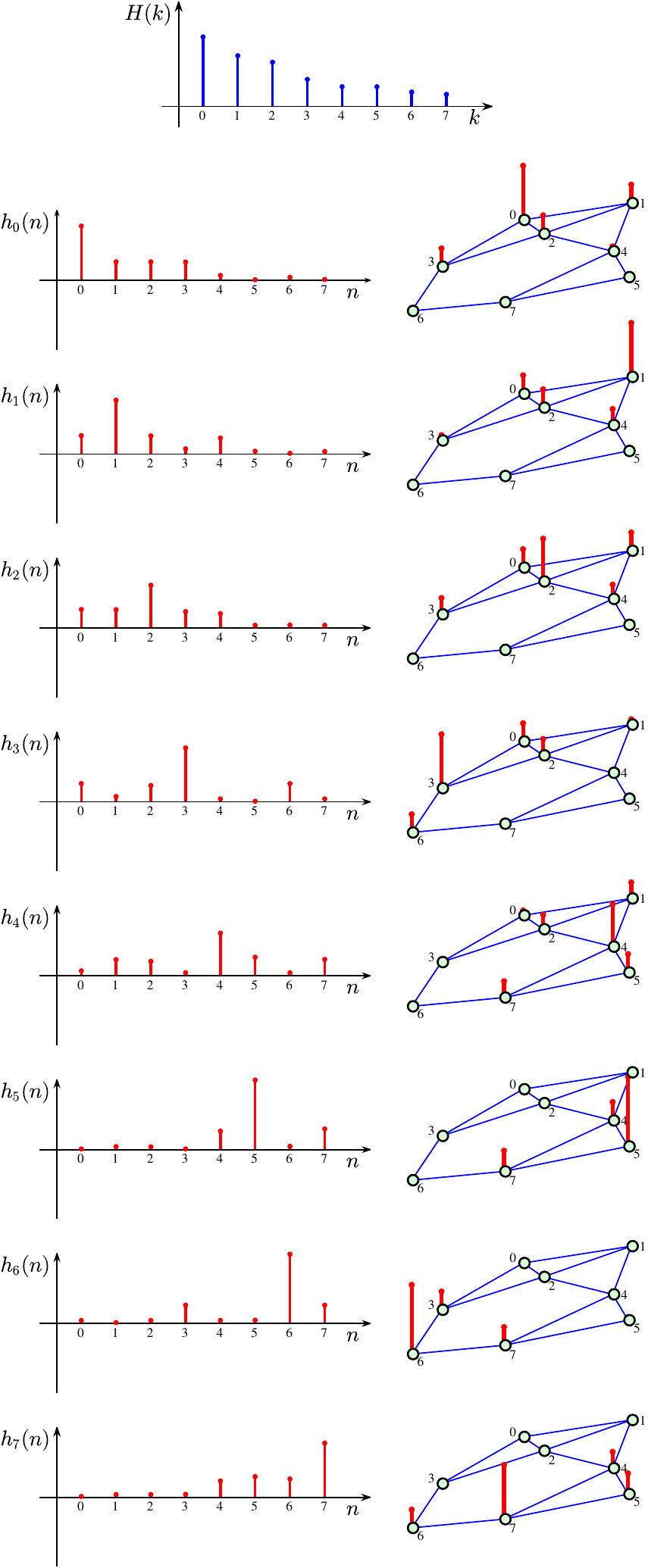}
\caption{An example of graph shift operator. \textit{Top}: The graph signal defined by its Laplacian GDFT, given by
	$H(k)=\exp (- 2 \lambda_k \tau)$. 
	\textit{Left and right column}: The graph signals $h_m(n)$ "shifted" for $m=0$ to $m=7$, calculated using $h_m(n)=T_m\{h(n)\}$ in (\ref{GSHGSH}). The shifted signal is shown both on the vertex index line (\textit{left}) and on the graph itself (\textit{right}).}
\label{shift_laplacian}
\end{figure}

\begin{Example} Consider a signal on graph from Fig. \ref{fig:sig-arb-graph}(a), which is defined by its graph Laplacian GDFT, given by
$$H(k)=\exp (- 2 \lambda_k \tau),$$
with $\tau=0.1573$. All shifted signals, $h_m(n)=T_m\{h(n)\}$, obtained using the shift operator in (\ref{GSHGSH}),
are shown in Fig. \ref{shift_laplacian}.
\end{Example}

\subsection{The $z$-transform of a Signal on a Graph} \label{GZT}

The relation between the graph signal shift operators, $T_m\{h(n)\}$ and $\mathcal{T}_m\{h(n)\}$, which are respectively used used to define   the generalized convolutions in (\ref{Huge_LConv}) and (\ref{Huge_LConv22}), can be established based on the definitions of  $H(\lambda_k)$ and  $H(k)$. Consider $H(\lambda_k)$, defined by (\ref{Huge_LConv}), as a graph discrete Fourier transform of signal $h(n)$. The samples of the graph signal $h(n)$ are then equal to the IGDFT of $H(\lambda_k)$, that is
$$h(n)=\sum_{k=0}^{N-1}H(\lambda_k)u_k(n)$$
while the system coefficients $h_n$, $n=0,1,\dots,M-1$, are related to $H(\lambda_k)$ by (\ref{Huge_LConv}), that is
$$H(\lambda_k)=h_0+h_1\lambda_k+\dots+h_{M-1}\lambda_k^{M-1}.$$
For $M=N$, the vector forms of the last two relations are
\begin{gather*}
[h(0),  \, \, h(1), \dots, h(N-1)]^T=\mathbf{U}H(\mathbf{\Lambda}) \\
H(\mathbf{\Lambda})=\mathbf{V}_\lambda [h_0, \, \, h_1, \dots, h_{N-1}]^T
\end{gather*}
so that the signal, $h(n)$, and the coefficients, $h_n$, can be related as  
\begin{equation}[h_0, \, \, h_1, \dots, h_{N-1}]^T=\mathbf{V}^{-1}_\lambda \mathbf{U}^T[h(0),  \, \, h(1), \dots, h(N-1)]^T. \label{Sig_for_Z}
\end{equation}

\begin{Remark} In classical DFT (the case of a directed circular graph and its adjacency matrix, when $\mathbf{U}^H$ should be used instead of $\mathbf{U}^T$), the signal samples, $h(n)$, which are obtained as the inverse DFT of $H(\lambda_k)$ and the system coefficients, $h_n$, are the same, since the eigenvalues are equal to the corresponding shift operators in the spectral domain, $\lambda_k=\exp(-j 2 \pi k/N)$ and $u_k(n)=\exp(j 2 \pi nk/N)/\sqrt{N}=\lambda_k^{-n}/\sqrt{N}$, with $h_n=h(n)/\sqrt{N}$ and 
	$$H(k)=\frac{1}{\sqrt{N}}\sum_{n=0}^{N-1}h(n)e^{-j 2 \pi nk/N}.$$
	 Therefore, for classical DFT analysis, the following relation holds
$$
\sqrt{N} \mathbf{V}_\lambda = (\mathbf{U}^H)^{-1}.   
$$
This relation is obvious from (\ref{VandMatr}) and $u^*_k(n)=\lambda_k^{n}/\sqrt{N}$, and  will be used to define \textit{the $z$-transform of a graph signal}. 
\end{Remark}

 \noindent\textbf{The $z$-transform of graph signals.} For a given graph signal $\mathbf{x}=[x(0), x(1), \dots, x(N-1)]^T$, following the reasoning as in (\ref{Sig_for_Z}), the coefficients of a system $[x_0 , x_1, \dots, x_{N-1}]^T$ which corresponds to a system transfer function that would have the same GDFT as the graph signal itself are 
$$[x_0 , x_1, \dots, x_{N-1}]^T=\mathbf{V}^{-1}_\lambda \mathbf{U}^T[x(0), x(1), \dots, x(N-1)]^T$$
or
$$[x_0 , x_1, \dots, x_{N-1}]^T=\mathbf{V}^{-1}_\lambda [X(0), X(1), \dots, X(N-1)]^T.$$
The graph $z$-transform of a signal $\mathbf{x}$ is therefore equal to the classic $z$-transform of coefficients $[x_0 , x_1, \dots, x_{N-1}]^T$, 
\begin{equation}
X(z^{-1})=\mathcal{Z}\{x_n\}=x_0+x_1 z^{-1} +\dots+x_{N-1}z^{-(N-1)} \label{H_in_the_GzTXX}
\end{equation}
so that the following  holds
$$Y(z^{-1})=H(z^{-1})X(z^{-1}) $$
The output signal, $y(n)$, can now be obtained as
$$[y(0), y(1), \dots, y(N-1)]^T= \mathbf{U}\mathbf{V}_\lambda [y_0, \, \, y_1,\dots, y_{N-1}]^T,$$
where the output graph signal, $y(n)$, results from  the inverse $z$-transform of the coefficients, $y_n$, that is
$$Y(z^{-1})=\mathcal{Z}\{y_n\}=y_0+y_1 z^{-1} +\dots+y_{N-1}z^{-(N-1)}.$$

\begin{figure}
	\centering
	\includegraphics[]{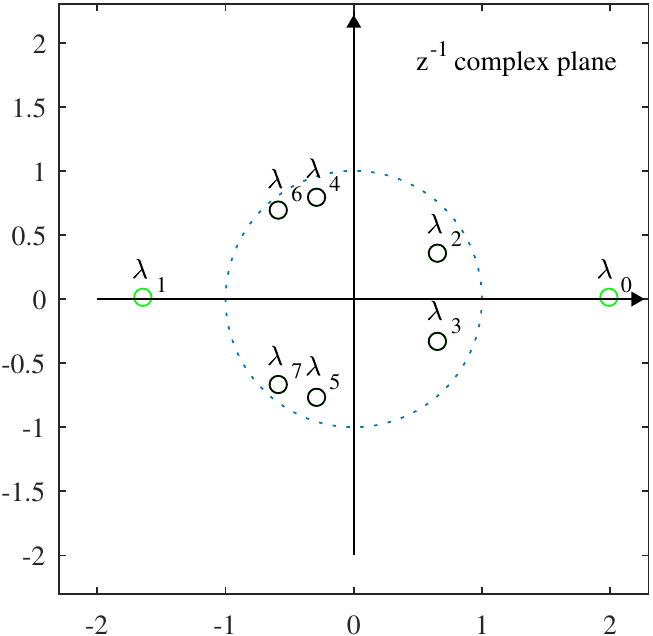}
	\caption{Complex eigenvalues of the adjacency matrix of a directed graph in {\color{blue}Fig. \ref{I-GSPb_ex1a}(b) in Part I}.}
	\label{Hilbert_Transform}
\end{figure}

The $z$-transform representation in the complex valued $z$-domain may be of interest when the eigenvalues are complex-valued, which occurs in the decomposition of adjacency matrices of undirected graphs. For example, for the graph from {\color{blue}Fig. \ref{I-GSPb_ex1a}(b) in Part I} and its adjacency matrix, the eigenvalues are shown in Fig. \ref{Hilbert_Transform}.

\noindent \textit{Definition:} The analytic graph signal, $X_a(k)$, and the graph Hilbert transform, $X_h(k)$, are defined in the spectral domain as
$$X_a(k)=\Big(1+\operatorname{sign}(\Im (\lambda_k))\Big)X(k)$$
$$X_h(k)=j\operatorname{sign}\Big(\Im (\lambda_k)\Big)X(k)$$ 
$$X(k)=X_a(k)+jX_h(k),$$ 
where $\Im (\lambda_k)$ denotes imaginary part of $\lambda_k$.
If these relations are applied to the standard DFT with $\lambda_k=e^{-j2\pi k/N}$ we would obtain the corresponding classical signal processing definitions. 

\subsection{Shift Operator in the Spectral Domain}

A shift operation in the spectral domain can be defined in the same way as the shift in the vertex domain. Consider a product of two graph signals, $x(n)y(n)$, defined on an undirected graph. The GDFT of this product then takes the form 
\begin{gather}
\mathrm{GDFT}\{x(n)y(n)\}=\sum_{n=0}^{N-1}x(n)y(n)u_k(n)=
\nonumber \\
\sum_{n=0}^{N-1}\sum_{i=0}^{N-1}X(i)u_i(n)y(n)u_k(n)= \sum_{i=0}^{N-1}X(i)Y_i(k), \nonumber
\end{gather}
where 
$$Y_i(k)=\sum_{n=0}^{N-1}y(n)u_i(n)u_k(n)$$
 can be considered as a shift of $Y(k)$ by $i$ spectral indices. 
 
 \begin{Remark}
 As desired, a shift by $i=0$ in the spectral domain produces the original value, $Y_0(k)=Y(k)$, up to a constant factor $1/\sqrt{N}$. This relation does not hold for the shift operators in the vertex domain. 
\end{Remark}

\subsection{Parseval's Theorem on a Graph}

Consider two graph signals, $x(n)$ and
$y(n)$, which are observed on an undirected graph and their spectra, $X(k)$ and $Y(k)$. Then, Parseval's theorem has the form 
\begin{equation}
\sum_{n=0}^{N-1}x(n)y(n)=\sum_{k=0}^{N-1}X(k)Y(k)
\end{equation}
and it holds for any two graph signals. 
 
To prove Parseval's theorem on graphs, consider
\begin{align}
\sum_{n=0}^{N-1}x(n)y(n)&=\sum_{n=0}^{N-1}\bigg[\sum_{k=0}^{N-1}X(k)u_k(n)\bigg]y(n)\notag \\
&=\sum_{k=0}^{N-1}X(k)\sum_{n=0}^{N-1}y(n)u_k(n),
\end{align}
to yield Parseval's equivalence between the energies in the original and spectral domains. It has been assumed that the graphs are undirected, so that $\mathbf{U}^{-1}=\mathbf{U}^{T}$ holds. This theorem is quite general and applies to both the graph Laplacian and the adjacency matrix based decompositions on undirected graphs.

\subsection{Optimal Denoising}\label{OptimaDenoising}

Consider a measurement, $\mathbf{x}$, composed of a slow-varying graph signal, $\mathbf{s}$, and a fast changing disturbance, $\boldsymbol{\varepsilon}$, to give
$$
\mathbf{x}=\mathbf{s}+\boldsymbol{\varepsilon}.
$$
The aim is to design a filter for disturbance suppression (denoising), the output of which is denoted by $\mathbf{y}=H(\mathbf{x})$.

The optimal denoising task may then be defined as a minimization of the objective function
\begin{equation}
J=\frac{1}{2}\Vert \mathbf{y}-\mathbf{x}\Vert_2^2 + \alpha \mathbf{y}^T \mathbf{L} \mathbf{y}. \label{LLLKKK}
\end{equation} 
Physically, the minimization of the first term $\frac{1}{2}\Vert \mathbf{y}-\mathbf{x}\Vert_2^2$ forces the output signal $\mathbf{y}$ to be as close as possible to the available observations $\mathbf{x}$, in terms of the energy of their Euclidean distance (minimum error energy), while the second term represent a measure of smoothness of  $\mathbf{y}$ (see Section \ref{sec:OrderinLaplacian}). This is also physically meaningful, as the original input, $\mathbf{s}$, was low-pass and smoother than the disturbance, $\boldsymbol{\varepsilon}$.  The parameter $\alpha$ provides a balance between the closeness of output, $\mathbf{y}$, to $\mathbf{x}$ and the output smoothness criterion.

To solve this minimization problem, we differentiate
$$
\frac{\partial J}{\partial \mathbf{y}^T} = \mathbf{y}-\mathbf{x} + 2\alpha \mathbf{L} \mathbf{y} =\mathbf{0}
$$
which results in 
$$
\mathbf{y}= (\mathbf{I} + 2\alpha \mathbf{L})^{-1} \mathbf{x}.
$$
The spectral domain form of this relation follows from $\mathbf{L}=\mathbf{U}^T\mathbf{\Lambda}\mathbf{U}$, $\mathbf{Y}=\mathbf{U}^T\mathbf{y}$, and $\mathbf{X}=\mathbf{U}^T\mathbf{x}$, to yield
$$
\mathbf{Y}= (\mathbf{I} + 2\alpha \mathbf{\Lambda})^{-1} \mathbf{X}. 
$$
The element-wise  transfer function of the above spectral input/output relation then takes the form
\begin{equation}
H(\lambda_k)=\frac{1}{1+2\alpha\lambda_k}. \label{TranfFGRFF}
\end{equation}
\begin{Remark} For a small $\alpha$, we have $H(\lambda_k) \approx 1$, that is, an all-pass behavior of (\ref{TranfFGRFF}), with  no signal smoothing, which yields $\mathbf{y} \approx \mathbf{x}$. On the other hand, for a large $\alpha$, $H(\lambda_k) \approx \delta(k)$. The resulting $ \mathbf{y} \approx const.$ is maximally smooth (a constant output, without any variation). 
\end{Remark}

\begin{Example}  The noisy signal from Fig. \ref{fig:MNE_fig_e} was filtered using the optimal  filter in (\ref{TranfFGRFF}) with $\alpha=1$, and the result is shown in Fig. \ref{MNE_fig_optimal_filt}. The achieved SNR was 19.16 dB. 

\begin{figure}[thpb]
	\centering
	\includegraphics[]{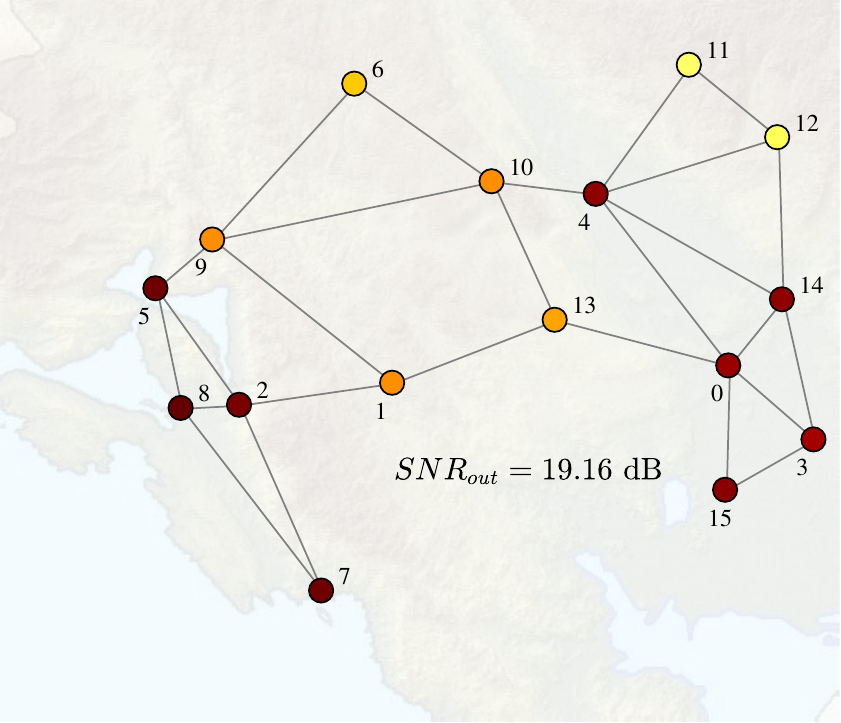}
	\caption{Graph signal denoising for a noisy signal from Fig. \ref{fig:MNE_fig_e}, which is filtered using an optimal  filter in (\ref{TranfFGRFF}), with $\alpha=1$.}
	\label{MNE_fig_optimal_filt}
\end{figure}
\end{Example}

\noindent\textbf{Other cost functions.} Among many possible alternatives, we will introduce two more cost functions for graph signal denoising, which exploit  different constraints imposed on the solution. 

Instead of enforcing the smoothness of  the output signal,  we may instead desire that its deviation from a linear form (that is, the signal, $\mathbf{y}$, which satisfies $\mathbf{L} \mathbf{y}=\mathbf{0}$) is as small as possible. This can be achieved with the cost function given by 
\begin{equation}
J=\frac{1}{2}\Vert \mathbf{y}-\mathbf{x}\Vert_2^2 + \alpha \Vert  \mathbf{L} \mathbf{y} \Vert^2_2=\frac{1}{2}\Vert \mathbf{y}-\mathbf{x}\Vert_2^2 + \alpha \mathbf{y}^T \mathbf{L}^2 \mathbf{y} \label{LLLLLK}
\end{equation}
which yields a closed form denoising solution
$$\mathbf{y}= (\mathbf{I} + 2\alpha \mathbf{L}^2)^{-1} \mathbf{x} $$
 with the corresponding element-wise spectral domain relation $
 H(\lambda_k)=1/(1+2\alpha\lambda^2_k).
 $ 

A combination of the two cost function forms in (\ref{LLLKKK}) and (\ref{LLLLLK}), may provide additional flexibility in the design of the filter transfer function, for example 
$$J=\frac{1}{2}\Vert \mathbf{y}-\mathbf{x}\Vert_2^2 + \alpha \mathbf{y}^T \mathbf{L} \mathbf{y}+ \beta \mathbf{y}^T \mathbf{L}^2 \mathbf{y}
$$ 
would yield the transfer function
$$
H(\lambda_k)=\frac{1}{1+2\alpha\lambda_k+2\beta\lambda^2_k}.
$$ 
This transfer function form can be further fine-tuned through the choice of the parameters $\alpha$ and $\beta$. For example, if we desire the component corresponding to $\lambda_1 \ne 0$ not to be attenuated, we would use $\alpha +\beta \lambda_1=0$. Such a cost function can be straightforwardly extended to produce a transfer function for $M$ unattenuated components.

\noindent\textbf{Sparsity promoting solutions.} Some applications require to promote the sparsity of the output graph signal, rather than its smoothness. Such solutions then naturally rest upon compressive sensing theory which requires the two-norm in the previous cost functions to be replaced with the norms that promote sparsity. Two examples of such cost functions are
\begin{equation}
J=\frac{1}{2}\Vert \mathbf{y}-\mathbf{x}\Vert_2^2 + \alpha \Vert  \mathbf{L} \mathbf{y} \Vert_p^p \label{SparPr1}
\end{equation}
and 
\begin{align}
J\!=\!\frac{1}{2}\! \sum_{n=0}^{N-1}\! (y(n)\!-\!x(n))^2\!\! +\! \alpha \sum_{n=0}^{N-1} \!\!\left(\sum_{m=0}^{N-1}W_{mn}(y(n)\!-y(m))^2\!\!\right)^{p/2}\label{SparPr2} 
\end{align}
with $0\le p\le 1$.

% {\color{red} While the minimization of $\mathbf{y}^T\mathbf{Ly}$ promotes the smoothness, the minimization of the sparsity of $Ly$ will promote constant (or linear y) with the smallest possible number of steps (nonzero values of Ly). }

\begin{Remark}
The zero-norm, $\ell_0$, with $p=0$, is the best in promoting sparsity, since for $p=0$ the second term in the cost function in (\ref{SparPr1}) counts (and minimizes) the number of nonzero  elements in $\mathbf{L} \mathbf{y}$. Minimization of the sparsity of $\mathbf{Ly}$ promotes constant (or linear) solutions for $\mathbf{y}$, with the smallest number of discontinuities (nonzero elements of vector $\mathbf{Ly}$). In the second cost function in (\ref{SparPr2}), the zero-norm promotes the smallest possible number of nonzero elements of the term $\sum_{m=0}^{N-1}W_{mn}(y(n)-y(m))^2$; this is also known as the total variation (TV) approach.	However, the minimization of such objective functions cannot be achieved in an analytic way, like in the standard MSE case of $p=2$.  
\end{Remark}

On the other hand, the choice of $p=1$ with one-norm, $\ell_1$, makes the above cost functions convex, allowing for gradient descend methods be used to arrive at the solution, while producing the same solution as with $p=0$, under some mild conditions. The $\ell_1$-norm serves as an analytic proxy to the $\ell_0$-norm \cite{kim2009ell_1}.

\subsection{Systems on a Graph Defined Using Random Walk Laplacian}

While common choices for the graph shift operator are: (i) adjacency matrix, $\mathbf{S}=\mathbf{A}$, and (ii) graph Laplacian,  $\mathbf{S}=\mathbf{L}$, normalized versions of the adjacency matrix,  graph Laplacian, $\mathbf{S}=\mathbf{D}^{-1/2}\mathbf{L}\mathbf{D}^{-1/2}$, and random walk (diffusion) matrix, $\mathbf{S}=\mathbf{D}^{-1}\mathbf{W}$, can also be used \cite{stankovic2019introduction,Eldar2017}. 
Various shift operators produce corresponding eigenvector (signal decomposition) bases, such as those  analyzed in Part I and given in Table \ref{tab:1a}.

 A generalized form 
of the output from a system on a graph can then be written as
\begin{equation}
\mathbf{y}=h_0 \mathbf{S}^0\, \mathbf{x}+h_1 \mathbf{S}^1\, \mathbf{x}+\dots+h_{M-1} \mathbf{S}^{M-1}\, \mathbf{x}=\sum_{m=0}^{M-1}h_m \mathbf{S}^m\, \mathbf{x},
\label{eq:filter-timeS}
\end{equation}
where,  by definition $\mathbf{S}^0=\mathbf{I}$, while $h_0$, $h_1$,  \ldots, $h_{M-1}$ are the system coefficients.

\begin{table}[tb]
	
	\centering
	\caption{Summary of graph spectral basis vectors.  }
	
	\renewcommand{\arraystretch}{1.25}
	\begin{tabular}{|l|l|}
		\hline\hline
		\textbf{Operator} & \textbf{Eigenanalysis}  \\[5pt]
		\hline\hline 
		Graph Laplacian & $\mathbf{L}\mathbf{u}_k=\lambda_k\mathbf{u}_k$  \\[5pt]
		\hline\hline
		Generalized eigenvectors    &   \\[0pt] 
		of graph Laplacian  &  $\mathbf{L}\mathbf{u}_k=\lambda_k\mathbf{D}\mathbf{u}_k$ \\[5pt]
		\hline\hline
		Normalized graph  Laplacian & $\mathbf{D}^{-\frac{1}{2}}\mathbf{L}\mathbf{D}^{-\frac{1}{2}}\mathbf{u}_k=\lambda_k\mathbf{u}_k$ \\[5pt]
		\hline\hline 
		Adjacency matrix & $\mathbf{A}\mathbf{u}_k=\lambda_k\mathbf{u}_k$ \\[5pt]
		\hline\hline 
		Normalized adjacency matrix & $\Big(\frac{1}{\lambda_{\max}}\mathbf{A}\Big)\mathbf{u}_k=\lambda_k\mathbf{u}_k$  \\[5pt]
		\hline\hline
	\end{tabular}
	\renewcommand{\arraystretch}{1}
	\label{tab:1a}	
\end{table}

An unbiased version of the random walk shift operator can also be employed in this context, defined as
\begin{equation}
{\bf S}=({\bf I}+{\bf D})^{-1}({\bf I}+{\bf W}), \label{eq:GRW_GSO}
\end{equation}
as it exhibits the desirable property of \textit{asymptotic signal energy preservation} \cite{Scalzo2019_3}. The shift operator in (\ref{eq:GRW_GSO}) can be derived under the assumption that the random graph signal, ${\bf x}$, follows the \textit{general random walk} (GRW) model, which exhibits the following properties:
\begin{enumerate}[label=\roman*)]
	\item \textit{Graph Markov property}, that is, the random process is dependent only of its shifted state,
	\begin{equation}
		P \left( {\bf x} \, \bigg| \, \bigcap_{m > 0} {\bf S}^{m}{\bf x}  \right) = P\left( {\bf x} \, \Big|  {\bf S}{\bf x}  \right) ; \label{eq:Markov_graph}
	\end{equation}
	\item \textit{Graph Martingale property}, whereby the conditional expectation of the random process is equal to its shifted state, which can be written as
	\begin{equation}
		E \left\{ {\bf x} \, \bigg| \, \bigcap_{m > 0} {\bf S}^{m}{\bf x} \right\} = {\bf S}{\bf x} . \label{eq:Martingale_graph}
	\end{equation}
\end{enumerate}
In this way, the random walk can be described by a \textit{Markov matrix}, ${\bf P} \in \mathbb{R}^{N \times N}$, with its $(m,n)$-th element defined as the transition probability, $P_{mn}$, of going from vertex $m$ to vertex $n$. By setting ${\bf S} = {\bf P}$, this shift operator is unbiased, since each row in ${\bf P}$ sums up to unity, i.e. ${\bf P}\,{\bf 1} = {\bf 1}$. Furthermore, owing to the graph Martingale condition in (\ref{eq:Martingale_graph}), the shift operator exhibits a dual role of the expectation operator, since ${\bf S}{\bf x} = {\bf P}{\bf x} = E\left\{{\bf x}\right\}$. With this result, it can also be proven that with an increase in the number of vertices, $N$, the shift operator is asymptotically power preserving (isometric), that is \cite{Scalzo2019_3}
\begin{equation}
\lim_{N \to \infty} \, \|{\bf S x}\|^{2} \,  = \, E\left\{\|{\bf x}\|^{2}\right\}.
\end{equation}
Therefore, the class of systems based on this graph shift also exhibits the following boundedness property
\begin{equation}
\lim_{N \to \infty} \, \|{\bf y}\|^{2} \, \leq \sum_{m=0}^{M-1} |h_{m}|^{2} E\left\{\|{\bf x}\|^{2}\right\}.
\end{equation}
The use of the Markov matrix as the shift operator was recently proposed in \cite{stankovic2019introduction,Eldar2017}, and the above analysis further justifies this concept.

In practice, the actual probabilities of vertex transition are often unknown but can be inferred from the available information of the graph topology, implied by the weight matrix, ${\bf W}$. In the limit, Donsker's theorem states that the GRW has a probability density which convergences to that of the \textit{Wiener process} \cite{Donsker1951,Billingsley1999,Durrett1996,Revuz1999}. In the graph setting, for a walker at a vertex $m$, the \textit{central limit theorem} \cite{Billingsley1995} asserts that after a sufficiently large number of independent steps, the probability of walker's position is Gaussian distributed, $P_{mn} \propto e^{-r_{mn}^{2}}$, where $r_{mn}$ is a measure of physical distance between vertices $m$ and $n$. Consequently, the elements of the GRW weight matrix, denoted by $\tilde{{\bf W}}$, are given by
\begin{equation}
\tilde{W}_{mn} = \begin{cases}
e^{-r_{mn}^{2}}, & (m,n) \in \mathcal{E},\\
1, & m=n,\\
0, & (m,n) \notin \mathcal{E}.
\end{cases} \label{eq:GRW_weights}
\end{equation}
Notice that in a probabilistic setting \textit{the vertices are implicitly self-connected}; to ensure that the transition probabilities sum up to unity, we therefore need to normalise the GRW weights to obtain $P_{mn} =  \tilde{W}_{mn}/ \tilde{D}_{mm}$. Notice that the standard weight matrix, ${\bf W}$, has zeros on the diagonal so that for $\tilde{{\bf W}}$ in (\ref{eq:GRW_weights}), ${\bf \tilde{{\bf W}}}=({\bf I}+{\bf W})$ and ${\bf \tilde{{\bf D}}}=({\bf I}+{\bf D})$. Therefore, this graph shift operator takes the form in (\ref{eq:GRW_GSO}).

\begin{Example}
Consider again the multi-sensor setup described in Section \ref{sec2}, and shown in Fig. \ref{fig:BSD_GRW}(a). The graph shift operator based on the GRW model was employed within a first order averaging system ($h_{0}=0$, $h_{1}=1$), as in (\ref{eq:filter-timeS}), to estimate the true temperature from the observed temperature field. The weight matrix elements, $W_{mn}=e^{-r_{mn}^{2}}$, were specified based on the Euclidean distance between vertices, $r_{mn}$, thereby accounting for the difference in latitude, longitude and altitute. The resulting denoised temperature field is illustrated in Fig. \ref{fig:BSD_GRW}(b) and demonstrates the attained increase in the SNR from $14.2 \, \text{dB}$ to $19.8 \, \text{dB}$, which results from the desirable unbiasedness and asymptotic power preservation properties of the shift operator.
	
\begin{figure}[ht]
	\centering
	\begin{minipage}{0.235\textwidth}
		\centering
		\includegraphics[width=0.9\textwidth, trim={6cm 13.7cm 0.5cm 3cm}, clip]{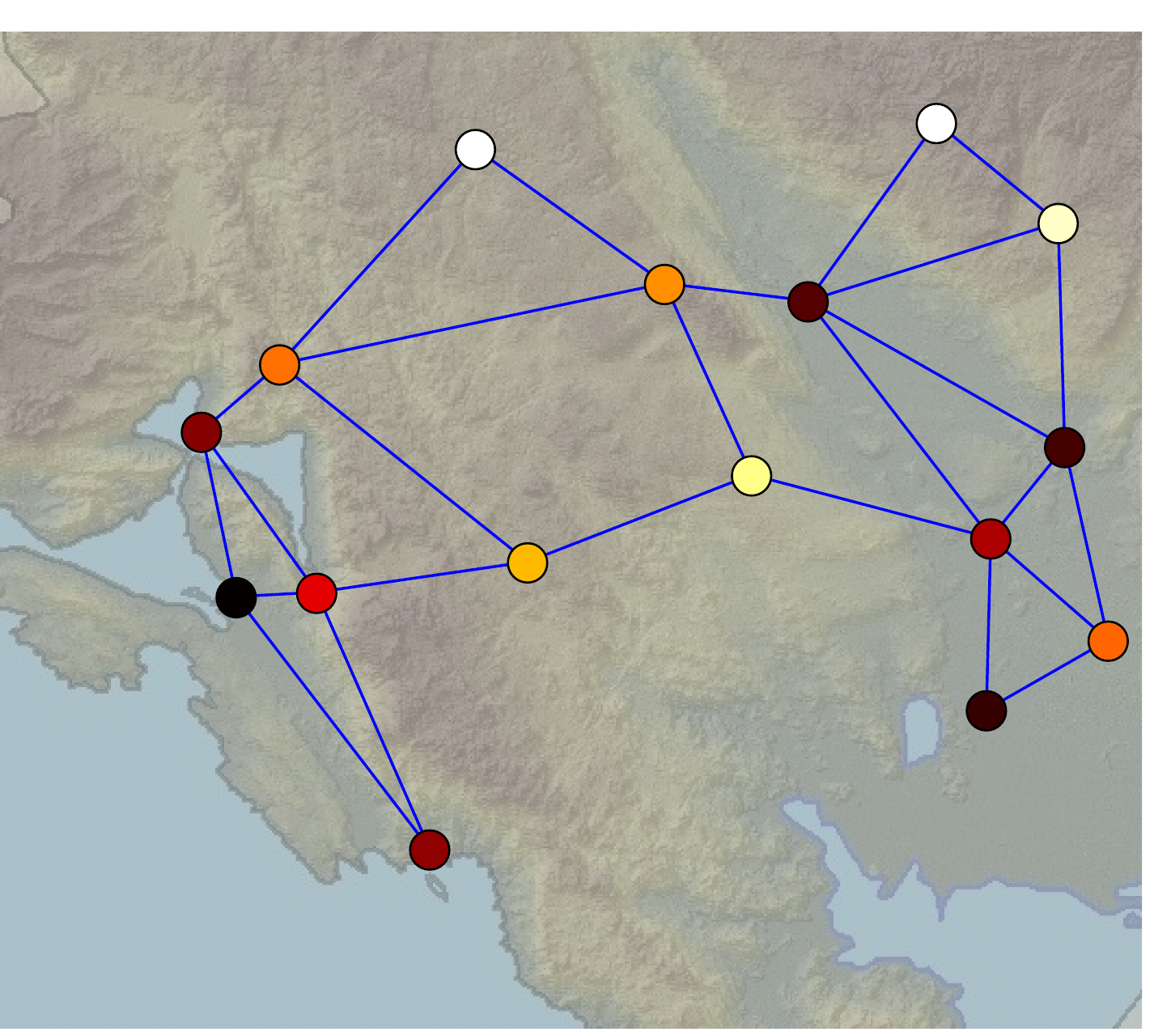} 
		{{\footnotesize (a) Observed field\\ ($\text{SNR} = 14.2 \, \text{dB}$).}}  
	\end{minipage}
	\begin{minipage}{0.235\textwidth}
		\centering
		\includegraphics[width=0.9\textwidth, trim={6cm 13.7cm 0.5cm 3cm}, clip]{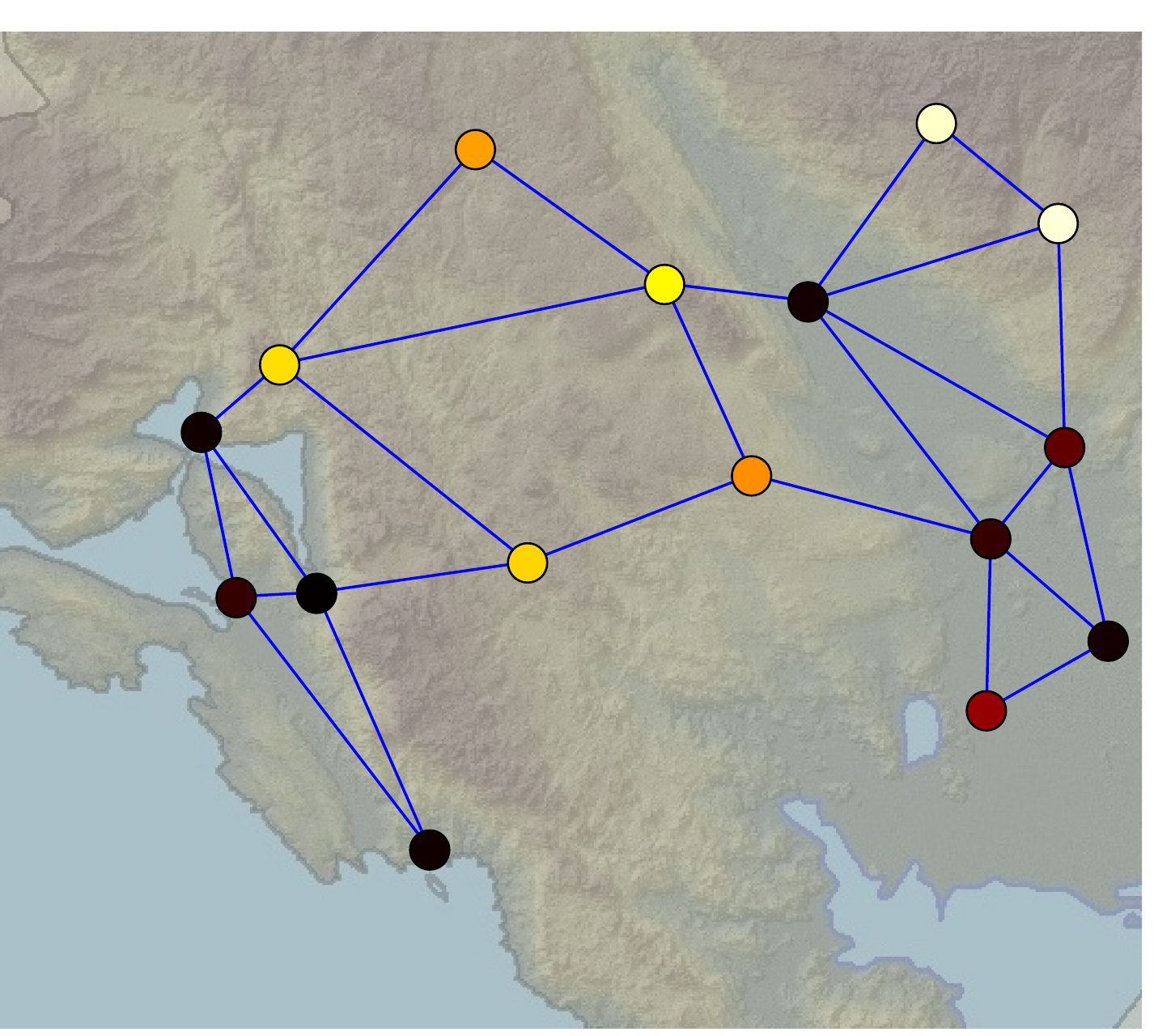} 
		{{\footnotesize (b) GRW local expectation\\ ($\text{SNR} = 19.8 \, \text{dB}$).}}
	\end{minipage}
\caption{Local average operator on a graph based on the generalized random walk (GRW) shift operator. The graph signal intensity is designated by the vertex color.} 
\label{fig:BSD_GRW}
\end{figure}
\end{Example}

\section{Subsampling, Compressed Sensing, and Reconstruction }
\label{sec4}
Graphs may comprise of a very large number of vertices, of the order of millions or even higher. The associated computational and storage issues bring to the fore the consideration of potential advantages of subsampling and compressive sensing  defined on graphs. We here present several basic approaches to subsampling, along with their relations to classical signal processing \cite{chen2015sampling,chen2015discrete,chen2015signal,chen2016signal,Tsitsvero2016,wang2015local,stankovictutorial,stankovic2015digital,narang2011downsampling,nguyen2015downsampling,narang2012perfect,segarra2015interpolation,marques2016sampling,anis2016efficient,behjat2016signal,Tanaka,sakiyama2014oversampled,Tremblay,leskovec2006sampling}.   

\subsection{Subsampling of Low-Pass Graph Signals}\label{SubLow}

For convenience, we shall start from the simplest case where  the considered graph signal is of a low-pass nature. Such a signal can be expressed as a linear combination of $K<N$  eigenvectors of the graph Laplacian which exhibit the lowest smoothness indices,
\begin{equation}
x(n)=\sum_{k=0}^{K-1}X(k)u_k(n), \,\, n=0,1,\dots,N-1. \label{ZeroValXInv}
\end{equation}
The GDFT domain coefficients of this ($K$-sparse) signal in the GDFT domain are of the following form 
\begin{equation}
\mathbf{X}=[X(0),X(1),\dots,X(K-1),0,0,\dots,0]^T. \label{ZeroValX}
\end{equation}

Recall that a graph signal is sparse in the GDFT domain if $K \ll N$. The smallest number of graph signal samples, $M$, needed to recover the sparse signal is therefore $M=K<N$. For stability of reconstruction, it is common to employ $K \le M < N$ graph signal samples. The vector of available graph signal samples will be referred to as the \textit{measurement vector}, and will be denoted by $\mathbf{y}$, while 
the set of vertices (a random subset of $\mathcal{V}=\{0,1,2,\dots,N-1\}$) over which the samples of  graph signal are available is denoted by
$$\mathbb{M}=\{n_1,n_2, \dots, n_M\}.$$
 The measurement matrix can now be defined using the IGDFT,  $\mathbf{x}=\mathbf{U}\,\mathbf{X},$ of which an element-wise form is given by (\ref{ZeroValXInv}).   
The equations in (\ref{ZeroValXInv}) corresponding to the available graph signal samples at vertices $n \in \mathbb{M}=\{n_1,n_2, \dots, n_M\}$ then define the system 
$$\left[
\begin{matrix} 
x(n_1)\\
x(n_2)\\
\vdots\\
x(n_M)\\
\end{matrix}
\right]
\!=\!
\left[
\begin{matrix} 
u_{0}(n_1)\!\! &\!\! u_{1}(n_1) \!\!\!\!& \dots\!\!\!\! & u_{N-1}(n_1) \\
u_{0}(n_2)\!\! &\!\! u_{1}(n_2)\!\!\!\! & \dots \!\!\!\!& u_{N-1}(n_2)\\
\vdots\!\! &\!\vdots\!\!\!\!&\!\!\!\!\ddots\!\!\!\!&\vdots\\
u_{0}(n_M)\!\! & \!\!u_{1}(n_M)\!\!\!\! & \dots \!\!\!\!& u_{N\!-\!1}(n_M)
\end{matrix}
\right]\!
\left[
\begin{matrix} 
X(0)\\
X(1)\\
\vdots\\
X(N-1)\\
\end{matrix}
\right],
$$
for which the matrix form is given by
\begin{equation}
\mathbf{y}=\mathbf{A}_{MN}\mathbf{X}, \label{MesBBB}
\end{equation}
where $\mathbf{A}_{MN}$ is the \textit{measurement matrix} and the \textit{ measurements vector} 
$$\mathbf{y}=[x(n_1), x(n_2),\dots,x(n_M)]^T$$
 consists of the available graph signal samples. 
In general, since  $M<N$ this system is underdetermined, and cannot be solved uniquely for $\mathbf{X}$  without additional constraints.

The assumption that the spectral representation of a signal contains a linear combination of only $K\le M$ slowest varying eigenvectors allows us to exclude the GDFT coefficients $X(K), X(K+1), \dots,X(N-1)$ in (\ref{ZeroValX}) since these are zero-valued and do not contribute to the formation of graph signal samples. With this in mind, the $M \times N$ system of equations in (\ref{MesBBB}) is reduced to the following $M \times K$ system 
$$\left[
\begin{matrix} 
x(n_1)\\
x(n_2)\\
\vdots\\
x(n_M)\\
\end{matrix}
\right]
\!=\!
\left[
\begin{matrix} 
u_{0}(n_1)\!\! &\!\! u_{1}(n_1) \!\!\!\!& \dots\!\!\!\! & u_{K-1}(n_1) \\
u_{0}(n_2)\!\! &\!\! u_{1}(n_2)\!\!\!\! & \dots \!\!\!\!& u_{K-1}(n_2)\\
\vdots\!\! &\!\vdots\!\!\!\!&\!\!\!\!\ddots\!\!\!\!&\vdots\\
u_{0}(n_M)\!\! & \!\!u_{1}(n_M)\!\!\!\! & \dots \!\!\!\!& u_{K\!-\!1}(n_M)
\end{matrix}
\right]\!
\left[
\begin{matrix} 
X(0)\\
X(1)\\
\vdots\\
X(K-1)\\
\end{matrix}
\right],
$$
or, in the matrix form
\begin{equation}
\mathbf{y}=\mathbf{A}_{MK}\mathbf{X}_K, \label{MesBBBK}
\end{equation}
where the definitions of the reduced measurement matrix $\mathbf{A}_{MK}$ and the reduced GDFT vector $\mathbf{X}_K$ are obvious.  
For $M=K$ independent measurements, this system can be solved uniquely, while for $M>K$ the system is typically overdetermined and the solution is found in the least squares (LS) sense, as \cite{stankovictutorial}
\begin{equation}\mathbf{X}_K=(\mathbf{A}^T_{MK}\mathbf{A}_{MK})^{-1}\mathbf{A}_{MK}^T\mathbf{y}=\textrm{pinv}(\mathbf{A}_{MK})\mathbf{y},\label{Recpinv}
\end{equation} 
where $\textrm{pinv}(\mathbf{A}_{MK})=(\mathbf{A}^T_{MK}\mathbf{A}_{MK})^{-1}\mathbf{A}_{MK}^T$ is the matrix pseudo-inverse of $\mathbf{A}_{MK}$. 

After $\mathbf{X}_K$  is calculated, all GDFT values follow directly  as $\mathbf{X}=[X(0),X(1),\dots,X(K-1),0,0,\dots,0]^T$, where the assumed zero values are added for $X(K)$, $X(K+1)$, $\dots$, $X(N-1)$. The graph  signal is then recovered at all vertices using $\mathbf{x}=\mathbf{U}\,\mathbf{X}$. 

\noindent\textbf{Recovery condition.} The signal reconstruction in (\ref{Recpinv}) is possible if the inverse $(\mathbf{A}^T_{MK}\mathbf{A}_{MK})^{-1}$ exists, which means that
\begin{equation}
\textrm{rank}(\mathbf{A}^T_{MK}\mathbf{A}_{MK})=K.\label{rankK}
\end{equation}
In terms of the matrix condition number, this requirement is equivalent to  
$$\textrm{cond}(\mathbf{A}^T_{MK}\mathbf{A}_{MK})<\infty,$$
that is, a nonsingular $\mathbf{A}^T_{MK}\mathbf{A}_{MK}$.

 \begin{Remark} For noisy measurements of graph signals, the noise in the reconstructed GDFT coefficients is directly related to the input noise and the matrix condition number. If we are able to choose the available signal sample positions (vertices), then the \textit{sampling strategy} would be to find the set of measurements so that these produce the condition number which is as close to unity as possible (for stability and reduced influence of noise).
 \end{Remark}

 \begin{Example} To demonstrate the principle of reconstruction from a reduced set of graph signal samples, consider the values of a graph signal at $M=3$ vertices, given by
$$\mathbf{y}=[x(0), x(2), x(6)]^T=[ 1.140,0.996, 0.563]^T,$$
as shown in Fig. \ref{fig:lowpass-cs} (upper panel). Assume that the graph signal is of a low-pass type, with $K=2$ lowest nonzero GDFT coefficients $X(0)$ and $X(1)$. 
The GDFT coefficients of this graph signal can then be reconstructed from 
\begin{equation}
\mathbf{y}=\mathbf{A}_{32}\mathbf{X}_2,
\end{equation}
that follows from the definition in (\ref{ZeroValXInv}) for the assumed available signal samples, $x(n)$, at the three  vertices $n=0$, $n=2$, and $n=6$, for two nonzero coefficients, $X(0)$ and $X(1)$,   
$$
\left[
\begin{matrix} 
	x(0)\\
	x(2)\\
	x(6)\\
\end{matrix}
\right]
=
\left[
\begin{matrix} 
	u_{0}(0) & u_{1}(0)\\
	u_{0}(2) & u_{1}(2)\\
	u_{0}(6) & u_{1}(6) \\
\end{matrix}
\right]
\left[
\begin{matrix} 
	X(0)\\
	X(1)\\
\end{matrix}
\right].
$$
The rank of the matrix $\mathbf{A}_{32}$ is 2. The corresponding matrix condition number is $\textrm{cond}(\mathbf{A}^T_{32}\mathbf{A}_{32})=4.33,$ while the reconstructed nonzero values of the GDFT are $X(0)=2$ and $X(1)=1,$ to yield the reconstructed graph signal $\mathbf{x}=\mathbf{U}\,\mathbf{X}$, with $\mathbf{X}=[2, 1,0,0,0,0,0,0]^T$,  as shown in Fig. \ref{fig:lowpass-cs} (lower panel).  

 \begin{figure}
 \centering
 \includegraphics[]{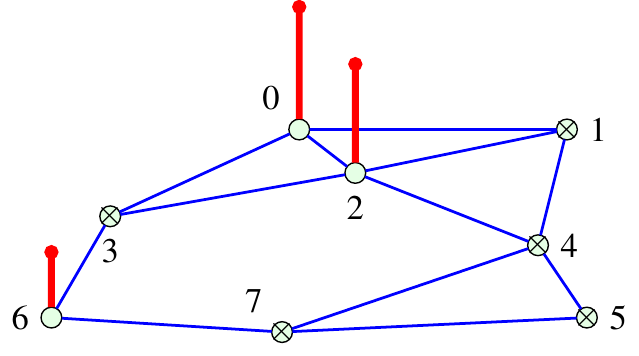}
 \hfill
 \includegraphics[]{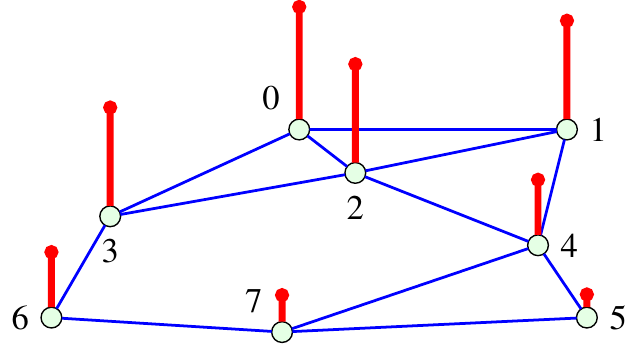}
  
 \caption{Illustration of the subsampling of a lowpass graph signal. Top: A graph signal with missing samples at vertices $1,3,4,5$ and $7$. Bottom: The reconstructed graph signal.}
 \label{fig:lowpass-cs}
 \end{figure}
\end{Example}

 \begin{Remark} For a directed circular graph, with the eigenvectors $u_k(n)=\exp(j2\pi nk/N)/\sqrt{N}$, the above downsampling and interpolation relations are identical to those in classical signal processing \cite{stankovic2015digital}.       
\end{Remark}

\subsection{Subsampling of Sparse Graph Signals}\label{SubSGS}
The subsampling of graph signals which are sparse in the GDFT domain will be next considered for the cases of both known and unknown positions of the nonzero GDFT coefficients.  
\subsubsection{Known Coefficient Positions in GDFT}\label{SubKnown}
The previous analysis holds not only for a low-pass type of the graph signal, $\mathbf{x}$, and its corresponding GDFT, $\mathbf{X}$, but also for case of GDFT, $\mathbf{X}$, with $K$ nonzero values at arbitrary, but known spectral positions, that is, 
$$
X(k)=0 \textrm{ for } k \notin \mathbb{K}=\{k_1,k_2,\dots,k_K\}.$$
Similar to (\ref{MesBBB}), the corresponding system of equations
\begin{equation}
\left[
\begin{matrix} 
x(n_1)\\
x(n_2)\\
\vdots\\
x(n_M)\\
\end{matrix}
\right]
\!=\!
\left[
\begin{matrix} 
u_{k_1}(n_1)\!\! &\!\! u_{k_2}(n_1) \!\!\!\!& \dots\!\!\!\! & u_{k_K}(n_1) \\
u_{k_1}(n_2)\!\! &\!\! u_{k_2}(n_2)\!\!\!\! & \dots \!\!\!\!& u_{k_K}(n_2)\\
\vdots\!\! &\!\vdots\!\!\!\!&\!\!\!\!\ddots\!\!\!\!&\vdots\\
u_{k_1}(n_M)\!\! & \!\!u_{k_2}(n_M)\!\!\!\! & \dots \!\!\!\!& u_{k_K\!}(n_M)
\end{matrix}
\right]\!
\left[
\begin{matrix} 
X(k_1)\\
X(k_2)\\
\vdots\\
X(k_K)\\
\end{matrix}
\right].\label{RndKnownposi}
\end{equation}  
of which the matrix form is $\mathbf{y}=\mathbf{A}_{MK}\mathbf{X}_K,$ is solved for the nonzero spectral values $X(k)$, $k \in \mathbb{K}$, in the same way as in the case of a low-pass signal presented in Section \ref{SubLow}.

\subsubsection{Support Matrices, Subsampling and Upsampling}
In graph signal processing literature, \textit{the subsampling problem is often defined using the so called support matrices}. Assume that a graph signal, $\mathbf{x}$, is subsampled in such way that it is available on a subset of vertices $n \in \mathbb{M}=\{n_1,n_2, \dots, n_M\}$, rather than on the full set of vertices. For this subsampled signal, we can define its  upsampled version, $\mathbf{x}_s$, by adding zeros at the vertices where the signal is not available. Using a mathematical formalism, the \textit{subsampled and upsampled} version, $\mathbf{x}_s$, of the original signal, $\mathbf{x}$, is then  
 \begin{equation}
 \mathbf{x}_s=\mathbf{B}\mathbf{x}, \label{xsxs}
 \end{equation}
 where \textit{the support matrix} $\mathbf{B}$ is an $N\times N$ diagonal matrix with ones at the diagonal positions which correspond to $\mathbb{M}=\{n_1,n_2, \dots, n_M\}$ and zeros elsewhere.  The subsampled and upsampled version, $\mathbf{x}_s$, of the signal $\mathbf{x}$ is obtained is such a way that the signal $\mathbf{x}$ is subsampled on a reduced set of vertices, and then upsampled by adding zeros at the original signal positions where the subsampled signal is not defined.
  
  Recall that in general a signal, $\mathbf{x}$, with $N$ independent values cannot be reconstructed from its $M<N$ nonzero values in $\mathbf{x}_s$, without additional constraints.  However, for graph signals which are also sparse in the GDFT domain, the additional constraint is that the signal, $\mathbf{x}$, has only $K \le M$ nonzero coefficients in the GDFT domain,  $\mathbf{X}= \mathbf{U}^T\mathbf{x}$, at  $k \in \mathbb{K}=\{k_1,k_2, \dots, k_K\}$, so that the relation
  $$\mathbf{X}=\mathbf{C}\mathbf{X}$$
   holds, where the support matrix $\mathbf{C}$ is an $N \times N$ diagonal matrix with ones at the diagonal positions which correspond to $\mathbb{K}=\{k_1,k_2, \dots, k_K\}$ and zeros elsewhere. Note the presence of the GDFT, $\mathbf{X}$, is on both sides of this equation, contrary to $\mathbf{x}_s=\mathbf{B}\mathbf{x}$ in (\ref{xsxs}).  The reconstruction formula then follows from $$\mathbf{x}_s=\mathbf{B}\mathbf{x}=\mathbf{B}\mathbf{U}^{}\mathbf{X}=\mathbf{B}\mathbf{U}^{}\mathbf{C}\mathbf{X}.$$
   as $\mathbf{X}=\mathrm{pinv}\big(\mathbf{B}\mathbf{U}^{}\mathbf{C}\big)\mathbf{x}_s$. The inversion $$\mathbf{X}=\mathbf{C}\mathbf{X}=\mathrm{pinv}\big(\mathbf{B}\mathbf{U}^{}\mathbf{C}\big)\mathbf{x}_s$$
     is possible for $K$ nonzero coefficients of $\mathbf{C}\mathbf{X}$  if the rank of $\mathbf{B}\mathbf{U}^{}\mathbf{C}$ is $K$ (if there are $K$ linearly independent equations), that is 
$$\mathrm{rank}(\mathbf{C})=K=\mathrm{rank}\big(\mathbf{B}\mathbf{U}^{}\mathbf{C}\big).
$$
This condition is equivalent to (\ref{rankK}) since the nonzero part of matrix $\mathbf{B}\mathbf{U}^{}\mathbf{C}$ is equal to $\mathbf{A}_{MK}$ in  (\ref{RndKnownposi}). 
\subsubsection{Unknown Coefficient Positions}\label{CS_Solu_UnKP}
The reconstruction problem is more complex if the positions of nonzero spectral coefficients $ \mathbb{K}=\{k_1,k_2,\dots,k_K\}$ are not known. This case has been addressed within standard compressive sensing theory and can be formulated as
\begin{equation}
\min \left\Vert \mathbf{X} \right\Vert _0 \text{ subject to } \mathbf{y}=\mathbf{A}_{MN}\mathbf{X}, \label{CSnorm0}
\end{equation}
where $\Vert \mathbf{X} \Vert_0$ denotes the number of nonzero elements in $\mathbf{X}$ ($\ell_0$ pseudo-norm).

While the ways to solve this minimization problem are manifold,  we here adopt  a  simple, two-step approach:

\begin{enumerate}
\item 
Estimate the positions $\mathbb{K}=\{k_1,k_2,\dots,k_K\}$ of the nonzero coefficients using $M > K$ signal samples,
% in one-step or iterative way.
\item
Reconstruct the nonzero coefficients of $\mathbf{X}$ at  the estimated positions $\mathbb{K}$, along with the signal $\mathbf{x}$ at all vertices, using the methods for the reconstruction with the known nonzero coefficient positions, described in Sections \ref{SubLow} and \ref{SubKnown}. The nonzero coefficients at positions  $\mathbb{K}$  are calculated as $\mathbf{X}_K=\textrm{pinv}(\mathbf{A}_{MK})\mathbf{y}$.
\end{enumerate}

The nonzero positions of the GDFT in Step 1 can be estimated through the projection of measurements (available signal samples), $\mathbf{y}$, on the measurement matrix 
$$
\mathbf{A}_{MN}=\left[
\begin{matrix} 
u_{0}(n_1) & u_{1}(n_1) & \dots & u_{N-1}(n_1) \\
u_{0}(n_2) & u_{1}(n_2) & \dots & u_{N-1}(n_2)\\
\vdots & \vdots & \ddots & \vdots \\
u_{0}(n_M) & u_{1}(n_M) & \dots & u_{N-1}(n_M)
\end{matrix}
\right]
$$
to give
\begin{equation}
\mathbf{X}_0=\mathbf{A}_{MN}^T\mathbf{y}, \label{iniialES}
\end{equation}
where the positions of $K$ largest values in $\mathbf{X}_0$ are used as an estimate of the nonzero positions, $\mathbb{K}$. This procedure can also be implemented in an iterative way \cite{stankovictutorial}, where
\begin{enumerate}[label=(\roman*)]
\item In the first iteration we assume $K=1$ and proceed to estimate the largest spectral component in the graph signal. Upon determining its position as $k_1=\mathrm{argmax}|\mathbf{A}_{MN}^{T}\mathbf{y}|$, the initially empty set of the nonzero positions becomes $\mathbb{K}=\{k_1\}$.   The reconstructed vector $\mathbf{y}_1=\mathbf{A}_1\mathbf{X}_1$, where $\mathbf{X}_1=\textrm{pinv}(\mathbf{A}_{M1})\mathbf{y}$, is then removed from the measurements, $\mathbf{y}$. In this case, the matrix $\mathbf{A}_{M1}$ is a column of the matrix $\mathbf{A}_{MN}$ defined by the index $k_1$. The difference $\mathbf{e}=\mathbf{y}-\mathbf{y}_1$ is used as the measurement vector in the next step.
	
\item  The position of the second largest spectral component in the graph signal is estimated by solving $k_2=\mathrm{argmax}|\mathbf{A}_{MN}^{T}\mathbf{e}|$. The set of nonzero positions now becomes $\mathbb{K}=\{k_1, k_2\}$. The first and the second component of the graph signal are now estimated as $\mathbf{X}_2=\textrm{pinv}(\mathbf{A}_{M2})\mathbf{y}$, where the matrix $\mathbf{A}_{M2}$ is a submatrix of the measurement matrix, $\mathbf{A}_{MN}$, which consists of the columns defined by the indices $k_1$ and $k_2$. The reconstructed vector $\mathbf{y}_2=\mathbf{A}_2\mathbf{X}_2$, is removed from the measurements, $\mathbf{y}$, with the error, $\mathbf{e}=\mathbf{y}-\mathbf{y}_2$, now acting as the new measurement vector.

\item  The procedure is iteratively repeated $K$ times or until the remaining measurement values in $\mathbf{e}$  are negligible. In the cases when the sparsity, $K$, is unknown, the procedure is iterated until $\|\mathbf{e}\|_2<\varepsilon$, where $\varepsilon$ is a predefined precision.

%This is the same as if we used zeros for all unavailable samples (norm-two solution of the problem).
\end{enumerate}

\begin{Example} Consider a sparse graph signal, of the sparsity degree $K=2$, measured at vertices $n=2, 3, 4, 5,$ and $7$, which takes the values
$$ \mathbf{y}=[    0.707,1.307,0.407,1.307,0.407]^T, $$
as shown in  Fig. \ref{fig:graph-cs} (top panel). Our task is to reconstruct the full signal, that is, to find the missing samples $x(0)$, $x(1)$, and $x(6)$. 

 The estimate positions of the nonzero elements in the GDFT, $\mathbf{X}$, the initial estimate,  $\mathbf{X}_0$, is calculated for given measurements,  $\mathbf{y}$,  according to (\ref{iniialES}). 
Because $K=2$, the positions of the two nonzero coefficients are estimated as positions of the two largest values in $\mathbf{X}_0$. In the considered case, $\mathbb{K}=\{0,3\}$, as shown in Fig. \ref{fig:graph-cs} (bottom panel).
The GDFT coefficients are then reconstructed for the sparsity degree $K=2$, as $\mathbf{X}_2=\textrm{pinv}(\mathbf{A}_{52})\mathbf{y}$, resulting in $X(0)=2$, $X(3)=1.2$, as illustrated in Fig. \ref{fig:graph-cs} (bottom--right). Finally, the reconstructed graph signal at all vertices, $\mathbf{x}=\mathbf{UX}$, is shown in the middle panel of Fig. \ref{fig:graph-cs}.

\begin{figure}
	\centering
	\includegraphics[]{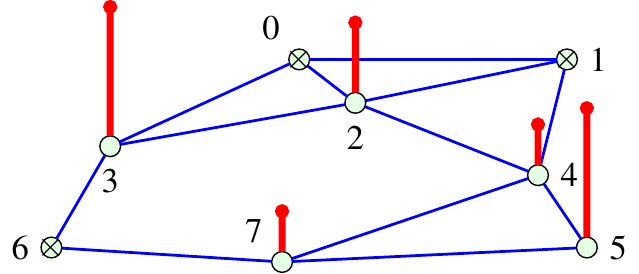} \hspace{5mm} (a)
	\hfill
	\includegraphics[]{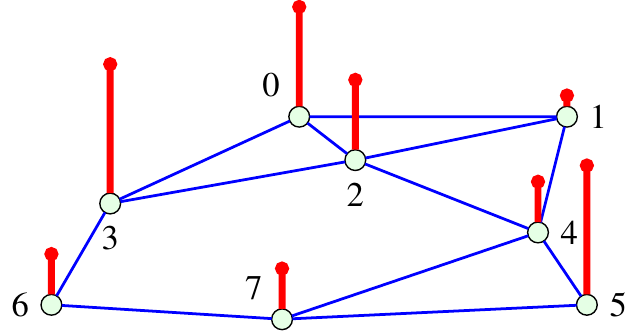} \hspace{5mm} (b)
	
	\vspace*{3mm}
	\includegraphics[]{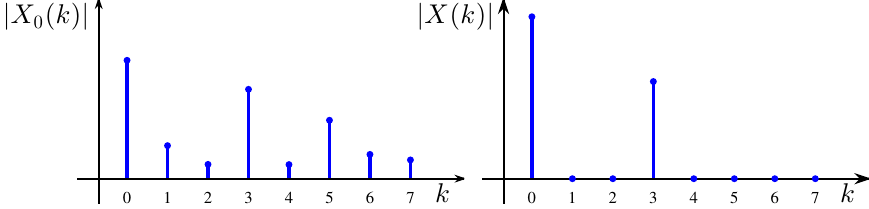}(c)
	
	\caption{Compressive sensing on graphs. (a) Available samples (measurements), $\mathbf{y}=[x(2),x(3),x(4),x(5),x(7)]^T$, with missing samples at $n=0,1,6$. (b) Reconstructed signal, $\mathbf{x}$, over the whole set of vertices. (c) Initial estimate of the GDFT, $X_0(k)$, (left), and the reconstructed sparse GDFT, $X(k)$, (right). }
	\label{fig:graph-cs}
\end{figure}

\end{Example}

\subsubsection{Unique Reconstruction Conditions}

As is the case with the standard compressive sensing problem, the initial  GDFT estimate, $\mathbf{X}_0$, will produce correct positions of the nonzero elements, $X(k)$, and the reconstruction will be unique, if 
$$K < \frac{1}{2}\left(1+\frac{1}{\mu}\right),$$
where $\mu$ is equal to the maximum value of the inner product among any two columns of the measurement matrix, $\mathbf{A}_{MN}$ ($\mu$ is referred to as the coherence index) \cite{stankovic2019intuitive}.

For illustration of the uniqueness of reconstruction, recall that a $K$-sparse signal can be written as $$x(n)=\sum_{i=1}^{K}X(k_i)u_{k_i}(n),$$
of which the  initial estimate in (\ref{iniialES}) is equal to $\mathbf{X}_0=\mathbf{A}_{MN}^T\mathbf{y} = \mathbf{A}_{MN}^T\mathbf{A}_{MN}\mathbf{X}$, or element-wise
$$X_0(k)=\sum_{i=1}^{K}X(k_i)\sum_{n \in \mathbb{M}}u_k(n)u_{k_i}(n)=\sum_{i=1}^{K}X(k_i) \mu(k,k_i),$$ 
where $\mathbb{M}=\{n_1,n_2, \dots, n_M\}$ and 
$$\mu(k,k_i)=\sum_{n \in \mathbb{M}}u_k(n)u_{k_i}(n).$$ 
If the maximum possible absolute value of $\mu(k,k_i)$ is denoted by $\mu=\max|\mu(k,k_i)|$ (coherence index of $\mathbf{A}_{MN}$) then, in the worst case scenario, the amplitude of the largest component, $X(k_i)$, (assumed with the normalized amplitude 1), will be reduced for the maximum possible influence of other equally strong (unity) components $1-(K-1)\mu$, and should be greater than the maximum possible disturbance at $k \ne k_i$, which is $K\mu$. From $1-(K-1)\mu>K\mu$, the unique reconstruction condition follows; see also \cite{stankovictutorial,stankovic2019intuitive}. 

In order to define other unique reconstruction conditions, we shall consider again the solution to $\mathbf{y}=\mathbf{A}_{MN}\mathbf{X}$ which assumes a minimum number of nonzero coefficients in $\mathbf{X}$. Assume that the sparsity degree $K$ is known, then a set of $K$ measurements would yield a possible solution, $\mathbf{X}_K$, for any combination of $K$ nonzero coefficients in $\mathbf{X}$. For another set of $K$ measurements, we would obtain another set of possible solutions, $\mathbf{X}_K$. Then, a common solution between these two sets of solutions would be the solution to our problem. For a unique solution,
there are no two different $K$-sparse solutions $\mathbf{X}_K^{(1)}$ and $\mathbf{X}_K^{(2)}$ if all possible matrices, $\mathbf{A}^T_{M2K}\mathbf{A}_{M2K}$, are nonsingular. 
Namely, both of these two different solutions would satisfy measurement equations, 
\begin{gather*}
\mathbf{A}_{M2K}
\left[
\begin{array}
[c]{cc}%
\mathbf{X}^{(1)}_K \\
\mathbf{0}_K   
\end{array}
\right]
=\mathbf{y}
\text{ and }  
\mathbf{A}_{M2K}
\left[
\begin{array}
[c]{cc}%
\mathbf{0}_K \\
\mathbf{X}^{(2)}_K   
\end{array}
\right]
=\mathbf{y},
\end{gather*}
where $\mathbf{A}_{M2K}=\left[
\begin{array}
[c]{cc}%
\mathbf{A}_{MK}^{(1)} &
\mathbf{A}_{MK}^{(2)}   
\end{array}
\right]$. Obviously, if we subtract these two matrix equations we get a zero-vector on the right-side and a nonzero solution for the resulting vector, 
$$\mathbf{X}_{2K}=
\left[
\begin{array}
[c]{cc}%
\mathbf{X}^{(1)}_K \\
-\mathbf{X}^{(2)}_K   
\end{array}
\right],$$
 requires the zero-valued determinant of  $\mathbf{A}_{M2K}$. The nonzero  determinant of  $\mathbf{A}_{M2K}$ guarantees that two such, nonzero solutions, $\mathbf{X}_K^{(1)}$ and $\mathbf{X}_K^{(2)}$, cannot exist. If all possible submatrices $\mathbf{A}_{M2K}$ of the measurement matrix $\mathbf{A}_{MK}$ are nonsigular, then two solutions of sparsity $K$ cannot exist, and the solution is unique.
The requirement that all reduced measurement matrices corresponding to a $2K$-sparse  $\mathbf{X}$ are nonsingular can be written in several forms, listed below 
$$\mathrm{det} \{\mathbf{A}^T_{M2K}\mathbf{A}_{M2K}\}=d_1 d_2 \cdots d_{2K} \ne 0$$
$$\mathrm{cond} \{\mathbf{A}^T_{M2K}\mathbf{A}_{M2K}\}=\frac{d_{\max}}{d_{\min}} \le \frac{1+\delta_{2K}}{1-\delta_{2K}} < \infty$$
$$1-\delta_{2K} \le d_{\min} \le \frac{\left\Vert \mathbf{A}_{M2K}\mathbf{X}_{2K}\right\Vert _{2}^{2}}{\left\Vert
	\mathbf{X}_{2K}\right\Vert _{2}^{2}}  \le d_{\max} \le 1+\delta_{2K}$$
where $d_i$ are the eigenvalues of $\mathbf{A}^T_{M2K}\mathbf{A}_{M2K}$, $d_{\min}$ is the minimum eigenvalue, $d_{\max}$ is the maximum eigenvalue, and $\delta_{2K}$ is the restricted isometry constant.
All these conditions are satisfied if $d_{\min}>0$ or $0 \le \delta_{2K}<1$.

Noisy data require robust estimators, and  thus more strict bounds on $d_{\min}$ and $\delta_{2K}$. For example, it has been shown that the condition $0\le \delta_{2K}<0.41$ will guarantee stable inversion of $\mathbf{A}^T_{M2K}\mathbf{A}_{M2K}$ and  consequently a robust reconstruction for noisy signals; in addition, this bound will allow for convex relaxation of the reconstruction problem \cite{candes2008restricted}. 
Namely, the previous problem, (\ref{CSnorm0}), can be solved using the convex relation from the norm-zero to a norm-one formulation given by
$$\min \left\Vert \mathbf{X} \right\Vert _1 \text{ subject to } \mathbf{y}=\mathbf{A}_{MN}\mathbf{X}.$$
The solutions to these two problem formulations are the same if the measurement matrix satisfies the previous conditions, with $0\le \delta_{2K}<0.41$. The signal reconstruction problem can now be solved using optimization techniques, such as gradient-based approaches or linear programming methods \cite{stankovictutorial,candes2008restricted}.  
 
\subsection{Measurements as Linear Combinations of Samples} 

It should be mentioned that if some spectrum coefficients of a graph signal are strongly related to only a few of the signal samples, then these signal samples may not be good candidates for the measurements. 

\begin{Example} Consider a graph with one of its eigenvectors  of the form close to $u_i(n)=\delta(n-m)$. This case is possible on graphs, in contrast to the classic DFT analysis where the basis functions are spread over all sensing instants (vertices). A similar scenario is also possible in  wavelet analysis or short time Fourier transforms, which also allow for some of the transform coefficients to be related to only a few of the signal samples. In the assumed simplified case, if a considered sparse signal contains a nonzero coefficient, $X(i)$, corresponding to $u_i(n)=\delta(n-m)$, then all information about $X(i)$ is contained in the graph signal sample $x(m)$ only. This is prohibitive to the principle of reduced number of samples, since an arbitrary set of available samples may not contain $x(m)$.  
	
	In classical and graph data analysis this class of problems  is solved by defining a more complex form of the measurements, $y(n)$, through linear combinations of all signal samples rather than the original samples themselves. In this way, each  measurement, $y(n)$, will contain information about all signal samples, $x(n)$, $n=0,2,\dots,N-1$.  
\end{Example}

Such measurements are  linear combinations of all signal samples, and are given by
$$\left[
\begin{matrix} 
y(1)\\
y(2)\\
\vdots\\
y(M)\\
\end{matrix}
\right]
\!=\!
\left[
\begin{matrix} 
b_{11}\!\! &\!\! b_{12} \!\!\!\!& \dots\!\!\!\! & b_{1N} \\
b_{21}\!\! &\!\! b_{12}\!\!\!\! & \dots \!\!\!\!& b_{2N}\\
\vdots\!\! &\!\vdots\!\!\!\!&\!\!\!\!\ddots\!\!\!\!&\vdots\\
b_{M1}\!\! & \!\!b_{M2}\!\!\!\! & \dots \!\!\!\!& b_{MN}
\end{matrix}
\right]\!
\left[
\begin{matrix} 
x(0)\\
x(1)\\
\vdots\\
x(N-1)\\
\end{matrix}
\right],
$$
or in a matrix form
 $$\mathbf{y}=\mathbf{B}_{MN}\mathbf{x}.$$
 The weighting coefficients for the measurements, $b_{mn}$, in the matrix, $\mathbf{B}_{MN}$,  may be, for example, drawn from a Gaussian random distribution. 
 
 For reconstruction, the sparsity of a graph signal, $\mathbf{x}$, should be again assumed in the GDFT domain. The relation of the measurement vector, $\mathbf{y}$, with this sparsity domain vector of coefficients, $\mathbf{X}$, is then given by
 $$\mathbf{y}=\mathbf{B}_{MN}\mathbf{x}=\mathbf{B}_{MN}\mathbf{UX}=\mathbf{A}_{MN}\mathbf{X}.$$
 The reconstruction is now obtained as a solution to
$$\min \left\Vert \mathbf{X} \right\Vert _0 \text{ subject to } \mathbf{y}=(\mathbf{B}_{MN}\mathbf{U})\mathbf{X}$$
or as a solution of the corresponding convex minimization problem, $$\min \left\Vert \mathbf{X} \right\Vert _1 \text{ subject to } \mathbf{y}=(\mathbf{B}_{MN}\mathbf{U})\mathbf{X},$$
as described in Section \ref{CS_Solu_UnKP}.

\subsection{Aggregate Sampling} 
A specific form of a linear combination of graph signals is referred to as \textit{aggregate sampling}. 

For clarity, we shall first establish an interpretation of sampling in classical signal processing through its graph counterpart -- sampling on a directed circular graph (Fig. \ref {fig:sig-on-graph}).  Consider  a graph signal, $\mathbf{x}$, at a vertex/instant $n$. If the signal is observed at this vertex/instant only, then its value is $y_0(n)=x(n)$. Upon applying the graph shift operator, we have $\mathbf{y}_1=\mathbf{A}\mathbf{x}$, then for the same vertex, $n$, we have $y_1(n)=x(n-1).$ If we continue this “shift and observe” operation on the directed circular graph $N$ times at the same vertex/instant, $n$, we will eventually have all signal values $x(n), x(n-1), \dots, x(n-N+1)$ observed at vertex $n$. 

To proceed with signal reconstruction, observe that if the shifts are stopped after $M<N$ steps, the available signal samples will be $x(n), x(n-1), \dots, x(n-M+1)$. From this reduced set of measurements/samples we can still recover the full graph signal, $\mathbf{x}$, using compressive sensing based reconstruction methods, if the appropriate reconstruction conditions are met.  

\noindent \textbf{Principle of aggregate sampling on arbitrary graph.} The same procedure can be applied to a signal observed in the same way on an arbitrary graph.
Assume that we observe the graph signal at  only one vertex, $n$, and obtain one graph signal sample 
$$y_0(n)=x(n),$$
which will be considered as the measurement $y(0)=y_0(n)$.

\begin{figure}[thpb]
	\centering
%	\includegraphics[]{MNE_fig_shift}(a))
	
%	\vspace{5mm}
	
%	\includegraphics[]{MNE_fig_shifted}(b)

	\includegraphics[]{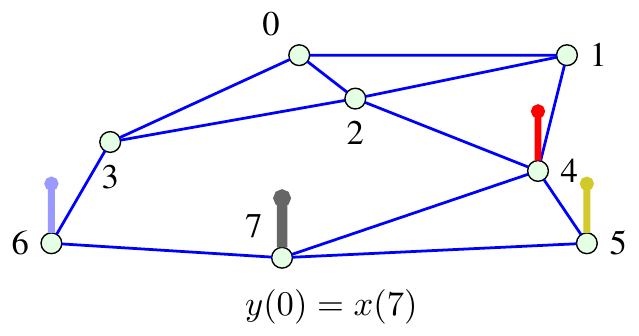}

	(a) signal $\mathbf{x}$

	\vspace{5mm}
	
	\includegraphics[]{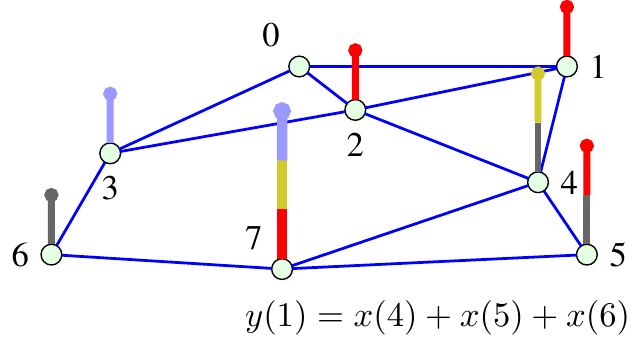}
	
	(b) shifted signal $\mathbf{Ax}$
	
	\caption{Principle of aggregate sampling. (a) A graph signal $\mathbf{x}$. (b) Its graph shifted version $\mathbf{Ax}$. For example, for a graph signal value observed at the vertex $n=7$ in the graph in (a) the measurement is $y(0)=x(7)$,  and the aggregate measurement at the same vertex, $n=7$, after the graph signal is shifted, is equal to  $y(1)=x(4)+x(5)+x(6)$ in (b). These two observations, $y(0)$ and $y(1)$, would be sufficient to reconstruct a signal whose sparsity degree is $K=2$ with nonzero values at the known spectral index positions, $k_1$ and $k_2$, if the reconstruction condition (\ref{rankK}) is satisfied for the matrix $\mathbf{A}_{MN}=\mathbf{B}_{MN}\mathbf{U}$ at the specified spectral index positions.}
	\label{fig:MNE_fig_shift}
\end{figure}

This graph signal may now be “graph shifted” to produce $\mathbf{y}_1=\mathbf{A}\mathbf{x}$. Recall that in a one-step signal shift on a graph, all signal samples will move by one step along the graph edges, as described in detail in Section \ref{graphshiftsec} and illustrated in Fig. \ref{fig:MNE_fig_shift}. The sample of a graph signal at vertex $n$ will now be a sum of all signal samples that have shifted to this vertex. Its value is obtained as an inner product of the $m$th row of the adjacency matrix, $\mathbf{A}$, and the original signal vector, $\mathbf{x}$. The value of graph shifted signal  at the vertex $n$, is therefore given by
$$y_1(n)=\sum_m A_{nm}x(m),$$
and represents a linear combination of some of the signal samples, which is now  considered as the measurement $y(1)=y_1(n)$.

One more signal shift on the graph yields
$$y_2(n)=\sum_m A^{(2)}_{nm}x(m),$$ 
where  $A^{(2)}_{nm}$ are the elements of matrix $\mathbf{A}^2=\mathbf{A}\mathbf{A}$ (see {\color{blue}  Property $M_2$ in Part I, Section \ref{I-section_graph_properties}}). 
Such an observed value, after two one step shifts, $y_2(n)$ at a vertex $n$, represents a new linear combination of some signal samples and will be considered as the measurement $y(2)=y_2(n)$.
 
If we proceed with shifts $M=N$ times, a system of $N$ linear equations, $\mathbf{y}=\mathbf{B}_{MN}\mathbf{x}$, is obtained from which  all signal values, $x(n)$, can be calculated. If we stop at $M<N$, the signal can still be recovered using compressive sensing based reconstruction methods if the signal is sparse and the reconstruction conditions are met.

Instead of $M$ signal samples (instants) at one vertex, we may use, for example, $P$ samples at vertex $n$ and $(M-P)$ samples from a vertex $m$. Other combinations of vertices and samples may be also used to obtain $M$ measurements and to fully reconstruct a signal.  

\subsection{Filter Bank on a Graph}

\begin{figure}
	\centering
	\includegraphics[scale=1]{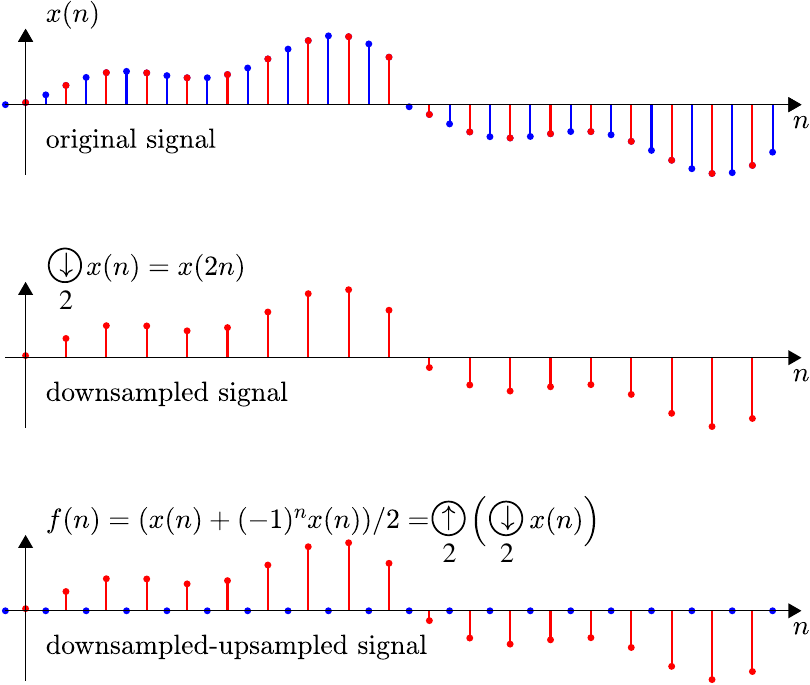}
	\caption{Principle of a signal, $x(n)$, downsampling and upsampling in the classical time domain.}
	\label{DOW_UP_ill}
\end{figure}

Subsampling and upsampling are the two standard operators used to alter the scale at which the signal is processed. Subsampling of a signal by a factor of 2, followed by the corresponding upsampling, can be described in classical signal processing by 
$$f(n)=\frac{1}{2}\Big(x(n)+(-1)^nx(n)\Big)=\frac{1}{2}\Big((1+(-1)^n)x(n)\Big),$$
as illustrated in Fig. \ref{DOW_UP_ill}.

This is the basic operation used in multiresolution approaches based on filter banks and can be extended to signals on graphs in the following way. Consider a graph with the set of vertices $\mathcal{V}$. Any set of vertices can be considered as a union of two disjoint subsets $\mathcal{E}$ and $\mathcal{H}$, such that $\mathcal{V}=\mathcal{E} \cup \mathcal{H}$ and $\mathcal{E} \cap \mathcal{H} = \emptyset$. The subsampling-upsampling procedure can then be performed in the following two steps:

\begin{enumerate}
	\item 
	Subsample the signal on a graph by keeping only signal values on the vertices $n \in \mathcal{E}$, while not altering the original graph topology, 
	\item 
	Upsample the graph signal by setting the signal values for the vertices  $n \notin \mathcal{E}$ to zero.
\end{enumerate}
This combined subsampling-upsampling operation produces a graph signal 
$$f(n)=\frac{1}{2}\Big(1+(-1)^{\beta_{\mathcal{E}}(n)}\Big)x(n),$$
where 
$$\beta_{\mathcal{E}}(n)=
\begin{cases}
0, & \text{ if } n \in \mathcal{E} \\
1, & \text{ if }n \in \mathcal{H}.
\end{cases}
$$
The values of the resulting graph signal, $f(n)$, are therefore $f(n)=x(n)$ if $ n \in \mathcal{E}$ and $f(n)=0$ elsewhere.

The vector form of the subsamped-upsampled graph signal, $f(n)$, which comprises all $n \in \mathcal{V}$, is given by
\begin{equation}\mathbf{f}=\frac{1}{2}
(\mathbf{x}+
\mathbf{J}
_{\mathcal{E}}
\mathbf{x})
=\frac{1}{2}(\mathbf{I}+\mathbf{J}_{\mathcal{E}})\mathbf{x}, \label{Fnsubup}
\end{equation}
where 
$\mathbf{J}_{\mathcal{E}}=\textrm{diag}((-1)^{\beta_{\mathcal{E}}(n)}),~n\in \mathcal{V}$. 

\begin{figure}
	\centering
	\includegraphics[scale=1]{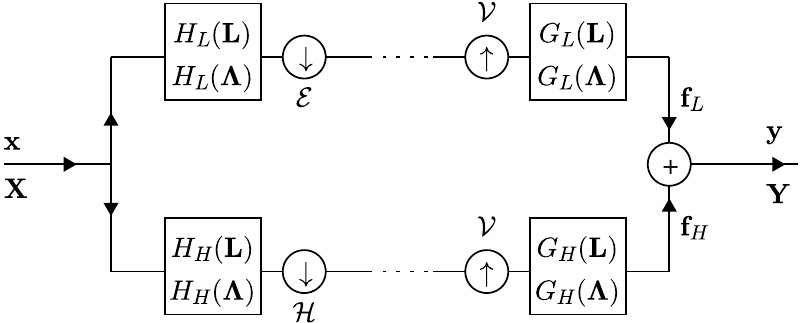}
	\caption{Principle of a filter bank for a graph signal.}
	\label{Filter_bank_Wavelet_GraphR}
\end{figure}

The focus of our analysis will be on the two-channel wavelet filter bank on a graph, shown in Fig. \ref{Filter_bank_Wavelet_GraphR}. As in the classical wavelet analysis framework for temporary signals, such a filter bank provides decomposition of a graph signal into the corresponding low-pass (smooth)  and  high-pass (fast-varying) constituents. The analysis side (left part of the system in Fig. \ref{Filter_bank_Wavelet_GraphR}) consists of two channels with filters characterized by the vertex domain operators $H_L(\mathbf{L})$ and $H_H(\mathbf{L})$, with the corresponding spectral domain operators $H_L(\mathbf{\Lambda})$ and $H_H(\mathbf{\Lambda})$. The operator $H_L(\mathbf{L})$ acts as a low-pass filter, transferring the low-pass components  of the graph signal, while the operator $H_H(\mathbf{L})$ does the opposite, acting as a high-pass filter. The low-pass filter, $H_L(\mathbf{H})$, is followed by a downsampling operator which keeps only the graph signal values, $\mathbf{x}$, at the vertices  $n \in \mathcal{E}$. Similarly, the high-pass filtering with the operator $H_H(\mathbf{L})$, is subsequently followed by a downsampling to the vertices $n \in \mathcal{H}$. These operations are crucial to alter the scale at which the graph signal is processed.
	
	The synthesis side (right part in Fig. \ref{Filter_bank_Wavelet_GraphR}), comprises the complementary upsampling and filtering operations, aiming to perform the graph signal reconstruction based on the upsampled versions, $\frac{1}{2}(\mathbf{I}+\mathbf{J}_{\mathcal{E}})H_L(\mathbf{L})\mathbf{x}$ and $\frac{1}{2}(\mathbf{I}+\mathbf{J}_{\mathcal{H}})H_H(\mathbf{L})\mathbf{x}$, of signals obtained on the filter bank analysis side. Therefore, upon performing the upsampling of these signals onto the original set of vertices, $\mathcal{V}$, by adding zeros to the complementary sets of vertices, filtering is performed by adequate low-pass, $G_L(\mathbf{L})$, and high-pass, $G_H(\mathbf{L})$, filters, to replace the zeros with meaningful values, as required for a successful reconstruction of the original signal. As in the classical wavelet analysis, to achieve the perfect (distortion-free) reconstruction it is necessary to conveniently design the analysis filters, $H_L(\mathbf{L})$ and $H_H(\mathbf{L})$, and the synthesis filters, $G_L(\mathbf{L})$ and $G_H(\mathbf{L})$, as well as to determine adequate downsampling and upsampling operators. 
	
	It will be shown that the spectral folding phenomenon, described by (\ref{I-bipartite_for_wavelets2}) in Part I, characterized by the specific spectral symmetry in the case of bipartite graphs, can be used to form the basis for the two-channel filter bank framework discussed in this Section.

Consider a graph signal, $\mathbf{x}$, and the filter-bank as in Fig. \ref{Filter_bank_Wavelet_GraphR}. If the graph signal, $\mathbf{x}$, passes through a low-pass analysis filter, $H_L(\mathbf{L})$, the output signal is $H_L(\mathbf{L})\mathbf{x}$. According to (\ref{Fnsubup}), the downsampled-upsampled form of the output signal, $H_L(\mathbf{L})\mathbf{x}$, is given by $\frac{1}{2}(\mathbf{I}+\mathbf{J}_{\mathcal{E}})H_L(\mathbf{L})\mathbf{x}$. After the syntheses filter, $G_L(\mathbf{L})$,  the graph signal output becomes 
\begin{equation} \mathbf{f}_L
=\frac{1}{2}G_L(\mathbf{L})(\mathbf{I}+\mathbf{J}_{\mathcal{E}})H_L(\mathbf{L})\mathbf{x}. \label{LoFL}
\end{equation}
The same holds for the high-pass part 
\begin{equation}\mathbf{f}_H
=\frac{1}{2}G_H(\mathbf{L})(\mathbf{I}+\mathbf{J}_{\mathcal{H}})H_H(\mathbf{L})\mathbf{x}, \label{HiFH}
\end{equation}
where $\mathbf{J}_{\mathcal{H}}=-\mathbf{J}_{\mathcal{E}}=\textrm{diag}((-1)^{1-\beta_{\mathcal{E}}(n)})$ and 
\begin{equation}\mathbf{J}_{\mathcal{H}}+\mathbf{J}_{\mathcal{E}}=\mathbf{0}. \label{HiFH0}
\end{equation}
The overall output is a sum of these two signals, as illustrated in Fig. \ref{Filter_bank_Wavelet_GraphR}, which after rearranging of terms gives 
\begin{gather} \mathbf{y}=\mathbf{f}_L+\mathbf{f}_H
=\frac{1}{2}(G_L(\mathbf{L})H_L(\mathbf{L})+
G_H(\mathbf{L})H_H(\mathbf{L}))\mathbf{x}+ \nonumber
\\
\frac{1}{2}(G_L(\mathbf{L})\mathbf{J}_{\mathcal{E}}H_L(\mathbf{L})
+G_H(\mathbf{L})\mathbf{J}_{\mathcal{H}}H_H(\mathbf{L}))\mathbf{x}.
\end{gather} 

The perfect reconstruction condition, $ \mathbf{y}= \mathbf{x}$, is then achieved if 
\begin{gather}
G_L(\mathbf{L}) H_L(\mathbf{L})+
G_H(\mathbf{L}) H_H(\mathbf{L})=2 \mathbf{I} \label{SigRecCond}, \\
G_L(\mathbf{L})\mathbf{J}_{\mathcal{E}}H_L(\mathbf{L})
-G_H(\mathbf{L})\mathbf{J}_{\mathcal{E}}H_H(\mathbf{L})=\mathbf{0}. \label{SigRecCondAL}
\end{gather}

\noindent\textbf{Spectral solution.} For the spectral representation of the filter-bank signals in the domain of Laplacian basis functions, we will use the decomposition of the graph Laplacian in the form
\begin{equation}\mathbf{F}=\mathbf{U}^T\mathbf{f}=\frac{1}{2}
(\mathbf{U}^T\mathbf{x}+
\mathbf{U}^T\mathbf{J}
_{\mathcal{E}}
\mathbf{x})=\frac{1}{2}
(\mathbf{X}+\mathbf{X}^{(alias)}),
\label{FnsubupLL}
\end{equation}
where $\mathbf{X}^{(alias)}=\mathbf{U}^T\mathbf{J}
_{\mathcal{E}}
\mathbf{x}$ is the aliasing spectral component. 

In the case of \textit{bipartite graphs}, the matrix operator $\mathbf{U}^T\mathbf{J}
_{\mathcal{E}}$ produces the transformation matrix $\mathbf{U}^T$ with reversed (left-right flipped) order of eigenvectors. This is obvious from (\ref{I-bipartite_for_wavelets2}) in Part I, since
\begin{align*}
\mathbf{U}^T\mathbf{J}_{\mathcal{E}}&=\begin{bmatrix}
\mathbf{u}_{{0}}  &
\mathbf{u}_{{1}}&
\dots & 
\mathbf{u}_{{N-1}}
\end{bmatrix}^T\mathbf{J}_{\mathcal{E}}
\\
&=\left[\begin{array}{c c  c  c}
~~\mathbf{u}_{{0\mathcal{E}}} & ~~\mathbf{u}_{{1\mathcal{E}}} &  & ~~\mathbf{u}_{{N-1\mathcal{E}}} \\
-\mathbf{u}_{{0\mathcal{H}}}&
-\mathbf{u}_{{1\mathcal{H}}} & \dotsm & -\mathbf{u}_{{N-1\mathcal{H}}}
\end{array}\right]^T\\
&=\begin{bmatrix}
\mathbf{u}_{{N-1}}  &
\mathbf{u}_{{N-2}}&
\dots & 
\mathbf{u}_{{0}}
\end{bmatrix}^T=\mathbf{U}^T_{\text{LR}}
\end{align*}
where $$\mathbf{u}_k=\begin{bmatrix}
\mathbf{u}_{k\mathcal{E}}\\\mathbf{u}_{k\mathcal{H}}
\end{bmatrix},~\mathbf{u}_{N-1-k}=\begin{bmatrix}
\mathbf{u}_{k\mathcal{E}}\\-\mathbf{u}_{k\mathcal{H}}
\end{bmatrix},~k=0,1,\dots N-1,$$ 
and 
$$\mathbf{U}_{\text{LR}}=\begin{bmatrix}
\mathbf{u}_{{N-1}}  &
\mathbf{u}_{{N-2}}&
\dots & 
\mathbf{u}_{{0}}
\end{bmatrix}$$
 is a left-right flipped version of the eigenvector matrix $$\mathbf{U}=\begin{bmatrix}
 \mathbf{u}_{{0}}  &
 \mathbf{u}_{{1}}&
 \dots & 
 \mathbf{u}_{{N-1}}
 \end{bmatrix}.$$
The element-wise form of equation (\ref{FnsubupLL}) is given by
$$F(k)=\frac{1}{2}(X(k)+X(N-1-k)).$$
For \textit{bipartite graphs and the normalized graph Laplacian}, we can write 
$$F(\lambda_k)=\frac{1}{2}(X(\lambda_k)+X(2-\lambda_k)).$$
The second term in $F(\lambda_k)$ represents an aliasing component of the GDFT of the original signal.

The spectral representation of (\ref{SigRecCond}) is obtained with a left-multiplication by $\mathbf{U}^T$ and a right-multiplication by $\mathbf{U}$, 
\begin{gather}
\mathbf{U}^TG_L(\mathbf{L})\mathbf{U}\mathbf{U}^T H_L(\mathbf{L})\mathbf{U}+
\mathbf{U}^TG_H(\mathbf{L})\mathbf{U}\mathbf{U}^T H_H(\mathbf{L})\mathbf{U}=2 \mathbf{I} \nonumber,
\end{gather} 
having in mind that we can add $\mathbf{U}^T\mathbf{U}=\mathbf{U}\mathbf{U}^T=\mathbf{I}$  between $G_L(\mathbf{L})$ and $H_L(\mathbf{L})$, and  between $G_H(\mathbf{L})$ and $H_H(\mathbf{L})$. 
Using the spectral domain definition of the transfer functions, $\mathbf{U}^T H_L(\mathbf{L}) \mathbf{U}=H_L(\mathbf{\Lambda})$, we get the spectral domain form of the reconstruction condition (\ref{SigRecCond}) as
\begin{gather}
G_L(\mathbf{\Lambda})H_L(\mathbf{\Lambda})+G_H(\mathbf{\Lambda})H_H(\mathbf{\Lambda})=2\mathbf{I}. \label{firstBPR}
\end{gather} 
For the aliasing part in equation (\ref{SigRecCondAL}), the left-multiplication is performed by $\mathbf{U}^T$, while the right-multiplication is done by $\mathbf{U}^T_{\text{LR}}$. The first term in (\ref{SigRecCondAL}) is then of the form
\begin{gather}
\mathbf{U}^TG_L(\mathbf{L})\mathbf{U}\mathbf{U}^T\mathbf{J}_{\mathcal{E}}H_L(\mathbf{L})\mathbf{U}_{\text{LR}} 
=\mathbf{U}^TG_L(\mathbf{L})\mathbf{U}\mathbf{U}^T_{\text{LR}}H_L(\mathbf{L})\mathbf{U}_{\text{LR}}\nonumber \\
=G_L(\mathbf{\Lambda})H_L^{(R)}(\mathbf{\Lambda}),
\end{gather}
since $\mathbf{U}^T\mathbf{J}_{\mathcal{E}}=\mathbf{U}^T_{\text{LR}}$ and $\mathbf{U}^T_{\text{LR}}\mathbf{U}_{\text{LR}}=\mathbf{I}$. The term $$H_L^{(R)}(\mathbf{\Lambda})=\mathbf{U}^T_{\text{LR}}H_L(\mathbf{L})\mathbf{U}_{\text{LR}}=H_L(2\mathbf{I}-\mathbf{\Lambda})$$
is just a reversed order version of the diagonal matrix $H_L(\mathbf{\Lambda})$, with diagonal elements $H_L(\lambda_{N-1-k})=H_L(2-\lambda_{k})$ instead of $H_L(\lambda_{k})$. 

The same holds for the second term in (\ref{SigRecCondAL}) which is equal to $G_H(\mathbf{L})\mathbf{J}_{\mathcal{H}}H_H(\mathbf{L})$, yielding the final spectral form of the aliasing condition in (\ref{SigRecCondAL}) as
\begin{gather}
G_L(\mathbf{\Lambda})H_L(2\mathbf{I}-\mathbf{\Lambda})-G_H(\mathbf{\Lambda})H_H(2\mathbf{I}-\mathbf{\Lambda})=\mathbf{0}. \label{firstBPR2} 
\end{gather} 
An element-wise solution to the system in (\ref{SigRecCond})-(\ref{SigRecCondAL}), for bipartite graphs and the normalized graph Laplacian,   according to (\ref{firstBPR})  and (\ref{firstBPR2}), reduces to 
\begin{gather}
G_L(\lambda_k)H_L(\lambda_k)+G_H(\lambda_k)H_H(\lambda_k)=2,\label{RecConDFB} \\
G_L(\lambda_k)H_L(2-\lambda_k)-G_H(\lambda_k)H_H(2-\lambda_k)=0 \label{AliConDFB}.
\end{gather}
 
\begin{Remark}
\textit{A quadratic mirror filter solution} would be such that for the designed transfer function of the low-pass analysis filter, $H_L(\lambda)$, the other filters are 
\begin{gather}
G_L(\lambda)=H_L(\lambda), \nonumber \\ 
H_H(\lambda)=H_L(2-\lambda), \nonumber \\
G_H(\lambda)=H_H(\lambda)=H_L(2-\lambda). \label{quadraticM}
\end{gather}
For this solution, the design equation is given by
\begin{equation}H_L^2(\lambda)+H_L^2(2-\lambda)=2,
\label{recCondQMF}
\end{equation}
while the aliasing cancellation condition,  (\ref{AliConDFB}), is always satisfied.  
\end{Remark}

An example of such a system would be an ideal low-pass filter, defined by  
$H_L(\lambda)=\sqrt{2}$ for $\lambda<1$ and  $H_L(\lambda)=0$ elsewhere. Since  $H_H(\lambda)=H_L(2-\lambda)$  holds for systems on bipartite graphs, this satisfies the reconstruction condition. For the vertex domain realization, an approximation of the ideal filter with a finite neighborhood filtering relation would be required.

\begin{Example}  Consider a simple form of the low-pass system 
		$$H_L^2(\lambda)=2-\lambda,$$
	which satisfies the design equation, $H_L^2(\lambda)+H_L^2(2-\lambda)=2.$ 
	 It also satisfies the condition that its form is of low-pass type for the  normalized Laplacian of bipartite graphs, $H_L^2(\lambda_0)=2-\lambda_0=2$, since $\lambda_0=0$, and $H_L^2(\lambda_{\max})=2-\lambda_{\max}=0$, as $\lambda_{\max}=2$. The vertex domain system operators which satisfy all four quadratic mirror analysis and synthesis filters in (\ref{quadraticM}), are \begin{gather*}
	 H_L(\mathbf{\Lambda})=\sqrt{\mathbf{2\mathbf{I}-\Lambda}}, \hspace{5mm} G_L(\mathbf{\Lambda})=H_L(\mathbf{\Lambda})=\sqrt{\mathbf{2\mathbf{I}-\Lambda}}, \\ H_H(\mathbf{\Lambda})=H_L(\mathbf{2I-\Lambda})=\sqrt{\mathbf{\Lambda}},  \hspace{5mm} G_H(\mathbf{\Lambda})=H_H(\mathbf{\Lambda})=\sqrt{\mathbf{\Lambda}}.
	 \end{gather*}
	 The spectral domain filtering form for the low-pass part of graph signal is then obtained from (\ref{LoFL}), as 
	\begin{gather*} \mathbf{F}_L=\mathbf{U}^T\mathbf{f}_L
	=\frac{1}{2}\mathbf{U}^TG_L(\mathbf{L})(\mathbf{I}+\mathbf{J}_{\mathcal{E}})H_L(\mathbf{L})\mathbf{x} \\
	=\frac{1}{2}\mathbf{U}^TG_L(\mathbf{L})\mathbf{U}\mathbf{U}^T(\mathbf{I}+\mathbf{J}_{\mathcal{E}})H_L(\mathbf{L})\mathbf{U}_{\text{LR}}\mathbf{U}^T_{\text{LR}}\mathbf{U}\mathbf{X} \\
	=\frac{1}{2}G_L(\mathbf{\Lambda})H_L(\mathbf{\Lambda})\mathbf{X}+\frac{1}{2}G_L(\mathbf{\Lambda})H_L(2\mathbf{I}-\mathbf{\Lambda})\mathbf{X}_{\text{UD}} 
	\end{gather*} 
	since $\mathbf{U}^T\mathbf{U}=\mathbf{I}$, $\mathbf{U}^T_{\text{LR}}\mathbf{U}_{\text{LR}}=\mathbf{I}$, $\mathbf{U}^T\mathbf{J}_{\mathcal{E}}=\mathbf{U}^T_{\text{LR}}$,  $\mathbf{U}^T_{\text{LR}}\mathbf{U}=\mathbf{I}_{\text{LR}}$, and $\mathbf{I}_{\text{LR}}\mathbf{X}=\mathbf{X}_{\text{UD}}$, where $\mathbf{I}_{\text{LR}}$ is an anti-diagonal (backward) identity matrix, and $\mathbf{X}_{\text{UD}}$ is the GDFT vector, $\mathbf{X}$, with elements flipped upside-down.   
	
	The same holds for the high-pass part in (\ref{HiFH}), to yield
	\begin{gather*} \mathbf{F}_H
	=\frac{1}{2}\mathbf{U}^TG_H(\mathbf{L})(\mathbf{I}+\mathbf{J}_{\mathcal{H}})H_H(\mathbf{L})\mathbf{x} \\
	=\frac{1}{2}G_H(\mathbf{\Lambda})H_H(\mathbf{\Lambda})\mathbf{X}-\frac{1}{2}G_H(\mathbf{\Lambda})H_H(2\mathbf{I}-\mathbf{\Lambda})\mathbf{X}_{\text{UD}}
	\end{gather*}  
and
$$\mathbf{F}_L+\mathbf{F}_H=\mathbf{X}.$$
\begin{figure}
	\centering
	\includegraphics[  width=3.5 cm]{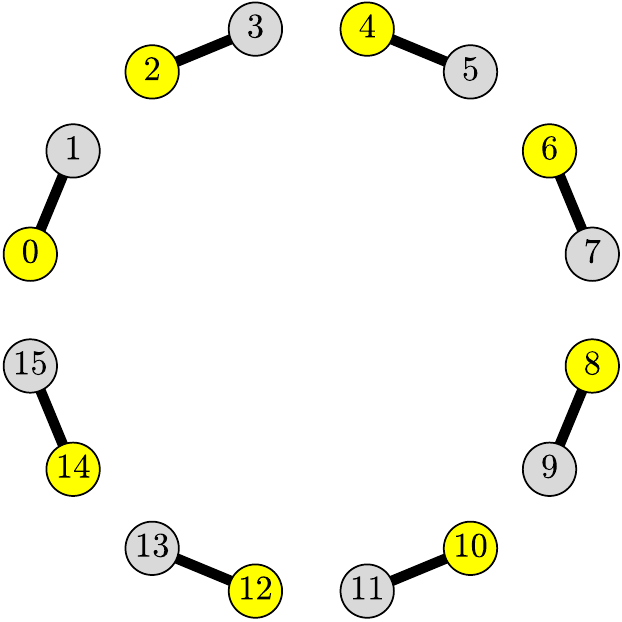} (a) \hfill
	\includegraphics[  width=3.5 cm]{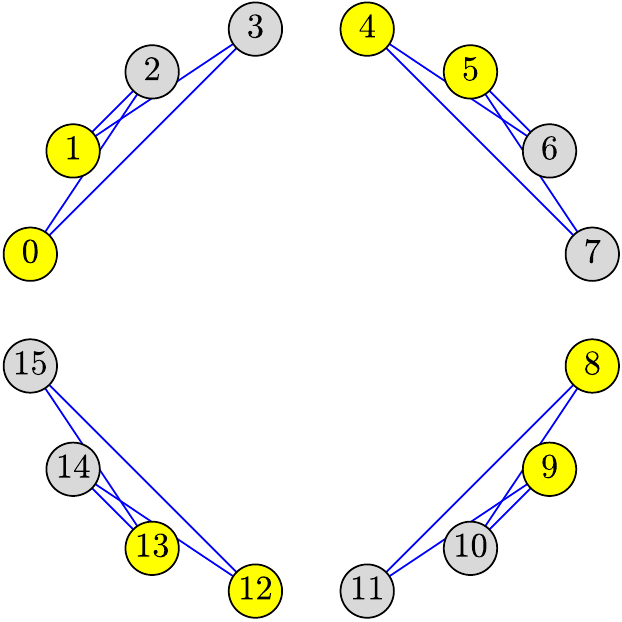} (b)
	
	\caption{Bipartite graph for the Haar wavelet transform   with $N=16$ vertices. (a) Vertices in yellow are used for the low-pass part of the signal and correspond to the set  $\mathcal{E}$, while the vertices in gray belong to the set  $\mathcal{H}$. This is the highest two-vertex resolution level for the Haar wavelet. (b) Graph for a four-vertex resolution level in the Haar wavelet. }
	\label{haar_gr_1}
\end{figure}

\begin{figure*}[hbtp]
	\begin{align}
	\mathbf{L}=
	{	\color{blue}
		\begin{matrix}
		\text{\footnotesize 0}\\
		\text{\footnotesize 2}\\
		\text{\footnotesize 4}\\
		\text{\footnotesize 6}\\
		\text{\footnotesize 8}\\
		\text{\footnotesize 10}\\
		\text{\footnotesize 12}\\
		\text{\footnotesize 14}\\
		\text{\footnotesize 1}\\
		\text{\footnotesize 3}\\
		\text{\footnotesize 5}\\
		\text{\footnotesize 7}\\
		\text{\footnotesize 9}\\
		\text{\footnotesize 11}\\
		\text{\footnotesize 13}\\
		\text{\footnotesize 15}
		\end{matrix}
	} 
	\left[
	\begin{array}{*{16}r}
	1 &  0    &  0    & 0 & 0 &  0    &  0    &  0    &  \cellcolor[gray]{0.9} -1    &  \cellcolor[gray]{0.9} 0    &  \cellcolor[gray]{0.9} 0    & \cellcolor[gray]{0.9} 0    &  \cellcolor[gray]{0.9} 0    & \cellcolor[gray]{0.9} 0 & \cellcolor[gray]{0.9} 0 & \cellcolor[gray]{0.9} 0  \\
	0    &  1 & 0 &  0    &  0    &  0    &  0    &   0    &  \cellcolor[gray]{0.9} 0    &  \cellcolor[gray]{0.9} -1    & \cellcolor[gray]{0.9} 0    &  \cellcolor[gray]{0.9} 0    & \cellcolor[gray]{0.9} 0 & \cellcolor[gray]{0.9} 0 & \cellcolor[gray]{0.9} 0  & \cellcolor[gray]{0.9} 0   \\
	0    & 0 &  1 &  0    &  0    & 0 &  0    & 0 &  \cellcolor[gray]{0.9} 0    &  \cellcolor[gray]{0.9} 0    & \cellcolor[gray]{0.9} -1    &  \cellcolor[gray]{0.9} 0    & \cellcolor[gray]{0.9} 0 & \cellcolor[gray]{0.9} 0 & \cellcolor[gray]{0.9} 0  & \cellcolor[gray]{0.9} 0   \\
	0 &  0    &  0    &  1 &  0    &  0    &  0    &  0    &  \cellcolor[gray]{0.9} 0    &  \cellcolor[gray]{0.9} 0    & \cellcolor[gray]{0.9} 0    &  \cellcolor[gray]{0.9} -1    & \cellcolor[gray]{0.9} 0 & \cellcolor[gray]{0.9} 0 & \cellcolor[gray]{0.9} 0  & \cellcolor[gray]{0.9} 0   \\
	0 &  0    &  0    &  0    &  1 &  0    &  0    &  0    &  \cellcolor[gray]{0.9} 0    &  \cellcolor[gray]{0.9} 0    & \cellcolor[gray]{0.9} 0    &  \cellcolor[gray]{0.9} 0    & \cellcolor[gray]{0.9} -1 & \cellcolor[gray]{0.9} 0 & \cellcolor[gray]{0.9} 0  & \cellcolor[gray]{0.9} 0   \\
	0    &  0    & 0 &  0    &  0    &  1 &  0   &  0    &  \cellcolor[gray]{0.9} 0    &  \cellcolor[gray]{0.9} 0    & \cellcolor[gray]{0.9} 0    &  \cellcolor[gray]{0.9} 0    & \cellcolor[gray]{0.9} 0 & \cellcolor[gray]{0.9} -1 & \cellcolor[gray]{0.9} 0  & \cellcolor[gray]{0.9} 0   \\
	0    &  0    &  0    &  0    &  0    &  0    &  1 &  0    &  \cellcolor[gray]{0.9} 0    &  \cellcolor[gray]{0.9} 0    & \cellcolor[gray]{0.9} 0    &  \cellcolor[gray]{0.9} 0    & \cellcolor[gray]{0.9} 0 & \cellcolor[gray]{0.9} 0 & \cellcolor[gray]{0.9} -1  & \cellcolor[gray]{0.9} 0   \\
	0    &  0    & 0 &  0    &  0    &  0    &  0    &  1 &  \cellcolor[gray]{0.9} 0    &  \cellcolor[gray]{0.9} 0    & \cellcolor[gray]{0.9} 0    &  \cellcolor[gray]{0.9} 0    & \cellcolor[gray]{0.9} 0 & \cellcolor[gray]{0.9} 0 & \cellcolor[gray]{0.9} 0  & \cellcolor[gray]{0.9} -1   \\
	\cellcolor[gray]{0.9} -1    &  \cellcolor[gray]{0.9} 0    & \cellcolor[gray]{0.9} 0    &  \cellcolor[gray]{0.9} 0    & \cellcolor[gray]{0.9} 0 & \cellcolor[gray]{0.9} 0 & \cellcolor[gray]{0.9} 0  & \cellcolor[gray]{0.9} 0   & 1 &  0    &  0    &  0    &  0    &  0    &  0    &  0     \\
	\cellcolor[gray]{0.9} 0    &  \cellcolor[gray]{0.9} -1   & \cellcolor[gray]{0.9} 0    &  \cellcolor[gray]{0.9} 0    & \cellcolor[gray]{0.9} 0 & \cellcolor[gray]{0.9} 0 & \cellcolor[gray]{0.9} 0  & \cellcolor[gray]{0.9} 0     &  1 & 0 &  0    &  0    &  0    &  0    &  0     \\
	\cellcolor[gray]{0.9} 0    &  \cellcolor[gray]{0.9} 0    & \cellcolor[gray]{0.9} -1    &  \cellcolor[gray]{0.9} 0    & \cellcolor[gray]{0.9} 0 & \cellcolor[gray]{0.9} 0 & \cellcolor[gray]{0.9} 0  & \cellcolor[gray]{0.9} 0   &  0    & 0 &  1 &  0    &  0    & 0 &  0    &  0     \\
	\cellcolor[gray]{0.9} 0    &  \cellcolor[gray]{0.9} 0    & \cellcolor[gray]{0.9} 0    &  \cellcolor[gray]{0.9} -1    & \cellcolor[gray]{0.9} 0 & \cellcolor[gray]{0.9} 0 & \cellcolor[gray]{0.9} 0  & \cellcolor[gray]{0.9} 0    &  0    &  0    &  0    &  1 & 0 &  0    &  0    &  0     \\
	\cellcolor[gray]{0.9} 0    &  \cellcolor[gray]{0.9} 0    & \cellcolor[gray]{0.9} 0    &  \cellcolor[gray]{0.9} 0    & \cellcolor[gray]{0.9} -1 & \cellcolor[gray]{0.9} 0 & \cellcolor[gray]{0.9} 0  & \cellcolor[gray]{0.9} 0    &  0    &  0    &  0    & 0 &  1 &  0    & 0 &  0     \\
	\cellcolor[gray]{0.9} 0    &  \cellcolor[gray]{0.9} 0    & \cellcolor[gray]{0.9} 0    &  \cellcolor[gray]{0.9} 0    & \cellcolor[gray]{0.9} 0 & \cellcolor[gray]{0.9} -1 & \cellcolor[gray]{0.9} 0  & \cellcolor[gray]{0.9} 0    &  0    &  0    & 0 &  0    &  0    &  1 &  0    &  0     \\
	\cellcolor[gray]{0.9} 0    &  \cellcolor[gray]{0.9} 0    & \cellcolor[gray]{0.9} 0    &  \cellcolor[gray]{0.9} 0    & \cellcolor[gray]{0.9} 0 & \cellcolor[gray]{0.9} 0 & \cellcolor[gray]{0.9} -1  & \cellcolor[gray]{0.9} 0   &  0    &  0    &  0    &  0    & 0 &  0    &  1 &  0     \\
	\cellcolor[gray]{0.9} 0    &  \cellcolor[gray]{0.9} 0    & \cellcolor[gray]{0.9} 0    &  \cellcolor[gray]{0.9} 0    & \cellcolor[gray]{0.9} 0 & \cellcolor[gray]{0.9} 0 & \cellcolor[gray]{0.9} 0  & \cellcolor[gray]{0.9} -1    &  0    &  0    &  0    &  0    &  0    &  0    &  0    &  1 
	\end{array}\label{LL16Haar}
	\right] \\
	{	\color{blue}
		\begin{array}{*{16}c}
		\text{\footnotesize 0 \hspace{1 mm}  } &
		\text{\footnotesize 2 \hspace{1 mm} }  &
		\text{\footnotesize 4 \hspace{1 mm}  }  &
		\text{\footnotesize 6 \hspace{1 mm} }  &
		\text{\footnotesize 8 \hspace{0 mm}  }  &
		\text{\footnotesize 10 \hspace{0mm}  }  &
		\text{\footnotesize 12 \hspace{0mm} }  &
		\text{\footnotesize 14 \hspace{0.5 mm} }  &
		\text{\footnotesize 1 \hspace{0.5 mm} }  &
		\text{\footnotesize 3 \hspace{0.5 mm} }  &
		\text{\footnotesize 5 \hspace{0.5 mm} }  &
		\text{\footnotesize 7 \hspace{0.5 mm} }  &
		\text{\footnotesize 9 \hspace{0 mm} }  &
		\text{\footnotesize 11 \hspace{0mm} }  &
		\text{\footnotesize 13 \hspace{0mm} }  &
		\text{\footnotesize 15 \hspace{0mm} }
		\end{array}\nonumber
	}
	\end{align} 
\end{figure*}

Therefore, after the one-step filter-bank based decomposition on a bipartite graph, we have a new low-pass signal, $\mathbf{f}_L$, for which the nonzero values are at the vertices in $\mathcal{E}$, and a high-pass signal,  $\mathbf{f}_H$, with nonzero values only on $\mathcal{H}$. Note that the high-pass operator on the graph signal is the graph Laplacian, $\mathbf{L}$, while the low-pass operator is  $\mathbf{2\mathbf{I}-L}$, which easily reduces to $|\mathbf{L}|$, for the normalized graph Laplacian used here. 

Another simple transfer function that satisfies the design equation (\ref{recCondQMF}) is $H_L(\lambda)=\sqrt{2}\cos(\pi\lambda/4)$. A similar analysis can also  be done for this transfer function and other functions defined by (\ref{quadraticM}). 

The considered transfer functions $H_L(\lambda)=\sqrt{2-\lambda}$  and $H_L(\lambda)=\sqrt{2}\cos(\pi\lambda/4)$  have several disadvantages, the most important being that they are not sufficiently smooth in the spectral domain at the boundary  interval points \cite{stankovic2015digital}.  In addition, although the graph Laplacian, $\mathbf{L}$, is commonly sparse (with a small number of nonzero elements in large graphs), the transfer function form $H_L(\mathbf{L})=\sqrt{\mathbf{2I-L}}$ is not sparse. This is the reason to use other forms which are sufficiently smooth toward the boundary points, along with their polynomial approximations, $H_L(\mathbf{\Lambda})=c_0\mathbf{\Lambda}+c_1\mathbf{\Lambda}^2+\dots+c_{M-1}\mathbf{\Lambda}^{M-1}$, with the coefficients $c_0,c_1,\dots,c_{M-1}$, that approximate $H_L(\lambda)$ and $H_H(\lambda)=H_L(2-\lambda)$ for each $\lambda=\lambda_k$, $k=0,1,\dots,N-1$. This topic will be addressed in detail on a general form of graphs in Section \ref{sec7}.

\end{Example} 

The classic time-domain Haar wavelet (and scale) functions are easily obtained for a bipartite graph, such that $\mathcal{E}={0,2,4,\dots,N-2}$ and $\mathcal{H}={1,3,5,\dots,N-1}$, with the adjacency/weighting matrix defined by the elements $A_{mn}=1$, for $(m,n)\in \{(0,1),(2,3),\dots,(N-2,N-1)\}$,  as shown in Fig. \ref{haar_gr_1}(a). This adjacency matrix has the block form as in {\color{blue}equation (\ref{I-BlckDL}), Part I}. The corresponding graph Laplacian is given in (\ref{LL16Haar}). Its eigenvectors are equal to the wavelet transform functions. The bipartite graph for the four-vertex resolution level in the Haar wavelet transform is shown in  Fig. \ref{haar_gr_1}(b).

\section{Time-Varying Signals on Graphs}
\label{sec5}
We shall denote a time-varying signal by $x_p(n)$, where $n$ designates the vertex index and $p$ the discrete-time index. For uniform sampling in time, the index $p$ corresponds to the time instant $p\Delta t$, where  $\Delta t$ is the sampling interval. In general, this type of data can be considered within the graph Cartesian product framework (given in {\color{blue}Property $M_{15}$, Section \ref{I-section_graph_properties}, Part I}).
The resulting graph $\mathcal{G}=(\mathcal{V},\mathcal{B})$ follows as a Cartesian product of the given graph  $\mathcal{G}_1=(\mathcal{V}_1,\mathcal{B}_1)$ and a simple path (or circular) graph $\mathcal{G}_2=(\mathcal{V}_2,\mathcal{B}_2)$ that corresponds to the classical uniformly samples time-domain axis. 

\begin{Example} A graph topology for a time varying signal on a graph is shown in {\color{blue} Part I, Fig. \ref{I-fig:cartesian}}, where the graph vertices are designated by $1,2,3,4,5$ and time instants are denoted as the $a,b,c$ vertices on the 	path graph. The resulting Cartesian product graph, for the analysis of this kind of signals, is shown in {\color{blue} Part I, Fig. \ref{I-fig:cartesian}}. 
\end{Example}

The adjacency matrix of a Cartesian product of two graphs is then given by
$$\mathbf{A}=\mathbf{A}_1 \otimes \mathbf{I}_{N_2}+\mathbf{I}_{N_1} \otimes \mathbf{A}_2=\mathbf{A}_1\oplus\mathbf{A}_2,$$
where $\mathbf{A}_1 $ is the adjacency matrix of the graph of interest $\mathcal{G}_1$, and $\mathbf{A}_2$ is the adjacency matrix  for the path or circular graph, $\mathcal{G}_2$, which designates the sampling grid, while $N_1$ and $N_2$ denote, respectively,
the number of vertices in $\mathcal{G}_1$ and $\mathcal{G}_2$.

We will next consider a simple and important example of a time-varying signal defined on graph in an iterative way, which designates the \textit{diffusion process on a graph}  in time.

\subsection{Diffusion on Graph and Low Pass Filtering}

Consider the diffusion equation 
$$\partial \mathbf{x} / \partial t = -\alpha \mathbf{Lx}.$$ 
Its discrete-time form, at a time instant $p$, may be obtained by using the backward difference approximation of the partial derivative ($\partial \mathbf{x} / \partial t \sim  \mathbf{x}_{p+1}-\mathbf{x}_{p} $),  and has the form
$$  \mathbf{x}_{p+1}-\mathbf{x}_{p}= -\alpha \mathbf{L}\mathbf{x}_{p+1}$$
or 
$ \mathbf{x}_{p+1}(\mathbf{I}+\alpha \mathbf{L})=\mathbf{x}_p$
to produce
$$ \mathbf{x}_{p+1}=(\mathbf{I}+\alpha \mathbf{L})^{-1}\mathbf{x}_p.$$

On the other hand, the forward difference approximation ($\partial \mathbf{x} / \partial t \sim  \mathbf{x}_{p}-\mathbf{x}_{p-1} $) to the diffusion equation yields 
$$  \mathbf{x}_{p+1}-\mathbf{x}_{p}= -\alpha \mathbf{L}\mathbf{x}_{p}$$
or 
$$  \mathbf{x}_{p+1}=(\mathbf{I} -\alpha \mathbf{L})\mathbf{x}_{p}.$$

It is interesting to note that these iterative forms lead to the \textit{minimization of the quadratic form of a graph signal}, $E_x=\mathbf{x}\mathbf{L}\mathbf{x}^T$, (see {\color{blue}Section \ref{I-section_smoothness_of_eigenvectors}, Part I}).  The minimum of this quadratic form can be found  based on the steepest descent method, whereby the signal value at a time instant $p$ is moving in the direction opposite to the gradient, toward the energy minimum position, with a step constant $\alpha$. The gradient of the quadratic form, $E_x=\mathbf{x}\mathbf{L}\mathbf{x}^T$, is $\partial E_x / \partial \mathbf{x}^T =2\mathbf{x}\mathbf{L}$, which results in an iterative procedure 
\begin{equation} \mathbf{x}_{p+1}=\mathbf{x}_{p}-\alpha \mathbf{L}\mathbf{x}_{p}=(\mathbf{I} -\alpha \mathbf{L})\mathbf{x}_{p}. \label{firsttaub}
\end{equation}
 This relation can be used for simple and efficient filtering of graph signals (with the aim to minimize $E_x$ as a measure of signal smoothness). The spectral domain relation follows immediately, and has the form   
$$  \mathbf{X}_{p+1}=(\mathbf{I} -\alpha \mathbf{\Lambda})\mathbf{X}_{p}$$
or for every individual component
$$  X_{p+1}(k)=(1 -\alpha \lambda_k)X_{p}(k).$$

Recall that the eigenvalues, $\lambda_k$, represent the index of smoothness for a spectral vector (eigenvector), $\mathbf{u}_k$, with a small $\lambda_k$ indicating smooth slow-varying elements of the eigenvectors; therefore, for low-pass filtering we should retain the slow-varying eigenvectors in a spectral representation of the graph signal.   Obviously, these slow-varying components will pass through this system since 
 $(1 -\alpha \lambda_k)$ is close to $1$ for small $\lambda_k$, while the fast-varying components with a larger $\lambda_k$, are attenuated. This iterative procedure will converge if $|1-\alpha\lambda_{\max}|<1$. 
 
 In a stationary state of a diffusion process, the trivial minimal energy solution is obtained when 
 $$  \lim_{p \to \infty}  X_{p+1}(k)=(1 -\alpha \lambda_k)^{p+1}X_{0}(k),$$ that is,
 all components $X_{p+1}(k)$ tend to $0$, except for the constant component, $X_{p+1}(0)$, for which $\lambda_0=0$. This component therefore defines the stationary state (maximally smooth solution). In order to avoid this effect in the processing of data on graphs, and to retain several low-pass components (eigenvectors) in the signal, the iteration process in (\ref{firsttaub}) can be used in alternation with   
\begin{equation}  \mathbf{x}_{p+2}=(\mathbf{I} +\beta \mathbf{L})\mathbf{x}_{p+1}. \label{secondtaub}
\end{equation}
This is the basis for Taubin’s $\alpha-\beta$ algorithm, presented next.
\subsection{Taubin’s $\alpha-\beta$ algorithm}
When the two iterative processes in (\ref{firsttaub}) and (\ref{secondtaub})  are used in a successive order, the resulting system on a graph is referred to as Taubin’s $\alpha-\beta$ algorithm. This algorithm is widely used for low-pass filtering of data on graphs, since it is very simple, and admits efficient implementation in the vertex domain. 
 
\noindent\textit{ Definition:} Taubin’s $\alpha-\beta$ algorithm is a two-step iterative algorithm for efficient low-pass data filtering on graphs. Its two steps are defined in a unified way as
 \begin{equation}
 \mathbf{x}_{p+2}=(\mathbf{I} +\beta \mathbf{L})(\mathbf{I} -\alpha \mathbf{L})\mathbf{x}_{p}.
 \label{two-step-iterative1}
  \end{equation}
 The corresponding element-wise transfer function spectral domain of the two iteration steps in (\ref{two-step-iterative1}) is given by
 $$H(\lambda_k)=(1 +\beta \lambda_k)(1 -\alpha \lambda_k).$$
 After $K$ iterations of this algorithm, the spectral domain transfer function can be written as  
 \begin{equation}
 H_K(\lambda_k)=\Big((1 +\beta \lambda_k)(1 -\alpha \lambda_k)\Big)^K. \label{TaubinK}
 \end{equation}
 For some values of $\alpha <\beta$, this system can be a good and computationally very simple approximation of a graph low-pass filter.
 
  \begin{figure}[h]
 	\centering
 	\includegraphics[]{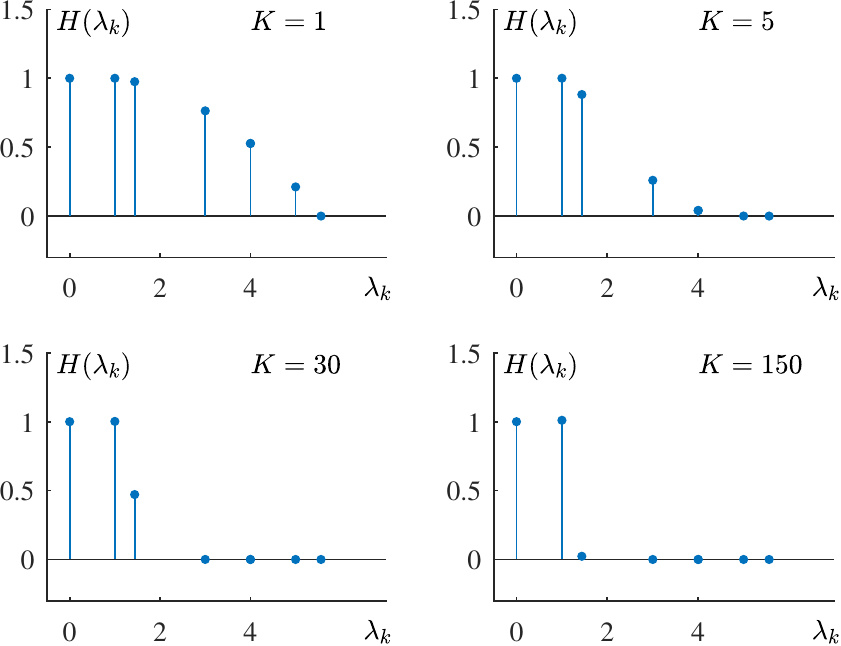}
 	\caption{Filter approximation in the spectral domain for a varying number of iterations, $K$, using Taubin's algorithm and the graph Laplacian matrix of the graph in Fig. \ref{fig:sig-arb-graph}.}
 	\label{Touring}
 \end{figure}

  \begin{Example} Consider the graph from {\color{blue} Fig.~\ref{fig:sig-arb-graph}(a)} and its graph Laplacian, $\mathbf{L}$. For the choice of parameters  $\alpha =0.1798$ and $\beta =0.2193$, the spectral transfer function in (\ref{TaubinK}) is shown in Fig. \ref{Touring} for the considered graph filter, and for the numbers of iterations in Taubin’s  algorithm $K=1, 5, 30$ and $150$. Observe how the transfer function, $H(\lambda_k)$, approaches the ideal low-pass form as the number of iterations, $K$, increases.  
  
 The task is next to low-pass filter the noisy signal from Fig. \ref{fig:filtering_Laplacian}(b), with the initial noisy signal is denoted by $\mathbf{x}_{0}$. Then $  \mathbf{x}_{1}=(\mathbf{I} -0.1545 \mathbf{L})\mathbf{x}_{0}$ is calculated using the corresponding graph Laplacian, followed by obtaining $  \mathbf{x}_{2}=(\mathbf{I} +0.1875 \mathbf{L})\mathbf{x}_{1}$. In the third and fourth iteration, the signal values $  \mathbf{x}_{3}=(\mathbf{I} -0.1545 \mathbf{L})\mathbf{x}_{2}$ and $  \mathbf{x}_{4}=(\mathbf{I} +0.1875 \mathbf{L})\mathbf{x}_{3}$ are calculated. This two-step iteration cycle is repeated $K=20$ times. The resulting signal is the same as the output of an ideal low-pass filter shown in Fig. \ref{fig:filtering_Laplacian}(c).
  %, as it can be expected from  Fig. \ref{Touring}(d).   
  
  Finally, the noisy signal from Fig. \ref{fig:MNE_fig_e} was filtered using  Taubin’s $\alpha-\beta$ algorithm, with  $\alpha=0.1$ and $\beta=0.1$, over $K=100$ iterations,  and the result is shown in Fig. \ref{MNE_fig_taub_filt}. Observe the reduced level of additive noise in the output.

\begin{figure}[thpb]
	\centering
	\includegraphics[]{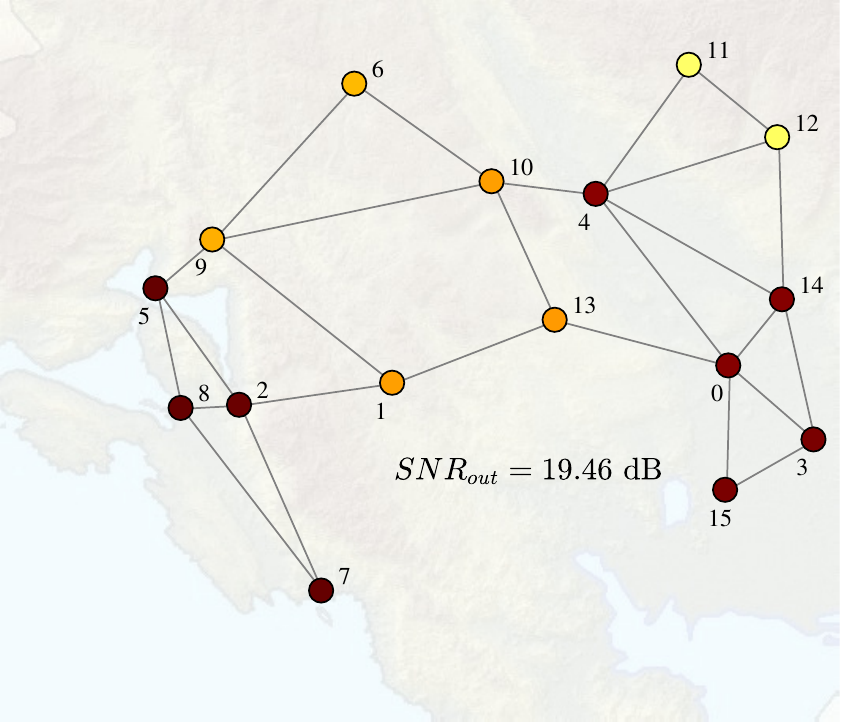}
	\caption{The noisy signal from Fig. \ref{fig:MNE_fig_e} was filtered using $K=100$ iterations of the Taubin two-step algorithm   with $\alpha=0.1$ and $\beta=0.1$.}
	\label{MNE_fig_taub_filt}
\end{figure}
\end{Example}
 
\section{Random Graph Signals}
\label{sec6}
This section extends the concepts of data analytics for deterministic signals on  graphs addressed so far, to introduce notions of random signals on graphs, their properties, and statistical graph-specific methods for their analysis. The main focus is on wide-sense stationary (WSS) data observed on graphs. In general, the stationarity of a signal is inherently related to the signal shift operator and its properties. 
We have already presented two approaches to define a shift on a graph (through the adjacency matrix and the graph Laplacian, and their spectral decompositions). These will be used, along with other general properties of WSS signals,  to define the conditions for wide sense stationarity of random signals on graphs  \cite{perraudin2017stationary,marques2017stationary,loukas2016stationary,chepuri2016subsampling,puy2016random,zhang2015graph}.  However the main obstacle toward extending the classical statistical data analytics to graphs is that the shift on a graph typically does not preserve signal energy (isometry property), that is, $\left\Vert \mathbf{Ax} \right\Vert^2_2 \ne \left\Vert \mathbf{x} \right\Vert_2^2.$ 

For completeness, we first provide a short review of WSS definitions in classical signal processing, together with their properties. 

\subsection{Review of WSS and related Properties for Random Signals in Standard Time Domain }

\noindent \textit{Definition:} A real-valued random signal, $x(n)$, is WSS in the standard time domain if its mean value is time-invariant, $\mu_x(n)= \mathrm{E}\{x(n)\}=\mu_x$, and its autocorrelation function is shift-invariant, that is,  $r_x(n,n-m)=\mathrm{E}\{x(n)x(n-m)\}=r_x(m)$. 

\begin{Remark}\label{WSS:R1} A random WSS time-domain signal, $x(n)$, can be considered as an output of a linear shift invariant system with impulse response, $h(n)$, which is driven by a white noise input, $\varepsilon(n)$, with $r_{\varepsilon}(n,m)=\delta(n-m)$. 
\end{Remark}

\begin{Remark} In classical time domain, the eigenvectors, $\mathbf{u}_k$, of the shift operator $y(n)=x(n-1)$, or in a matrix form $\mathbf{y}=\mathbf{Ax}$, are the DFT basis functions, with $\mathbf{A}=\mathbf{U}\mathbf{\Lambda}\mathbf{U}^H$. This property is discussed in detail and proven in  {\color{blue} Part I, Section \ref{I-Section_DFT_basis_functions}, equations (\ref{I-DFTdiffeq})-(\ref{I-DFT_baisis}).}
\end{Remark}

\begin{Remark} For a random signal, its DFT  $\mathbf{X}=\mathbf{U}^H\mathbf{x}$ is also a random signal with the power spectrum matrix  $\mathbf{P}_x=\mathrm{E}\{\mathbf{X}\mathbf{X}^H\}$, where  $\mathbf{U}^H$ is the DFT transformation matrix. For WSS signals, the matrix $\mathbf{P}_x$ is  diagonal and has the power spectral density (PSD) as its diagonal values
$$p_x(k)=\mathrm{DFT}\{r_x(n)\}=\mathrm{E}\{|X(k)|^2\}.$$
\end{Remark}

\begin{Remark}\label{WSS:R2} For WSS random signals, their correlation matrix,  $\mathbf{R}_x=\mathrm{E}\{\mathbf{x}\mathbf{x}^T\}$, is  diagonalizable with the same transform matrix,  $\mathbf{U}$, which defines the DFT, $\mathbf{X} {\overset{def}{=}} \mathbf{U}^H\mathbf{x}$, with $\mathbf{x} {\overset{def}{=}} \mathbf{U}\mathbf{X}$. The proof follows from
\begin{align}
\mathbf{R}_x&=\mathrm{E}\{\mathbf{x}\mathbf{x}^T\}
=\mathrm{E}\{\mathbf{U}\mathbf{X}(\mathbf{U}\mathbf{X})^H\}\nonumber\\&=
\mathbf{U}\mathrm{E}\{\mathbf{X}\mathbf{X}^H\}\mathbf{U}^H=\mathbf{U}\mathbf{P}_x\mathbf{U}^H, 
\end{align}      
and the fact that $\mathbf{P}_x$ is a diagonal matrix for WSS signals.
\end{Remark}

The properties of the WSS signals in classical analyses, presented in this subsection,  will be used  next to define the corresponding properties of \textit{random signals on undirected graphs}.

\subsection{Adjacency Matrix Based Definition of GWSS}\label{AdjMGWSSS}

Consider a real-valued white noise signal  on a graph, $\boldsymbol{\varepsilon}=\{\varepsilon(n)\}$. Following Remark \ref{WSS:R1}, a signal $\mathbf{x}$ on the graph is graph wide sense stationary (GWSS) if it can be considered an output of a linear shift invariant system on a graph,  
$H(\mathbf{A})=\sum_{m=0}^{M-1}h_m \mathbf{A}^m$,    
which is driven by a white noise input, $\boldsymbol{\varepsilon}$, that is
$$\mathbf{x}=H(\mathbf{A})\boldsymbol{\varepsilon}.$$

\begin{Remark} The autocorrelation matrix, $\mathbf{R}_{x}=\mathrm{E}\{\mathbf{x}\mathbf{x}^T\}$, of a GWSS signal is  diagonalizable with the eigenmatrix of the adjacency matrix, $\mathbf{A}$, since (\textit{cf.} Remark \ref{WSS:R2}) 
\begin{gather}
\mathbf{A}=\mathbf{U}\mathbf{\Lambda}\mathbf{U}^{-1}=\mathbf{U}\mathbf{\Lambda}\mathbf{U}^T \nonumber \\
\mathrm{E}\{\mathbf{x}\mathbf{x}^T\}=\mathbf{U}\mathbf{P}_x\mathbf{U}^T, \label{RxDiagoniz}
\end{gather}
where $\mathbf{P}_x$ is a diagonal matrix.
The values on the diagonal of matrix $\mathbf{P}_x$ can be comprised into the vector $\mathbf{p}_x$, which represents the PSD of a graph signal, $\mathbf{x}$, $p_x(k)=\mathrm{E}\{|X(k)|^2\}$.  
\end{Remark}

To prove this property  for a signal $\mathbf{x}=H(\mathbf{A})\boldsymbol{\varepsilon}$, consider $$\mathbf{R}_{x}=\mathrm{E}\{\mathbf{x}\mathbf{x}^T\}=\mathrm{E}\{H(\mathbf{A})\boldsymbol{\varepsilon}\Big(H(\mathbf{A})\boldsymbol{\varepsilon}\Big)^T\}=H(\mathbf{A})H^T(\mathbf{A}),
$$
since $\mathrm{E}\{\boldsymbol{\varepsilon}\boldsymbol{\varepsilon}^T\}=\mathbf{I}$. Using $H(\mathbf{A})=\mathbf{U}^TH(\mathbf{\Lambda})\mathbf{U}$, we obtain 
$$
\mathbf{R}_x=\mathbf{U}^T|H(\mathbf{\Lambda})|^2\mathbf{U},
$$
which concludes the proof that the matrix $
\mathbf{P}_x$ is diagonal
$$
\mathbf{P}_x=|H(\mathbf{\Lambda})|^2,
$$
with the diagonal elements equal to the PSD of signal $\mathbf{x}$,
$$p_x(k)=|H(\lambda_k)|^2.$$ 

The periodogram of  a graph signal can be estimated using $K$ realizations of the random signal, denoted by $\mathbf{x}_i$,  and is equal to the diagonal elements of  the matrix 
$$\hat{\mathbf{P}}_x=\frac{1}{K}\sum_{i=1}^K|\mathbf{X}_i|^2=
\mathbf{U}^T\frac{1}{K}\sum_{i=1}^K(\mathbf{x}_i\mathbf{x}_i^T)\mathbf{U}.$$

Consider a system on a graph, with a spectral domain transfer function $H(\mathbf{\Lambda})$. Assume that the input signal to this system is GWSS, with PSD $p_x(k)$. The PSD of the output graph signal, $y(n)$, is then given by
   $$p_y(k)=|H(\lambda_k)|^2p_x(k).$$
This expression is conformal with the output power of a standard linear system.

\subsection{Wiener filter on a graph}
Consider a real-valued graph signal, $\mathbf{s}$, which serves as an input to a linear shift-invariant system on an undirected graph, to yield a noisy output
$$
\mathbf{x}=\sum_{m=0}^{M-1}h_m\mathbf{A}^m\mathbf{s}+\boldsymbol{\varepsilon}.
$$
In the spectral domain, this system is described by
$$\mathbf{X}=H(\mathbf{\Lambda})\mathbf{S}+\mathbf{E},$$
where $\mathbf{E}$ is the GDFT of the noise, $\boldsymbol{\varepsilon}$.

Assume that the signal and noise are statistically independent, and that the noise is a zero-mean GWSS random signal. 
The aim is to find the system function of the optimal  filter, $G(\mathbf{\Lambda})$, such that its output $\mathbf{Y}=G(\mathbf{\Lambda})\mathbf{X}$, estimates the GDFT of the input, $\mathbf{S}$, in the least squares sense. This condition can be expressed as 
$$
e^2=\mathrm{E}\{\left\Vert \mathbf{S}-\mathbf{Y} \right\Vert _2 ^2\}=\mathrm{E}\{\left\Vert \mathbf{S}-G(\mathbf{\Lambda})\mathbf{X} \right\Vert _2 ^2\}.
$$
Upon setting the derivative of $e^2$  with respect to the elements of $G(\mathbf{\Lambda})$ to zero, we arrive at
$$
2\mathrm{E}\{(\mathbf{S}-G(\mathbf{\Lambda})\mathbf{X})\mathbf{X}^T\}=\mathbf{0},
$$
which results in the system function of the graph Wiener filter in the form
\begin{align*}
G(\mathbf{\Lambda})&=\frac{\mathrm{E}\{\mathbf{S}\mathbf{X}^T\}}
{\mathrm{E}\{\mathbf{X}\mathbf{X}^T\}}
=\frac{\mathrm{E}\{\mathbf{S}(H(\mathbf{\Lambda})\mathbf{S}+\mathbf{E})^T\}}
{\mathrm{E}\{(H(\mathbf{\Lambda})\mathbf{S}+\mathbf{E})(H(\mathbf{\Lambda})\mathbf{S}+\mathbf{E})^T\}}
\\&=\frac{H(\mathbf{\Lambda})\mathbf{P}_s}{H^2(\mathbf{\Lambda})\mathbf{P}_s+\mathbf{P}_{\varepsilon}}
\end{align*}
or element-wise 
$$
G(\lambda_k)=\frac{H(\lambda_k)p_s(k)}{H^2(\lambda_k)p_s(k)+E(k)}.
$$
When the noise is not present, the elements of the vector $\mathbf{E}$ are zero-valued, $E(k)=0$ for all $k$, and the graph inverse filter (introduced in Section \ref{inversesystem}) directly follows.

\begin{Remark} 
The above expressions for the graph Wiener filter are conformal with the standard frequency domain Wiener filter, given by 
$$
G(\omega)=\frac{P_s(\omega)}{P_s(\omega)+P_{\varepsilon}(\omega)},
$$
which again demonstrates the generic nature of Graph Data Analytics.
\end{Remark}

\subsection{Spectral Domain Shift Based Definition of GWSS}
Consider an $m$-step shift on a graph defined using
 the graph filter response
\begin{equation}
\mathcal{T}_m\{h(n)\}=h_m(n)=\sum_{k=0}^{N-1}H(\lambda_k)u_k(m)u_k(n).
\end{equation}
The matrix form of this relation is given by
\begin{equation}
\mathcal{T}_h=H(\mathbf{L}) =\sum_{m=0}^{M-1}h_m \mathbf{L}^m=\mathbf{U} H(\mathbf{\Lambda}) \mathbf{U}^T, \label{OPHSpec}
\end{equation}
where $\mathcal{T}_m\{h(n)\}$ are the elements of $\mathcal{T}_h$.

Note that the graph filter response function is well localized on a graph. Namely, if we use, for example, the $(M-1)$-neighborhood of a vertex $n$, within a filtering function of order $M$ defined by $H(\mathbf{\Lambda})$, then only the vertices within this neighborhood are used in the calculation of graph filter response. From (\ref{OPHSpec}), we see that the localization operator acts in the spectral domain and associates the corresponding shift to the vertex domain. 

\noindent\textit{Definition:} A random graph signal, $x(n)$, is GWSS if its autocorrelation function is invariant with respect to the shift, $\mathcal{T}_m\{r_x(n)\}$, that is $$r_{x}(m)=\mathrm{E}\{x(n)x(n-m)\}=\mathcal{T}_m\{r_x(n)\}.$$
Similar to (\ref{RxDiagoniz}), the autocorrelation matrix, $\mathbf{R}_{x}$, of a GWSS signal is  diagonalizable based on the matrix of eigenvectors of the graph Laplacian $\mathbf{L}$, whereby
\begin{gather}
\mathbf{L}=\mathbf{U}\mathbf{\Lambda}\mathbf{U}^T.
\end{gather}

For the basic autocorrelation we use
$$ 
\mathbf{R}_x=\mathbf{U}P_x(\mathbf{\Lambda})\mathbf{U}^T
$$
so that
$$
\mathcal{T}_m\{r_x(n)\}=\sum_{k=0}^{N-1}p_x(\lambda_k)u_k(m)u_k(n)
$$
where 
$$P_x(\mathbf{\Lambda})=\mathbf{U}\mathbf{R}_x\mathbf{U}^T$$ 
is a diagonal matrix.

\subsection{Isometric Shift Operator} 
Another possible approach may be based on the shift operator defined as 
$\mathcal{T}_m=\exp(j\pi \sqrt{\mathbf{L}/\rho})$, where $\rho$ is the upper bound on the eigenvalues, $\rho=\max_k\{\lambda_k\}$,  \cite{girault2015stationary,girault2015translation}.
Physically, this operator casts the eigenvalues of the Laplacian, $\mathbf{L}$, onto a unit circle, thus preserving in this way the isometry property, since
\begin{equation}
\mathcal{T}_h=\exp(j\pi \sqrt{\mathbf{L}/\rho}) =\mathbf{U} \exp(j\pi \sqrt{\mathbf{\Lambda}/\rho}) \mathbf{U}^T.
\end{equation}
The property $f(\mathbf{L})=\mathbf{U} f(\mathbf{\Lambda}) \mathbf{U}^H$ was used above. 
Observe that for real-valued eigenvalues,  $\lambda_k$, all eigenvalues of the matrix $\exp(j\pi \sqrt{\mathbf{\Lambda}/\rho})$ reside on the unit circle, with  the frequency $0\le \omega_k = \pi \sqrt{\lambda_k/\rho} \le \pi$ being associated to the eigenvector $\mathbf{u}_k$.

However, with this setup the corresponding GDFT bases are not necessarily orthogonal for a general graph structure. In turn, the work in \cite{Scalzo2019_4} drew inspiration from the classical definition of the unitary shift operator acting on Hilbert spaces \cite{Fillmore1974} to define an isometric shift operator with orthogonal GDFT bases. Since the adjacency matrix, ${\bf A}$, provides the minimal information required to fully reflect the connectivity structure arising from the graph topology, and therefore to define the most elementary graph shift, the task of determining an isometric graph shift operator can be formulated as finding the matrix ${\bf S}$ closest to ${\bf A}$ in a Hilbert space, that is,
\begin{equation}
\begin{split}
\max_{{\bf S}} \quad & \langle{\bf S},{\bf A}\rangle = \mathrm{tr} \left\{ {\bf S}{\bf A}^{T} \right\} \notag \\
\textnormal{s.t.} \quad &  {\bf S}^{T}{\bf S} = {\bf S}{\bf S}^{T} = {\bf I} 
\end{split} \label{eq:optimization}
\end{equation}
This can be achieved analytically by evaluating the singular value decomposition of ${\bf A}$, given by ${\bf A} = {\bf U} {\boldsymbol \Sigma}  {\bf V}^{T}$ where ${\bf U}, {\bf V} \in \mathbb{R}^{N \times N}$ are respectively the left and right matrix of singular vectors, and ${\boldsymbol \Sigma} \in \mathbb{R}^{N \times N}$ is the diagonal matrix of singular values. In this way, the backward shift operator can be expressed as 
\begin{equation}
{\bf S} = {\bf U}{\bf Q}{\bf V}^{T} \label{eq:GSO_sym_orth}
\end{equation}
where 
\begin{equation}
{\bf Q} = \left[\begin{array}{cc}
{\bf I}_{(N-1) \times (N-1)} & {\boldsymbol 0}_{(N-1) \times 1} \\
{\boldsymbol 0}_{1 \times (N-1)} & \det({\bf U}{\bf V}^{T})
\end{array}\right]
\end{equation}
ensures that $\det({\bf S})=1$, so as to produce a \textit{proper rotation} matrix. The resulting matrix, ${\bf S}$, is called the \textit{symmetric orthogonalization} of the matrix ${\bf A}$, and is unique \cite{Lowdin1950,Lowdin1970}. The solution is also closely related to the orthogonal Procrustes problem \cite{Schonemann1966} and the Kabsch algorithm \cite{Kabsch1976}. 

\begin{Example}
	Consider the directed graph in Fig. \ref{fig:unitary_GSO}(a) which has $N=8$ vertices in the set $\mathcal{V}=\{0,1,2,3,4,5,6,7\}$. The backward and forward shifted versions of the signal in Fig. \ref{fig:unitary_GSO}(b) were evaluated using both the elementary shift matrix, ${\bf A}$, and the proposed isometric graph shift operator, ${\bf S}$ in (\ref{eq:GSO_sym_orth}). Notice that, as desired, the signal energy was preserved when employing the isometric graph shift operator, ${\bf S}$, while the energy of the signals shifted through ${\bf A}$ increased.

\begin{figure}[thpb]
	\centering
	\begin{minipage}{0.235\textwidth}
		\centering
		\includegraphics[width=0.8\textwidth, trim={7.2cm 22cm 7.4cm 1.9cm}, clip]{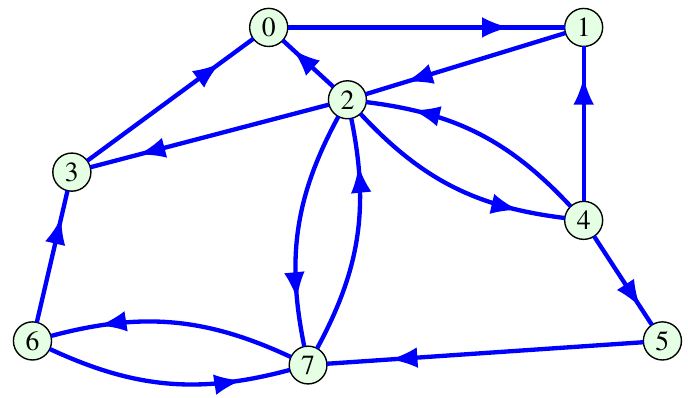} \\
		{{\footnotesize (a) Directed graph structure.}}
		\label{fig:structure}
	\end{minipage}
	\hfill
	\begin{minipage}{0.235\textwidth}
		\centering
		\includegraphics[width=0.9\textwidth, trim={4cm 12cm 4cm 11cm}, clip]{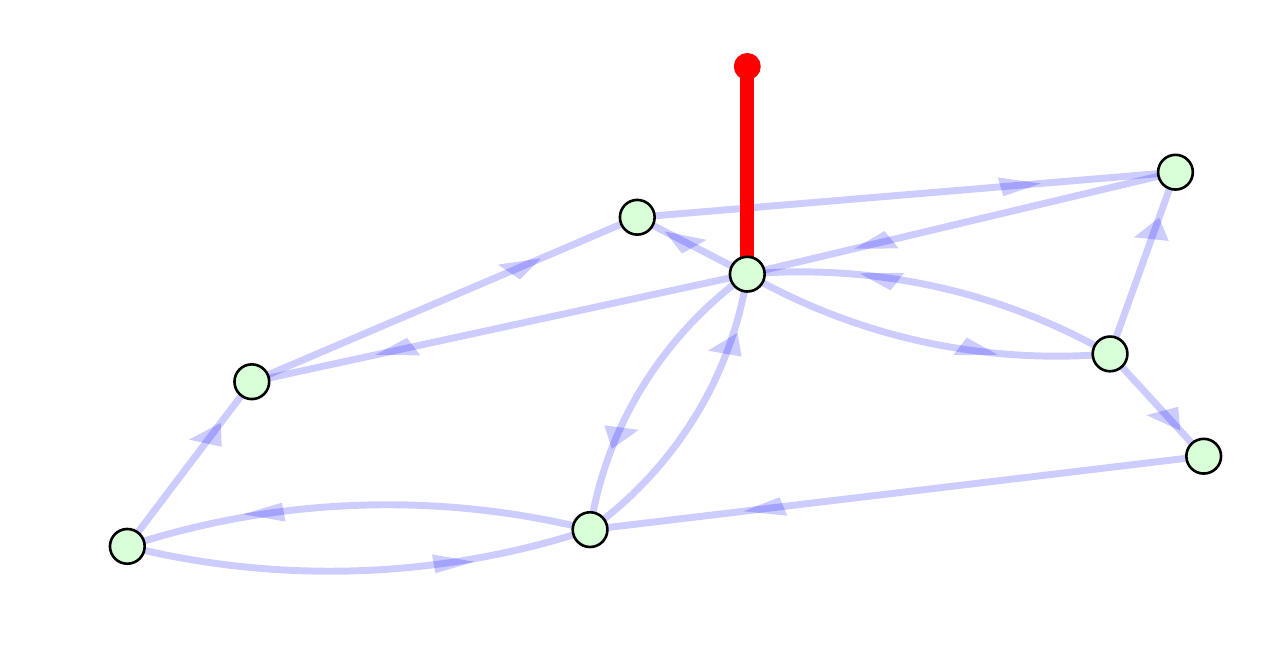} 
		{\footnotesize (b) Graph signal, ${\bf x}$.}    
		\label{fig:orig}
	\end{minipage}
	\vspace{-0.4cm}
	\vskip\baselineskip
	\centering
	\begin{minipage}{0.235\textwidth}
		\centering		
		\includegraphics[width=0.9\textwidth, trim={4cm 12.8cm 4cm 10.5cm}, clip]{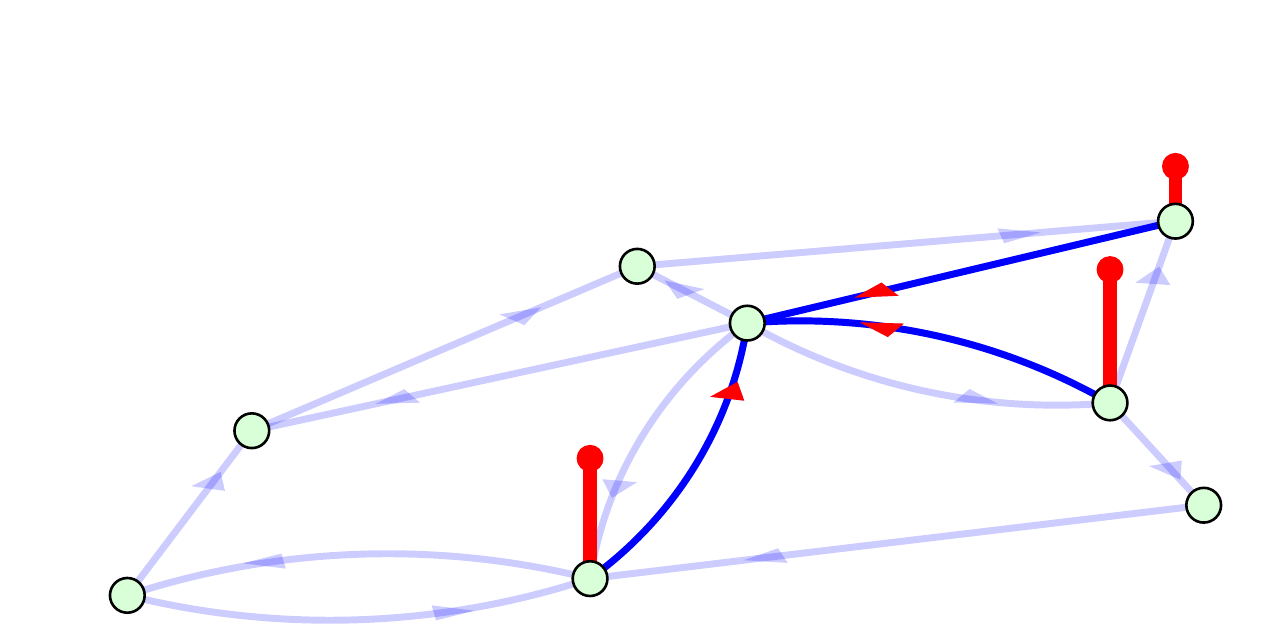}
		{\footnotesize (c) Backward shifted signal, ${\bf S}{\bf x}$.}
		\label{fig:Sbshift}
	\end{minipage}
	\hfill
	\begin{minipage}{0.235\textwidth}  
		\centering 
		\includegraphics[width=0.9\textwidth, trim={4cm 12.8cm 4cm 10.5cm}, clip]{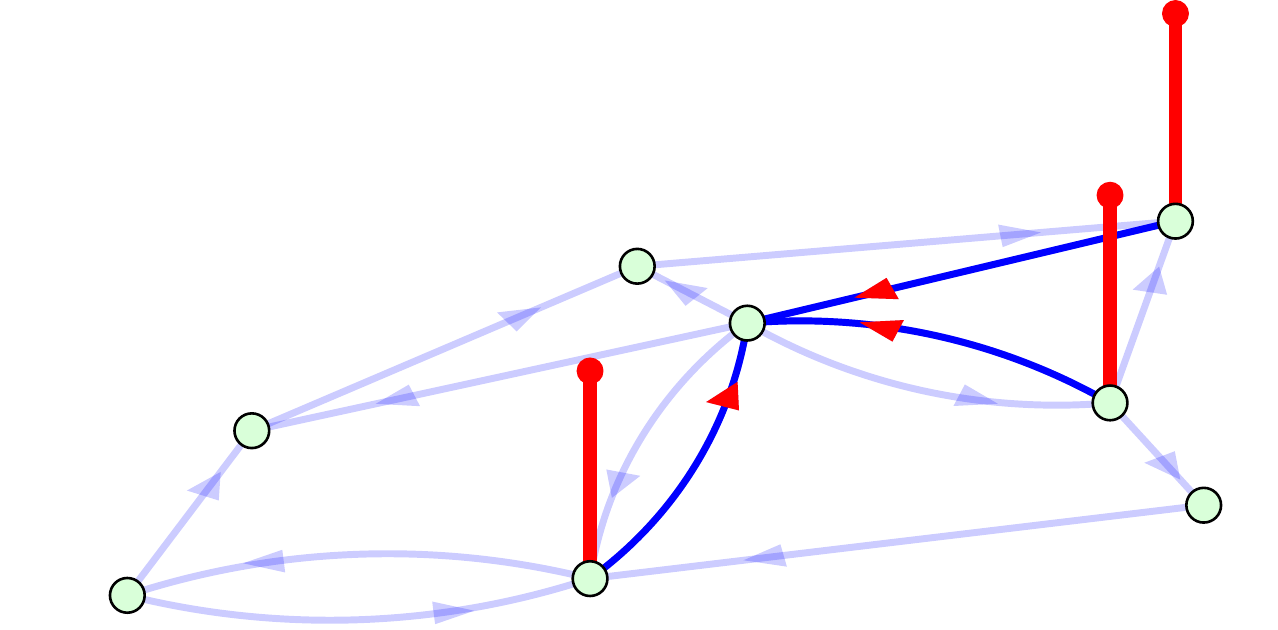} 
		{\footnotesize (d) Backward shifted signal, ${\bf A}{\bf x}$.}
		\label{fig:Abshift}
	\end{minipage}
	\vspace{-0.4cm}
	\vskip\baselineskip
	\begin{minipage}{0.235\textwidth}   
		\centering 
		\includegraphics[width=0.9\textwidth, trim={4cm 12.8cm 4cm 10.9cm}, clip]{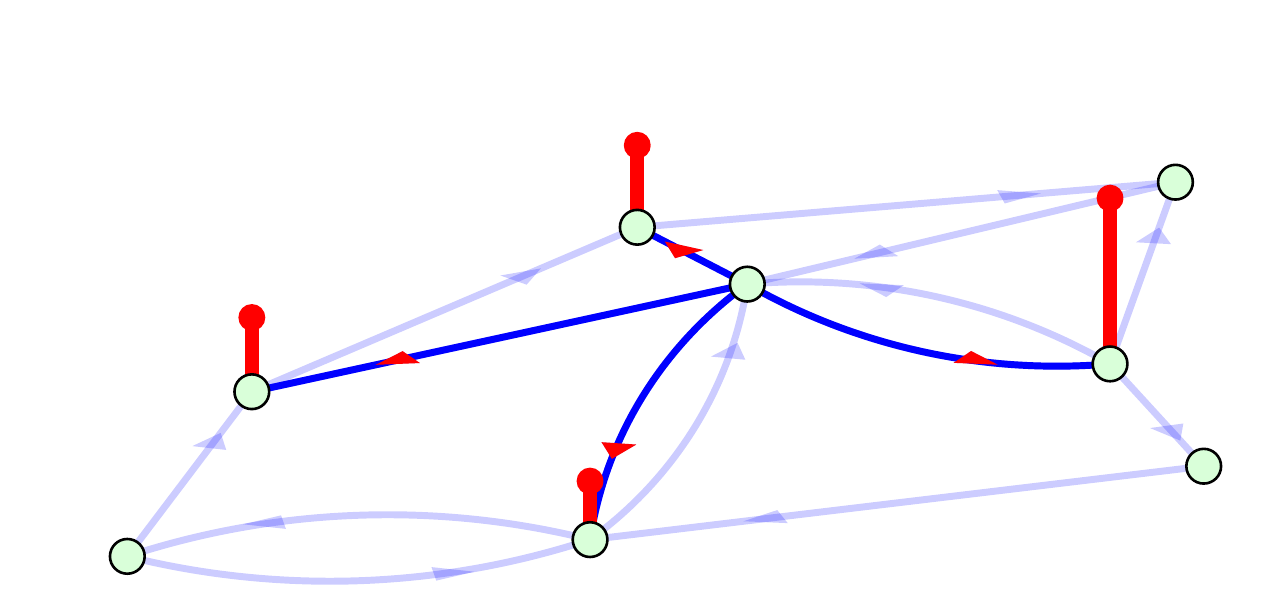} 
		{\footnotesize (e) Forward shifted signal, ${\bf S}^{T}{\bf x}$.}
		\label{fig:Sfshift}
	\end{minipage}
	\hfill
	\begin{minipage}{0.235\textwidth}   
		\centering 
		\includegraphics[width=0.9\textwidth, trim={4cm 12.8cm 4cm 10.9cm}, clip]{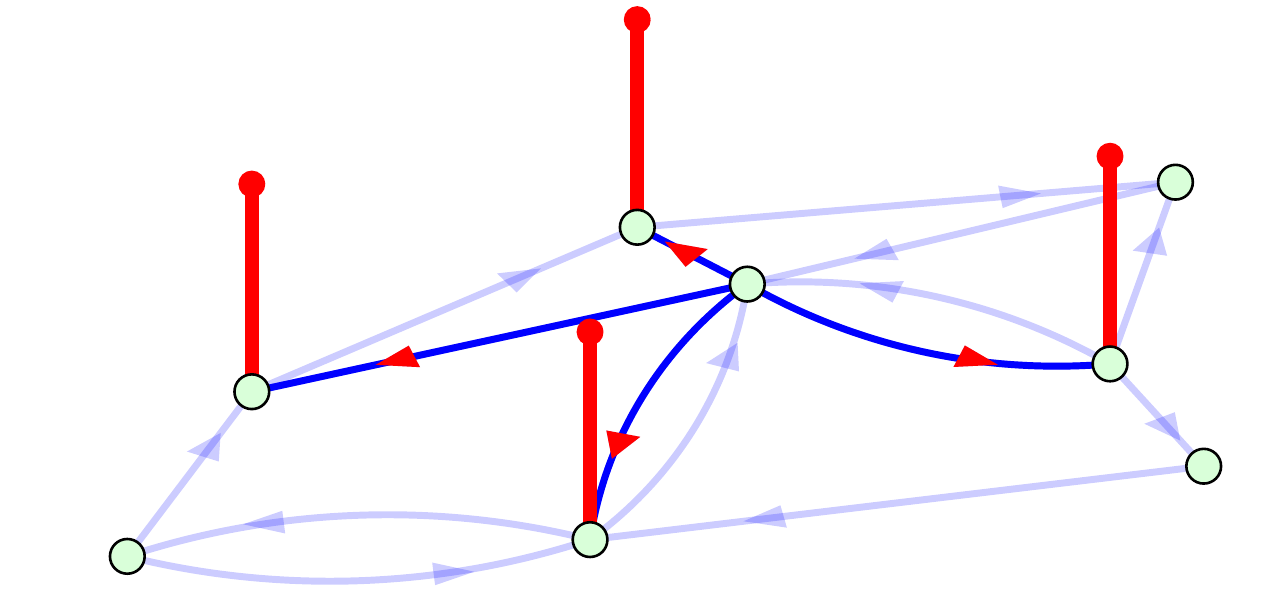} 
		{\footnotesize (f) Forward shifted signal, ${\bf A}^{T}{\bf x}$.}  
		\label{fig:Afshift}
	\end{minipage}
	\vspace{-0.5cm}
	\caption{Graph signal shifts by the proposed isometric shift operator, ${\bf S}$, evaluated on a directed graph. (a) Directed graph structure. (b) A simple graph signal, ${\bf x}$. (c) Backward shifted version of ${\bf x}$, given by ${\bf S}{\bf x}$. (d) Backward shifted version, ${\bf A}{\bf x}$. (e) Forward shifted version, ${\bf S}^{T}{\bf x}$. (f) Forward shifted version, ${\bf A}^{T}{\bf x}$. The red arrows indicate the movement of the pulse at vertex $n=2$ towards vertices connected with (i) outgoing blue arrows in (a) for a forward shift and (ii) to vertices connected with incoming blue arrows in (a) for a backward shift.}  
	\label{fig:unitary_GSO}
\end{figure}
\end{Example}

\section{Vertex-Frequency Representations}
\label{sec7}

 Oftentimes in practical applications concerned with large graphs, we may not be interested in the analysis 
 of the entire graph signal, but rather in its local behavior. Indeed, the Big Data paradigm has revealed the possibility of using smaller and localized subsets of the available information to enable reliable mathematical analysis and local characterization of subsets of data of interest \cite{sandryhaila2014big}.  Our aim in this section is to characterize the localized graph signal behavior simultaneously in the vertex-frequency domain, in a natural analogy with  classical time-frequency analysis \cite{stankovic2014time,cohen1995time,boashash2015time}. 

It is important to note that, while the concept of  window functions for signal localization  has been extended 
to signals defined on graphs \cite{shuman2016vertex, shuman2012windowed,zheng2016multi,tepper2016short,stankovic2017vertex},  such extensions are not straightforward, since, owing to inherent properties of graphs as irregular but interconnected domains, even an operation which is very simple in classical time-domain analysis, like the time shift, 
cannot be straightforwardly generalized to graph signal domain. This has resulted in several approaches 
to the definition of the graph shift operator, and much ongoing research in this domain \cite{shuman2016vertex, shuman2012windowed,zheng2016multi,tepper2016short,stankovic2017vertex}.

A common approach to signal windowing in the graph domain is to utilize the eigenspectrum of a graph signal to obtain  window function 
for each graph vertex \cite{shuman2013emerging}. Another possibility is to define the 
window support as a local neighborhood for each vertex \cite{stankovic2017vertex}.
In either case, the localization window is defined based on a set of vertices that contain the 
current vertex, $n$, and all vertices that are close  in some sense to the vertex $n$, that is, a neighborhood of vertex $n$. In this monograph, special attention is devoted to the class of local graph Fourier transform approaches which can be implemented in the vertex domain, since this domain often offers a basis for numerically efficient analysis in the case of very large graphs.  

Notice that, as in 
classical signal analysis, a localization window should be narrow enough so as to provide 
good localization of signal properties, but at the same time wide enough to produce high 
resolution in the spectral domain. 

With vertex-frequency analysis serving as a key to graph signal estimation, filtering, and efficient representation, two  forms of the local graph Fourier transform inversion are considered here, while the inversion condition is defined within the frames framework, that is, based on the  analysis of energy of the graph spectrogram. A relation between the graph wavelet transform and the local graph Fourier transform implementation and its inversion  is also established. 

\begin{Remark}
The energy versions of the vertex-frequency representations are also considered, as these representations can be implemented without a localization window, and they can serve as estimators of the local smoothness index. 
\end{Remark}

The reduced interference vertex-frequency distributions, which satisfy the marginal property and localize graph signal energy in the vertex-frequency domain are also defined, and are subsequently related to classical time-frequency analysis, as a special case.

Consider a graph with $N$ vertices, $n \in\mathcal{V}=\{0,1,\dots, N-1\}$, which are connected with edges whose weights are $W_{mn}$. 
Spectral analysis of graphs is most commonly based on the eigendecomposition of the graph Laplacian, $\mathbf{L}$, or the adjacency matrix, $\mathbf{A}$. By default, we shall assume the decomposition of the graph Laplacian, $\mathbf{L}$, if not stated otherwise.

\subsection{Localized Graph Fourier Transform (LGFT)}\label{s:VLS}

The localized graph Fourier transform (LGFT), denoted by $S(m,k)$, can be considered as an extension of the standard time-localized (short-time) Fourier transform (STFT), and can be  calculated as the GDFT
of a signal, $x(n)$, multiplied by an appropriate vertex localization window function, $h_m(n)$, to yield
\begin{equation}
S(m,k)=\sum_{n=0}^{N-1} x(n)h_m(n)\; u_{k}(n). \label{VFSPEC}
\end{equation}
In general, it is desired that a graph window function, $h_m(n)$, should be such that it 
localizes the signal content around the vertex $m$. To this end, its values should be close to $1$ at vertex $m$ and vertices in its close neighborhood, while it should approach to $0$ for vertices that are far from  vertex $m$.  For an illustration of the concept of localization window on a graph see Fig. \ref{VF_windows_LVS}, panels (a) and (c).

The localized GDFT in (\ref{VFSPEC}) admits a matrix notation, $\mathbf{S}$, and contains all elements, $S(m,k)$, $m=0,1,\dots,N-1$, $k=0,1,\dots,N-1$. The columns of $\mathbf{S}$ which correspond to a vertex $m$ are given by 
$$\mathbf{s}_m=\mathrm{GDFT}\{x(n)h_m(n)\}= \mathbf{U}^T\mathbf{x}_{m},$$ where  $\mathbf{x}_{m}$ is the vector of which the elements, $x(n)h_m(n)$, are equal to the graph signal samples, $x(n)$, multiplied by the window function, $h_m(n)$, centered at the vertex $m$, {while matrix $\mathbf{U}$ is composed of the eigenvectors $\mathbf{u}_k$, with elements $u_k(n)$, $k=0,1,\dots,N-1$, of the graph Laplacian as its columns}. 

\smallskip
\noindent\textbf{Special cases:} 
\begin{itemize}
	\item 
	For $h_m(n)=1$, the localized vertex spectrum is  equal to the standard 
	spectrum, 
	$S(m,k)=X(k)$, in (\ref{VFSPEC}) for each $m$; this means that no vertex localization is 
	performed. 
	\item 
	If $h_m(m)=1$ and $h_m(n)=0$ for $n\ne m$, the localized vertex 
	spectrum is equal to the graph signal, $S(m,0)=x(m)/\sqrt{N}$, for $k=0$. 
\end{itemize}

In the following, we outline  ways to create vertex domain windows with desirable localization characteristics, and address two methods for defining graph localization window functions, $h_m(n)$:
\begin{itemize} 
	\item Spectral domain definition of windows, $h_m(n)$, which are defined using their spectral basic function. The spectral domain definition of the window is shown to be related  to the wavelet transform. 
	\item Vertex domain window definitions, with one method  bearing a direct relation to the spectral analysis of the graph window, while the other method represents a  purely vertex domain formulation.  
\end{itemize}

\begin{figure*}
	\centering
	
	\includegraphics[]{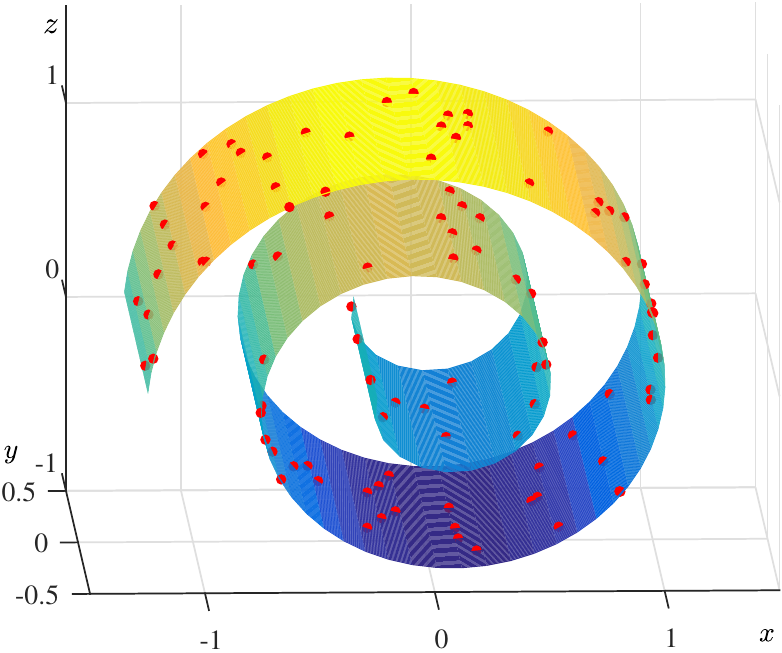}\hspace{2mm}(a)\hspace{5mm}
	\includegraphics[]{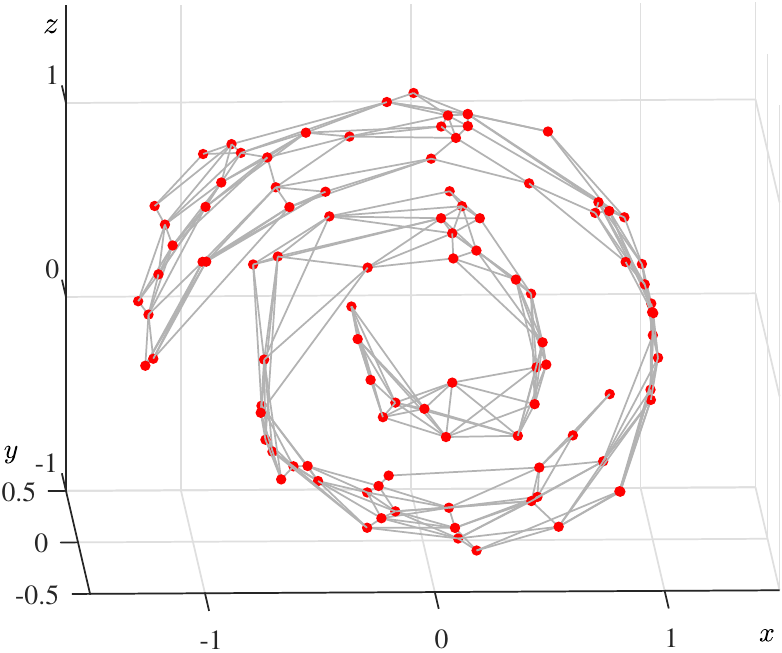}\hspace{2mm}(b)
	
	\vspace{10mm}
	
	\includegraphics[]{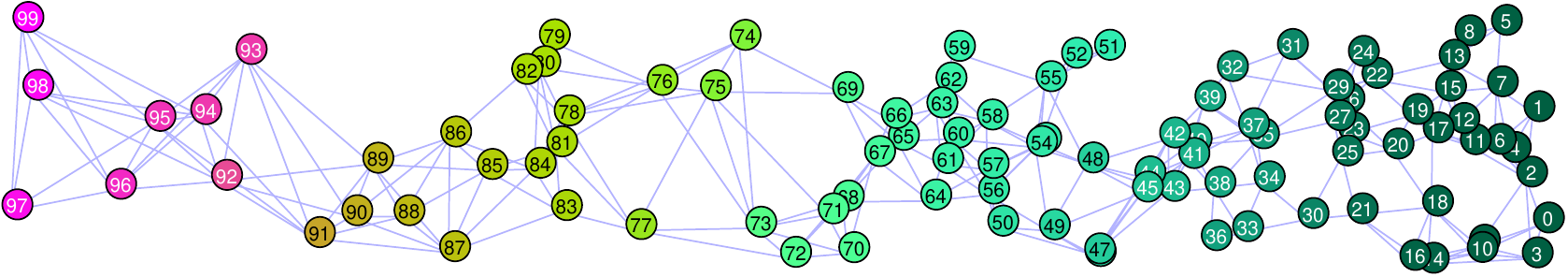}\hspace{7mm}(c)
	
	\vspace{10mm}
	
	\includegraphics[]{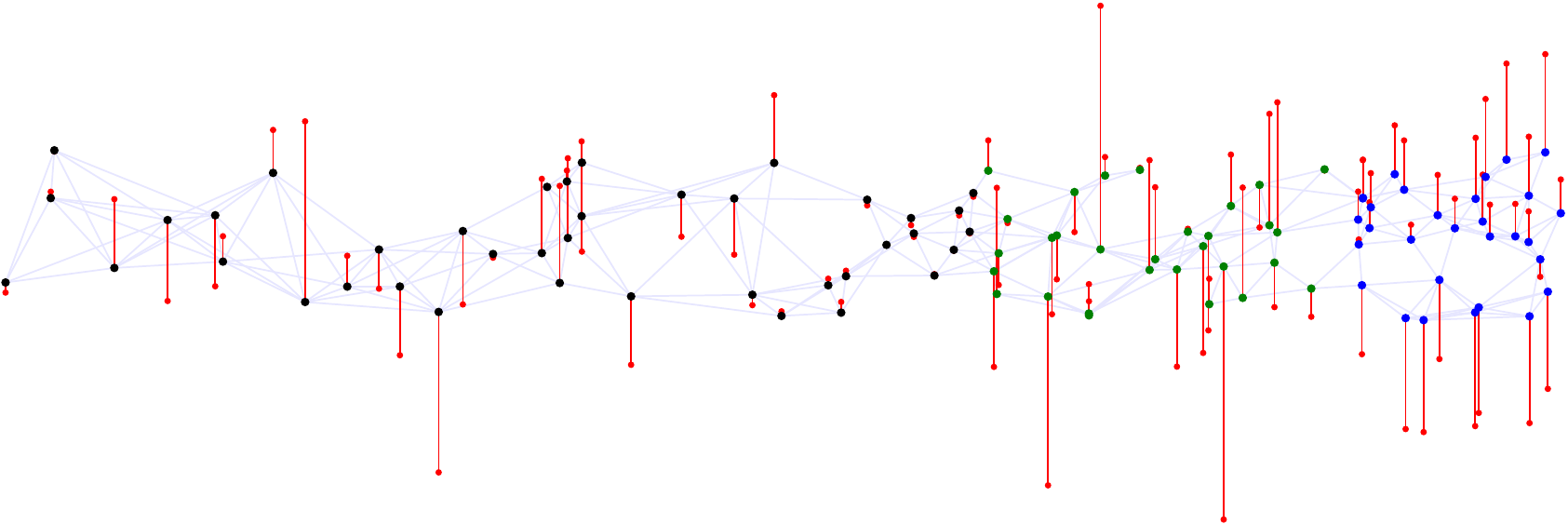}\hspace{7mm}(d)
	
	\caption{Concept of a signal on a graph. (a) Vertices on a three-dimensional  manifold Swiss roll surface. (b) A graph representation on the Swiss roll manifold. (c) Two-dimensional presentation of the three-dimensional graph from (b), with vertex colors defined by the three smoothest graph Laplacian eigenvectors $u_1(n)$, $u_2(n)$, and $u_3(n)$. (d) A signal observed on the graph in (c), which is composed of three Laplacian eigenvectors (signal components). The supports of these three components  are designated by different vertex colors. The vertex-frequency representations are then assessed based on their ability to accurately resolve and localize these three graph signal components. }
	\label{VF_graph3ab}
\end{figure*}

\subsubsection{Windows Defined in the GDFT Domain}

The basic function of a window, $h(n)$, can be conveniently defined in the spectral domain, for example, in the form
\begin{equation}
H(k)=C
\exp (-\lambda_k \tau),
\label{win-spect}
\end{equation}
where $C$ denotes the “window amplitude” and $\tau>0$ is a 
constant which determines the window width in the spectral domain. Notice that the graph shifted and “modulated” versions of this window are straightforwardly  obtained using the generalized convolution of graph signals, defined in Section \ref{Sec:ConvSigGRR}.  The graph-shifted window in the vertex domain is then defined by the IGDFT of  $H(k)u_k(m)$, to give the window localized at the vertex $m$, denoted by $h_m(n)$, as in (\ref{signShiftOngr}), in the form
\begin{equation}
h_m(n)=h(n)*\delta_m(n)=\sum_{k=0}^{N-1}H(k)u_k(m)u_k(n).
\label{win-spect1}
\end{equation}
 An example of two windows obtained in this way is given in Fig. \ref{VF_windows_LVS}(a), (b). 
 
 Observe that the exponential function in (\ref{win-spect}) corresponds to a Gaussian window in  classical analysis (thus offering the best time-frequency concentration \cite{stankovic2014time,cohen1995time,boashash2015time}),  since  graph signal processing on a 	path graph reduces to classical signal analysis. In this case, the eigenvalues of the graph Laplacian, $\lambda$, may be related to the frequency, $\omega$, in  classical signal analysis as $\lambda \sim \omega^2$. 

\noindent\textbf{ Properties of graph window functions.} The  graph window which is localized at the vertex $m$, and defined by (\ref{win-spect1}), satisfies the following properties:
\begin{enumerate}[label=\subscript{W}{{\arabic*}}:] 
	\item Symmetry, $h_m(n)=h_n(m)$, which follows from the definition in (\ref{win-spect1}).
	\item A sum of all  coefficients of a localized window, $h_m(n)$, is equal to $H(0)$, since
	\begin{gather*}
	\sum_{n=0}^{N-1}h_m(n)=\sum_{k=0}^{N-1}H(k)u_k(m)\sum_{n=0}^{N-1}u_k(n)\\
	=
	\sum_{k=0}^{N-1}H(k)u_k(m)\delta(k)\sqrt{N}=H(0),
	\end{gather*}
	with $\sum_{n=0}^{N-1}u_k(n)=\delta(k)\sqrt{N}$, following from the definition of the eigenvectors, $u_k(n)$.
	\item
	The Parseval theorem for  $h_m(n)$ has the form
	\begin{equation}
	\sum_{n=0}^{N-1}|h_m(n)|^2=\sum_{k=0}^{N-1}|H(k)u_k(m)|^2.
	\end{equation}
\end{enumerate}
These properties  will be used in the sequel in the inversion analysis of the LGFT.

Based on the above properties, the LGFT can now be written as 
\begin{gather}
S(m,k)=\sum_{n=0}^{N-1} x(n)h_m(n)\; u_{k}(n) \label{LGFTDEF}\\
=\sum_{n=0}^{N-1} \sum_{p=0}^{N-1}x(n)H(p)u_p(m)u_p(n) \; u_{k}(n). \label{LGFTDef1}
\end{gather}
The modulated (frequency shifted) version of the window centered at a vertex $m$ and for a spectral index $k$ will be referred to as the \textit{vertex-frequency kernel}, $\mathcal{H}_{m,k}(n)$, which is defined as
\begin{equation}
\mathcal{H}_{m,k}(n)=h_m(n)u_k(n)=\Big( \sum_{p=0}^{N-1}H(p)u_p(m)u_p(n)  \Big) u_k(n). \label{KernelGFDFTmk}
\end{equation}
Using the kernel notation, it becomes obvious that the LGFT in (\ref{LGFTDef1}), for a given vertex $m$ and a spectral index $k$, physically represents  a projection of a graph signal, $x(n)$, onto the graph kernel, $\mathcal{H}_{m,k}(n)$, that is,
\begin{equation}
S(m,k)=\langle\mathcal{H}_{m,k}(n),x(n)\rangle=\sum_{n=0}^{N-1}\mathcal{H}_{m,k}(n)x(n).\label{KernelGFDFTDEF}
\end{equation}

\begin{Remark}
	The classical STFT, a basic tool in time-frequency analysis, can be obtained as a special case of the GDFT when the graph is directed and circular. For this type of graph, the eigendecomposition  of the adjacency matrix produces complex-valued eigenvectors of the form $u_k(n)\sqrt{N} =\exp(j 2 \pi nk/N)$. Then, having in mind the \textit{complex nature of these eigenvectors}, 	
	\begin{gather}
	S(m,k)
	=\sum_{n=0}^{N-1} \sum_{p=0}^{N-1}x(n)H(p)u_p^*(m)u_p(n) \; u_{k}^*(n), \nonumber
	\end{gather}
	the value of $S(m,k)$ in (\ref{LGFTDEF})  becomes the standard STFT, that is
	\begin{gather}
	S(m,k)=\frac{1}{N^{3/2}}\sum_{n=0}^{N-1} \sum_{p=0}^{N-1}x(n)H(p)e^{j -\frac{2 \pi}{N} mp}e^{j  \frac{2 \pi}{N} np}e^{-j  \frac{2 \pi}{N} nk}, \nonumber \\
	=\frac{1}{N}\sum_{n=0}^{N-1} x(n)h(n-m)e^{-j 2 \pi nk/N},
	\end{gather}
	where $h(n)$ is the inverse DFT of $H(k)$.
\end{Remark}

\begin{Example} To illustrate the principle of  local vertex-frequency representations, consider the graph and the graph signal from Fig. \ref{VF_graph3ab}.
	A graph with $N=100$ vertices, randomly placed on the so called Swiss roll surface, is shown in Fig. \ref{VF_graph3ab}(a). The vertices are connected with edges whose weights are defined as $W_{mn}=\exp(-r_{mn}^2/\alpha)$, where $r_{mn}$ is the Euclidean  distance between the vertices $m$ and $n$, measured along the Swiss roll manifold, and $\alpha$ is a constant. Small weight values were hard-thresholded to zero, in order to reduce the number of edges associated with each vertex to only a few  strongest ones. The so produced graph is shown  in Fig. \ref{VF_graph3ab}(b), and its two-dimensional presentation in Fig. \ref{VF_graph3ab}(c). Vertices are ordered so that the values of the Fiedler eigenvector, $u_1(n)$, are nondecreasing.  
	
	A signal on this graph was created so as to be composed of parts of three Laplacian eigenvectors. For the subset, $\mathcal{V}_1$, of all vertices, $\mathcal{V}$,  which comprises the vertices with indices from $m=0$ to $m=29$, the eigenvector with the spectral index $k=8$ was used. 
	For the subset, $\mathcal{V}_2$,  with the vertex indices from $m=30$ to $m=59$, the signal was equal to the eigenvector $u_{66}(n)$, that is, with $k=66$. The remaining vertices form the vertex subset $\mathcal{V}_3$, and the signal on this subset was equal to the eigenvector with the spectral index $k=27$. The amplitudes of these eigenvectors were scaled too.

	Consider now the  
	vertex-frequency localization kernels, 
	$$\mathcal{H}_{m,k}(n)=h_m(n)u_k(n),$$ 
	shown in Fig. \ref{VF_windows_LVS}. The constant eigenvector, $u_0(n)=1/\sqrt{N}$, was used in the panel shown in Fig. \ref{VF_windows_LVS}(a) at $m=34$. In this case, the localization window, $h_{34}(n)$, is shown since $\mathcal{H}_{34,0}(n)=h_{34}(n)/\sqrt{N}$. The illustration is repeated in the panel in Fig. \ref{VF_windows_LVS}(c) for the vertex  $m=78$. The frequency shifted version of these two vertex-domain kernels, shown in Figs. \ref{VF_windows_LVS}(a) and (c), are given respectively in Figs. \ref{VF_windows_LVS}(b) and (d), where $\mathcal{H}_{m,20}(n)=h_m(n)u_{20}(n)$ is shown for $m=34$ and $m=78$, respectively.

	Next, the vertex-frequency representation, $S(n,k)$, using the LGFT and the localization window defined in the spectral domain is shown in Fig. \ref{VF_graph3g}. From this representation, we can clearly identify the three constituent signal components, within their intervals of support. The marginal properties, such as the projections of $S(n,k)$ onto the vertex index axis and the spectral index axis, are also clearly distinguishable. From the marginal properties, we can conclude that the considered graph signal in hand is spread over all vertex indices, while its spectral localization is dominated by the three spectral indices which correspond to the three components of the original graph signal. In an ideal case of  vertex-frequency analysis, these marginals should  respectively  be equal to $|x(n)|^2$ and $|X(k)|^2$, which is not the case here.

	\begin{figure*}
		\centering
		\includegraphics[]{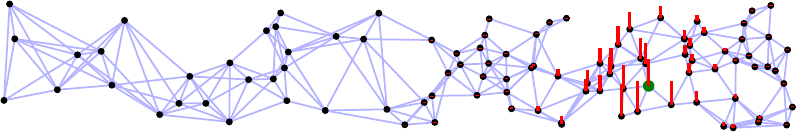}(a)
		\hspace{2mm}
		\includegraphics[]{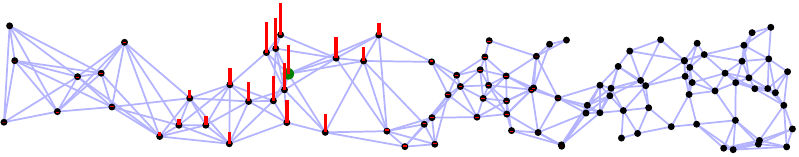}(b)
		
		$\mathcal{H}_{34,0}(n)=h_{34}(n)u_{0}(n) \sim h_{34}(n)$ \hspace{35mm} 
		$\mathcal{H}_{78,0}(n)=h_{78}(n)u_{0}(n) \sim h_{78}(n)$
		
		\vspace{8mm}
		
		\includegraphics[]{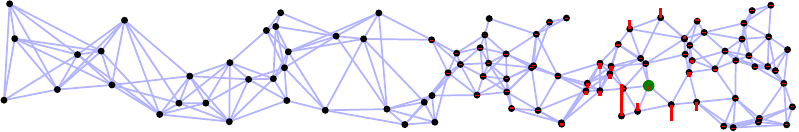}(c)
		\hspace{2mm}
		\includegraphics[]{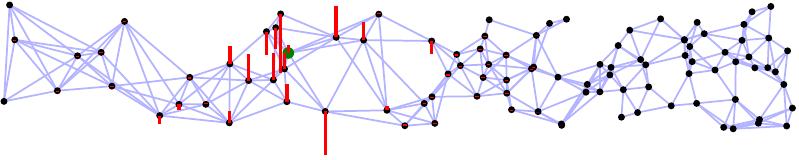}(d)
		
		$\mathcal{H}_{34,20}(n)=h_{34}(n)u_{20}(n)$ \hspace{55mm} 
		$\mathcal{H}_{78,20}(n)=h_{78}(n)u_{20}(n)$
		
		\vspace{4mm}
		
		\includegraphics[scale=0.9]{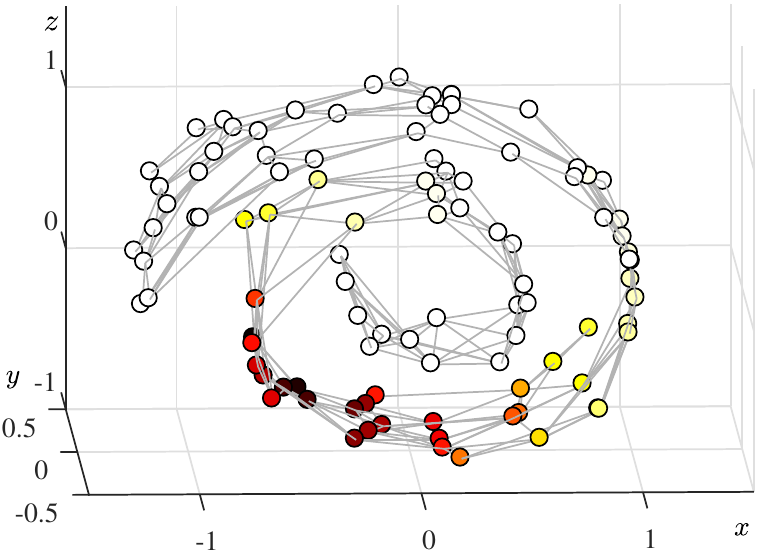} \hspace{6mm} (e)
		\hspace{6mm}
		\includegraphics[scale=0.9]{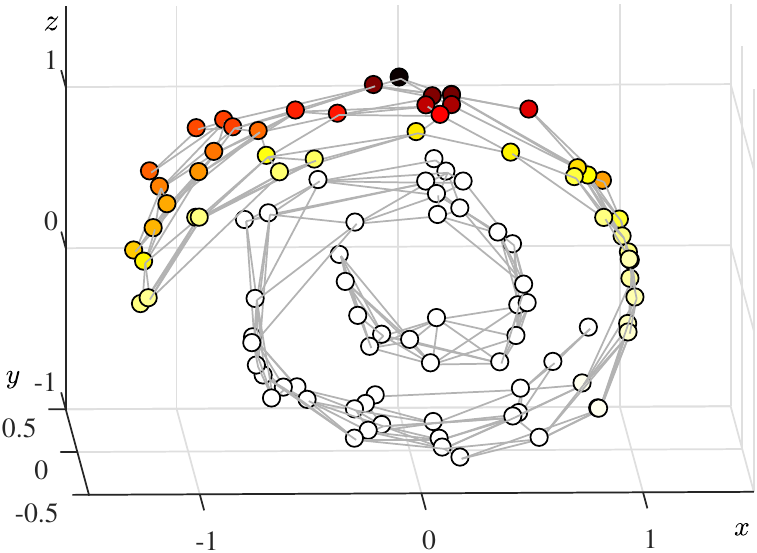} \hspace{8mm} (f)
		
		$\mathcal{H}_{34,0}(n)=h_{34}(n)u_{0}(n) \sim h_{34}(n)$ \hspace{35mm} 
		$\mathcal{H}_{78,0}(n)=h_{78}(n)u_{0}(n) \sim h_{78}(n)$

		\caption{Illustration of localization kernels, $\mathcal{H}_{m,k}(n)=h_m(n) u_{k}(n)$, for vertex-frequency analysis based on \textit{spectral domain defined windows} within the local graph Fourier transform, $S(m,k)=\sum_{n=0}^{N-1} x(n)\mathcal{H}_{m,k}(n)$. (a) Localization kernel $\mathcal{H}_{34,0}(n)=h_{34}(n)u_{0}(n) \sim h_{34}(n)$, for a constant eigenvector, $u_0(n)=1/\sqrt{N}$, centered at the vertex $m=34$. (b) The same localization kernel as in (a) but centered at the vertex $m=78$. (c) Localization kernel, $\mathcal{H}_{34,20}(n)=h_{34}(n) u_{20}(n)$, centered at the vertex $m=34$ and frequency shifted by $u_{20}(n)$. Notice that the variations in kernel amplitude indicate the effects of modulation of the localization window, $h_m(n)$. (d) The same localization kernel as in (c), but centered at the vertex $m=78$.
			(e) Three-dimensional representation of the kernel $\mathcal{H}_{34,0}(n)=h_{34}(n)u_{0}(n)$. (f) Three-dimensional representation of the kernel $\mathcal{H}_{78,0}(n)=h_{78}(n)u_{0}(n)$.
		}
		\label{VF_windows_LVS}
	\end{figure*}

	\begin{figure}
		\centering
		\includegraphics[scale=0.9]{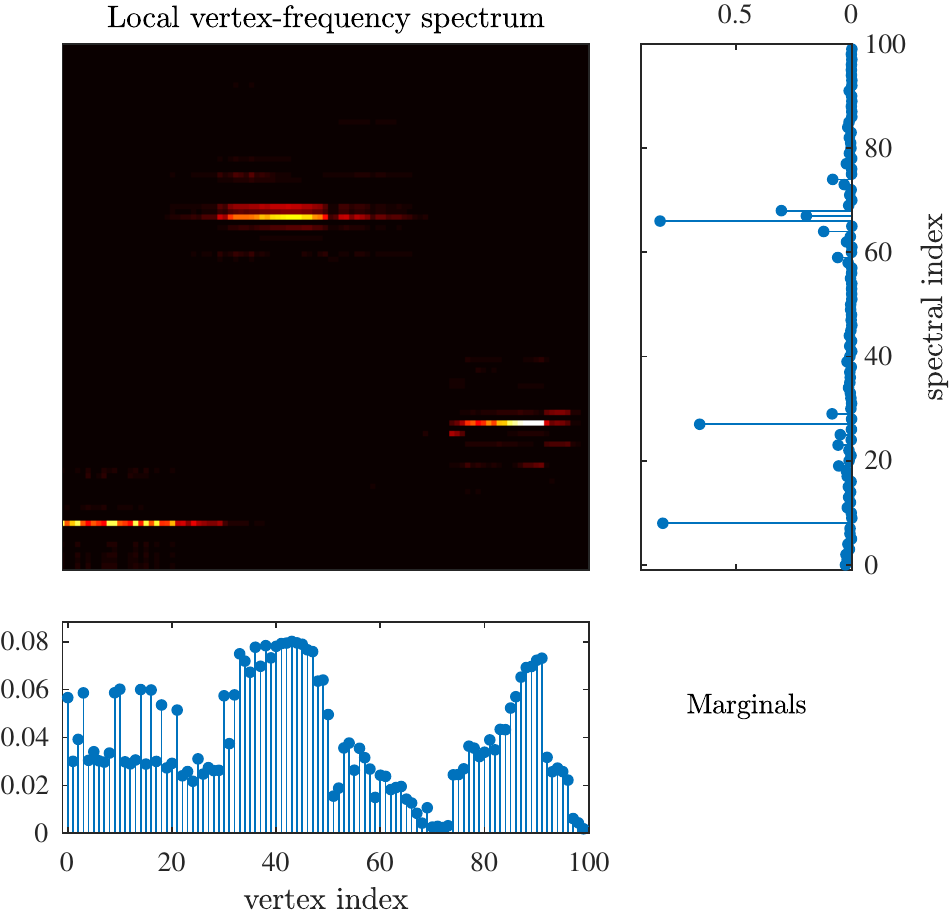}
		\caption{Local vertex-frequency spectrum calculated using the LGFT and the vertex-frequency localized kernels defined in the spectral domain, as in (\ref{KernelGFDFTmk}). From this representation, observe that the graph signal consists of three distinct components located at spectral indices $k=8$, $k=66$, and $k=27$, with the corresponding vertex index subsets $\mathcal{V}_1$, $\mathcal{V}_2$, and $\mathcal{V}_3$, where $\mathcal{V}_1 \cup \mathcal{V}_2 \cup \mathcal{V}_3=\mathcal{V}$. The marginal (vertex and spectrum-wise) properties are shown in the panels to the right and below the vertex-frequency representation. Observe that, while the graph signal is spread across all vertices, its spectral content is localized at the three spectral indices which correspond to the constituent signal components. In an ideal case of vertex-frequency analysis, these marginals should  be respectively equal to $|x(n)|^2$ and $|X(k)|^2$.   }
		\label{VF_graph3g}
	\end{figure}
	
\end{Example}

\subsubsection{Spectral Domain Localization of the LGFT}

Recall that the classical STFT admits frequency localization in the spectral domain; this is achieved based on the DFT of the original signal and a spectral domain window. For graph signals, we may also adapt this approach to perform signal localization in the spectral domain, whereby the LGFT is obtained as an inverse GDFT of $X(p)$ that is localized by a spectral domain window, $H(k-p)$, which is centered around spectral index $k$, that is   
\begin{equation}
S(m,k)=\sum_{p=0}^{N-1} X(p)H(k-p)\; u_{p}(m). \label{FDLGFT}
\end{equation}
Note that this form of the LGFT can be entirely implemented in the graph spectral domain. The spectral domain LGFT form in (\ref{FDLGFT}) can be implemented using band-pass transfer functions,  $H_k(\lambda_p)=H(k-p)$, as
\begin{equation}
S(m,k)=\sum_{p=0}^{N-1} X(p)H_k(\lambda_p)\; u_{p}(m). \label{FDLGFTLam}
\end{equation}
\begin{Remark}
 Recall that the classical time-frequency analysis counterpart of (\ref{FDLGFT}) is \cite{stankovic2014time} $$S(m,k)=\frac{1}{\sqrt{N}}\sum_{p=0}^{N-1} X(p)H(k-p)e^{j \frac{2 \pi}{N} mp}.$$  
\end{Remark}

\subsubsection{LGFT Realization with Band-Pass Functions}
Assume that the GDFT of the localization window, $h_m(n)$, corresponds to the transfer function of a band-pass system on a graph, centered at an eigenvalue, $\lambda_k$, and around it, and that it is defined in the form of a polynomial given by
\begin{equation}
H_k(\lambda_p)=h_{0,k}+h_{1,k}\lambda_p+\dots+h_{M-1,k}\lambda_p^{M-1}, \label{Huge_LConvFFF}
\end{equation}
with $(M-1)$ as the polynomial order and $k=0,1,\dots,K$, where $K$ is the number of spectral bands. 

The vertex shifted version of the window, $h_m(n)$, has the GDFT of the form,  $\mathrm{GDFT}\{h(n)*\delta_m(n)\} = H(p)u_p(m)$.  Therefore, the inverse GDFT of $H_k(\lambda_p)u_p(m)$ represents a vertex domain kernel, where $H_k(\lambda_p)$ is centered at the spectral index $k$ by definition, while $u_p(m)$ corresponds to the shift in the vertex domain which centers the window at the vertex $m$. In other words, this kernel, centered around the spectral index $k$ and  vertex $m$, is defined as
\begin{equation}
\mathcal{H}_{m,k}(n)=\sum_{p=0}^{N-1}H_k(\lambda_p)u_p(m)u_p(n). \label{kernelFR}
\end{equation}

\begin{Remark}
	It is important to emphasize crucial difference between the vertex-frequency kernels in (\ref{KernelGFDFTmk}) and (\ref{kernelFR}). The kernel in (\ref{KernelGFDFTmk}) is defined based on the low-pass transfer function $H(k)$, such as in (\ref{win-spect}),  appropriately shifted in the vertex domain and the spectral domain, to be centered at a vertex $m$ and at a spectral index $k$. This is achieved involving adequate modulation terms $u_k(n)$ and $u_p(m)$. The transfer function in the kernel given by (\ref{kernelFR}), $H_k(\lambda_p)$, is  centered at $k$ by definition  (\ref{Huge_LConvFFF}). Hence, it is needed to perform  the spectral  modulation only, by $u_p(n)$, in order to center the kernel, $\mathcal{H}_{m,k}(n)$, at a vertex $m$. Therefore, the main difference between the kernels in  (\ref{KernelGFDFTmk}) and (\ref{kernelFR}) is that the spectral shift in (\ref{KernelGFDFTmk}) is achieved by a  modulation in the vertex domain using $u_k(n)$, while in (\ref{kernelFR}) the kernel is directly shifted (defined as a pass-band function) in the spectral domain.

\noindent\textbf{Classical time-frequency domain kernel.} To additionally clarify the previous two forms of kernels, we will observe their special cases for a circular directed graph and write the kernels in the classical time-frequency domain. 

The kernel defined by (\ref{KernelGFDFTmk}) uses low-pass function $H(k)$ and assumes the following form
	\begin{gather*}
		\mathcal{H}_{m,k}(n)=\frac{1}{N^{3/2}}\sum_{p=0}^{N-1}H(p)e^{-j\frac{2\pi}{N}mp}e^{j\frac{2\pi}{N}np}e^{-j\frac{2\pi}{N}kn}\\
		=\frac{1}{N}h(n-m)e^{-j\frac{2\pi}{N}kn}=\frac{1}{\sqrt{N}}h_k(n-m),
	\end{gather*}
	which is shifted for $m$ in time, and modulated by the  $k$th eigenvector elements  $u^*_k(n)=e^{-j\frac{2\pi}{N}kn}/\sqrt{N}$, to achieve centering around the spectral index $k$.

The classical time-frequency domain form of the kernel in (\ref{kernelFR}) is given by
\begin{gather*}
\mathcal{H}_{m,k}(n)=\frac{1}{N}\sum_{p=0}^{N-1}H_k(\lambda_p)e^{-j\frac{2\pi}{N}mp}e^{j\frac{2\pi}{N}np}\\=\frac{1}{N}\sum_{p=0}^{N-1}H(p-k)e^{-j\frac{2\pi}{N}mp}e^{j\frac{2\pi}{N}np}=\frac{1}{\sqrt{N}}h_k(n-m),
\end{gather*}
where $h_k(n-m)$ is the temporary shifted version of $h_k(n)=\mathrm{IGDFT}\{H_k(\lambda_p)\}=\mathrm{IDFT}\{H(k-p)\}$, which corresponds to the already frequency shifted (band-pass) transfer function $H_k(\lambda_p)=H(p-k)$.
\end{Remark}

In the case of kernel (\ref{kernelFR}), the local vertex-frequency transform for a vertex, $m$, and a spectral index, $k$, becomes
\begin{gather}
S(m,k)=\sum_{n=0}^{N-1} \mathcal{H}_{m,k}(n)x(n) \nonumber \\  
\!=\!\!\!\sum_{n=0}^{N-1} \sum_{p=0}^{N-1} x(n)H_k(\lambda_p)u_p(m)u_p(n) 
\! = \!\!\! \sum_{p=0}^{N-1} X(p)H_k(\lambda_p)u_p(m).\label{PKerLGFT}
\end{gather}
The relation (\ref{PKerLGFT}) can be written in a vector form as
\begin{align}
\mathbf{s}_k&=\mathbf{U}H_k(\mathbf{\Lambda}) \mathbf{U}^T  \mathbf{x}=H_k(\mathbf{L}) \mathbf{x}=\sum_{p=0}^{M-1}h_{p,k} \mathbf{L}^p\, \mathbf{x}, \label{PolyVert}
\end{align} 
where $\mathbf{s}_k$ is the  column vector with elements $S(m,k)$, $m=0,1,\dots,N-1$, and the property of the eigendecomposition of a matrix polynomial is used in  derivation. \textit{The number of bands (shifted transfer functions, $H_k(\lambda_p)$, $k=0,1,\dots,K$) is equal to $K+1$ and is not related to the total number of indices, $N$. }

\begin{Example}\label{EX:simple_dec}  Consider the simplest decomposition into a low-pass and high-pass part of a graph signal, with $K=1$. In this case, the two values, $k=0$ and $k=1$, represent respectively the low-pass part and high-pass part of the graph signal. Such a decomposition can be achieved using the graph Laplacian with $h_{0,0}=1$, $h_{0,1}=-1/\lambda_{\max}$, and $h_{1,0}=0$, $h_{1,1}=1/\lambda_{\max}$, where the coefficients are chosen so as to form a simple linearly decreasing function of $\lambda_p$ for the low-pass, and a linearly increasing function of $\lambda_p$ for the high-pass, in the corresponding transfer functions. These low-pass and high-pass transfer functions are respectively given by
	\begin{align}
	H_0(\lambda_p)=(1-\frac{\lambda_p}{\lambda_{\max}}),  \,\,\, \,\,\,
	H_1(\lambda_p)=\frac{\lambda_p}{\lambda_{\max}}, \notag
	\end{align}
	which leads to the vertex domain implementation of the LGFT in the form
	\begin{align}
	\mathbf{s}_0=(\mathbf{I}-\frac{1}{\lambda_{\max}}\mathbf{L})\, \mathbf{x},  \,\,\, \,\,\,
	\mathbf{s}_1=\frac{1}{\lambda_{\max}}\mathbf{L}\, \mathbf{x}. \notag
	\end{align}
	
	To improve the spectral resolution, we can employ the same transfer function, but divide the low-pass part into its low-pass and high-pass part. The same can be performed for the high-pass part, to obtain $$\mathbf{s}_{00}=\Big(\mathbf{I}-\frac{\mathbf{L}}{\lambda_{\max}}\Big)^2 \mathbf{x}, \,\,\,\, \mathbf{s}_{01}=2\Big(\mathbf{I}-\frac{\mathbf{L}}{\lambda_{\max}}\Big)\frac{\mathbf{L}}{\lambda_{\max}} \mathbf{x}, \,\,\,\, \mathbf{s}_{11}=\frac{\mathbf{L}^2}{\lambda^2_{\max}} \mathbf{x}.$$  
	The factor 2 appears in the new middle pass-band, $\mathbf{s}_{01}$, since the low-high-pass and the high-low-pass components are the same. 
	
	A division into $(K+1)$ bands would correspond to the terms of a binomial form $$\Big((\mathbf{I}-\mathbf{L}/\lambda_{\max})+\mathbf{L}/\lambda_{\max}\Big)^K \, \mathbf{x},$$ with the corresponding transfer functions in the vertex domain given by
	$$H_k(\mathbf{L})={K \choose k}\Big(\mathbf{I}-\frac{1}{\lambda_{\max}}\mathbf{L}\Big)^{K-k}\Big(\frac{1}{\lambda_{\max}}\mathbf{L}\Big)^k.$$
\end{Example}  
\begin{Example}\label{EX:Tharf} Consider the transfer functions $H_k(\lambda_p)$, $p=0,1,\dots,N-1$, $k=0,1,\dots,K$ in the spectral domain, corresponding to the binomial form terms for $K=25$, which are shown in Fig. \ref{test_bands_a}(a). These functions are used for the LGFT calculation at vertex indices $m=0,1,\dots,N-1$ in the $k=0,1,\dots,K$ bands for the graph and signal from Fig. \ref{VF_graph3ab}. Since the bands are quite spread out, the resulting LGFT is also spread along the frequency axis. The frequency concentration can be improved by reassigning the values of $S(m,k)$ to the position of their maximum value along the frequency band index, $k$, for each vertex index, $m$. The so reassigned LGFT values are given in Fig. \ref{test_bands_b}.  
	
	\begin{figure}
		\centering
		\includegraphics[]{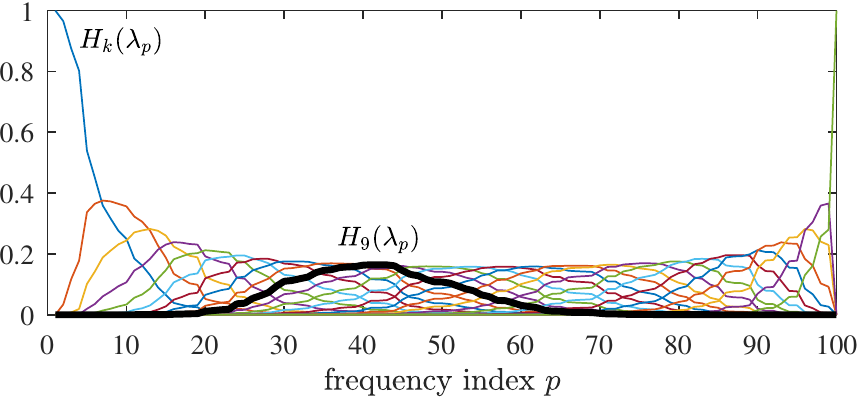}(a)
		
		\vspace{5mm}
		
		\includegraphics[]{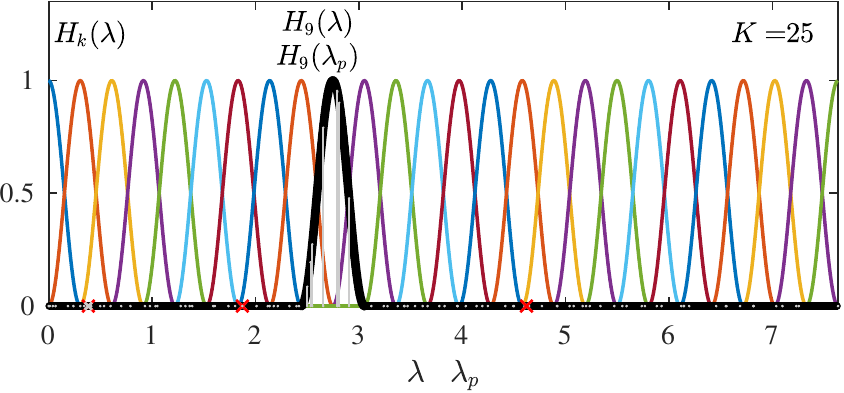}(b)
		
		\vspace{5mm}
		
		\includegraphics[]{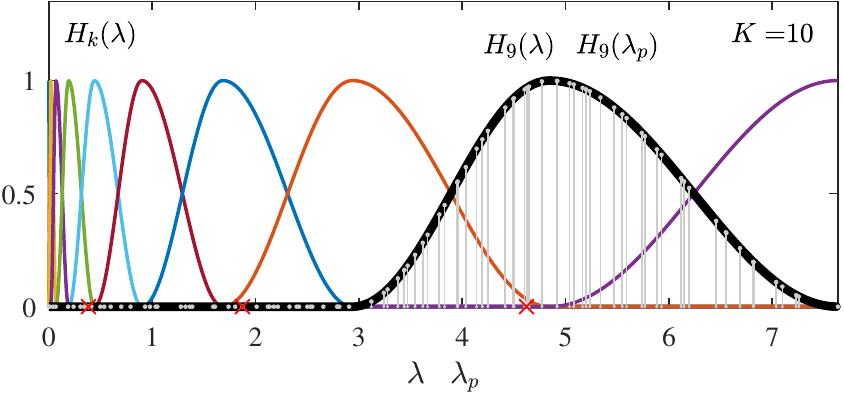}(c)
		\caption{Exemplar of transfer functions in the spectral domain. (a) The spectral domain transfer functions $H_k(\lambda_p)$, $p=0,1,\dots,N-1$, $k=0,1,\dots,K$ which correspond to the binomial form terms for $K=50$. (b) The transfer functions $H_k(\lambda_p)$, $p=0,1,\dots,N-1$, $k=0,1,\dots,K$ which correspond to the raised cosine (Hann) window form for $K=25$. (c) The spectral index-varying (wavelet-like) transfer functions $H_k(\lambda_p)$, $p=0,1,\dots,N-1$, $k=0,1,\dots,K$  which correspond to the raised cosine (Hann) window form for $K=13$.  
			The transfer function $H_9(\lambda)$ is designated by the thick black line for each considered domain, while its discrete values at $\lambda_p$, $H_9(\lambda_p)$, are shown in gray, in panels (b) and (c).
		}
		\label{test_bands_a}
	\end{figure}
	
	\begin{figure}
		\centering
		\includegraphics[]{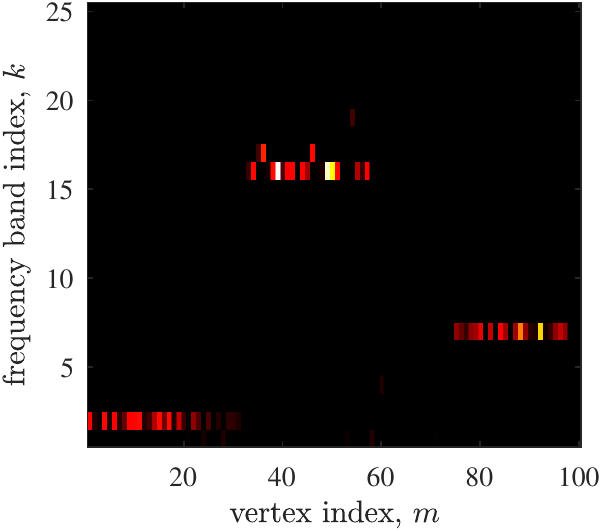}(a)
		
		\vspace{5mm}
		
		\includegraphics[]{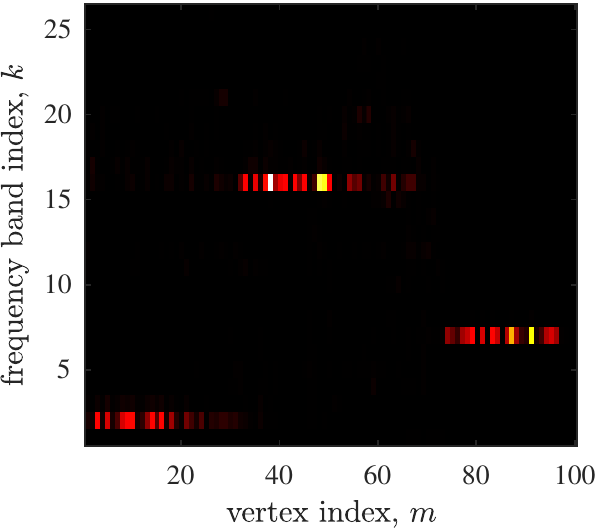}(b)
		
		\vspace{5mm}
		
		\includegraphics[]{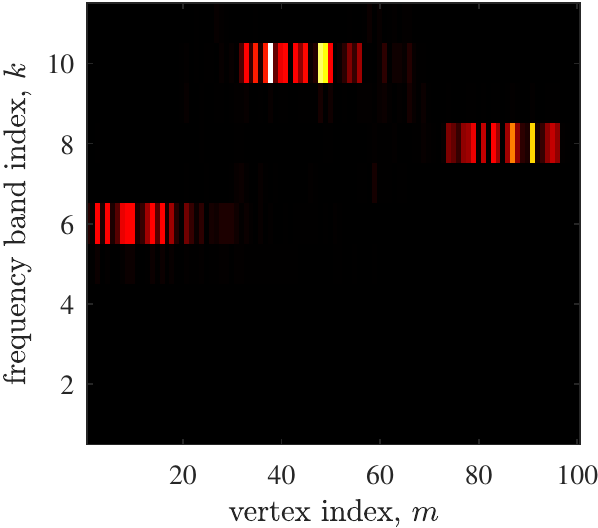}(c)
		\caption{Vertex-frequency representation of a three-component signal in Fig. \ref{VF_graph3ab}(d). (a) The LGFT of the signal from Fig \ref{VF_graph3ab}(d), calculated using the transfer functions for frequency selection given in Fig. \ref{test_bands_a}(a). The LGFT values, $S(m,k)$, were reassigned to the position of its maximum value along the frequency band index, $k$, for each vertex index, $m$. (b) The LGFT of the signal from Fig \ref{VF_graph3ab}(d), calculated using the transfer functions for frequency selection given in Fig. \ref{test_bands_a}(b). The LGFT values, $S(m,k)$, were reassigned to the positions of their maximum values along the frequency band index, $k$, for each vertex index, $m$. (c) The LGFT of the signal from Fig \ref{VF_graph3ab}(d), calculated using the wavelet-like transfer functions for frequency selection given in Fig. \ref{test_bands_a}(c). 
			%(d) Vertex-frequency representation from Fig. \ref{test_bands_b}(c) with the eigenvalue index, $p$, axis instead of the frequency band index, $k$.
		}
		\label{test_bands_b}
	\end{figure}
	
	%\begin{figure}
	%       \centering
	%       \includegraphics[]{test_bands_c}
	%       \caption{XXX bands c}
	%       \label{test_bands_c}
	%\end{figure}  

\begin{figure}
	\centering
	\includegraphics[]{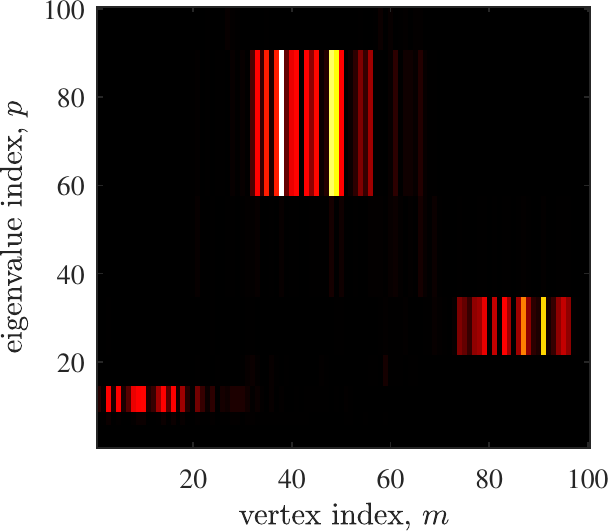} 
	\caption{Vertex-frequency representation from Fig. \ref{test_bands_b}(c) with the axis of the eigenvalue index, $p$,  instead of the frequency band index, $k$. The same value of LGFT, $S(m,k)$, is assigned to each spectral index, $p$, when $\lambda_p \in (\frac{a_k+b_k}{2},\frac{b_k+c_k}{2}]$, and without any scaling.}
	\label{test_bands_bC}
\end{figure}

\begin{figure}
	\centering
\includegraphics[]{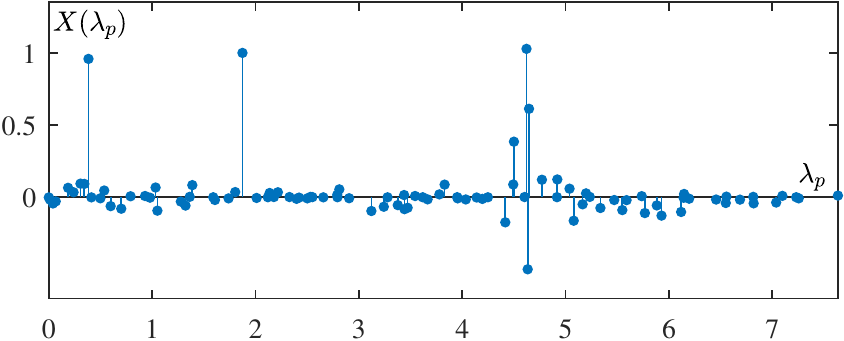}(a)

\vspace{5mm}

	\includegraphics[]{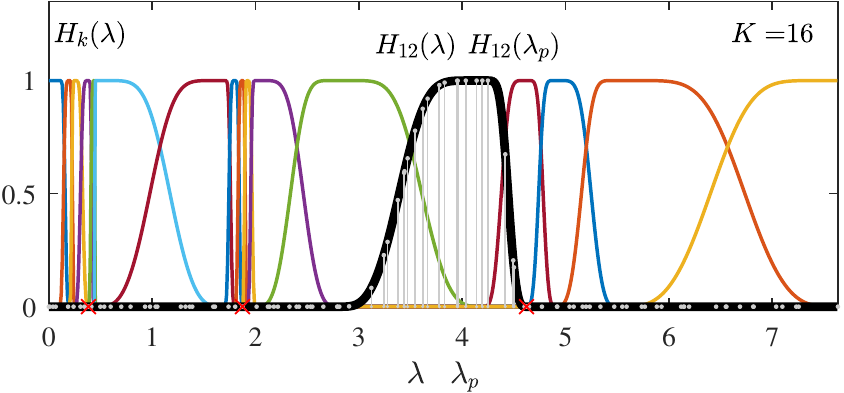}(b)
	
		\vspace{5mm}
		
	\includegraphics[]{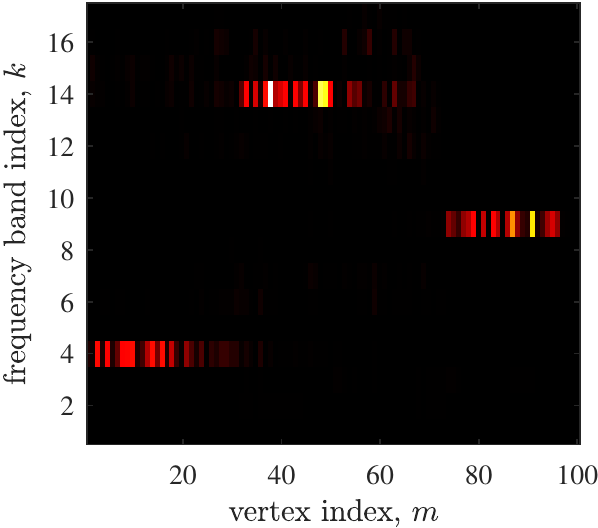}(c)
	
		\vspace{5mm}
		
	\includegraphics[]{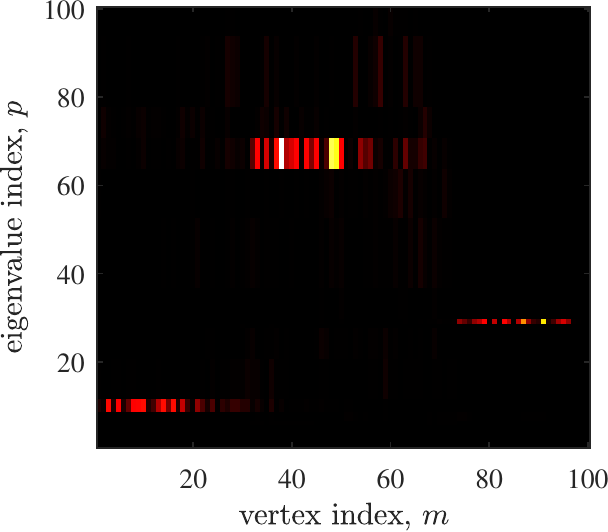}(d)
	\caption{A graph signal and transfer functions in the spectral domain for a signal adaptive LGFT. (a) Graph signal in the spectral domain, $X(p)$, as a function of the eigenvalues, $\lambda_p$. (b) The spectral domain transfer functions $H_k(\lambda_p)$, $p=0,1,\dots,N-1$, $k=0,1,\dots,K$ which satisfy the condition $\sum_{k=0}^{K}H_k^2(\lambda_p)=1$, with $K=16$. (c) The LGFT of the signal from Fig \ref{VF_graph3ab}(d), calculated using the transfer functions for frequency selection given in (b). (d) Vertex-frequency representation from (c) with the eigenvalue (spectral) index, $p$, axis instead of the frequency band index, $k$. The same value of LGFT, $S(m,k)$, is assigned to each spectral index, $p$, when $\lambda_p \in (\frac{a_k+b_k}{2},\frac{b_k+c_k}{2}]$, without any scaling. }
	\label{test_bands_bCSA}
\end{figure}

\end{Example}

Of course, any band-pass function, $H_k(\mathbf{\Lambda})$, can be used in (\ref{FDLGFTLam}) or (\ref{PKerLGFT}) to produce the LGFT in the form
\begin{align}
\mathbf{s}_k&=\mathbf{U}H_k(\mathbf{\Lambda}) \mathbf{U}^T  \mathbf{x}=H_k(\mathbf{L}) \mathbf{x}. \label{GenerFunVert}
\end{align}
Commonly used examples of such band-pass functions are the spline or raised cosine (Hann window) functions. We will next use the general form of the shifted raised cosine functions as the transfer functions, defined by  
\begin{equation}
H_k(\lambda)=\begin{cases}
	\sin^2\bigg(\frac{\pi}{2}\frac{a_k}{b_k-a_k}(\frac{\lambda}{a_k}-1)\bigg), \text{ for } a_k<\lambda \le b_k \\
	\cos^2\bigg(\frac{\pi}{2}\frac{b_k}{c_k-bk}(\frac{\lambda}{b_k}-1)\bigg), \text{ for } b_k<\lambda \le c_k \\
	0 , \text{ elsewhere, }
\end{cases} 
\label{defHannWGen}
\end{equation} 
where $(a_k,b_k]$ and $(b_k,c_k]$, $k=1,2,\dots,K$, define the spectral bands for $H_k(\mathbf{\Lambda})$. For uniform bands within $0\le \lambda \le \lambda_{\max}$, the intervals can be defined by 
\begin{gather}
a_k=a_{k-1}+\frac{\lambda_{\max}}{K} \nonumber \\
b_k=a_k+\frac{\lambda_{\max}}{K} \nonumber \\
c_k=a_k+2\frac{\lambda_{\max}}{K} \label{intervalsabc}
\end{gather} 
with $a_1=0$ and $\lim_{\lambda\to 0}(a_1/\lambda)=1$. 
The initial transfer function, $H_0(\lambda)$, is defined using only $0=b_0\le \lambda \le c_0=\lambda_{\mathrm{max}}/K$, while the last transfer function, $H_K(\lambda)$, is defined using the interval $a_K < \lambda \le b_K=\lambda_{\max}$  in (\ref{defHannWGen}).  

The raised cosine transfer function satisfy the following condition
\begin{equation}
\sum_{k=0}^{K}H_k(\lambda_p)=1.
\end{equation}
 The conditions for graph signal reconstruction from the LGFT will be discussed  in Section \ref{GFTInversion}.

\begin{Example}\label{EX:Tharf2}
The shifted raised cosine functions, defined by  (\ref{defHannWGen}) and (\ref{intervalsabc}),  are shown in Fig. \ref{test_bands_a}(b) for the graph from Fig. \ref{VF_graph3ab},  for $K=25$. These functions are used for the LGFT calculation of the graph signal from Fig. \ref{VF_graph3ab} at the vertex indices $m=0,1,\dots,N-1$, and in $(K+1)$ spectral bands, $k=0,1,\dots,K$. The absolute LGFT values are given in Fig. \ref{test_bands_b}(b). Spectral resolution depends on the number of bands $K$, with a larger number of spectral bands resulting in a higher spectral resolution.  
\end{Example}

\begin{Example} The experiment from Examples \ref{EX:Tharf} and \ref{EX:Tharf2}  is repeated with varying bounds of the spectral intervals in the raised cosine transfer functions $H_k(\lambda_p)$, $p=0,1,\dots,N-1$, $k=0,1,\dots,K$. The spectral index-varying (wavelet-transform like) form of the raised cosine transfer functions $H_k(\lambda_p)$, $p=0,1,\dots,N-1$, $k=0,1,\dots,K$, is defined by the interval bounds 
$\lambda_{\max}\Big(\big(1.5+p\big)/11.5\Big)^5$, for $p=0,1,2,\dots,10$,
\begin{gather*}
a_k \in \{0,0.004, 0.02,0.07, 0.19, 0.44,    0.9,    1.7,  2.9\}, \\
b_k \in \{0.004, 0.02,0.07, 0.19, 0.44,    0.9,    1.7,  2.9, 4.8\}, \\
c_k \in \{0.02,0.07, 0.19, 0.44,    0.9,    1.7,  2.9, 4.8,7.63\}, \\
k=1,2,\dots,9,
\end{gather*} 
and depicted in Fig. \ref{test_bands_a}(c).  The LGFT values, $S(m,k)$, calculated with the  so-obtained transfer functions, $H_k(\lambda_p)$,  are shown in Fig. \ref{test_bands_b}(c). In order to illustrate the change of resolution in this case, the LGFT was reassigned to each eigenvalue $\lambda_p$, $p=0,1,\dots,N-1$, and shown in Fig. \ref{test_bands_bC}. As in classical wavelet transform, the spectral resolution is lower for the higher spectral indices.   
\end{Example} 

\subsubsection{Signal Adaptive LGFT}

The spectral graph wavelet-like transform is just an example of  varying  spectral transfer functions in the LGFT, where the spectral resolution is the highest (spectral wavelet functions narrowest) for small values of the smoothness index, $\lambda_p$. The spectral resolution decreases as the spectral wavelet functions become wider for large smoothness index values, Fig. \ref{test_bands_a}(c). In general, the change of resolution may be arbitrary and signal adaptive, for example, the resolution may be higher for the spectral intervals of $\lambda$ which are rich in signal components and lower within the intervals where there are no signal components. 

Before introducing an example with a signal adaptive LGFT, we will modify  the transfer functions, $H_k(\lambda_p)$, in (\ref{defHannWGen}) to satisfy the condition
\begin{equation}
\sum_{k=0}^{K}H_k^2(\lambda_p)=1.
\end{equation}
as this will be important for the frame-based LGFT inversion. 

Notice that a simple transformation of the transfer functions, $H_k(\lambda_p) \rightarrow H^2_k(\lambda_p)$, would allow for the condition $\sum_{k=0}^{K}H^2_k(\lambda_p)=1$ to hold instead of $\sum_{k=0}^{K}H_k(\lambda_p)=1$. This means that a simple removal of squares in the sine and cosine functions in (\ref{defHannWGen}) would produce a form to satisfy the  condition $\sum_{k=0}^{K}H_k^2(\lambda_p)=1.$ Both of these conditions will be used in Section \ref{GFTInversion} in various approaches to the graph signal reconstruction from the LGFT.

By removing the squares in the sine and cosine functions in (\ref{defHannWGen}), their first derivative loses continuity in $\lambda$ at the end interval points. In order to preserve continuous derivatives, the arguments in the sine and cosine functions can be mapped by a polynomial,
$$v_x(x)=x^4(35-84x+70x^2-20x^3), \text{ for } 0 \le x \le 1,$$
with $
v_x(0)=0$ and $v_x(1)=1$.
In this way, we arrive at the Meyer wavelet-like transfer functions \cite{meyer1992wavelets} for the LGFT calculation, given by
\begin{equation}
H_k(\lambda)=\begin{cases}
\sin\bigg(\frac{\pi}{2}v_x\Big(\frac{a_k}{b_k-a_k}(\frac{\lambda}{a_k}-1)\Big)\bigg), \text{ for } a_k<\lambda \le b_k\\
\cos\bigg(\frac{\pi}{2}v_x\Big(\frac{b_k}{c_k-b_k}(\frac{\lambda}{b_k}-1)\Big)\bigg), \text{ for } b_k<\lambda \le c_k\\
0 , \text{ elsewhere. }
\end{cases} 
\label{MeyerdefHannWGen}
\end{equation} 
The initial transfer function, $k=0$, and the last transfer function, $k=K$, are calculated using only the half of the interval, as explained after the spectral band definition in relation (\ref{intervalsabc}).

\begin{Example}
	The transfer functions of the form defined in (\ref{MeyerdefHannWGen}) are used with signal adaptive intervals. These intervals are defined in such a way that they are small (fine) around $\lambda$, where a significant signal spectral content is detected, and are big (rough) around $\lambda$  where the signal spectral content is low, as in Fig. \ref{test_bands_bCSA}(a) and (b). The intervals are narrow (with a high resolution) around the three signal components at $\lambda=0.38$, $\lambda=1.87$, and $\lambda=4.62$. Vertex-frequency representation with these transfer functions is shown in Fig. \ref{test_bands_bCSA}(c) and (d) with the spectral band index, $k$, and the assigned eigenvalue (spectral) index, $p$, as a spectral axis. Fine intervals around the spectral signal components allowed for high spectral resolution representation, as in  Fig. \ref{test_bands_bCSA}(c), with a smaller number of transfer functions $K+1=17$.  A wider interval width for the third component resulted in a lower spectral resolution than in the case of the other two components.  		
\end{Example}

\subsubsection{Polynomial LGFT Approximation}
Bandpass LGFT functions, $H_k(\lambda),~k=0,1,\dots,K$, of the form (\ref{defHannWGen}) or (\ref{MeyerdefHannWGen}) can be implemented using the Chebyshev finite $(M-1)$-order polynomial approximation,  $\bar{P}_{k,M-1}(\lambda)$, $k=0,1,\dots,K$, of the form
\begin{equation}
\bar{P}_{k,M-1}(\lambda)=\frac{c_{k,0}}{2}+\sum_{m=1}^{M-1}c_{k,m}\bar{T}_m(\lambda).
\end{equation}
This leads to the vertex domain implementation of the spectral LGFT form, given by
$$\mathbf{s}_k=\bar{P}_{k,M-1}(\mathbf{L})\mathbf{x},$$
for $k=0,1,2,\dots,K,$ with
\begin{gather}
\bar{P}_{k,M-1}(\mathbf{L})=\frac{c_{k,0}}{2}+\sum_{m=1}^{M-1}c_{k,m}\bar{T}_m(\mathbf{L}), \label{plylgft} \\
=h_{0,k}\mathbf{I}+h_{1,k}\mathbf{L}+h_{2,k}\mathbf{L}^2+\dots+h_{(M-1),k}\mathbf{L}^{M-1} \nonumber
\end{gather}
as discussed in Section \ref{Plyappsystem} and shown in Table \ref{tablecoeffLGHT}. The polynomial form in (\ref{plylgft}) uses only  the
 $(M-1)$-neighborhood in calculation of the LGFT for each considered vertex, without the need for eigendecomposition analysis, thus significantly reducing the computational cost.

\begin{figure}
	\centering
	\includegraphics[]{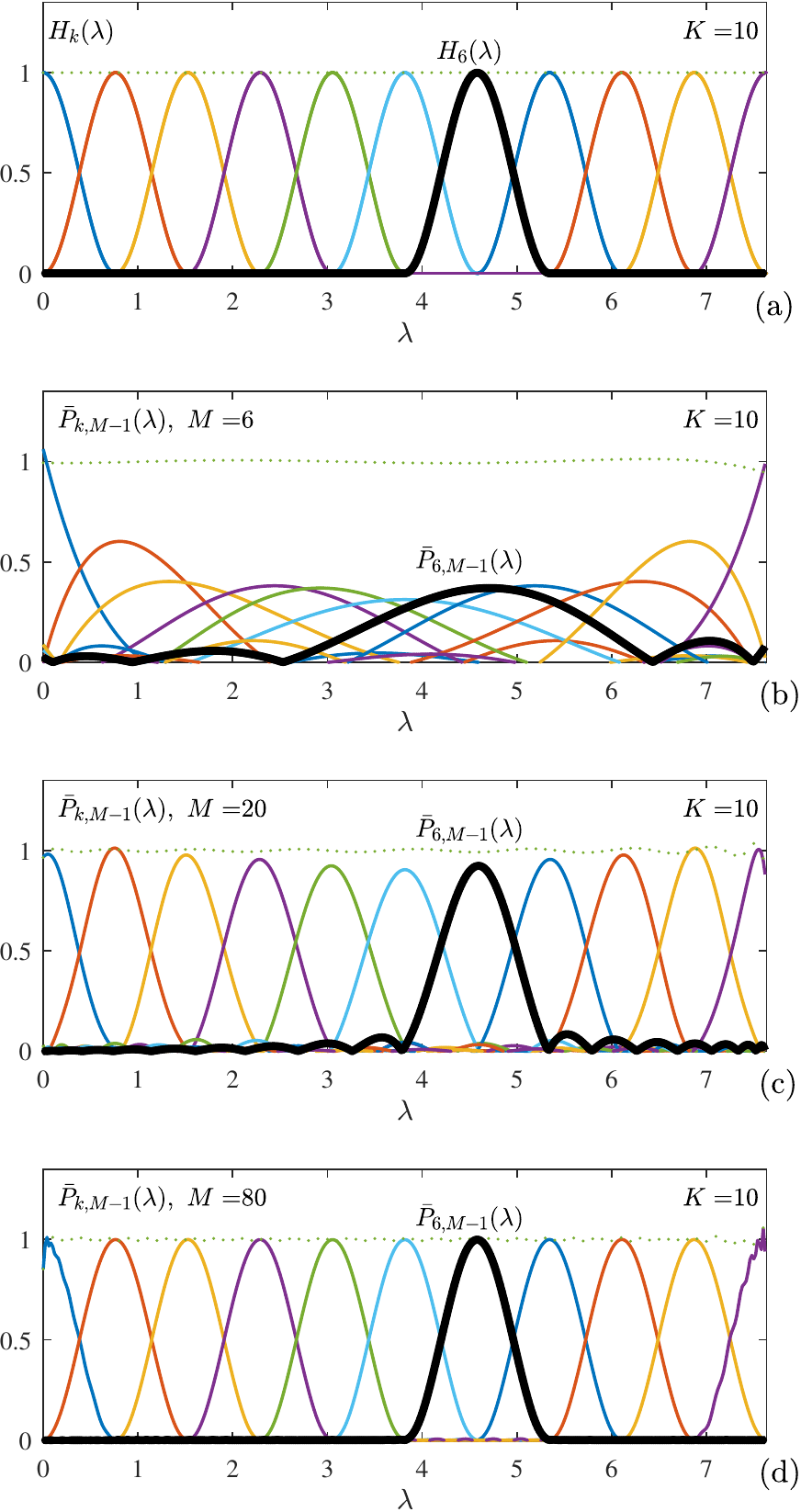}

	\caption{Chebyshev approximation of LGFT transfer functions, which correspond to the raised cosine window in the spectral domain. 
		(a) Original transfer functions $H_k(\lambda),~k=0,1,\dots,K$, for $K=10$. The  dotted horizontal line  designates $\sum_{k=0}^{K}H_k(\lambda)$. (b) Polynomial Chebyshev approximations, $\bar{P}_{k,M-1}(\lambda),k=0,1,\dots,K$, with $M=6$.  (c)  Polynomial Chebyshev approximations, $\bar{P}_{k,M-1}(\lambda),k=0,1,\dots,K$, with $M=20$. (d)  Polynomial Chebyshev approximations, $\bar{P}_{k,M-1}(\lambda),k=0,1,\dots,K$, with $M=80$. The dotted horizontal  line designates $\sum_{k=0}^{K}\bar{P}_{k,M-1}(\lambda)$, which is close to $1$ in all considered approximations, thus guaranteeing stable transform invertibility.
		Transfer function $H_6(\lambda)$ and approximations, $\bar{P}_{6,M-1}(\lambda)$, are designated by the thick black line.
	}
	\label{LGFT_Chebyshev}
\end{figure}

\begin{figure}
	\centering
	\includegraphics[]{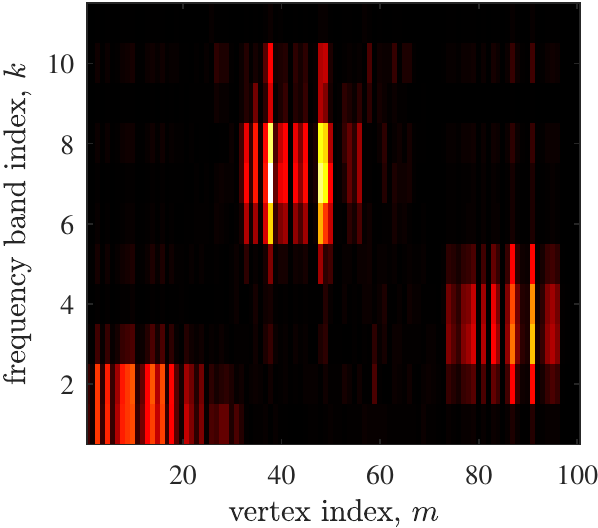}(a)
	
	\vspace{5mm}
	
	\includegraphics[]{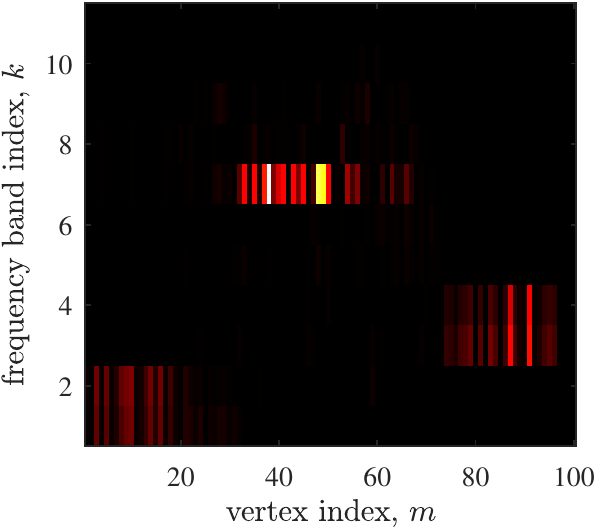}(b)
	
	\vspace{5mm}
	
	\includegraphics[]{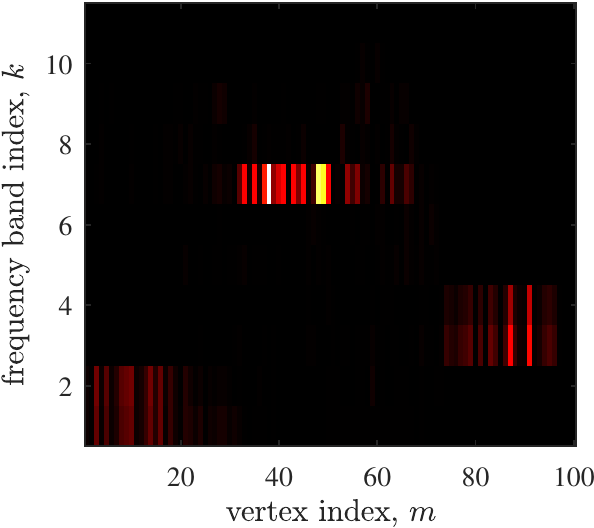}(c)
	\caption{Vertex-frequency representation of a three-component signal in Fig. \ref{VF_graph3ab}(d). The LGFT is based on raised cosine (Hann window) like bandpass transfer functions for frequency selection, with  $K=10$, approximated using the Chebyshev polynomials of various order, as shown in Fig. \ref{LGFT_Chebyshev} (b), (c), (d).  
		(a) The LGFT of the signal from Fig \ref{VF_graph3ab}(d), calculated using the Chebyshev polynomial approximation  of transfer functions given in Fig. \ref{LGFT_Chebyshev} (b), with $M=6$.  
		(b) The LGFT of the signal from Fig \ref{VF_graph3ab}(d), calculated using the transfer functions for frequency selection given in Fig. \ref{test_bands_a}(c), with $M=20$.  
		(c)  The LGFT of the signal from Fig \ref{VF_graph3ab}(d), calculated using the transfer functions for frequency selection given in Fig. \ref{test_bands_a}(d), with $M=80$. Low resolution in (a) can be directly related with low $M=6$ used in approximation in Fig. \ref{LGFT_Chebyshev} (b). The resolution is considerably improved for $M=20$.
		}
	\label{chebyshev_LGFT}
\end{figure}

\begin{Example}\label{ExLGFTCHP}
	Consider the shifted transfer functions, $H_k(\lambda),~k=0,1,\dots,K$, defined by  (\ref{defHannWGen}) and (\ref{intervalsabc}), shown in Fig. \ref{LGFT_Chebyshev}(a), for $K=10$. Functions $H_k(\lambda)$ satisfy $\sum_{k=0}^{K}H_k(\lambda)=1$, which is numerically confirmed and designated by the horizontal dotted line in \ref{LGFT_Chebyshev}(a). Each individual transfer function,  $H_k(\lambda)$, is approximated using the Chebyshev polynomial, $\bar{P}_{k,M-1},k=0,1,\dots,K$, as detailed in Section \ref{Plyappsystem}, with three polynomial orders defined by $M=6$, $M=20$ and $M=80$. These polynomial approximations are shown in Fig. \ref{LGFT_Chebyshev}(b), (c) and (d). In each considered case, summations $\sum_{k=0}^{K}\bar{P}_{k,M-1}(\lambda)$ are calculated. It can be observed that for different values of $M$, the summations in all considered cases are very close to 1, thus guaranteeing numerically stable invertibility of the LGFT, as discussed later. 
	
	\begin{table}
	\centering
	\caption{Coefficients, $h_{i,k}$, $i=0,1,\dots,M-1$, $k=0,1,\dots,K$,  for the polynomial calculation of the LGFT, $\mathbf{s}_k$, of a signal, $\mathbf{x}$ , in various spectral bands, $k$, shown in Fig. \ref{LGFT_Chebyshev}(b). The obtained LGFT of the three-component signal from Fig. \ref{VF_graph3ab}(d) is given in Fig. \ref{chebyshev_LGFT}(a). \vspace*{2mm}}
	\small
	\setlength{\tabcolsep}{3pt}

	\begin{tabular}{crrrrrr}
	\multicolumn{7}{c}{$\mathbf{s}_k=(h_{0,k}\mathbf{I}+h_{1,k}\mathbf{L}+h_{2,k}\mathbf{L}^2+h_{3,k}\mathbf{L}^3+h_{4,k}\mathbf{L}^4+h_{5,k}\mathbf{L}^5) \mathbf{x}$} \\
	\midrule
$k$ & $h_{0,k}$ & $h_{1,k}$ & $h_{2,k}$ & $h_{3,k}$ & $h_{4,k}$ & $h_{5,k}$ \\
\midrule
  0 & $ 1.062$ & $-1.925$ & $ 1.168$ & $-0.3115$ & $ 0.03776$ & $-0.001702$ \\
  1 & $-0.002$ & $ 1.773$ & $-1.655$ & $ 0.5357$ & $-0.07250$ & $ 0.003508$ \\
  2 & $-0.154$ & $ 1.016$ & $-0.601$ & $ 0.1295$ & $-0.01155$ & $ 0.000349$ \\
  3 & $ 0.005$ & $-0.301$ & $ 0.621$ & $-0.2674$ & $ 0.04200$ & $-0.002225$ \\
  4 & $ 0.089$ & $-0.748$ & $ 0.869$ & $-0.3042$ & $ 0.04217$ & $-0.002040$ \\
  5 & $ 0.060$ & $-0.381$ & $ 0.319$ & $-0.0704$ & $ 0.00461$ & $ 0.000000$ \\
  6 & $-0.024$ & $ 0.277$ & $-0.430$ & $ 0.2055$ & $-0.03570$ & $ 0.002040$ \\
  7 & $-0.076$ & $ 0.598$ & $-0.714$ & $ 0.2814$ & $-0.04292$ & $ 0.002225$ \\
  8 & $-0.027$ & $ 0.159$ & $-0.122$ & $ 0.0198$ & $ 0.00177$ & $-0.000349$ \\
  9 & $ 0.087$ & $-0.699$ & $ 0.868$ & $-0.3662$ & $ 0.06140$ & $-0.003508$ \\
 10 & $-0.026$ & $ 0.220$ & $-0.293$ & $ 0.1333$ & $-0.02435$ & $ 0.001536$ \\
 \bottomrule
	\end{tabular}
\label{tablecoeffLGHT}
	\end{table}

	The so obtained approximations of transfer functions,  $H_k(\lambda)$, are used  for the LGFT based vertex-frequency analysis. Absolute LGFT values, calculated for the three-component graph signal from Fig. \ref{VF_graph3ab}(d), are shown in Fig. \ref{chebyshev_LGFT}(a),(b) and (c), for $M=6$, $M=20$ and $M=80$. Low resolution in Fig. \ref{chebyshev_LGFT}(a) is directly related to the imprecise and very wide (with a low spectral resolution) approximation of the spectral transfer functions for $M=6$, in Fig. \ref{LGFT_Chebyshev} (b). Notice that high values of the polynomial order, $(M-1)$, increase calculation complexity and require wide vertex neighborhood in the calculation of the LGFT.

	 Based on the analysis of calculation complexity in Section \ref{Plyappsystem} we may conclude that an order of $KMN_{\mathbf{L}}$ of arithmetic operations is needed to calculate the LGFT in the vertex domain, with $(K+1)$ spectral bands, using a polynomial whose order is $(M-1)$. The number of nonzero elements in the graph Laplacian is denoted by $N_{\mathbf{L}}$.   
	% for the graph and signal from Fig. \ref{VF_graph3ab},  for $K=25$.
\end{Example}

\subsubsection{The Spectral Graph Wavelet Transform}

As in classical signal processing, wavelet coefficients can be defined as a projection of a graph signal onto the wavelet kernel functions. Assume that the basic form for the wavelet definition in the spectral domain  is a band-pass function, $H(\lambda)$. The wavelet in spectral domain then represents a scaled version of $H(\lambda)$ in scale $s_i$, $i=1,2,\dots,K$, and is denoted by $H(s_i\lambda)$.  The  wavelet kernel is already, by definition, of a high-pass form. Now, in the same way as in the case of the kernel form of the LGFT in (\ref{PKerLGFT}), 
the graph wavelet transform (spectral graph wavelet transform -- SGWT) is defined using the wavelet kernel, $\psi_{m,s_i}(n)$, 
\begin{equation}
\psi_{m,s_i}(n)= \sum_{p=0}^{N-1}H(s_i\lambda_p)u_p(m)u_p(n), \label{Waveletmk}
\end{equation}
which corresponds to the LGFT kernel, $\mathcal{H}_{m,k}(n)$, defined in (\ref{kernelFR}).  This yields the wavelet coefficients given by
\begin{gather*}
W(m,s_i)= \sum_{n=0}^{N-1} \psi_{m,s_i}(n)x(n)= \\
\sum_{n=0}^{N-1} \sum_{p=0}^{N-1}  H(s_i\lambda_p)x(n)u_p(m)u_p(n)
= \sum_{p=0}^{N-1}  H(s_i\lambda_p)X(p)u_p(m).
\end{gather*}

	The wavelet coefficients may be interpreted as the IGDFT of $ H(s_i\lambda_p)X(p)$, that is
	\begin{equation}
	W(m,s_i)= \mathrm{IGDFT}\{H(s_i\lambda_p)X(p)\}. \label{WaveletmkIDGDFT}
	\end{equation}

\begin{Remark}\label{IndexinWTSca}	
	We will use the notation $H(s_i\lambda)=H_i(\lambda)$ 	
	with the corresponding matrix function form $H_i(\boldsymbol{\Lambda})$. Notice that this scale-based indexing is opposite to the classical frequency band indexing. 
	The largest scale for $H(s_1 \lambda)$, $1<s_1 \lambda \leq M$,  is obtained for the smallest $s_1$,  $1/s_1 <\lambda \leq M/s_1$, where $M>1$ is the coefficient of the scale changes, which will be explained later. The associated spectral wavelet transfer function, $H(s_1\lambda)=H_1(\lambda)$, corresponds to the highest frequency band. The wavelet transfer function in scale $s_K$,   $H(s_K\lambda)=H_K(\lambda)$, is associated with the lowest frequency band. 
	Notation for the spectral scale function (low-pass transfer function complementary to $H(s_K\lambda)$ within the lowest spectral interval) is  $G(\lambda)$. The spectral scale function, $G( \lambda)$, plays the role of low-pass transfer function with spectral index 0 in the LGFT. Therefore,  $K$ spectral wavelet transfer functions $H(s_i\lambda)$, $i=1,2,\dots,K$, along with the scale function $G(\lambda)$, cover exactly $K+1$ spectral bands as in the LGFT case.
\end{Remark}	
	 According  to (\ref{eq:filterLF-time}), we can write
		\begin{equation}
		\mathbf{w}_i= H_i(\mathbf{L})\mathbf{x}, 
		\end{equation}
		where $\mathbf{w}_i$ a column vector with elements $W(m,s_i)$, $m=0,1,\dots,N-1$.
		
	If $H_i(\lambda)=H(s_i\lambda)$ can be approximated by a polynomial in $\lambda$, $H_i(\lambda) \approx P_i(\lambda)$, then the relation
	\begin{equation}
	\mathbf{w}_i \approx P_i(\mathbf{L})\mathbf{x}, 
	\end{equation}
	follows, where $P_i(\mathbf{L})$ is
	a polynomial in the graph Laplacian (see Section \ref{Plyappsystem} and Example \ref{ExLGFTCHP}).
	
\begin{figure}
		\centering
		\includegraphics[]{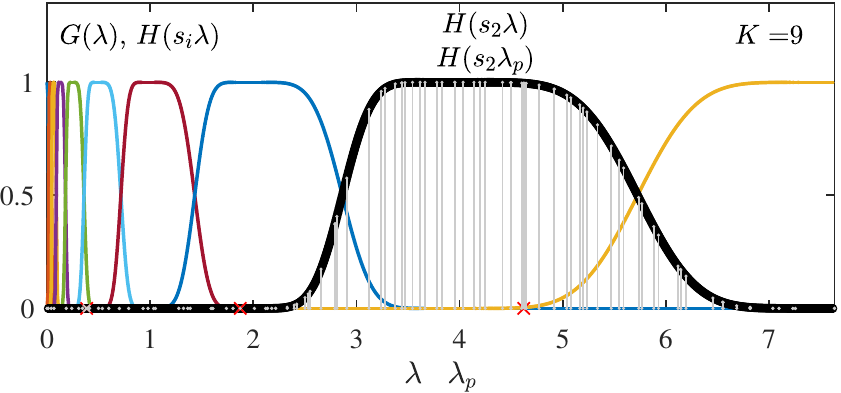}(a)
		
		\vspace{5mm}
		
			\includegraphics[]{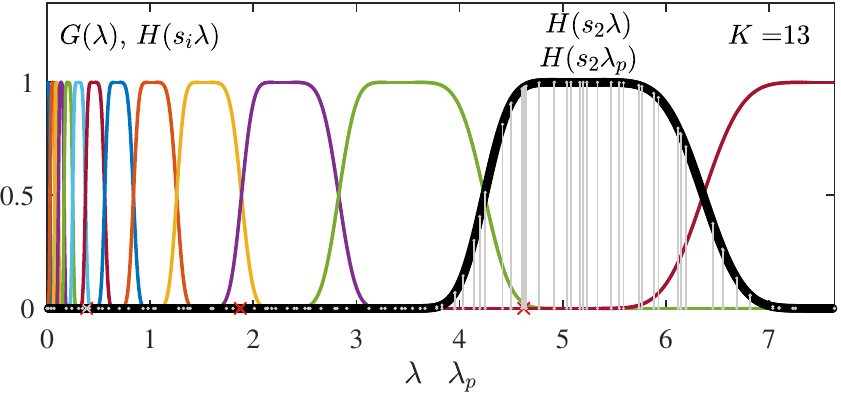}(b)
		
		\vspace{5mm}
		
		\includegraphics[]{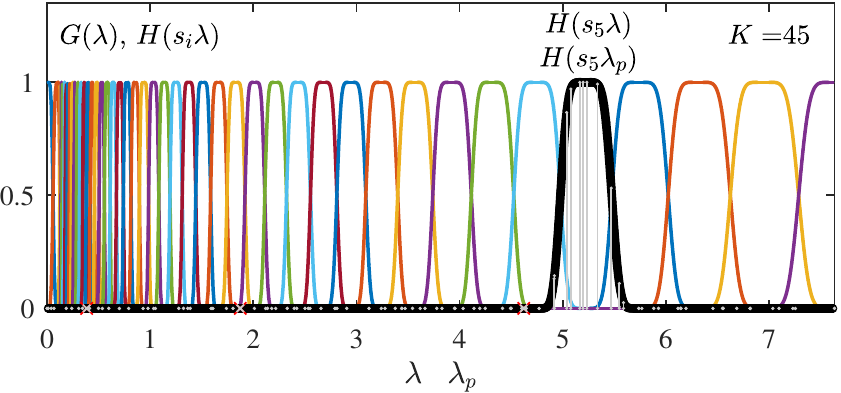}(c)
		\caption{Exemplars of  Meyer wavelet functions (acting as transfer functions in the wavelet transform), shown in the spectral domain. (a) Band-pass Meyer wavelet functions  $H(s_i\lambda)$, $i=1,2,\dots,K$ and the low-pass scale function $G(\lambda)$, for $K=9$ and  $M=2$. (b)  Band-pass Meyer wavelet functions  $H(s_i\lambda)$, $i=1,2,\dots,K$ and the low-pass scale function $G(\lambda)$, for $K=13$ and $M=3/2$. (c) Band-pass Meyer wavelet functions  $H(s_i\lambda)$, $i=0,1,\dots,K$ and the low-pass function $G(\lambda)$, for $K=45$ and $M=10/9$. Transfer functions $H(s_2\lambda)$, $H(s_2\lambda)$, $H(s_5\lambda)$ are designated by the tick black line, for each of the considered setups in (a), (b) and (c), respectively; their values at $\lambda_p$ are shown in gray.}
		\label{test_wavelet_meyer_bands_a}
	\end{figure}

\begin{figure}
		\centering
		\includegraphics[]{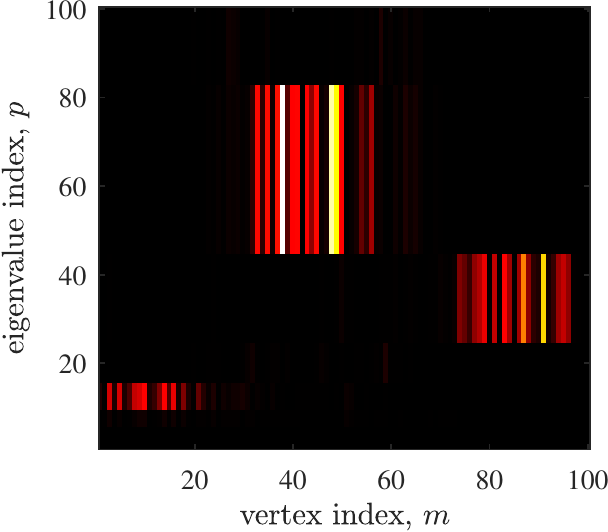}(a)
		
		\vspace{5mm}
		
		\includegraphics[]{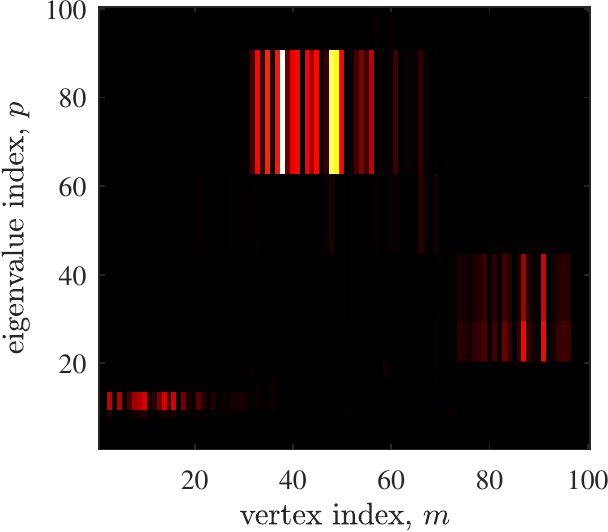}(b)
		
		\vspace{5mm}
		
		\includegraphics[]{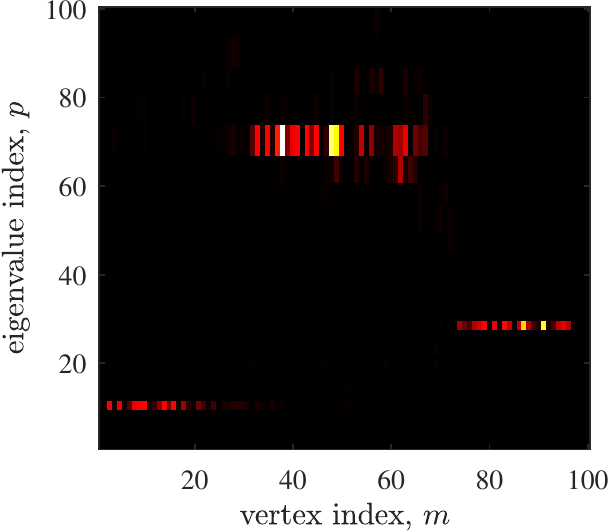}(c)
		\caption{
			Vertex-frequency representation of a three-component signal in Fig. \ref{VF_graph3ab}(d). (a) The Meyer wavelet transform of the signal from Fig \ref{VF_graph3ab}(d), calculated using the transfer functions for frequency selection given in Fig. \ref{test_wavelet_meyer_bands_a}(a). (b)  The Meyer wavelet transform of the signal from Fig \ref{VF_graph3ab}(d), calculated using the transfer functions for frequency selection given in Fig. \ref{test_wavelet_meyer_bands_a}(b). (c) The Meyer wavelet transform  of the signal from Fig \ref{VF_graph3ab}(d), calculated using  the Meyer wavelet transform transfer functions for frequency selection given in Fig. \ref{test_wavelet_meyer_bands_a}(c). Wavelet values were reassigned to spectral indices, $p$, in order to illustrate the change in resolution. The same value of SGWT, $W(m,k)$, is assigned to each spectral index, $p$, when $\lambda_p \in (\frac{a_k+b_k}{2},\frac{b_k+c_k}{2}]$, without any scaling. }
		\label{test_wavelet_meyer_bands_b}
	\end{figure}

	\begin{Example}\label{DefLGFTWVT}
	The wavelet transform (vertex-scale) representation of a three-component signal in Fig. \ref{VF_graph3ab}(d), obtained using the Meyer-like graph wavelet in the spectral domain, $\lambda$, will be illustrated  here.  As in classical wavelet transform, the wavelet in the first scale should correspond to the high-pass transfer function with nonzero values in the interval $\lambda_{\max}/M < \lambda \le \lambda_{\max}$, where $M>1$ is the coefficient of the scale changes. In classical wavelet transforms the dyadic scheme with $M=2$ is commonly used. \textit{The scale based indexing is opposite to the classical frequency indexing, where large indices indicate the high frequency content.} The  Meyer-like graph wavelet in the first scale is defined by \cite{meyer1992wavelets,leonardi2013tight}   	
\begin{gather*}
H(s_1\lambda)=\begin{cases}
\sin\bigg(\frac{\pi}{2}v_x\Big(q(s_1\lambda-1)\Big)\bigg), \text{ for } 1<s_1\lambda \le M, \\
0 , \text{ elsewhere. }
\end{cases} 
\end{gather*}
For $2\le i \le K$ the Meyer-like graph wavelet is given by
\begin{gather*}
	H(s_i\lambda)=\begin{cases}
	\sin\bigg(\frac{\pi}{2}v_x\Big(q(s_i\lambda-1)\Big)\bigg), \text{ for } 1<s_i\lambda \le M\\
	\cos\bigg(\frac{\pi}{2}v_x\Big(q(\frac{s_i\lambda}{M}-1)\Big)\bigg), \text{ for } M<s_i\lambda \le M^2\\
	0 , \text{ elsewhere, }
	\end{cases} \label{defMeyerW}
	\end{gather*}	 
	where $q=1/(M-1)$. The initial interval is defined by $s_1=M/\lambda_{\max}$,  so that $1<s_i\lambda \le M$ corresponds to $\lambda_{\max}/M<\lambda \le \lambda_{\max}$, while the other interval bounds are defined using a geometric sequence of scale factors, 
	$$s_i=s_{i-1}M=s_1M^{i-1}=\frac{1}{\lambda_{\max}}M^{i}.$$ 
	Observe that the larger the scale factor $s_i$ (and the scale index $i$), the narrower the transfer function, $H(s_i\lambda)$, while the progression coefficient is 
	$$M=(q+1)/q>1.$$ 
	In classical wavelet transforms the dyadic scheme with $M=2$ is commonly used. 
	 The last value of the scale factor, $s_K=M^K/\lambda_{\max}/M$, is defined by $K$ and indicates how close the last wavelet transfer function is to $\lambda=0$.
	
	The polynomial function, $v_x(x)$, is defined by 
		\begin{gather}
		v_x(x)=x^4(35-84x+70x^2-20x^3), \text{ for } 0 \le x \le 1, \text{ with} \nonumber\\ 
		v_x\Big(q(0)\Big)=v_x(0)=0, \text{ } v_x\Big(q(M-1)\Big)=v_x(1)=1.  \label{MeyerPoly}
		\end{gather}   
	
	 The wavelet transfer functions, 
	 $$H_i(\lambda)=H(s_i\lambda),$$ 
	 are of a band-pass type. The main property (condition for the reconstruction) is that the wavelet functions in two successive scales satisfy the following property
	 	\begin{gather*}H^2_i(\lambda)+H^2_{i+1}(\lambda) \\=\cos^2\bigg(\frac{\pi}{2}v_x\Big(q(\frac{s_i\lambda}{M}-1)\Big)\bigg)+\sin^2\bigg(\frac{\pi}{2}v_x\Big(q(\frac{s_i\lambda}{M}-1)\Big)\bigg)=1,
	 	\end{gather*}
	 	 within
	$$M<s_i\lambda \le M^2.$$
This property implies $\sum_{i=1}^K H^2(s_i \lambda) = 1$ 
	for all $\lambda$ except in the last interval, $s_K\lambda \in [0,M^2]$.
	To handle the low-pass spectral components (the interval for $\lambda$ closest to $\lambda=0$), the low-pass type \textit{scale function}, $G(\lambda))$, is added in the form
	$$
	G(\lambda)=\begin{cases}
	1, \text{ for } 0 \leq\lambda \le M/s_K=\lambda_{\max}/M^{K-1}\\
	\cos\bigg(\frac{\pi}{2}v_x\Big(q(\frac{s_K\lambda}{M}-1)\Big)\bigg), \text{ for } M<s_K\lambda \le M^2\\
	0 , \text{ elsewhere. }
	\end{cases}
	$$
	\end{Example}	
	 \begin{Remark}
		  The number of wavelet transfer functions, $K$, does not depend on the other wavelet parameters. A large value of $K$ will only increase the number of intervals and the resolution (producing smaller width of the first interval defined by $\lambda_{\max}/M^{K-1}$) toward $\lambda \to 0$, as shown in Fig. \ref{test_wavelet_meyer_bands_a}(a), (b), and (c). 
		  \end{Remark}	
		\begin{Remark}
		The wavelet transfer functions, $H(s_i \lambda)$, including the low-pass scale function, $G(\lambda)$, defined in Example \ref{DefLGFTWVT} satisfy the relation $$
		\sum_{i=1}^K H^2(s_i \lambda)+G^2(\lambda) =1.
		$$
		\end{Remark}	
	\begin{Example}
	For $q=1$, $M=2$, and  $K=9$ the Meyer wavelet functions are given in Fig. \ref{test_wavelet_meyer_bands_a}(a).  The Meyer wavelet functions for $q=3$, $M=4/3$, $K=13$ and  $q=9$, $M=10/9$, $K=45$ are shown in Fig. \ref{test_wavelet_meyer_bands_a}(b) and (c). The vertex-frequency representation of the signal from Fig. \ref{VF_graph3ab} using these three sets of wavelet transfer functions are shown in Fig. \ref{test_wavelet_meyer_bands_b}(a),(b), and (c).    
	\end{Example}
	
%	 The last (largest) interval for $M=2$ and nonzero values of $h_K(\lambda)=h_9(\lambda)$ is 
%	$$\frac{\lambda_{\max}}{M} < \lambda \le \lambda_{\max},$$  
%	while  the wavelet transfer function $h_{K-1}(\lambda)=h_8(\lambda)$ is nonzero for the intervals 
%	\begin{equation}\frac{\lambda_{\max}}{M^2} < \lambda \le \frac{\lambda_{\max}}{M} \text{ and } \frac{\lambda_{\max}}{M} < \lambda \le \lambda_{\max}, \label{limitsG9}
%	\end{equation}
%	with the function, $h_8(\lambda)$, values within these intervals as in (\ref{defMeyerW}).
%	The wavelet transfer function, $h_{K-1}(\lambda)=h_{8}(\lambda)$, corresponding to the interval for $\lambda$ given in (\ref{limitsG9}), with $\lambda_{\max}=7.63$,  is shown in Fig. \ref{test_wavelet_meyer_bands_a}(a) by a thick line.

	\noindent\textbf{Polynomial SGWT approximation.}  Chebyshev approximation of the wavelet functions, $H(s_i\lambda)=H_i(\lambda)$, in the form  
	\begin{equation}
	\bar{P}_{i,M-1}(\lambda)=\frac{c_{i,0}}{2}+\sum_{m=1}^{M-1}c_{i,m}\bar{T}_m(\lambda),
	\end{equation} 
	can be used for the vertex domain wavelet transform implementation
	\begin{gather*}
	\bar{P}_{i,M-1}(\mathbf{L})=\frac{c_{i,0}}{2}+\sum_{m=1}^{M-1}c_{i,m}\bar{T}_m(\mathbf{L}), \\
	i=0,1,2,\dots,K
	\end{gather*} 	
	 using only the $(M-1)$-neighborhood of each considered vertex, and without any graph Laplacian eigendecomposition analysis.
	The Chebyshev polynomials can be calculated recursively, as in (\ref{recurChebPol}), with a change of variables	and the recursive implementation as described in detail in Examples \ref{Cheb_GRDesignEx} and \ref{ExLGFTCHP}.
	
\subsubsection{Windows Defined Using the Vertex Neighborhood}

In order to show that the window, $h_m(n)$, which is localized at a vertex $m$ can also be defined using the vertex neighborhood, recall that the distance, $d_{mn}$, between vertices $m$ and $n$ is equal to the length of the shortest walk from vertex $m$ to vertex $n$, and that $d_{mn}$ takes integer values. Then, the window function can be defined as a function of vertex distance, in the form
$$
h_m(n)=g(d_{mn}),
$$ 
where $g(d)$ corresponds to any basic window function in classical signal processing.  For example, we can use the Hann window, given by
$$
h_m(n)=\frac{1}{2}\Big(1+\cos(\pi d_{mn}/D)\Big), \text{ for } 0\le d_{mn} < D,
$$
where $D$ is the assumed window width. 

For convenience, window functions for every vertex can be calculated in a matrix form as follows:
\begin{itemize}
	\item
	For the vertices for which the distance is $d_{mn}=1$, window functions  are defined trough an adjacency (neighborhood one) matrix $\mathbf{A}_1=\mathbf{A}$. In other words, the vertices which belong to the one-neighborhood of a vertex, $m$, are indicated by unit-value elements in the $m$th row of the adjacency matrix $\mathbf{A}$ (in unweighted graphs).  In weighed  graphs, the corresponding adjacency matrix $\mathbf{A}$ can be obtained from the weighting matrix  $\mathbf{W}$ as $\mathbf{A}=\operatorname{sign}( \mathbf{W})$.
	\item  
	Window functions for vertices $m$ and $n$, for which the distance is $d_{mn}=2$ are defined by the  matrix
	$$
	\mathbf{A}_2=(\mathbf{A} \odot \mathbf{A}_1 ) \circ (\mathbf{1}-\mathbf{A}_1) \circ (\mathbf{1}-\mathbf{I}),
	$$
	where the symbol $\odot$ denotes the logical (Boolean) matrix product, $\circ$ is the Hadamard (element-by-element) product, and $\mathbf{1}$ is a matrix with all elements equal to 1. The nonzero elements of the $m$th row of the matrix $\mathbf{A} \odot \mathbf{A}_1$ then designate the vertices that are connected to the vertex $m$ with walks of length $K=2$ or lower.  It should be mentioned that the element-by-element multiplication of  $(\mathbf{A} \odot \mathbf{A}_1 )$ by matrix $(\mathbf{1}-\mathbf{A}_1)$ removes the vertices connected with walks of length $1$, while the multiplication by $(\mathbf{1}-\mathbf{I})$ removes the diagonal elements from $(\mathbf{A} \odot \mathbf{A}_1 )$. 
	\item
	For $d_{mn}=d\ge 2 $,  we arrive at a recursive relation for the calculation of a matrix which will give the information about the vertices separated by the distance $d$. Such a matrix has the form
	\begin{equation}
	\mathbf{A}_{d}= (\mathbf{A} \odot \mathbf{A}_{d-1} ) \circ (\mathbf{1}-\mathbf{A}_{d -1})\circ (\mathbf{1}-\mathbf{I}). \label{AdDEF}
	\end{equation}
\end{itemize}

The window matrix for an assumed graph window width, $D$, can now be defined as
$$ 
\mathbf{P}_{D}= g(0)\mathbf{I}+g(1)\mathbf{A}_1+\dots+g(D-1)\mathbf{A}_{D-1},  
$$
so that a graph signal which is localized around a vertex $m$, may be formed based on this matrix, as 
$$
x_m(n)=h_m(n)x(n)=P_D(n,m)x(n).
$$

The LGFT representation of a graph signal, $x(n)$, then becomes
\begin{equation}
S(m,k)=\sum_{n=0}^{N-1} x(n)h_m(n)\; u_{k}(n)=\sum_{n=0}^{N-1} x(n)P_D(n,m)\; u_{k}(n),
\end{equation}
with the vertex-frequency kernel given by 
\begin{equation}
\mathcal{H}_{m,k}(n)=h_m(n)u_k(n)=P_D(n,m) u_{k}(n). \label{KernelGFDFTmkA}
\end{equation}
This allows us to arrive at the matrix form of the LGFT, given by
\begin{equation}
\mathbf{S}=\mathbf{U}^{T} \Big(\mathbf{P}_D \circ  [\mathbf{x},\  \mathbf{x},\ \ldots ,\ \mathbf{x}]\Big),
\end{equation}
where $[\mathbf{x},\  \mathbf{x},\ \ldots, \ \mathbf{x}]$ is an $N\times N$ matrix,  the columns of which are the signal vectors, $\mathbf{x}$.

For a rectangular function $g(d)=1$, and for any $d<D$, the LGFT can be calculated recursively with respect to the window width, $D$, as
\begin{equation}
\mathbf{S}_D = \mathbf{S}_{D-1} + \mathbf{U}^{T} \Big(\mathbf{A}_{D-1} \circ [\mathbf{x},\  \mathbf{x},\ \ldots, \ \mathbf{x}]\Big).
\end{equation}

\begin{Example} Consider the local vertex-frequency representation of the signal from Fig. \ref{VF_graph3ab}, using vertex domain defined windows. 
		The  localization kernels, $\mathcal{H}_{m,k}(n)=h_m(n)u_k(n)$, are shown in Fig. \ref{VF_windows_neigh} for two vertices and two spectral indices. Observe that for the spectral index $k=0$, the localization kernel is proportional to the localization function $h_m(n)$, given in Fig.  \ref{VF_windows_neigh}(a) and (c) for the vertices  $m=34$ and  $m=78$. Frequency modulated forms of these localization functions are shown in Figs. \ref{VF_windows_neigh}(b) and (d), for the same vertices and $k=20$. 
	
	A vertex domain window is next used to analyze  the graph signal  from Fig. \ref{VF_graph3ab}. The vertex-frequency representation, $S(n,k)$, obtained with the LGFT and the vertex domain  localization window is given in Fig. \ref{VF_graph3h}. Again, we can observe three constituent graph  signal components in three distinct vertex regions. The marginals of  $S(n,k)$ are also shown in the right and bottom panels.  
	
	\begin{figure*}
		\centering
		\includegraphics[]{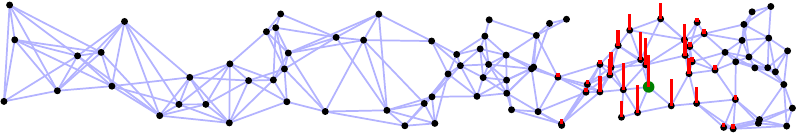}(a)
		\hspace{2mm}
		\includegraphics[]{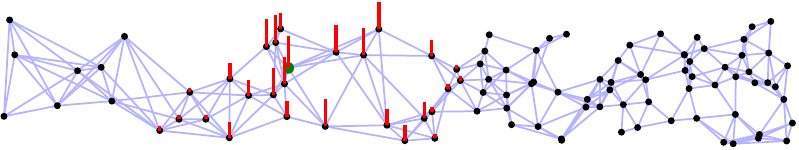}(b)
		
		$\mathcal{H}_{34,0}(n)=h_{34}(n)u_{0}(n) \sim h_{34}(n)$ \hspace{35mm} 
		$\mathcal{H}_{78,0}(n)=h_{78}(n)u_{0}(n) \sim h_{78}(n)$
		
		\vspace{8mm}
		
		\includegraphics[]{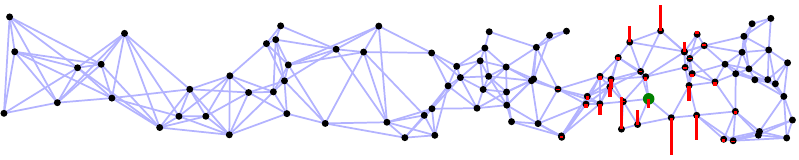}(c)
		\hspace{2mm}
		\includegraphics[]{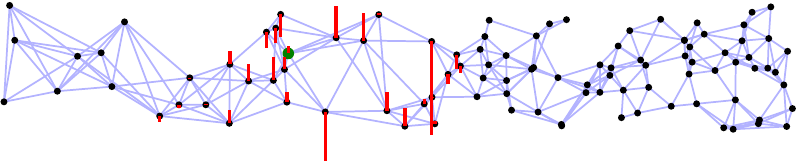}(d)
		
		$\mathcal{H}_{34,20}(n)=h_{34}(n)u_{20}(n)$ \hspace{55mm} 
		$\mathcal{H}_{78,20}(n)=h_{78}(n)u_{20}(n)$
		
		\vspace{4mm}
		
		\includegraphics[scale=0.9]{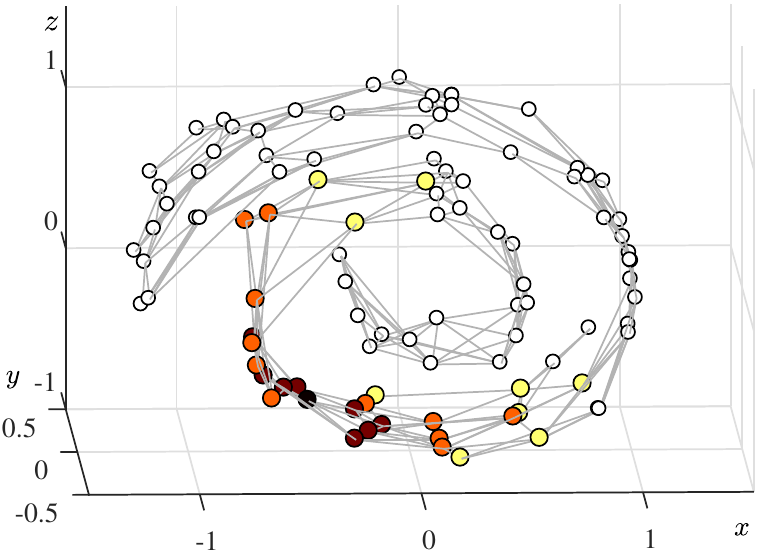} \hspace{6mm} (e)
		\hspace{6mm}
		\includegraphics[scale=0.9]{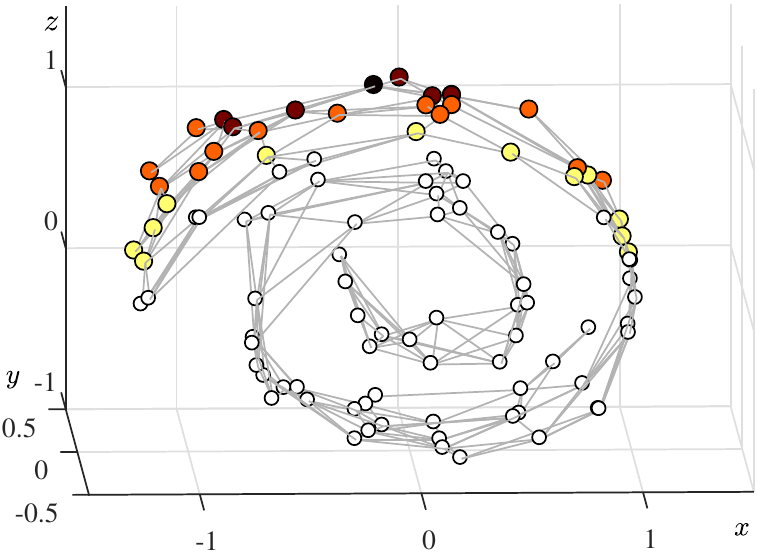} \hspace{8mm} (f)

		$\mathcal{H}_{34,0}(n)=h_{34}(n)u_{0}(n) \sim h_{34}(n)$ \hspace{35mm} 
		$\mathcal{H}_{78,0}(n)=h_{78}(n)u_{0}(n) \sim h_{78}(n)$

		\caption{Localization kernels for vertex-frequency analysis, $\mathcal{H}_{m,k}(n)=h_m(n) u_{k}(n)$, for the case of \textit{vertex domain defined windows} in the local graph Fourier transform, $S(m,k)=\sum_{n=0}^{N-1} x(n)\mathcal{H}_{m,k}(n)$. (a) Localization kernel $\mathcal{H}_{34,0}(n)=h_{34}(n)u_{0}(n) \sim h_{34}(n)$, for a constant eigenvector, $u_0(n)=1/\sqrt{N}$, centered at the vertex $m=34$. (b) The same localization kernel as in  (a), but centered at the vertex $m=78$. (c) Localization kernel, $\mathcal{H}_{34,20}(n)=h_{34}(n) u_{20}(n)$, centered at the vertex $m=35$ and frequency shifted by $u_{20}(n)$. Observe the variations in kernel amplitude, which indicate a modulation of the localization window, $h_m(n)$. (d) The same localization kernel as in (c), but centered at the vertex $m=78$. (e) Three-dimensional representation of the kernel $\mathcal{H}_{34,0}(n)=h_{34}(n)u_{0}(n)$. (f) Three-dimensional representation of the kernel $\mathcal{H}_{78,0}(n)=h_{78}(n)u_{0}(n)$.
		}
		\label{VF_windows_neigh}
	\end{figure*}
	
	\begin{figure}
		\centering
		\includegraphics[scale=0.9]{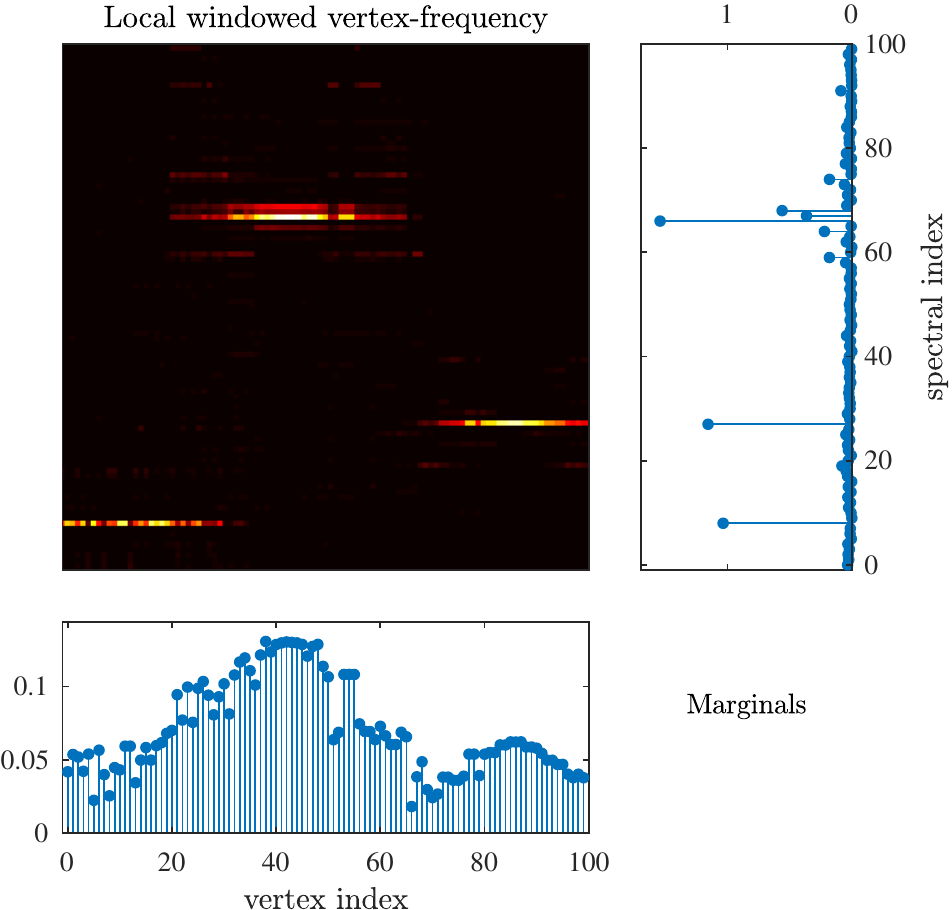}
		\caption{Local vertex-frequency spectrum calculated using the LGFT and vertex neighborhood windows, as in (\ref{KernelGFDFTmkA}). This representation immediately shows that the graph signal consists of three components located at spectral indices $k=8$, $k=66$, and $k=27$, with  the corresponding vertex indices in their respective vertex subsets $\mathcal{V}_1$, $\mathcal{V}_2$, and $\mathcal{V}_3$, where $\mathcal{V}_1 \cup \mathcal{V}_2 \cup \mathcal{V}_3=\mathcal{V}$. The marginal properties are also given in the panels to the right and below the vertex-frequency representation, and they differ from the ideal ones given respectively by $|x(n)|^2$ and $|X(k)|^2$.
		}
		\label{VF_graph3h}
	\end{figure}
	
\end{Example}

\begin{Remark}\textbf{Directed graphs.}  
The vertex neighborhood, as a set of vertices that can be reached from the considered vertex by a walk whose length is at most $D$, may be also defined  on directed graphs. In this case, this approach corresponds to one-sided windows in classical signal analysis. 

If we want to define two-sided window, then we should also include all vertices from which we can reach the considered vertex by walk whose length is at most $D$.
This means that 
for a directed graph we should assume that vertices with distance  $d_{mn}=1$ form the considered vertex $m$ are the vertices from which we can reach vertex $m$ with walk of length 1. In this case $\mathbf{A}_1 = \mathbf{A} + \mathbf{A}^T$ where addition is logical operation (Boolean OR). The matrix $\mathbf{A}_2$ is $$\mathbf{A}_2 = (\mathbf{A} \odot \mathbf{A} + \mathbf{A}^T \odot \mathbf{A}^T) \circ (\mathbf{1}- \mathbf{I}) \circ (\mathbf{1}-\mathbf{A}_1).$$ 

%$$\mathbf{A}_2 = (\mathbf{A} \odot \mathbf{A} + \mathbf{A}^T \odot \mathbf{A}^T) \circ \overline{\mathbf{I}} \circ \overline{\mathbf{A}}$$

This procedure could be continued for walks up to the desired maximal length $D$.

For a circular directed graph in this way, we will get the classical STFT with symmetric window.
\end{Remark}

\subsubsection{Window Parameter Optimization}

The concentration of local vertex spectrum representation can be measured using 
the normalized one-norm \cite{stankovic2001measure}, as 
\begin{equation}
\mathcal{M}=\frac{1}{F}\displaystyle \sum_{m=0}^{N-1}\sum_{k=0}^{N-1} 
|S(m,k)|=\frac{1}{F} \Vert \mathbf{S} \Vert_1,
\end{equation}
where 
$$F=\Vert \mathbf{S} \Vert_F= \sqrt{\displaystyle \sum_{m=0}^{N-1}\sum_{k=0}^{N-1} 
	|S(m,k)|^2}$$
 is the Frobenius norm of matrix $\mathbf{S}$. Alternatively, any other norm $\Vert \mathbf{S} \Vert_p^p$, with $0 \le p \le 1$ can be used instead of $\Vert \mathbf{S} \Vert_1$. Recall that norms with $p$ close to $0$ are noise sensitive, while the norm with $p=1$ is the only convex norm, which hence allows for gradient based optimization \cite{stankovic2001measure}.     

\begin{Example}
	The concentration measure, $\mathcal{M}(\tau)= \Vert \mathbf{S} \Vert_1 / \Vert \mathbf{S} \Vert_F$, for the signal from Fig. \ref{VF_graph3ab}, the window given in (\ref{win-spect}), and for  various $\tau$ is shown in Fig. \ref{VF_graph3_OPT}, along with the optimal vertex frequency representation. This representation is similar to that shown in Fig. \ref{VF_graph3g}, where an empirical value of $\tau=3$ was used, with the same localization window and kernel form. 
	
	The optimal $\tau$ can be obtained in  only a few steps through the iteration 
	$$\tau_k=\tau_{k-1}-\alpha \Big(\mathcal{M}(\tau_{k-1})-\mathcal{M}(\tau_{k-2})\Big),$$ 
	{with $\alpha$ a step-size parameter.}
	\begin{figure}
		\centering
		\hspace*{1cm}\includegraphics[]{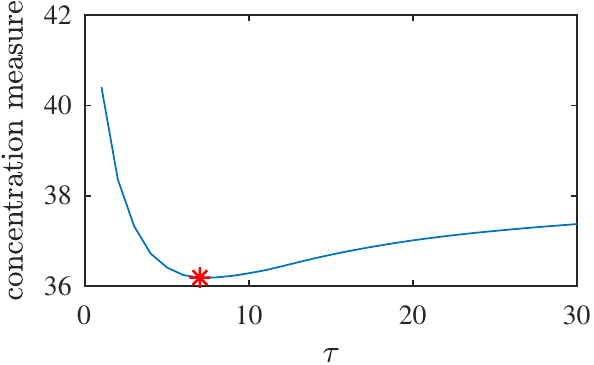}\hspace*{1cm} (a) 
		\vspace{3mm}
		
		\includegraphics[scale=0.9]{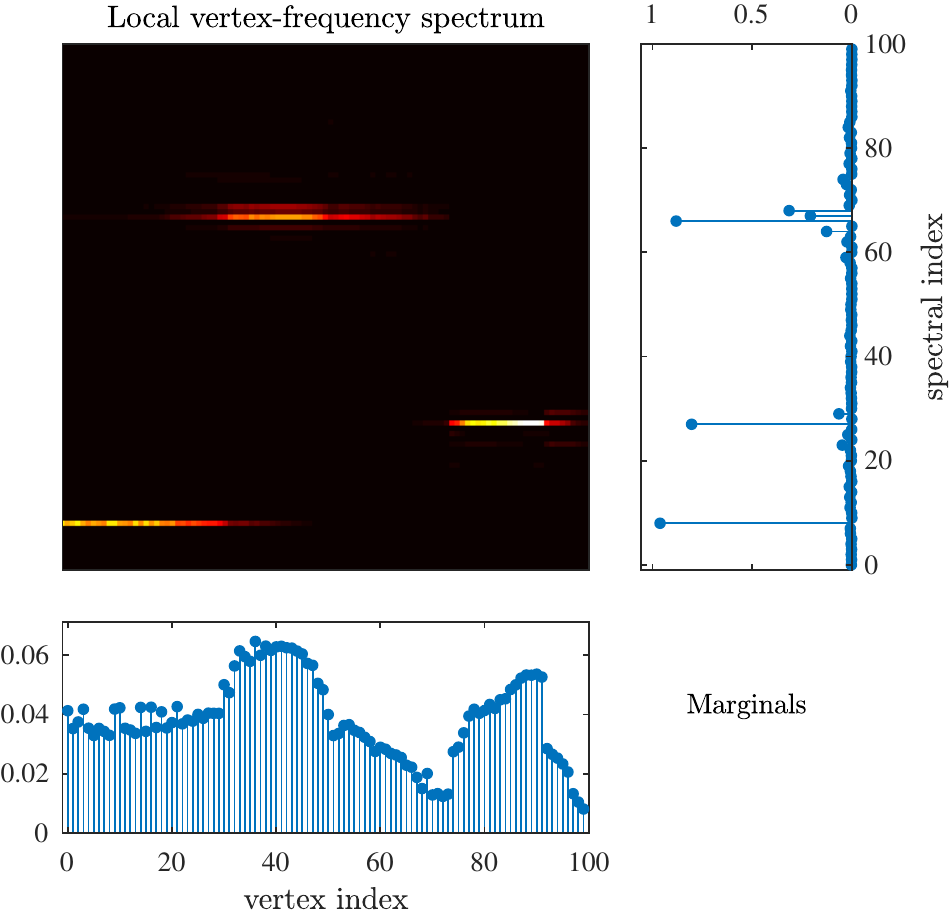}\hspace{-4mm}(b)
		
		\caption{Principle of the optimization of localization window. (a) Measure of the concentration of graph spectrogram for a varying  spectral domain window parameter $\tau$. (b) The corresponding optimal vertex-frequency representation, calculated with $\tau=7$, together  with its marginals.}
		\label{VF_graph3_OPT}
	\end{figure}
	
\end{Example}

The optimization of parameter $\tau$ can also be achieved trough  graph uncertainty principle based techniques \cite{Tsitsvero2016,Agaskar}.

\subsection{Inversion of the LGFT}\label{GFTInversion}

The inversion relation of the LGFT, calculated using any of the presented localization (window) forms, will next be considered in a unified way; the two approaches for the LGFT inversion here are: (i) inversion by summation of LGFT and (ii) kernel based inversion.

\subsubsection{Inversion by the Summation of the LGFT}
The reconstruction of a graph signal, $x(n)$, from its local spectrum,  
$S(m,k)$, can be performed through an inverse GDFT of (\ref{LGFTDEF}), based on the graph windowed signal
\begin{equation}
x(n)h_m(n)=\sum_{k=0}^{N-1} S(m,k)\, u_{k}(n)
\end{equation}
followed by a summation 
over all vertices, $m$, to yield
\begin{equation}
x(n)=\frac{1}{\sum_{m=0}^{N-1}h_m(n)} \sum_{m=0}^{N-1} \sum_{k=0}^{N-1} S(m,k)u_k(n).
\end{equation}

\begin{Remark}
	If the windows, $h_m(n)$, for every vertex, $n$, satisfy the condition
	$$\sum_{m=0}^{N-1}h_m(n)=1,$$ 
	then the reconstruction does not depend on the vertex index, $n$, or in other words such reconstruction is vertex independent. This becomes clear from 
	\begin{equation}
	x(n)=\sum_{m=0}^{N-1} \sum_{k=0}^{N-1} S(m,k)u_k(n)=\sum_{k=0}^{N-1} X(k)u_k(n),
	\end{equation}
	where $$X(k)=\sum_{m=0}^{N-1} S(m,k)$$
	is a projection of the LGFT onto the spectral index axis. 
	For windows obtained using the generalized graph shift in (\ref{AdDEF}), this conditions is always satisfied since $H(0)=1$.
\end{Remark}

The condition  $\sum_{m=0}^{N-1}h_m(n)=1$ can be enforced by normalizing the elements of the matrix $\mathbf{A}_d$, $d=1,2,\ldots,D-1$ in (\ref{AdDEF}), prior to the calculation of matrix $\mathbf{P}_D$, in such a way that the sum of each of its columns is equal to $1$, which allows us to arrive at $$\sum_{m=0}^{N-1}h_m(n)=\sum_{m=0}^{N-1}P_D(n,m)=\sum_{d=1}^{D-1}g(d)=const.$$

In general, the local vertex spectrum, $S(m,k)$, can also be calculated over a reduced set of vertices, $m \in \mathcal{M} \subset \mathcal{V}$. In this case, the summation over $m$ in the reconstruction formula should be executed over only the vertices $m \in \mathcal{M}$, while a vertex-independent reconstruction is achieved if $\sum_{m \in \mathcal{M}}h_m(n)=1$.

\subsubsection{Inversion of the LGFT with Band-Pass Functions}
For the LGFT, defined in (\ref{PolyVert}) as $
\mathbf{s}_k=\sum_{p=0}^{M-1}h_{p,k} \mathbf{L}^p\mathbf{x}$, the inversion is obtained by a summation over all spectral index shifts, $k=0,1,\dots,K$, that is
\begin{gather}
\sum_{k=0}^{K}\mathbf{s}_k=\sum_{k=0}^{K}\sum_{p=0}^{N-1}h_{p,k} \mathbf{L}^p\mathbf{x} 
=\sum_{k=0}^{K}H_k(\mathbf{L})  \mathbf{x}=\mathbf{x}, \label{unenerINV}
\end{gather}
if
$
\sum_{k=0}^{K}H_k(\mathbf{L})=\mathbf{I}. 
$
This condition is equivalent to the following spectral domain form 
\begin{equation}\sum_{k=0}^{K}H_k(\mathbf{\Lambda})=\mathbf{I} \label{ConDHH}
\end{equation} 
since   $\mathbf{U}\sum_{k=0}^{K}H_k(\mathbf{\Lambda})\mathbf{U}^T=\mathbf{I}$ and $\mathbf{U}^T\mathbf{U}=\mathbf{I}$. The condition in (\ref{ConDHH}) is used to define the transfer functions in Fig. \ref{test_bands_a}.

\subsubsection{Kernel-Based Inversion}
Another approach to the inversion of the local vertex spectrum, $S(m,k)$,  follows the Gabor expansion framework \cite{stankovic2014time}, whereby the local vertex spectrum, $S(m,k)$,  is projected back to the vertex-frequency localized kernels, $\mathcal{H}_{m,k}(n)$. The inversion for two forms of the LGFT, defined as in (\ref{LGFTDef1}) and (\ref{PKerLGFT}), will be analyzed. 

\noindent(a) For the LGFT defined in (\ref{LGFTDef1}), the sum of all of its projections to the localized kernels, $\mathcal{H}_{m,k}(n)$, is
\begin{gather}
\sum_{m=0}^{N-1}\sum_{k=0}^{N-1} S(m,k)\mathcal{H}_{m,k}(n)\!=\!\sum_{m=0}^{N-1} \! \bigg( \sum_{k=0}^{N-1}S(m,k)h_m(n)u_k(n) \bigg)
\nonumber \\
=\sum_{m=0}^{N-1} \bigg( \sum_{i=0}^{N-1}\underset{k \to i}{\text{IGDFT}}\{S(m,k)\}\underset{k \to i}{\text{IGDFT}}\{h_m(n)u_k(n)\} \bigg) \nonumber \\
=\sum_{m=0}^{N-1}\sum_{i=0}^{N-1}[x(i)h_m(i)][h_m(n)\delta(n-i)] \nonumber
\\ = \sum_{m=0}^{N-1}x(n)h^2_m(n) 
=x(n)\sum_{m=0}^{N-1}h^2_m(n), \label{GGABINV}
\end{gather}
where IGDFT denotes the inverse GDFT transform.  Parseval's theorem for graph signals
$$
\sum_{n=0}^{N-1} x(n)y(n) = \sum_{k=0}^{N-1} X(k)Y(k) 
$$
was used in the derivation. In this form of the LGFT all possible spectral shifts, $k=0,1,\dots,N-1$, are used.

The inversion formula for the local vertex spectrum, $S(m,k)$, which yields the original graph signal, $x(n)$, then becomes
\begin{equation}
x(n)=\frac{1}{\sum_{m=0}^{N-1}h^2_m(n)}\sum_{m=0}^{N-1}\sum_{k=0}^{N-1}S(m,k)\mathcal{H}_{m,k}(n). \label{GGABINV1}
\end{equation}
\begin{Remark}
	This kind of kernel-based inversion is vertex-invariant if the sum over all vertices, $m$, is invariant with respect to  $n$ and is equal to 1, that is 
	\begin{equation}
	\sum_{m=0}^{N-1}h^2_m(n)=1. \label{unener}
	\end{equation}	
	If the LGFT, $S(m,k)$, is calculated over a reduced set of vertices, $m \in \mathcal{M} \subset \mathcal{V}$, then the vertex independent reconstruction condition becomes $\sum_{m \in \mathcal{M}}h_m^2(n)=1$.
\end{Remark}

\noindent(b) For the  LGFT with spectral shifted spectral windows, defined in (\ref{PKerLGFT}), the kernel based inversion is of the form
\begin{gather}
x(n)=\sum_{m=0}^{N-1}\sum_{k=0}^{K}S(m,k)\mathcal{H}_{m,k}(n) \label{PKerLGFTInv}
\end{gather}
if the following condition
\begin{equation}\sum_{k=0}^{K}H^2_k(\lambda_p)=1 \label{PKerLGFTInv2} 
\end{equation}
is satisfied for all $\lambda_p$, $p=0,1,2,\dots,N-1$.

The inversion formula in (\ref{PKerLGFTInv}), with  condition  (\ref{PKerLGFTInv2}), follows from
\begin{gather}
\sum_{m=0}^{N-1}\sum_{k=0}^{K}S(m,k)\mathcal{H}_{m,k}(n)\\
=\sum_{m=0}^{N-1}\sum_{k=0}^{K}
\sum_{p=0}^{N-1} X(p)H_k(\lambda_p)u_p(m) \sum_{l=0}^{N-1} H_k(\lambda_l)u_l(m)u_l(n) \nonumber.
\end{gather}
 Since $\sum_{m=0}^{N-1}u_p(m)u_l(m)=\delta(p-l)$, the last expression reduces to the graph signal, $x(n)$,
\begin{gather}
\sum_{k=0}^{K}
\sum_{p=0}^{N-1} X(p)H_k(\lambda_p) H_k(\lambda_p)u_p(n)=x(n),
\end{gather}
if the transfer functions, $H_k(\lambda_p)$, $k=0,1,\dots,K$, satisfy the condition in (\ref{PKerLGFTInv2}) for all $\lambda_p$.

\subsubsection{Vertex-Varying Filtering} 

Filtering in the vertex-frequency domain may be implemented using a vertex-frequency support function, $B(m,k)$. The filtered LGFT is then given by
$$
S_f(m,k)=S(m,k) B(m,k),
$$
and the filtered signal, $x_f(n)$, is obtained by the inversion of $S_f(m,k)$ using the above mentioned inversion methods.
The filtering support function, $B(m,k)$, can be obtained, for example, by thresholding noisy values of the local vertex spectrum, $S(m,k)$.

\begin{figure}
		\centering
	\includegraphics[]{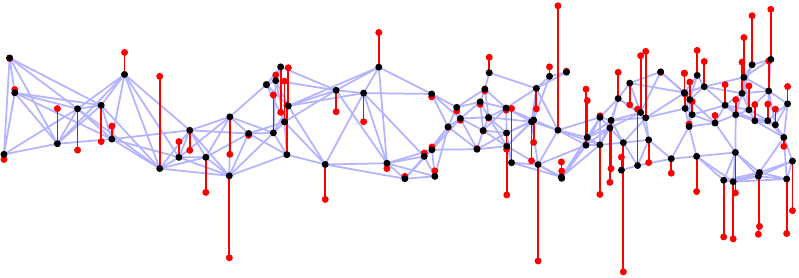}(a)
 
		\vspace{5mm} 
 
	\includegraphics[]{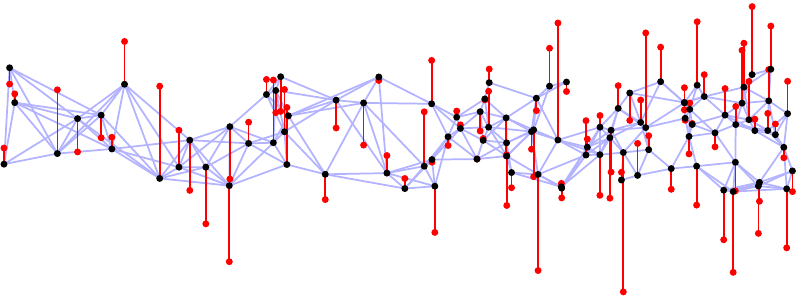}(b)
	
	\vspace{5mm}
	
	\includegraphics[]{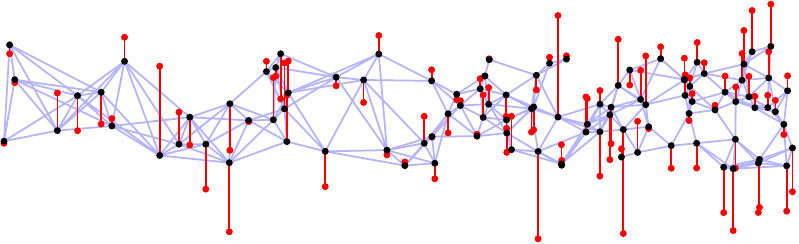}(c)

	\caption{Vertex-varying filtering of a graph signal. (a) The original graph signal, $x(n)$, from Fig. \ref{VF_graph3ab} (d). (b) The graph signal, $x(n)$, corrupted by an additive white Gaussian noise, at $SNR_{in}=5.3$ dB. (c) The graph signal, $x_f(n)$, after vertex-varying filtering based on thresholding of the LGFT of noisy graph signal, $S(m,k)$, with the final signal-to-noise ratio $SNR_{out}=10.36$ dB.}
	\label{lgft_bandpass_inversion}
\end{figure}

\begin{Example}
	Consider the graph signal, $x(n)$, from Fig. \ref{VF_graph3ab}(d), also shown in Fig. \ref{lgft_bandpass_inversion} (a), and its version corrupted by an additive white Gaussian noise,  at the signal-to-noise ratio of $SNR_{in}=5.3$ dB, given in Fig. \ref{lgft_bandpass_inversion} (b). The LGFT,   $S(m,k)$  of the noisy graph signal is calculated according to (\ref{PKerLGFT}), using shifted bandpass spectral transfer functions, $H_k(\lambda_p)$, $k=0,1,\dots,K$, $p=0,1,\dots,N-1$, given by (\ref{defHannWGen}) without squares ($H_k(\lambda_p) \rightarrow H^2_k(\lambda_p)$), which allows  $\sum_{k=0}^{K}H^2_k(\lambda_p)=1$ to hold, instead of $\sum_{k=0}^{K}H_k(\lambda_p)=1$. In this way, the condition for the inversion (\ref{PKerLGFTInv2}) is satisfied. The transfer functions, $H_k(\lambda_p)$, otherwise correspond to those shown in Fig. \ref{test_bands_a} (b) with $K=25$. 
	
	The vertex-varying filtering is performed using $
	S_f(m,k) = \allowbreak S(m,k) B(m,k)
	$ for $m=0,1,\dots,N-1$, $k=0,1,\dots,K$, with a simple thresholding-based filtering support function
	$$
	B(m,k)=\begin{cases}
	0, \text{ for } |S(m,k)|<T\\
	1 , \text{ otherwise, }
	\end{cases}
	$$
$m=0,1,\dots,N-1$, $k=0,1,\dots, K$,	with the threshold $T=0.09$ set empirically. The output graph signal, $x_f(n)$,  is obtained using the inversion relation in (\ref{PKerLGFTInv}) for the filtered LGFT, $S_f(m,k)$, and shown in Fig.~\ref{lgft_bandpass_inversion}(c). The achieved output SNR was $SNR_{out}=10.36$ dB.
	
\end{Example}

\subsection{Uncertainty Principle for Graph Signals}
In classical signal analysis, the purpose of a window function is  to enhance signal localization in the joint time-frequency domain. However, the uncertainty principle prevents the ideal localization in both time and frequency. Indeed, in the classical DFT analysis the uncertainty principle states that
\begin{equation}
\Vert \mathbf{x}\Vert_0 \Vert \mathbf{X} \Vert_0 \ge N, \label{uncertainityDFT}
\end{equation}
or in other words, that the product of the number of nonzero signal  values, $\Vert \mathbf{x}\Vert_0$, and the number of its nonzero DFT coefficients, $\Vert \mathbf{X}\Vert_0$, is greater or equal than the total number of signal samples $N$; they cannot  simultaneously assume small values. 

To arrive at the \textit{uncertainty principle for graph signals}, consider a graph signal, $\mathbf{x}$, and its spectral transform, $\mathbf{X}$, in a domain of orthonormal basis functions, $u_k(n)$.  Then, the uncertainty principle states that  \cite{Tsitsvero2016,Agaskar, elad2001generalized,perraudin2018global}
\begin{equation}
\Vert \mathbf{x}\Vert_0 \Vert \mathbf{X} \Vert_0 \ge \frac{1}{\max_{k,m} \{|u_k(m)|^2\}}. \label{uncertainity}
\end{equation}
This form of the uncertainty principle is generic, and indeed for the basis functions  $u_k(n)=\frac{1}{\sqrt{N}}\exp(j2\pi nk/N)$, the standard DFT uncertainty principle form in (\ref{uncertainityDFT}) follows. 

\begin{Remark}
Note, however, that in graph signal processing, the eigenvectors/basis functions can assume quite different forms than in the standard DFT case. For example, when one vertex is loosely connected with other vertices, then $\max \{|u_k(m)|^2\} \rightarrow 1$  and even $\Vert \mathbf{x}\Vert_0 \Vert \mathbf{X} \Vert_0 \ge 1$ is possible for the uncertainty condition in (\ref{uncertainity}). \textbf{This means that, unlike the classical Fourier transform-based time and frequency domains, a graph signal can be well localized in both the vertex and the spectral domains. }
\end{Remark}

\begin{Example}
	For the graph shown in Fig. \ref{VF_graph3ab}, we have $$\max_{k,m} \{|u_k(m)|^2\} = 0.8713$$
	 which indicates that even $\Vert \mathbf{x}\Vert_0 \Vert \mathbf{X} \Vert_0 \ge 1.1478$ is possible. In other words,  a graph signal for which the number of nonzero samples, $x(n)$, in the vertex domain is just two, will not violate the uncertainty principle even if it has just one nonzero GDFT coefficient, $X(k)$. 
\end{Example}

\subsection{Graph Spectrogram and Frames}

Based on (\ref{LGFTDEF}), the \textit{graph spectrogram} can be defined as 
\begin{equation}
|S(m,k)|^2=\Big|\sum_{n=0}^{N-1} x(n)h_m(n)\; u_{k}(n)\Big|^2.
\end{equation}
Then, according to Parseval's theorem, the \textit{vertex marginal property}, which is a projection of $|S(m,k)|^2$ onto the vertex index axis, is  given by 
\begin{gather*}
\sum_{k=0}^{N-1}|S(m,k)|^2=\sum_{k=0}^{N-1}S(m,k)\sum_{n=0}^{N-1} x(n)h_m(n)\; u_{k}(n) \\
=\sum_{n=0}^{N-1} |x(n)h_m(n)|^2,
\end{gather*}
which would be equal to the signal power, $|x(m)|^2$, at the vertex $m$, if $h_m(n)=\delta{(m-n)}$. Since this is not the case, the vertex marginal property of the graph spectrogram is equal to the power of the graph signal in hand, smoothed by the window, $h_m(n)$.

\noindent\textbf{Energy of graph spectrogram.} For the total energy of graph spectrogram, we consequently have
\begin{equation}
\sum_{m=0}^{N-1}\sum_{k=0}^{N-1}|S(m,k)|^2=\sum_{n=0}^{N-1} \Big(|x(n)|^2\sum_{m=0}^{N-1}|h_m(n)|^2\big)\Big).
\end{equation}
If 
$\sum_{m=0}^{N-1}|h_m(n)|^2=1$ for all $n$, then the spectrogram on the graph is \textit{energy unbiased} (statistically consistent with respect to the energy), that is
\begin{equation}
\sum_{m=0}^{N-1}\sum_{k=0}^{N-1}|S(m,k)|^2=\sum_{n=0}^{N-1} |x(n)|^2=||\mathbf{x}||^2=E_x.
\end{equation}

\noindent\textbf{The LGFT viewed as a frame.} A set of functions, $S(m,k)$, 
is called \textit{a frame} for the expansion of a graph signal, $\mathbf{x}$, if 
$$A ||\mathbf{x}||^2 \le \sum_{m=0}^{N-1} |S(m,k)|^2 \le B ||\mathbf{x}||^2,$$
where $A$ and $B$ are positive constants. If $A=B$, the frame is termed \textit{Parseval's tight frame} and the signal can be recovered 
as $$x(n)=\frac{1}{A}\sum_{m=0}^{N-1}\sum_{k=0}^{N-1}S(m,k)h_m(n)u_k(n).$$ The constants $A$ and $B$ govern the numerical stability of recovering the original signal $\mathbf{x}$ from the coefficients $S(m,k)$. 

The conditions for two forms of the LGFT, defined as in (\ref{LGFTDef1}) and (\ref{PKerLGFT}), to represent frames will be analyzed next:

\smallskip

\noindent \textbf{(a)} The LGFT, defined as in (\ref{LGFTDef1}), is a frame, since in this case Parseval's theorem holds \cite{behjat2016signal,hammond2011wavelets,sakiyama2014oversampled,girault2015stationary}, that is
\begin{equation}
\sum_{m=0}^{N-1}|h_m(n)|^2=\sum_{k=0}^{N-1}|H(k)|^2|u_k(n)|^2,
\end{equation}
which allows us to write
\begin{equation}
\frac{1}{N}H^2(0)\le \sum_{m=0}^{N-1}|h_m(n)|^2 \le \max_{m,k}|u_k(n)|^2\sum_{k=0}^{N-1}|H(k)|^2 = \gamma^2 E_h,  \label{FramLGFT}
\end{equation}
where $\gamma=\max_{m,k}|u_k(n)|$ and $$E_h=\sum_{k=0}^{N-1}|H(k)|^2.$$

By multiplying both sides of the above inequalities by $||\mathbf{x}||^2$, we arrive at
\begin{equation}
\frac{1}{N}H^2(0)||\mathbf{x}||^2\le \sum_{m=0}^{N-1}\sum_{k=0}^{N-1}|S(m,k)|^2\le ||\mathbf{x}||^2\gamma^2 E_h.
\end{equation}
A frame is termed  a \textit{tight frame} if the equality in (\ref{FramLGFT}) holds, that is,  if $$\sum_{m=0}^{N-1}|h_m(n)|^2=1,$$
what is the same condition as in (\ref{unener}).
\smallskip

\noindent \textbf{(b)} The LGFT defined in (\ref{PKerLGFT}) is a tight frame if 
\begin{gather}
\sum_{k=0}^{K}\sum_{m=0}^{N-1} |S(m,k)|^2= 
\sum_{k=0}^{K} \sum_{p=0}^{N-1} |X(p)H_k(\lambda_p)|^2=E_x,\label{TFPKerLGFT}
\end{gather}
where Parseval's theorem for the $S(m,k)$ as the GFT of $X(p)H_k(\lambda_p)$ was used to yield 
$$\sum_{m=0}^{N-1} |S(m,k)|^2=\sum_{p=0}^{N-1} |X(p)H_k(\lambda_p)|^2.$$
This means that the LGFT in (\ref{PKerLGFT}) is a tight frame if
$$\sum_{k=0}^{K}|H_k(\lambda_p)|^2=1 \text{ for } p=0,1,\dots,N-1.$$
This condition is used to define transfer functions in Fig. \ref{test_bands_a}(b) and (c). 

From (\ref{TFPKerLGFT}), it is straightforward to conclude that the graph spectrogram energy is bounded with 
\begin{gather}
A E_x\le \sum_{k=0}^{K}\sum_{m=0}^{N-1} |S(m,k)|^2 \le B E_x,\label{TFPKerLGFTBoun}
\end{gather}
where $A$ and $B$ are respectively the minimum and the maximum of value of $$g(\lambda_p)=\sum_{k=0}^{K} |H_k(\lambda_p)|^2.$$ 

\subsubsection{Graph Wavelet Transform Inversion}
The wavelet inversion formula
\begin{gather}
x(n)= \sum_{n=0}^{N-1} \sum_{i=0}^{K} \psi(n,s_i)W(n,s_i)
\end{gather}
can be derived in the same way and under the same condition as in (\ref{PKerLGFTInv})-(\ref{PKerLGFTInv2}), where a set of discrete scales for the wavelet calculation, denoted by $s \in \{s_1, s_1,\dots,s_{K}\}$, is assumed, and $\psi(n,s_0)$ is used as a notation for the\textit{ scale function}, $\phi(n)$, whose spectral transfer function is $G(\lambda)$, as explained in  Remark \ref{IndexinWTSca}. 
In the same way as in the LGFT case, it can be shown that the wavelet transform represents a frame with 
\begin{gather}
A||\mathbf{x}||^2 \le \sum_{n=0}^{N-1} \sum_{i=0}^{K} |W(n,s_i)|^2 \le  B||\mathbf{x}||^2,
\end{gather}
where \cite{leonardi2013tight,hammond2019spectral,behjat2019spectral}
\begin{gather*}
A=\min_{0\le\lambda\le \lambda_{\max}}{g(\lambda)}, \\
B=\max_{0\le\lambda\le \lambda_{\max}}{g(\lambda)},
\end{gather*}
and the function $g(\lambda)$ is defined by
\begin{gather*}
g(\lambda)=\sum_{i=1}^{K} H^2(s_i\lambda)+G^2(\lambda).
\end{gather*}
The low-pass scale function, $G(\lambda)$, is added in the reconstruction formula, since all $H(s_i\lambda)=0$ for $\lambda=0$, as explained in Example \ref{DefLGFTWVT} and Remark \ref{IndexinWTSca}. It should be mentioned that the spectral functions  of the wavelet transform, $H(s_i\lambda)$, form Parseval's frame if
$$
g(\lambda)=1.
$$
Since the number of wavelet transform coefficients, $W(n,s_i)$, for each $n$ and $i$, is greater than the number of signal samples, $N$, this representation is redundant, and this redundancy allows us to implement the transform through a fast algorithm, rather than using the explicit computation of all wavelet coefficients \cite{hammond2019spectral,behjat2019spectral}. Indeed, for large graphs, it can be computationally too complex to compute the full eigendecomposition of the graph Laplacian. A common way to avoid this computational burden is to use a polynomial approximation schemes for $H(s_i\lambda)$, $i=1,2,\dots,K$, and $G(\lambda)$. One such approach is the truncated Chebyshev polynomial approximation method which is based on  the application of the continuous spectral window functions  with  Chebyshev polynomials, which admit order-recursive calculation (see Section \ref{Plyappsystem} and Example \ref{ExLGFTCHP}). If, for a given scale, $s_i$, the wavelet function is approximated by a polynomial in the Laplacian, $P_i(\mathbf{L})$, then the wavelet transform can  be efficiently calculated using
\begin{equation}
\mathbf{w}_i= P_i(\mathbf{L})\mathbf{x},
\end{equation}
where $\mathbf{w}_i$ a column vector with elements $W(m,s_i)$, $m=0,1,\dots,N-1$. Note that this form corresponds to the LGFT form in (\ref{PolyVert}).

\subsection{Vertex-Frequency Energy Distributions}

The energy of a general signal is usually defined as
\begin{gather*}
E=\sum_{n=0}^{N-1}x^2(n)=\sum_{n=0}^{N-1}x(n)\sum_{k=0}^{N-1}X(k)u_k(n).
\end{gather*}
This expression can be rearranged into 
\begin{gather*}
E=\sum_{n=0}^{N-1}\sum_{k=0}^{N-1}x(n)X(k)u_k(n)=\sum_{n=0}^{N-1}\sum_{k=0}^{N-1}E(n,k),
\end{gather*}
where for each vertex, the vertex-frequency energy distribution, $E(n,k)$, is defined by \cite{stankovic2018vertex, stankovic2019vertex}
\begin{align}
E(n,k) =x(n)X(k)u_k(n)   =\sum_{m=0}^{N-1}x(n)x(m)u_k(m)u_k(n) \label{VFE}.
\end{align} 

\begin{Remark}
	The definition in (\ref{VFE}) corresponds to the Rihaczek distribution in classical time-frequency analysis \cite{stankovic2014time,cohen1995time,boashash2015time}. Observe that based on the Rihaczek distribution and the expression in (\ref{VFE}),  we may obtain a vertex-frequency representation even without a localization window. This very important property  is  also the main advantage (along with the concentration improvement) of classical time-frequency distributions with respect to the spectrogram and STFT based time-frequency representations. 
\end{Remark}

\textit{The marginal properties} of the vertex-frequency energy distribution, $E(n,k)$, are defined as its projections onto the spectral index axis, $k$, and the vertex index axis, $n$, to give 
$$
\sum_{n=0}^{N-1}E(n,k)=|X(k)|^2
\quad \text{    and    } \quad
\sum_{k=0}^{N-1}E(n,k)=x^2(n),
$$
which correspond respectively to the squared spectra, $|X(k)|^2$, and the signal power, $x^2(n)$, of the graph signal, $x(n)$.   

\begin{Example} Fig. \ref{VF_graph3f} shows the vertex-frequency distribution, $E(n,k)$, of the graph signal from Fig. \ref{VF_graph3ab}, together with its marginal properties. The marginal properties are satisfied up to the computer precision. Observe also that the localization of energy is better than in the cases obtained with the localization windows in Figs. \ref{VF_graph3g}, \ref{VF_graph3h}, and \ref{VF_graph3_OPT}. Importantly, the distribution, $E(n,k)$,  does not employ a localization window.  
	
	\begin{figure}
		\centering
		\includegraphics[scale=0.9]{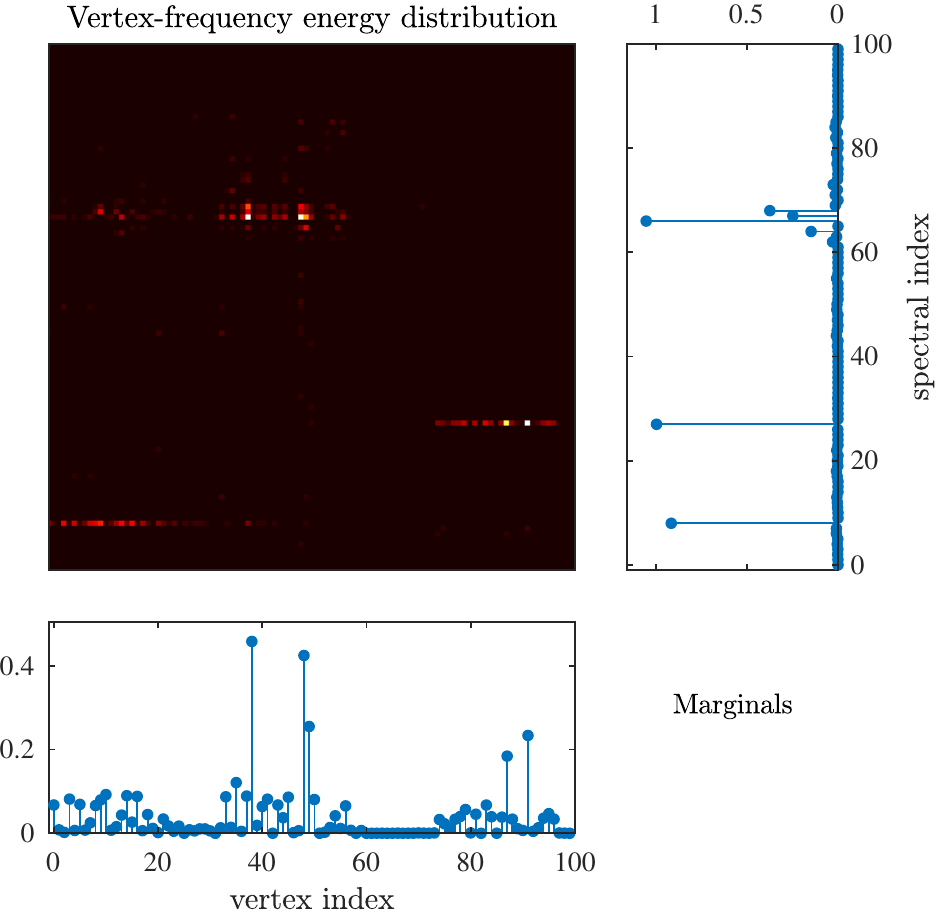}
		\caption{Vertex-frequency energy distribution for the graph signal whose vertex-frequency representation is given in Fig. \ref {VF_graph3g}. No localization window was used here.}
		\label{VF_graph3f}
	\end{figure}
\end{Example}

\subsubsection{Smoothness Index and Local Smoothness} 

The \textit{ smoothness index}, $l$, in graph signal processing plays the role of \textit{frequency}, $\omega$, in classical spectral analysis. For a graph signal, $\mathbf{x}$, the smoothness index is defined as the  \textit{Rayleigh quotient} of the matrix $\mathbf{L}$ and vector $\mathbf{x}$, that is ({\color{blue}see Section \ref{I-section_smoothness_of_eigenvectors}, Part I)}
\begin{equation}
l=\frac{\mathbf{x}^T\mathbf{L}\mathbf{x}}{\mathbf{x}^T\mathbf{x}} \ge 0. \label{LqfCut}
\end{equation}
\begin{Remark}
	The expression in (\ref{LqfCut}) indicates that the smoothness index can be considered as a measure of the rate of change of a graph signal. \textit{Faster changing signals (corresponding to high-frequency signals) have larger values of the smoothness index}. The maximally smooth graph signal is then a constant signal, $x(n)=c$, for which the smoothness index is $l=0$.

	In the mathematics literature, the inverse of the smoothness index is known as the \textit{curvature} ($\text{curvature} \sim 1/l$). While larger values of the smoothness index correspond to graph signals with larger rates of change (less smooth graph signals), the larger values of curvature would indicate smoother graph signals.   
\end{Remark}

Notice that the smoothness index for an eigenvector, $\mathbf{u}_k$, of the graph Laplacian, $\mathbf{L}$,  is equal to its corresponding eigenvalue, $\lambda_k$, that is
\begin{equation}
\frac{\mathbf{u}_k^T\mathbf{L}\mathbf{u}_k}{\mathbf{u}_k^T\mathbf{u}_k}=\lambda_k, \label{LqfCutEV}
\end{equation}
since by definition $\mathbf{L}\mathbf{u}_k= \lambda_k \mathbf{u}_k$.

\begin{Remark}
	If the above eigenvectors are the classical Fourier transform basis functions, then the smoothness index corresponds to the squared frequency of the considered basis function, $\lambda_k \sim \omega^2_k$, while the curvature corresponds to the squared period in harmonic signals.   
\end{Remark}    

This makes it possible to define the local smoothness  index  for a vertex $n$, $\lambda(n)$, in analogy with the standard instantaneous frequency, $\omega(t)$, at an instant $t$,   as \cite{dakovic2019local} 
\begin{equation}
\lambda(n)=\frac{\mathcal{L}_x(n)}{x(n)},
\label{local-smooth}
\end{equation}
where it was assumed that $x(n)\ne 0$ and $\mathcal{L}_{x}(n)$ are the elements of the vector $\mathbf{Lx}$.

The properties of the local smoothness include: 

\begin{enumerate}
	
	\item
	The local smoothness index, $\lambda(n)$, for a monocomponent signal 
	$$x(n)=\alpha u_{k}(n),$$
	is vertex independent, and is equal to the global smoothness index, $\lambda_k$, since 
	$$\mathcal{L}_{x}(n) = \alpha \mathcal{L}_{u_k}(n)=\alpha \lambda_k u_k(n).$$ 
	In the standard time-domain signal analysis, this property means that the instantaneous frequency of a sinusoidal signal is equal to its global frequency. 
	
	\item 
	Assume a piece-wise monocomponent signal 
	$$
	x(n)=\alpha_i u_{k_i}(n)\text{ for } n \in \mathcal{V}_i, \quad i=1,2,\ldots,M,
	$$
	where  $\mathcal{V}_i \subset \mathcal{V}$ are the subsets of the vertices such that $\mathcal{V}_i \cap \mathcal{V}_j=\emptyset$ for $i\ne j$, $\mathcal{V}_1 \cup \mathcal{V}_2 \cup \cdots \cup \mathcal{V}_M=\mathcal{V}$, that is, every vertex belongs to only one subset, $\mathcal{V}_i$. Given the monocomponent nature of this signal, within each subset, $\mathcal{V}_i$, the considered signal is proportional to the eigenvector, $u_{k_i}(n)$.
	
	Then, for each interior vertex, $n\in\mathcal{V}_i$, i.e., a vertex whose neighborhood lies in the same set, $\mathcal{V}_i$, the local smoothness index is given by
	\begin{equation}
	\lambda(n)=\frac{\alpha_i \mathcal{L}_{u_{k_i}}(n)}{\alpha_i u_{k_i}(n)}=\lambda_{k_i}.
	\label{loc-smooth}
	\end{equation}
	\item
	An ideally concentrated vertex-frequency distribution (ideal distribution) can be defined as 
	$$I(n,k) \sim |x(n)|^2 \delta\Big(\lambda_k-[\lambda(n)]\Big),$$
	whereby it is assumed that the local smoothness index is rounded to the nearest eigenvalue.
	
	This distribution can also be used as a local smoothness estimator, since for each vertex, $n$, the maximum of $I(n,k)$ is positioned at $\lambda_k = \lambda(n)$. An estimate of the spectral index at a vertex, $n$, denoted by $\hat{k}(n)$,  is then obtained as
	$$
	\hat{k}(n)=\arg \max_k \{I(n,k)\},
	$$
	so that the estimated local smoothness index becomes $\hat{\lambda}(n)=\lambda_{\hat{k}(n)}$.
	This type of estimator is widely used in classical time-frequency analysis \cite{stankovic2014time,cohen1995time,boashash2015time}.

	\item       
	
	\noindent\textbf{Local smoothness property.}  The vertex-frequency distribution, $E(n,k)$, satisfies the local smoothness property if  
	\begin{equation}
	\frac{\sum_{k=0}^{N-1} \lambda_k E(n,k)}{\sum_{k=0}^{N-1} E(n,k)}=\lambda(n).\label{LSPROP}
	\end{equation} 
	In that case, the centers of masses of the vertex-frequency distribution along the spectral index axis, $k$, should be exactly at $\lambda=\lambda(n)$, and can be used as an unbiased estimator of this graph signal parameter.  
	
\end{enumerate}

\begin{Example}
	The vertex-frequency distribution, defined by $E(n,k)=x(n)X(k)u_k(n)$, satisfies the local smoothness property in (\ref{LSPROP}), since 
	\begin{gather*}
	\frac{  {\sum_{k=0}^{N-1}} \lambda_k E(n,k)}{ {\sum_{k=0}^{N-1}} E(n,k)}=\frac{ {\sum_{k=0}^{N-1}} \lambda_k x(n)X(k)u_k(n)}{ {\sum_{k=0}^{N-1}} x(n)X(k)u_k(n)}
	=\frac{ \mathcal{L}_{x}(n) }{ x(n)}=\lambda(n).
	\end{gather*}
	The above relation follows from the fact that $\sum_{k=0}^{N-1} \lambda_k X(k)u_k(n)$ are the elements of the IGDFT of $\lambda_k X(k)$. Upon employing the matrix form of the IGDFT of $\mathbf{\Lambda} \mathbf{X}$, we have  $\mathbf{U}\mathbf{\Lambda} \mathbf{X}= \mathbf{U}\mathbf{\Lambda}  (\mathbf{U}^T \mathbf{U}) \mathbf{X} = (\mathbf{U}\mathbf{\Lambda}  \mathbf{U}^T) (\mathbf{U} \mathbf{X})= \mathbf{L}\mathbf{x}$. With the notation, $\mathcal{L}_{x}(n)$, for the elements of $\mathbf{L}\mathbf{x}$,  we next obtain 
	$$\sum_{k=0}^{N-1} \lambda_k X(k)u_k(n)= \mathcal{L}_{x}(n).$$
	
	The local smoothness index for the graph signal from Fig. \ref{VF_graph3ab} is shown in Fig. \ref{VF_graph3d}.
	
	\begin{figure}
		\centering
		\includegraphics[scale=0.9]{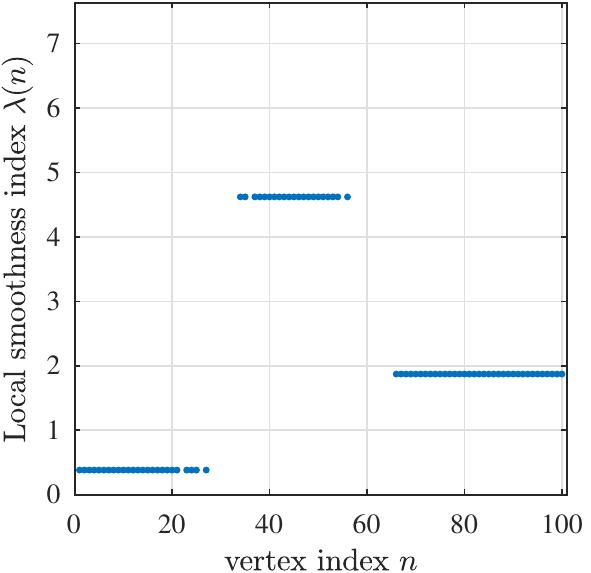}
		\caption{Local smoothness index, $\lambda(n)$, of the graph signal from Fig. \ref {VF_graph3ab}. }
		\label{VF_graph3d}
	\end{figure}
\end{Example}

\subsubsection{Reduced Interference Distributions (RID) on Graphs}
{In order to emphasize the close relations with classical time-frequency analysis, in this subsection we will use the complex-sensitive notation for eigenvectors and spectral vectors.} The frequency domain definition of the  energy distribution in (\ref{VFE}) is given by 
\begin{align}
E(n,k) & =x(n)X^*(k)u^*_k(n)  
=\sum_{p=0}^{N-1}X(p)X^*(k)u_p(n)u^*_k(n) \nonumber.
\end{align} 
Then, the general form of a graph distribution can be defined with the help of a kernel $\phi(p,k,q)$, as \cite{stankovic2018reduced}
\begin{gather}
G(n,k)=\sum_{p=0}^{N-1}\sum_{q=0}^{N-1}X(p)X^*(q)u_p(n)u^*_q(n)\phi(p,k,q). \label{GENRID}
\end{gather}
Observe that for $\phi(p,k,q)=\delta(q-k)$, the graph Rihaczek distribution in (\ref{VFE}) follows, while the unbiased energy condition $\sum_{k=0}^{N-1}\sum_{n=0}^{N-1}G(n,k)=E_x$ is satisfied if  $$\sum_{k=0}^{N-1}\phi(p,k,p)=1.$$

The so obtained distribution, $G(n,k)$, may also satisfy the vertex and frequency marginal properties, as elaborated bellow.
\begin{itemize}
	\item 
	The \textit{vertex marginal property} is satisfied if
	\begin{gather*}
	\sum_{k=0}^{N-1}\phi(p,k,q)=1.
	\end{gather*}
	This is obvious from 
	$$\sum_{k=0}^{N-1}G(n,k)=\sum_{p=0}^{N-1}\sum_{q=0}^{N-1}X(p)X^*(q)u_p(n)u^*_q(n)=|x(n)|^2.$$
	
	\item 
	The \textit{frequency marginal property} is satisfied if
	\begin{gather*}
	\phi(p,k,p)=\delta(p-k).
	\end{gather*}
	Then, the sum over all vertex indices produces  
	\begin{gather*}
	\sum_{n=0}^{N-1}G(n,k)=\sum_{p=0}^{N-1}|X(p)|^2\phi(p,k,p)=|X(k)|^2,
	\end{gather*} 
	since 
	$\sum_{n=0}^{N-1}u_p(n)u^*_q(n)=\delta(p-q)$, that is, the eigenvectors are orthonormal.
\end{itemize}

\subsubsection{Reduced Interference Distribution Kernels}

A straightforward extension of classical time-frequency kernels to graph signal processing would be naturally based upon exploiting the relation $\lambda \sim \omega^2$, together with an appropriate exponential kernel normalization.  

\begin{figure}
	% \centering
	\includegraphics[scale=0.9]{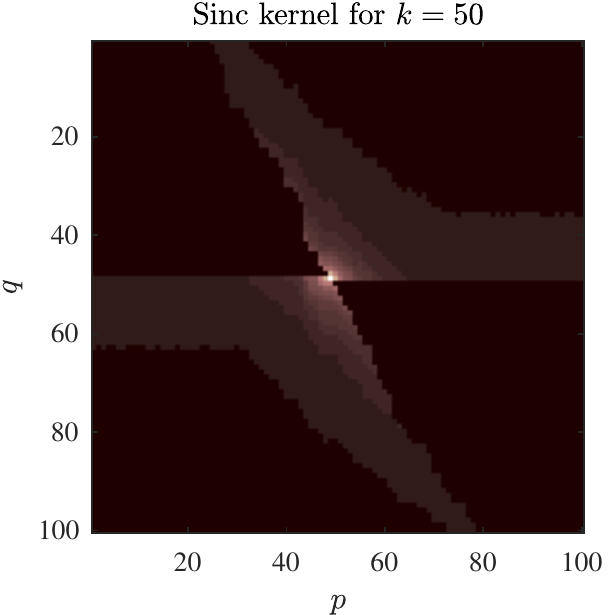}
	\caption{The sinc kernel of the reduced interference vertex-frequency distribution in the frequency domain. }
	\label{kerel}
\end{figure}

The simplest reduced interference kernel in the frequency-frequency shift domain,  which would satisfy the marginal properties, is the \textit{sinc kernel}, given by
\begin{gather*}
\phi(p,k,q)=\begin{cases}
\frac{1}{1+2|p-q|}, & \text{for } |k-p| \le |p-q|, \\
0, & \text{otherwise,}
\end{cases}
\end{gather*}
which is is shown in Fig. \ref{kerel} at the frequency shift corresponding to $k=50$. 

\begin{Example}
	The sinc kernel was used for a vertex-frequency representation of the signal from Fig. \ref{VF_graph3ab}(d), with the results shown in  Fig. \ref{VF_graph3_RID}. This representation is a smoothed version of the energy vertex-frequency distribution in Fig. \ref{VF_graph3f}, whereby both (vertex and frequency) marginals are preserved. 
	
	\begin{figure}
		\centering
		\includegraphics[scale=0.9]{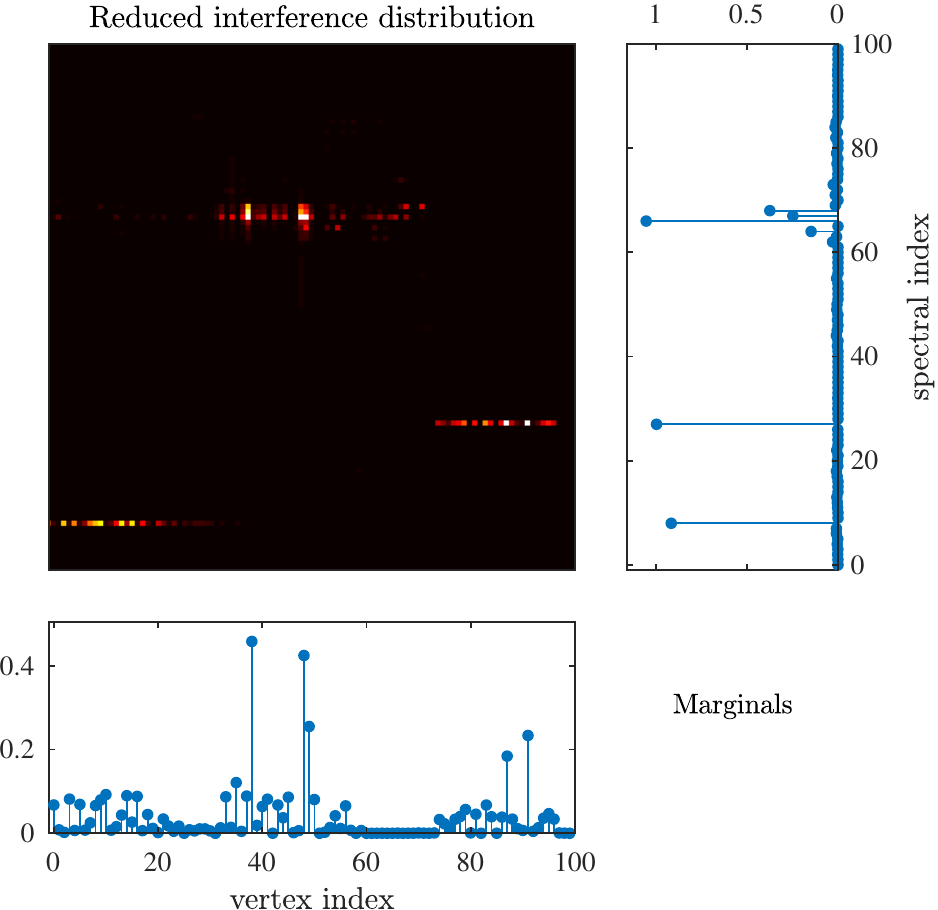}
		\caption{Reduced interference vertex-frequency distribution of a signal whose vertex-frequency representation is given in Fig. \ref {VF_graph3g}. The marginal properties are  given in the panels to the right and below the vertex-frequency representation, and are equal to their corresponding ideal forms given by $|x(n)|^2$ and $|X(k)|^2$. }
		\label{VF_graph3_RID}
	\end{figure}
\end{Example}

\begin{Remark} \textbf{Marginal properties of  graph spectrogram.} A general vertex-frequency distribution can be written for the vertex-vertex shift domain as a dual form of (\ref{GENRID}), to yield
	\begin{gather}
	G(n,k)=\sum_{m=0}^{N-1}\sum_{l=0}^{N-1}x(m)x^*(l)u_k(m)u^*_k(l)\varphi(m,n,l),
	\end{gather}
	where $\varphi(m,n,l)$ is the kernel in this domain (the same mathematical form as for the frequency-frequency shift domain kernel).  
	The frequency marginal is then satisfied if $
	\sum_{n=0}^{N-1}\varphi(m,n,l)=1$ holds, while the vertex marginal is met if $
	\varphi(m,n,m)=\delta(m-n).$ The relation of this distribution with the vertex domain spectrogram (\ref{VFSPEC}) is simple, and is given by \begin{gather*}\varphi(m,n,l)=h_n(m)h^*_n(l).
	\end{gather*}
	However, this kernel cannot satisfy both the frequency and vertex marginal properties, while the unbiased energy condition $\sum_{n=0}^{N-1}\varphi(m,n,m)=1$ reduces to (\ref{unener}).
\end{Remark}

\begin{Remark} \textbf{ Classical time-frequency analysis}  follows as a special case from the general form of graph distributions in (\ref{GENRID}),  if the considered graph is a directed circular graph. This becomes obvious upon recalling that the adjacency matrix eigendecomposition produces complex-valued eigenvectors of the form $u_k(n)=\exp(j 2 \pi nk/N)/\sqrt{N}$. With  the kernel choice
	$$\phi(p,k,q)=\phi
	(p-q,k-p)=\sum_{n=0}^{N-1} c(p-q,n)
	e^{-j \frac{2\pi n k}{N}}e^{j \frac{2\pi n p}{N}}$$
	in (\ref{GENRID}),  the classical (Rihaczek based) Cohen class of distributions directly follows, where $c(k,n)$ is the distribution kernel in the ambiguity domain \cite{stankovic2014time,cohen1995time,boashash2015time}.   
\end{Remark}
 
\section{Conclusion}

 Fundamental ideas of graph signals and their analysis have been introduced starting from an intuitive multisensor estimation example, frequently considered in traditional data analytics. The concept of systems on graphs has been defined using graph signal shift operators, which generalize the fundamental signal shift concepts in traditional signal processing. In part II of our monograph, the Graph Discrete Fourier Transform (GDFT) has been at the core of the spectral domain representation of graph signals and systems on graphs, and has been  defined based on both the adjacency matrix and graph Laplacian. These spectral domain representations have been used as the basis to introduce graph signal filtering concepts. Methods for the design of the graph filters have been presented next, including those based on the polynomial approximation.  Various ideas related to the sampling of graph signals, and particularly, the challenging topic of the subsampling, have also been addressed in this part of the monograph. This is followed by conditions for the recovery of signals on graphs, from a reduced number of samples.  The concepts of time-varying signals on graphs and  basic definitions related to random graph signals have also been reviewed. 
 
 While traditional approaches for graph analysis, clustering and segmentation consider only graph topology and spectral properties of graphs, when dealing with signals on graphs, localized analyzes should employed in order to consider both  data on graphs and the graph topology. Such a unified approach to define and implement graph signal localization methods, which takes into account both the data on graph and the corresponding graph topology, is at the core of vertex-frequency analysis. Like in classical time-frequency analysis, the main research efforts have been devoted to  linear representations of the graph signals which include a localization window for enhanced signal discrimination. Several methods for the definition of localization widows in the spectral and vertex domain have been addressed in Part II of this  monograph. Optimization of the window parameters, uncertainty principle, and inversion methods have also been discussed. Following classical time-frequency analysis, energy forms of vertex-frequency energy and reduced interference distributions, which do not use localization windows, have also been considered, together with their role as an estimator of the local smoothness index.

\section{Bibliography}
%\nocite{*}
\bibliographystyle{ieeetr}

\bibliography{graph-signal-processing}

\vspace*{3cm}	

\centering	
\Large{Graph Signal Processing -- Part III: \\ Learning Graph Topology}

\vspace*{5mm}

\normalsize 
L. Stankovic, D. Mandic, M Dakovic, \\ M. Brajovic, B. Scalzo, A. G. Constantinides
	
\end{document}